\pdfoutput=1

\documentclass[a4paper,11pt]{article}
\usepackage{jheppub}

\usepackage{bm}

\usepackage{color}

\usepackage{amsmath}
\usepackage{amssymb}
\usepackage{graphicx}
\usepackage{slashed}
\usepackage{soul}
\usepackage{lscape}
\usepackage{xcolor}
\usepackage{multirow}
\usepackage{placeins,hyperref}
\usepackage[separate-uncertainty=true]{siunitx}
\usepackage{silence}
\usepackage{caption}
\usepackage{subcaption}
\usepackage{mathrsfs}
\usepackage[version=3]{mhchem}
\usepackage{amsmath, amssymb, amscd, amsthm, amsfonts}
\usepackage{wrapfig}

\usepackage[italic]{hepnames}
\usepackage{gensymb}
\usepackage{textcomp}
\usepackage{booktabs}
\usepackage{float}
\usepackage{adjustbox}

\usepackage{xcolor} 
\usepackage{tikz} 
\usetikzlibrary{arrows.meta}

\usepackage{cleveref}
\usepackage{textgreek}
\usepackage{overpic}

\counterwithin{figure}{section}

\usepackage{lineno}

\begin{document}

\title{\centering Mineral Detection of Neutrinos and Dark Matter 2026 \\ Proceedings}


\author[1]{Alexey~Elykov,}
\affiliation[1]{Institute for Astroparticle Physics, Karlsruhe Institute of Technology, 76021 Karlsruhe, Germany}

\author[2]{Patrick~Stengel;}
\affiliation[2]{Jo\v{z}ef Stefan Institute, Jamova 39, 1000 Ljubljana, Slovenia}

\author[3]{Natsue~Abe,}
\affiliation[3]{Center for Mathematical Science and Advanced Technology (MAT), Japan Agency for Marine-Earth Science and Technology (JAMSTEC), Yokohama, Kanagawa 236-0001, Japan}

\author[4]{Daniel~Ang,}
\affiliation[4]{Quantum Technology Center, University of Maryland, College Park, MD 20742, USA}

\author[5]{Lorenzo~Apollonio,}
\affiliation[5]{INFN Milano, via Celoria 16 20133, Milano, Italy}

\author[6]{Levente~Balogh,}
\affiliation[6]{Department of Mechanical and Materials Engineering, Queen's University, 130 Stuart Street, Kingston, ON, K7L 2V9, Canada}

\author[7]{Laura~Baudis,}
\affiliation[7]{Department of Physics, University of Zurich, Winterthurerstrasse 190, 8057 Zurich, Switzerland}

\author[4]{Chinmay~Bharathulwar,}

\author[4]{Priyanshu~Bhattacharya,}

\author[8]{Yilda~Boukhtouchen,}
\affiliation[8]{Department of Physics, Engineering Physics, and Astronomy, Queen's University, 64 Bader Lane, Kingston, Ontario, K7L 2S8, Canada}

\author[8]{Joseph~Bramante,}

\author[9]{Vincent~Breton,}
\affiliation[9]{Laboratoire de Physique de Clermont, 4 Av. Blaise Pascal, 63170 Aubi\`{e}re, France}

\author[8]{Andrew~Buchanan,}

\author[10]{Jens~Burkhart,}
\affiliation[10]{MBI - Marietta Blau Institute for Particle Physics of the Austrian Academy of Sciences, 1010 Vienna, Austria}

\author[5]{Lorenzo~Caccianiga,}

\author[11]{Andrew Calabrese-Day,}
\affiliation[11]{Department of Physics, University of Michigan, Ann Arbor, MI 48103, USA}

\author[4]{Mason~Camp,}

\author[12]{Jordan~Chapman,}
\affiliation[12]{Virginia Tech National Security Institute, Blacksburg, Virginia 24060, USA}

\author[4]{Anson~Cook,}

\author[13,14]{Reza~Ebadi,}
\affiliation[13]{Department of Physics and Astronomy, The Johns Hopkins University, Baltimore, MD 21218, USA}
\affiliation[14]{Department of Physics and Astronomy, University of Delaware, Newark, DE 19716, USA}

\author[15]{Denis~Erkal,}
\affiliation[15]{School of Mathematics and Physics, University of Surrey, Guildford GU2 7XH, UK}

\author[16,17,18]{Katherine~Freese,}
\affiliation[16]{Department of Physics, University of Texas at Austin, 2515 Speedway, Austin, TX 78712, USA}
\affiliation[17]{The Oskar Klein Centre, Department of Physics, Stockholm University, AlbaNova, SE-106 91 Stockholm, Sweden}
\affiliation[18]{Nordita, Stockholm University and KTH Royal Institute of Technology, Hannes Alfvéns väg 12, SE-106 91 Stockholm, Sweden}

\author[8,19]{Audrey~Fung,}
\affiliation[19]{Asia Pacific Center for Theoretical Physics, Postech, Pohang 37673, Korea}

\author[20]{Shota~Futamura,}
\affiliation[20]{Institute for Space-Earth Environmental Research, Nagoya University, Furo-cho, Chikusaku, Nagoya, Aichi, 464-8601, Japan}

\author[5]{Claudio~Galelli,}

\author[8]{Kevin~Gao,}

\author[21]{Peter~W.~Graham,}
\affiliation[21]{Leinweber Institute for Theoretical Physics, Department of Physics, Stanford University, Stanford, CA 94305, USA}

\author[11]{Thomas~Haddock,}

\author[20]{Minako~Hashiguchi,}

\author[8]{Alexander~Hayes,}

\author[22]{Adam~A.~Hecht,}
\affiliation[22]{Department of Nuclear Engineering, University of New Mexico, Albuquerque, NM 87131, USA}

\author[23]{Samuel~C.~Hedges,}
\affiliation[23]{Center for Neutrino Physics, Virginia Tech, Blacksburg, VA 24061, USA}

\author[3]{Shigenobu~Hirose,}

\author[7]{Luisa~M.~H\"{o}tzsch,}

\author[23]{Patrick~Huber,}

\author[24]{Yohei~Igami,}
\affiliation[24]{Department of Geology and Mineralogy, Kyoto University, Kyoto, Japan}

\author[12,25,26]{Vsevolod~Ivanov,}
\affiliation[25]{Department of Physics, Virginia Tech, Blacksburg, Virginia 24061, USA}
\affiliation[26]{Center for Quantum Information Science and Engineering, Virginia Tech, Blacksburg, VA 24061, USA}

\author[7]{Florian~J\"{o}rg,}

\author[27]{Ayuki~Kamada,}
\affiliation[27]{Institute of Theoretical Physics, Faculty of Physics, University of Warsaw, ul. Pasteura 5, PL-02-093 Warsaw, Poland}

\author[20]{Takenori~Kato,}

\author[3]{Yoji~Kawamura,}

\author[1]{Katharina~Kehl,}

\author[28]{Chris~Kelso,}
\affiliation[28]{Department of Physics and Astronomy, University of North Florida, 1 UNF Dr., Jacksonville, FL 32250, USA}

\author[10]{Holger~Kluck,}

\author[11]{Emilie~M.~LaVoie-Ingram,}

\author[30]{Matthew~Leybourne,}
\affiliation[30]{Department of Geological Sciences and Geological Engineering, Queen's University, Kingston, Canada}

\author[4]{Gavishta~H.~M.~Liyanage,}

\author[6]{Thalles~Lucas,}

\author[25]{Brenden~A.~Magill,}

\author[31]{Paolo~Magnani,}
\affiliation[31]{Gran Sasso Science Institute, viale F. Crispi 7, 67100 L'Aquila, Italy}

\author[8]{Jennika~McIntosh,}

\author[32]{Naoki~Mizutani,}
\affiliation[32]{Department of Physics, Toho University, 2-2-1 Miyama, Funabashi, Chiba, Japan}

\author[33]{Kohta~Murase,}
\affiliation[33]{Department of Physics, The Pennsylvania State University, University Park, PA 16802, USA}

\author[32]{Tatsuhiro~Naka,}

\author[34]{Lina~Necib,}
\affiliation[34]{Physics Department and Kavli Institute for Astrophysics and Space Research, Massachusetts Institute of Technology, Cambridge, MA 02139, USA}

\author[11]{Pranav~Parvathaneni,}

\author[28]{Andre~Peterson,}

\author[25]{Zachary~S.~C.~Picker,}

\author[28]{Rabeya~Rabu,}

\author[14]{Harikrishnan~Ramani,}

\author[8,35,36]{Anupam~Ray,}
\affiliation[35]{Arthur B. McDonald Canadian Astroparticle Physics Research Institute, 64 Bader Lane, Queen's University, Kingston, Ontario, Canada}
\affiliation[36]{Perimeter Institute for Theoretical Physics, Waterloo, ON N2J 2W9, Canada}

\author[25]{Morteza~Roostaeinia,}

\author[11]{Hannah~Ross,}

\author[37]{Issei~Saikyo,}
\affiliation[37]{Particle Physics Laboratory, Department of Physics, Faculty of Science, Toho University, 2-2-1 Miyama, Funabashi, Chiba 274-8510, Japan}

\author[38]{Lukas~Scherne,}
\affiliation[38]{Physikalisches Institut, Universität Freiburg, Herrmann-Herder Str. 3 79104 Freiburg im Breisgau, Germany}

\author[4]{Maximilian~Shen,}

\author[39]{Aaron~Shugar,}
\affiliation[39]{Department of Art History and Art Conservation, Queen's University, Kingston, Ontario, Canada}

\author[11]{Joshua~Spitz,}

\author[40]{Kai~Sun,}
\affiliation[40]{Department of Materials Science and Engineering, University of Michigan, Ann Arbor, MI 48103, USA}

\author[4]{Jiashen~Tang,}

\author[41]{Erwin~H.~Tanin,}
\affiliation[41]{Department of Physics, Stanford University, Stanford, California 94305, USA}

\author[16]{Dionysios~P.~Theodosopoulos,}

\author[42]{Yoichi~Usui,}
\affiliation[42]{College of Earth Sciences and Civil Engineering, Kanazawa University, Kakuma, Kanazawa 9201192, Japan}

\author[43]{Pieter~Vermeesch,}
\affiliation[43]{Department of Earth Sciences, University College London, Gower Street, London WC1E 6BT, UK}

\author[8]{Aaron~Vincent,}

\author[4]{Ronald~Walsworth,}

\author[44]{Jin-Wei~Wang,}
\affiliation[44]{School of Physics, University of Electronic Science and Technology of China, Chengdu 611731, China}

\author[45]{David~Waters,}
\affiliation[45]{Department of Physics \& Astronomy, University College London, WC1E 6BT, UK}

\author[46]{Samuel~S.~Y.~Wong,}
\affiliation[46]{Department of Physics, University of Washington, Seattle, WA 98195, USA}

\author[11]{Audrey~Wu,}

\author[28]{Gregory~Wurtz,}

\author[47]{Wen~Yin,}
\affiliation[47]{Department of Physics, Tokyo Metropolitan University, Minami-Osawa, Hachioji-shi, Tokyo 192-0397, Japan}

\author[34]{Xiuyuan~Zhang,}

\author[11]{Zhexian~Zhang}

\emailAdd{alexey.elykov@kit.edu}
\emailAdd{patrick.stengel@ijs.si}


\abstract{The fourth ``Mineral Detection of Neutrinos and Dark Matter'' (MD$\nu$DM'26) meeting was held April 14–17, 2026 in Karlsruhe, Germany, hosted by the Institute for Astroparticle Physics (IAP) at Karlsruhe Institute of Technology (KIT).
This meeting was the fourth edition of the MD$\nu$DM workshops, continuing the success of previous meetings that were held at the Yokohama Institute for Earth Sciences JAMSTEC in Japan (2025); the Center for Neutrino Physics at Virginia Tech in Arlington, USA (2024); and the Institute for Fundamental Physics of the Universe (IFPU) in Trieste, Italy (2022).
These proceedings detail the contributions that were presented during MD$\nu$DM'26, illustrating the unprecedented progress in theoretical, computational and experimental studies towards the realization of the concept of mineral detectors.
Mineral detectors represent an emerging particle detection concept that has risen in prominence in recent years due to the advent of modern computational and high-resolution microscopy techniques.
Natural and synthetic crystals are capable of retaining microscopic damage features induced by nuclear recoils, which could be then read out with a variety of micrometer and nanometer resolution microscopy techniques.
On laboratory time scales mineral detectors could be employed for reactor neutrino monitoring and dark matter detection, with the potential to measure the directions as well as the energies of the induced nuclear recoils.
Uniquely, ancient natural crystals (so-called paleo-detectors) that have been recording nuclear recoils over geological timescales could be used for studying astrophysical neutrinos, cosmic rays, dark matter and heavy exotic particles, as well as the variation of their fluxes over our Galaxy's lifetime.
The research field of mineral detectors is highly interdisciplinary, combining complementary expertise in particle and astroparticle physics, condensed matter physics, materials science, geoscience, optics and robotics and application of AI/ML for data analysis.
In recent years the international MD$\nu$DM community has been successfully tackling the challenges associated with realizing the concept of mineral detectors, opening the pathway towards a fully fledged experimental program and potential future discoveries.
}

\maketitle
\flushbottom

\section*{Preface}

\addcontentsline{toc}{section}{Preface}


The emerging field of mineral detection aims to investigate particles both within and beyond the Standard Model (SM) through the interactions of such particles with atoms comprising the crystal lattice of natural mineral or synthetic crystal detectors. Damage to the crystal lattice of the detector must persist on timescales sufficiently long to be read out by one or more microscopy techniques in a controlled laboratory experiment or, in the case of a geological mineral detector, to be excavated and read out after recording particle interactions for up to $\sim 1 \,$Gyr. Mineral detection is thus a highly interdisciplinary field, combining expertise and tools from high energy and condensed matter physics, as well as from materials science and geoscience. Growing interest has led to an increasing number of groups from institutions across Europe, North America and Asia to investigate various experimental and theoretical aspects of mineral detection. 

The Mineral Detection of Neutrinos and Dark Matter (MD$\nu$DM) 2026 workshop was hosted by the Karlsruhe Institute of Technology (KIT) in Germany in April 2026\footnote{\href{https://indico.kit.edu/event/5425/}{\url{https://indico.kit.edu/event/5425/}}}. MD$\nu$DM 2026 was the fourth edition of the workshop, previously held at the Yokohama Institute for Earth Sciences JAMSTEC in Japan (2025)\footnote{\href{https://indico.ijs.si/event/2583/}{\url{https://indico.ijs.si/event/2583/}}}; the Center for Neutrino Physics at Virginia Tech in Arlington, USA (2024)\footnote{\href{https://indico.phys.vt.edu/event/62/}{\url{https://indico.phys.vt.edu/event/62/}}}; and the Institute for Fundamental Physics of the Universe (IFPU) in Trieste, Italy (2022)\footnote{\href{https://agenda.infn.it/event/32181/}{\url{https://agenda.infn.it/event/32181/}}}. MD$\nu$DM'26 was the most highly attended workshop yet, with 28 speakers from 22 institutions presenting updates on experimental progress towards the realization of mineral detectors and new theoretical investigations of mineral detector sensitivity to particle interactions. These proceedings are intended to give a snapshot of the progress in the field of mineral detection in the $\sim 1$ year period since the previous workshop. The growth and progress of the field over the prior years can be found in previous MD$\nu$DM proceedings~\cite{baum:2024eyr,MDDM2025Proceedings} and the whitepaper~\cite{baum_mineral_2023}.

Experimental work on mineral detection has primarily focused on the identification and calibration of suitable microscopy techniques and can be broadly characterized by the type of crystal lattice damage that these techniques are designed to read out. The initial experiments which established mineral detection were carried out several decades ago in searches for damage tracks, enlarged by chemical etching in geological mica samples, due to the transit of magnetic monopoles across the detector~\cite{price:1986ky} and nuclear recoils induced by weak-scale dark matter scattering~\cite{collar:1994mj,snowden-ifft:1995zgn,collar:1995aw,Snowden-Ifft:1996dug}. While the read out of the former uses optical microscopy (OM) to read out micro-scale monopole tracks and the latter uses atomic force microscopy (AFM) to read out nano-scale nuclear recoil tracks, the underlying mechanism creating the damage tracks is assumed to be similar. 

Several experimental approaches featured in these proceedings directly build upon the track-based read out techniques implemented in the initial mineral detector experiments, taking advantage of more recent advances in microscopy and sample preparation (for example, see Ref.~\cite{Boukhtouchen:2026rfz}). Sec.~\ref{sec:PRImuS} proposes to implement plasma etching of high energy nuclear recoil tracks induced by cosmogenic muons in the Earth's crust and Sec.~\ref{sec:Toho_NAKA} describes a technique for the chemical etching of tracks induced by heavy exotic particles in meteorites, both to be read out in olivine using OM. Sec.~\ref{sec:Queens} discusses the calibration of X-ray fluorescencence (XRF) imaging to detect micro-scale melt-tracks induced by heavy composite dark matter interactions in mica without etching. For the readout of nano-scale nuclear recoil tracks, Sec.~\ref{sec:KIT_IAP} discusses using  optical profilometry (OP) to detect etch pits from particle interactions accumulated over either laboratory or geological timescales in NaCl, olivine and mica, while Sec.~\ref{sec:DMICA_directional_signature} proposes to look for the directional signal of weak-scale dark matter over geological timescales using OP for chemically etched mica.   

Microscopy techniques which can potentially read out low energy recoil tracks induced by weak-scale dark matter scattering or coherent elastic neutrino nucleus scattering (CE$\nu$NS) without the need for etching are also being investigated (for example, see Ref.~\cite{TrackWidthPaper}). While chemical or plasma etching helps to increase the detectability of tracks, the etching process can slow down the throughput of microscopy techniques which need to read out significant quantities of mineral samples in rare event searches. The high resolution read out of low energy nuclear recoil tracks without etching also gives additional insight into the mechanisms for track formation. Sec.~\ref{sec:Michigan} discusses how Scanning Transmission Electron Microscopy (STEM) has been used to characterize damage tracks in ion irradiated natural olivine without etching. Sec.~\ref{sec:UNF} describes how electron backscatter diffraction (EBSD) imaging is used to read out crystal damage in ion irradiated olivine and quartz. A collaborative effort to image ion irradiated samples with x-ray microscopy techniques at synchrotron light sources is also mentioned in Sec.~\ref{sec:Michigan} and Sec.~\ref{sec:UNF}. 
 
Complementary to track-based read out techniques are those which use OM to identify optically active vacancy clusters, so-called color centers, which can be produced by nuclear recoil damage induced by particle interactions~\cite{Tilley2014}. Depending on the electronic structure of a given cluster of anionic vacancies, color centers absorb and emit light at specific frequencies and can be read out in crystals which are sufficiently transparent using techniques such as light-sheet microscopy~\cite{marshall_high-precision_2022,vladimirov:2023,araujo2025nuclear}. 
Sec.~\ref{sec:UMD} discusses the read out of nitrogen-vacancy centers induced in synthetic diamond by the interactions of weak-scale dark matter, using a light-sheet quantum diamond microscope (LS-QDM) to reconstruct the nuclear recoil direction, as well as energy. 
Sec.~\ref{sec:UZH_Bedretto} proposes both a weak-scale dark matter search and a measurement of reactor neutrino CE$\nu$NS in artificial LiF crystals using Selective Plane Illumination Microscopy (mesoSPIM) operating in an underground microscopy facility. 
Sec.~\ref{sec:UMD} (Sec.~\ref{sec:Hedges_Talk}) also proposes the use of optical microscopy to read out color centers arising from particle interactions in geological diamond (olivine) samples.

Simulations of nuclear recoil damage in crystals are essential tools for understanding the sensitivity of controlled laboratory experiments and geological mineral detector searches to particle interactions. Both experimental and theoretical studies of mineral detectors typically incorporate one or a combination of simulation codes for particle transport and damage formation. Higher energy particle interactions with matter can be modeled with particle transport codes such as \texttt{GEANT4}~\cite{GEANT4:2002zbu}, while the crystal damage due to the associated lower energy nuclear recoils has typically been simulated with binary collision approximation codes such as \texttt{SRIM}~\cite{zieglersrim2010}. More detailed simulations of low energy nuclear recoil damage in crystals, accounting for solid-state effects, are possible using classical molecular dynamics simulation codes such as \texttt{LAMMPS}~\cite{lammps2022}, which is described in Sec.~\ref{sec:Kluck} for cryogenic \ce{Al_2O_3}-based calorimeter experiments. First principles quantum molecular dynamics codes such as \texttt{VASP}~\cite{VASP1,VASP2} are discussed in Sec.~\ref{sec:VTQM} to simulate the formation of color centers in LiF-based detectors and metastable intrinsic defects in Germanium-based detectors. 

Recent theoretical studies of the sensitivities of mineral detectors to the interactions of neutrinos~\cite{baum:2019fqm,jordan:2020gxx,Tapia-Arellano:2021cml,Baum:2022wfc,Yamasaki:2026ynn}, cosmic rays~\cite{caccianiga_2024,Galelli_2026} and weak-scale (or lighter) dark matter~\cite{baum:2018tfw,drukier:2018pdy,Edwards:2018hcf,Baum:2021jak,baum:2021chx,Bramante_2022,Fung:2025cub,zhang2025darknesscrustsearchingtruly,Wang:2026you,Theodosopoulos:2026ehn,Chu:2026skp,hedges2026calorimetricapproachpaleodetectiondark,Graham:2026ivn,Theodosopoulos:2026eym} have explored the scientific potential of mineral detectors and motivate many of the experimental efforts presented in these proceedings. Expanding on the types of dark matter models previously considered, Sec.~\ref{sec:UT_Austin} projects the sensitivity of mineral detectors to weak-scale dark matter interactions with nuclei in a generalized non-relativistic effective field theory framework, while Sec.~\ref{sec:Higgsino} investigates how mineral detectors can probe the specific interactions of weak-scale thermal relic Higgsino dark matter with nuclei in the minimal supersymmetric standard model. Sec.~\ref{sec:MIT_XZ} discusses the sensitivity of mineral detectors to weak-scale dark matter in small-scale subhalos as the the solar system rotates around the Milky Way over geological timescales. For dark matter with masses below the weak scale, Sec.~\ref{sec:UESTC} describes the sensitivity of mineral detectors to dark matter boosted to relativistic energies by interactions with cosmic rays in the Milky Way dark matter halo.

Additional theoretical studies have investigated mineral detector signatures from the interactions of ultra-heavy dark matter, as well as other exotic phenomena. The long exposure times of mineral detectors over geological timescales potentially allows mineral detectors to be sensitive to ultra-heavy dark matter, which has a flux too small to be detected by conventional dark matter detection experiments operating on laboratory timescales~\cite{SinghSidhu:2019znk,ebadi:2021cte,Acevedo:2021tbl}. Mineral detectors have also been proposed to search for signatures of charged Planck-scale relics, including primordial black holes~\cite{Lehmann:2019zgt}, as well as signatures of proton decay~\cite{Baum:2024sst} and cosmic walls~\cite{yin:2025wuv}. Following earlier experimental studies of mineral detection searches for magnetic monopoles, Sec.~\ref{sec:Kamada} proposes mineral detector searches for minicharged magnetic monopoles, which arise in models with extended dark sectors. Sec.~\ref{sec:CosmicWalls} updates previous theoretical studies of mineral detector searches for cosmic walls with a detailed calculation of the nuclear recoil momentum and angular distributions.  

The contributions to the proceedings of MD$\nu$DM 2026 summarize the ongoing work in the field of mineral detection and the exciting scientific possibilities for future mineral detector experiments. Mineral detectors have now reached a new stage of development. Multiple grants have been awarded for mineral detector research, including \$1.25M from the Gordon and Betty Moore Foundation and \$1.2M from the U.S. National Science Foundation. International collaborations working on various aspects of mineral detectors have formed between researchers at institutions in at least 14 countries across Europe, North America and Asia~\cite{Feder2024PaleoDetectors}. In the coming years, the advanced microscopy techniques currently in development to enable imaging of nano- and micro-scale damage features will increasingly be applied to read out synthetic crystal and geological mineral samples, testing the wide variety of physical phenomena which mineral detectors are uniquely suited to discover. We look forward to the continued progress in the field of mineral detectors and the next MD$\nu$DM workshop scheduled for late May/early June 2027, hosted by Queen's University in Canada.

\acknowledgments
We thank the Karlsruhe Institute of Technology (KIT) for hosting the MD$\nu$DM'26 workshop in April 2026.
We also would like to especially thank Sabine Bucher, Dr. Klaus Eitel and Prof. Dr. Kathrin Valerius for their support and help with organizing the workshop. We also thank Yoji Kawamura for the careful review of these proceedings. 

\clearpage

\section{Optical Imaging of Crystals for Mineral Detector Applications at KIT}\label{sec:KIT_IAP}

Authors: {\it Katharina~Kehl and Alexey~Elykov}
\vspace{0.1cm} \\
Karlsruhe Institute of Technology, Institute for Astroparticle Physics, Germany
\vspace{0.3cm}

\subsection{Introduction}
Artificial and natural crystals acting as passive nuclear track detectors offer a new and exciting path for detecting standard model and exotic particles, such as neutrinos, particle dark matter candidates and particles originating from cosmic rays. 
Use of ancient natural crystals, may reveal the nature of fluxes of those particles over enormous timescales, thus helping us to answer some of the most fundamental questions about the nature of our Universe.

Karlsruhe Institute of Technology (KIT) is one of the major scientific research institutes in Europe with interdisciplinary expertise in a wide range of research fields, including material sciences, geology, microscopy, electronics, robotic systems, particle and astroparticle physics.
At the Institute for Astroparticle Physics (IAP) at KIT, we are performing a series of experimental, simulation and imaging studies towards the realization of crystal detectors on laboratory timescales and as so-called paleo-detectors.

With support from the KIT Elementary Particle and Astroparticle Physics Center (KCETA), we have begun to develop a custom open-source high-throughput FPGA-GPU-based data acquisition system (DAQ) for imaging optically active colour-centres with mesoSPIM \cite{mesospim}.
These efforts will be coupled with the development of a robotic system for placement and orientation of samples on mesoSPIM, thus aiming to develop a fully automated DAQ prototype for serial crystal imaging, data acquisition, reduction and processing. 
Another key area of research focuses on microscopy imaging of irradiated and blank crystal samples with electron microscopy and nanotomography.
This work is performed in collaboration with microscopy experts from the Institute of Nanotechnology (INT) and the Laboratory for Electron Microscopy (LEM).

Lastly, the following sections are devoted to the latest area of research at KIT-IAP, that was motivated by the pioneering research at INFN Milan \cite{Galelli_2026}, and focuses on the use of in-house KEYENCE VK-X3000 optical profiler for high-resolution topographical imaging of crystals.
This imaging technique and the developed machine-learning (ML) based analysis could be used to detect and study particle-induced tracks in the form of etch pits on chemically or plasma-etched crystal surfaces \cite{snowden-ifft:1995zgn, MDDM2025Proceedings, Galelli_2026}.

\subsection{High-resolution Topographical Imaging of Crystal Surfaces}

Tracks from particle interactions that were accumulated in ancient crystals over millions of years may allow us to not only detect signatures left by particle dark matter and neutrinos, but also to detect cosmic ray-induced tracks and study the changes in the cosmic ray flux over our galaxy’s lifetime \cite{Galelli_2026}.
However, these latent tracks are not directly imaged; instead, chemical or plasma etching could be used to preferentially dissolve regions of the crystal lattice containing particle-induced defects, producing etch pits that serve as signatures of the underlying particle interactions.
Together with colleagues from INFN Milan we are working on a pilot study aimed at large-area high-resolution topographical imaging of etched and unaltered NaCl, olivine and mica samples.
The main goal of our studies is the development of the experimental procedures and analysis pipelines for detecting naturally occurring and particle-induced features.\\

\begin{figure}[htbp]
\centering
\begin{subfigure}[t]{0.47\textwidth}
  \centering
  \includegraphics[width=0.7\linewidth]{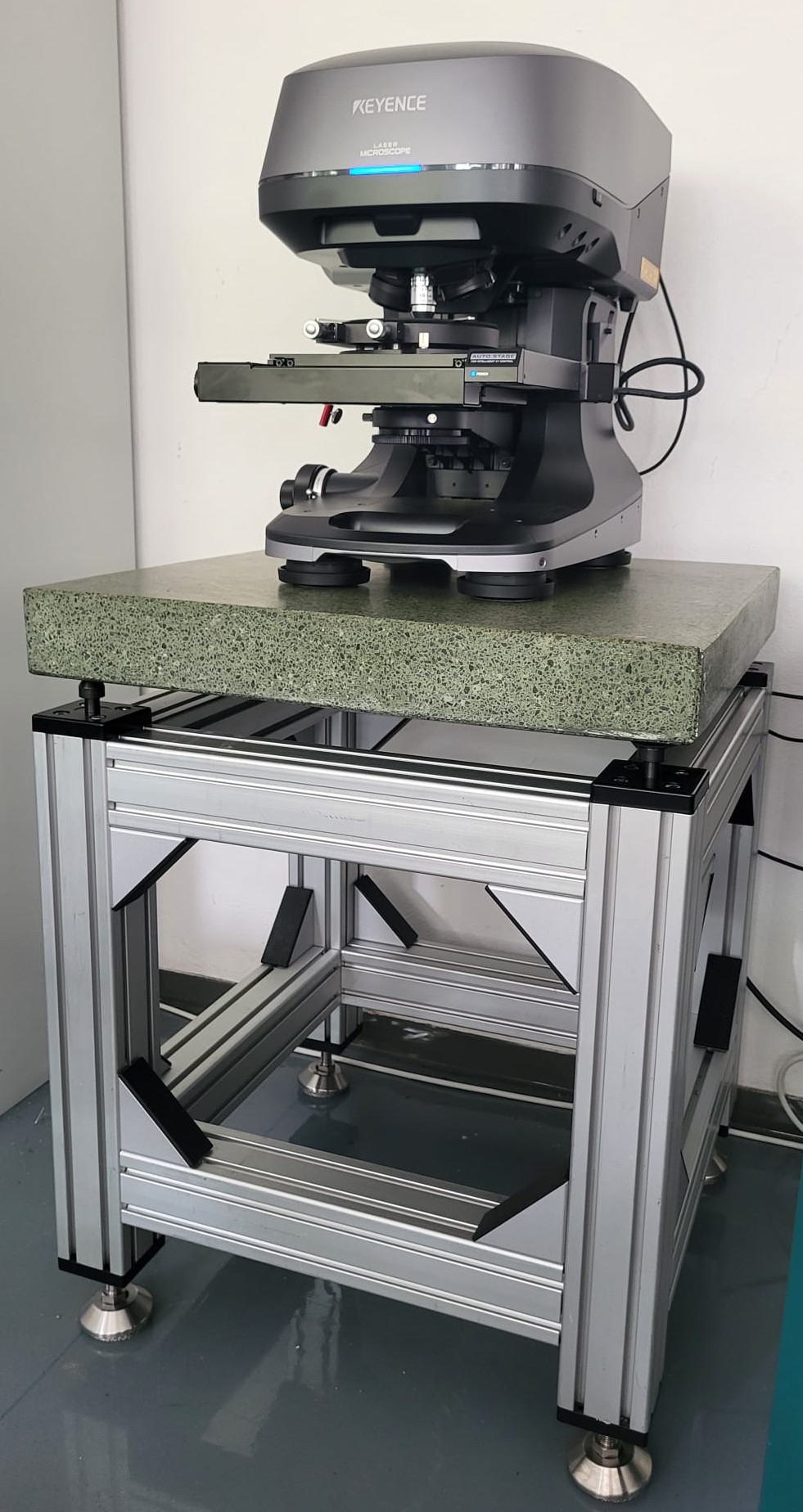}
  \caption{Optical profiler KEYENCE VK-X3000 on custom vibration-reducing table.}
  \label{fig:OpticalProfilerSetup}
\end{subfigure}\hfill
\begin{subfigure}[t]{0.47\textwidth}
  \centering
  \includegraphics[width=0.745\linewidth]{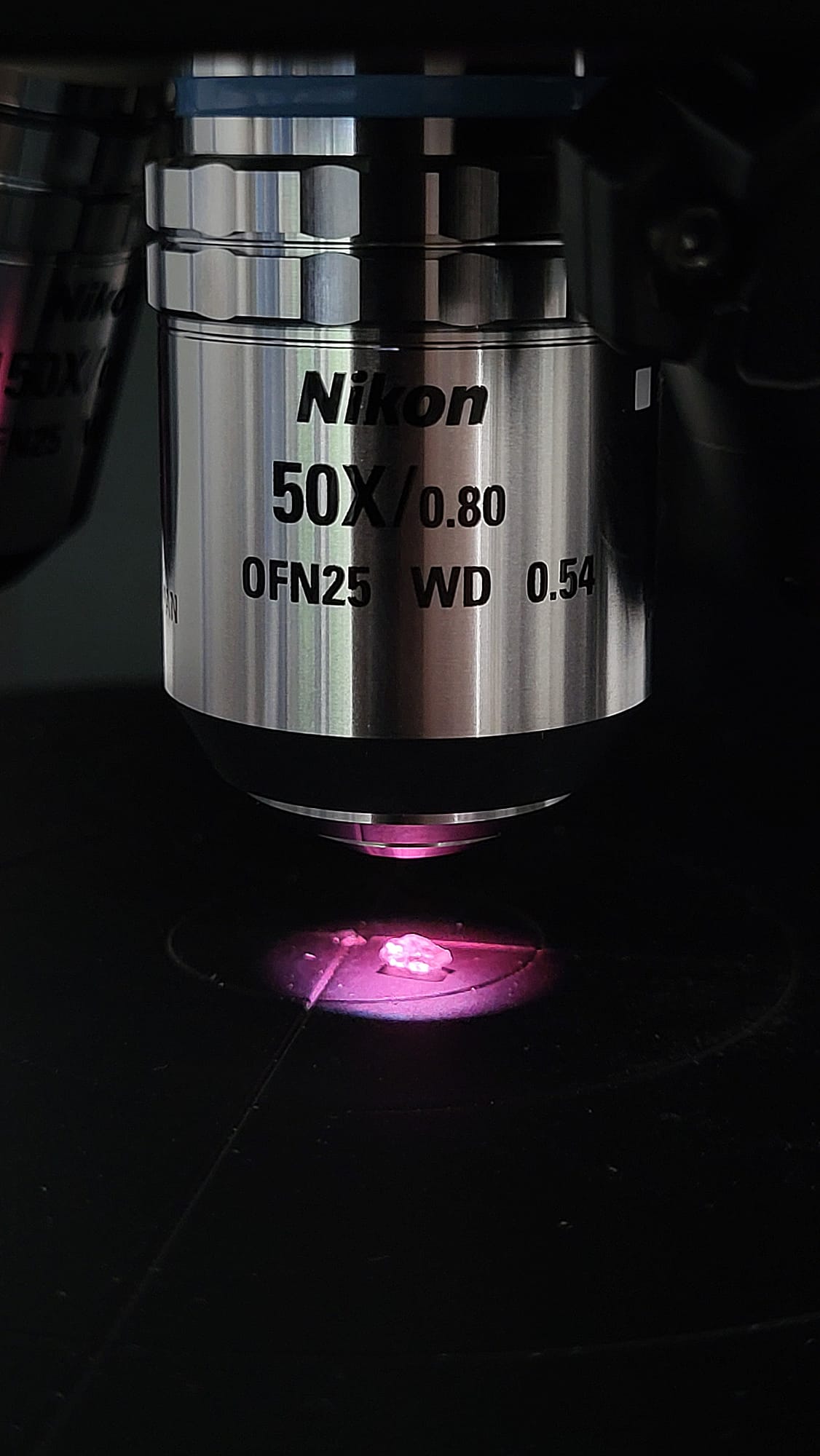}
  \caption{NaCl sample being imaged with KEYENCE VK-X3000 in laser confocal mode with 50x magnification. Objective is raised above its actual working distance for illustrative purposes.}
  \label{fig:SampleUnderMicroscope}
\end{subfigure}
\caption[Setup]{The microscopy setup at KIT-IAP.}
\label{fig:pictures_setup}
\end{figure}

\subsection{Optical Profiler Properties}
\label{sec:ops}
The KEYENCE VK-X3000, pictured in Figure \ref{fig:OpticalProfilerSetup}, is an optical profiler that offers several imaging modes; focus variation, white light interferometry and laser confocal. 
The available objectives for the latter mode have magnifications of 2.5x, 10x, 20x, 50x, and being the most recent addition, 150x. 
For our purposes, and as illustrated in Figure \ref{fig:SampleUnderMicroscope}, measurements are performed in laser confocal mode, as it provides the best resolution on the smallest scales.
The optical profiler can perform 3D surface imaging at high magnifications through automated scanning without contacting the sample's surface, thereby leaving it completely undamaged.
The resulting height map, according to the manufacturer, has a maximum z-resolution of 0.01 nm and a maximum scan area of 50 mm x 50 mm.

Data acquisition in laser confocal mode is performed using the automatic measurement mode.
This operation mode was found to be just as accurate as, or even more accurate than, manual measurement with respect to objective range and brightness settings while also being more time-efficient when working with a large number of images. 
In this operation mode, lens magnification and the measurement parameters are automatically determined and set by the KEYENCE software.

An essential feature of the optical profiler is the so-called image stitching.
This functionality, enables the automated acquisition and combination of adjacent images into a single larger image, thus enabling serial imaging of large-scale areas.
The maximum number of stitched images that can be acquired in one measurement in laser confocal mode is 560, which corresponds to an area of around 4 mm$^2$ using the highest-magnification objective. 

To acquire measurements in this mode, only the scanned area needs to be specified, alongside the exclusion of images where the sample is not visible or which are not of interest due to extreme unevenness of the samples' surface. 
The results of the measurement are then stored as a stitched image.
The individual images are also stored in a dedicated directory, in the manufacturer's own data format `.vk4', or in the newer version with additional metadata, `.vk6'.

However, the default stitched images produced by the optical profiler software are of limited use, as above the size of 25,160,256 pixels, corresponding to roughly 1350 x 1350 µm, the resolution of the combined image is automatically degraded.
Additionally, the transition regions between images exhibit a particularly significant resolution degradation. 
To address the aforementioned challenges we have developed an alternative post-acquisition image stitching method that will be detailed in section \ref{sec:image_processing_pipeline}

\subsection{Imaging Workflow}
\label{sec:imaging_workflow}

To reliably and consistently image large areas with the optical profiler we have established a dedicated imaging workflow.
First, an overview image with 2.5x magnification is taken of every side of the sample, providing a reference for sample orientation and measurement repeatability. 
Then, a stitched surface height measurement of each side, covering the entire sample is acquired at 10x and 20x magnification. 
These images showcase the overall topography of the sample and help to identify particularly flat areas for subsequent high-resolution imaging. 
Images of the identified flat areas are then taken with 50x and 150x magnification in automated stitching mode. 
Subsequently, the individual raw images are used for feature detection and are also combined into a larger image without resolution loss with the use of custom developed software, as described in section \ref{sec:image_processing_pipeline}.

While developing the imaging workflow, we have also identified several limitations, which could be addressed by further refinement of the imaging process, development of dedicated sample holders and optimization of the samples surfaces.
A frequently encountered problem is that the working distance of the 150x objective is only 1.5 mm and the actual vertical clearance can appear even smaller because the lens housing extends below the objective barrel. 
Some samples therefore cannot be imaged with the highest possible magnification and thus resolution, as either the surface roughness or sample inclination prevents the objective from being brought sufficiently close without damaging the sample or the lens.

Moreover, the measurement time in stitching mode is also quite long, requiring approximately an hour to acquire a single overall image composed of 20 individual images.
Reproducibility of the imaging process represents another challenge, as without a dedicated sample holder it is difficult to position the crystal on the stage in the exactly same manner each time it is being imaged. 
Lastly, changes in ambient lighting and external vibrations should be minimized as they were found to constitute a major disruptive factor during the imaging process.

\subsection{Image Processing Pipeline}
\label{sec:image_processing_pipeline}
As stated before, we are interested in achieving the highest possible resolution in every spatial dimension while scanning areas as large as possible with the optical profiler. 
While for overview purposes, an image covering the whole crystal surface could be beneficial, for machine-learning-based feature identification, working with individual smaller-scale images could be faster and more efficient. 
Hence, we developed a Python-based code \footnote{Available at: \url{https://github.com/KIT-Dark-Matter/PaleoDM_ML_Development}} capable of reading and processing the raw proprietary binary .vk4 and .vk6 files.
The file-reading portion of the code is largely based on the work of Yiming Zhang.\footnote{Repository: \url{https://github.com/duserzym/Keyence_VKX3000_data_processing}}

Building upon this foundation, we have developed custom software for combining the individual images acquired with the stitching function of the optical profiler. 
It is only necessary to specify the path to the folder containing the individual images and the software will automatically iterate over them and arrange them at their correct positions within a global grid.
As a subsequent image-processing step, the height map is extracted from each image data file and undergoes tilt correction, by fitting a plane using the least-squares method and subtracting this fitted plane from the data.
Feature contrast is further enhanced by applying a Gaussian filter from scikit-image and clipping the height data to a range where large structures are alleviated and small features become more prominent.
Finally, the processed image data is visualized as a two-dimensional height map, with the colour scale representing the surface height.
Following this procedure, we performed the high-resolution imaging and analysis of multiple mineral samples, with some of them presented here. 
It should be noted that this image processing pipeline represents a preliminary implementation and further refinement and optimization are ongoing.

\begin{figure}[htbp]
   \centering
   \includegraphics[width=0.6\textwidth]{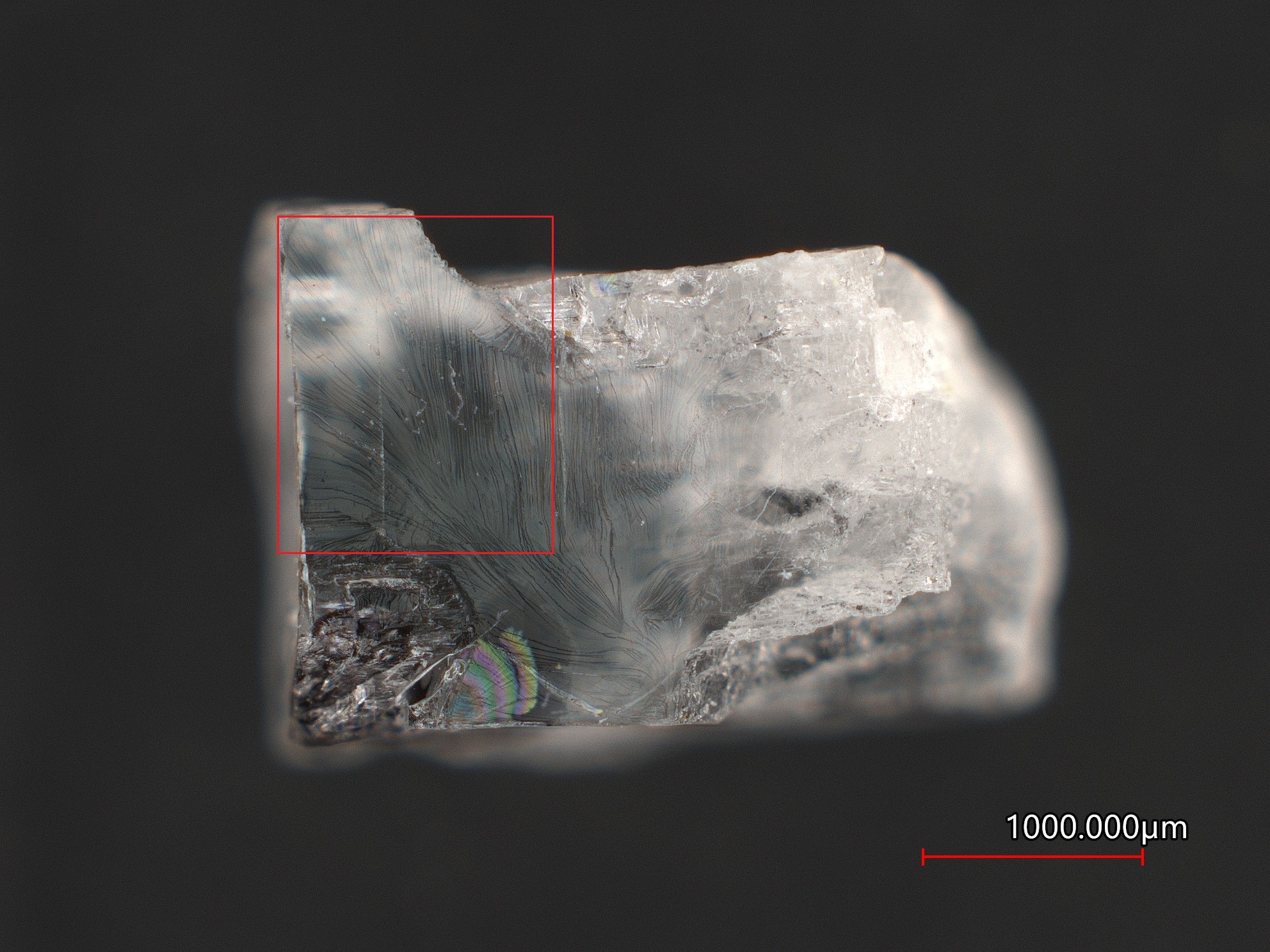}
   \caption{NaCl sample provided by Dr. Claudio Galelli. The red box indicates the area showcased in the stitched image in the subsequent figures.}
   \label{fig:NaCl_whole}
\end{figure}

\subsection{Imaging Results}
\label{sec:preliminary_results}
\begin{figure}[htbp]
\centering
\begin{subfigure}[t]{0.5\textwidth}
  \centering
  \includegraphics[width=1.0\linewidth]{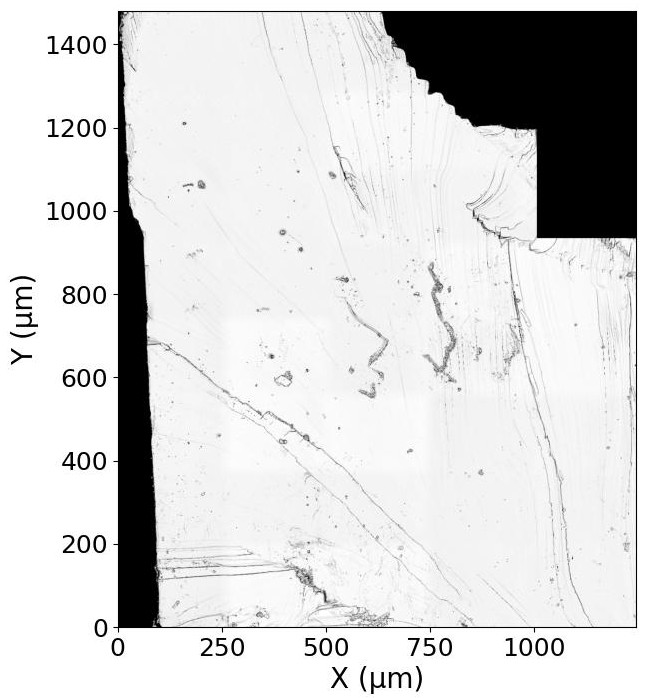}
  \caption{Optical overview image, that was automatically stitched by KEYENCE software.}
  \label{fig:NaCl_optical}
\end{subfigure}
\vspace{1em}
\begin{subfigure}[t]{0.49\textwidth}
  \centering
  \includegraphics[width=1.22\linewidth]{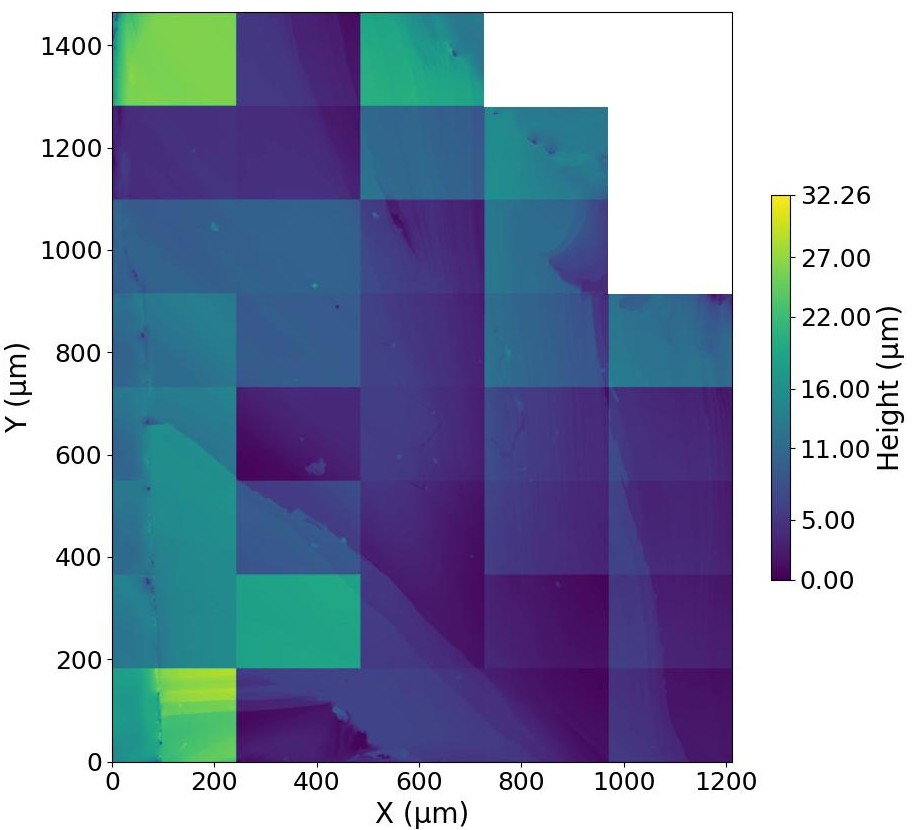}
  \caption{Unprocessed, height map image stitched by our custom-developed software.}
  \label{fig:NaCl_uncorr}
\end{subfigure}\hfill
\begin{subfigure}[t]{0.5\textwidth}
  \centering
  \includegraphics[width=1.3\linewidth]{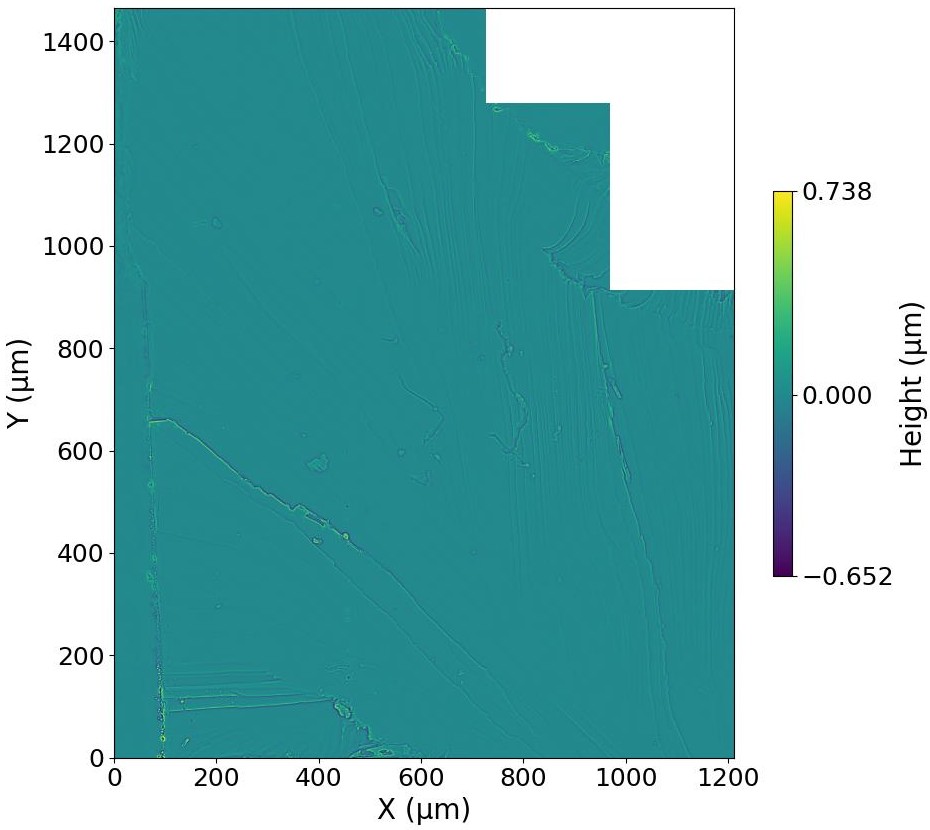}
  \caption{Processed, custom stitched height map, where each individual tile was tilt corrected and enhanced via image processing.}
  \label{fig:NaCl_corr}
\end{subfigure}\hfill
\caption[NaCl]{Stitched images of a NaCl sample, imaged with 50x magnification.}
\end{figure}

To demonstrate the capabilities of the developed processing pipeline, a NaCl sample shown in Figure \ref{fig:NaCl_whole}, was imaged at different magnifications.
The stitched optical image produced by the software is presented in Figure \ref{fig:NaCl_optical}. 
It provides an overview of the sample and illustrates the low-resolution transition regions between adjacent images, which become visible upon closer inspection. 
The corresponding uncorrected, manually stitched height map, shown in Figure \ref{fig:NaCl_uncorr}, exhibits significant height offsets between individual images. 
These height offsets are mitigated during the tilt correction of the data and the individual tiles further undergo image processing for feature enhancement, resulting in the height map shown in Figure \ref{fig:NaCl_corr}.

\begin{figure}[htbp]
\centering
\begin{subfigure}[t]{0.49\textwidth}
  \centering
  \includegraphics[width=1.0\linewidth]{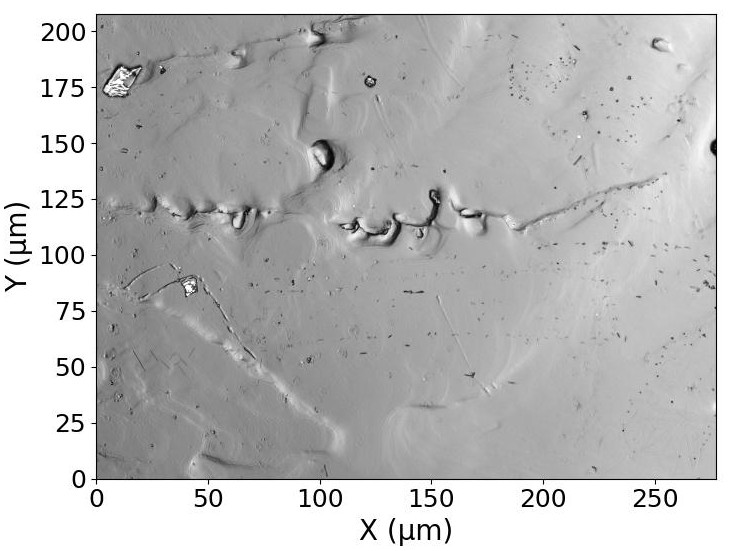}
  \caption{Optical overview image.}
  \label{fig:NaCl_single_50x_refl}
\end{subfigure}
\begin{subfigure}[t]{0.49\textwidth}
  \centering
  \includegraphics[width=1.275\linewidth]{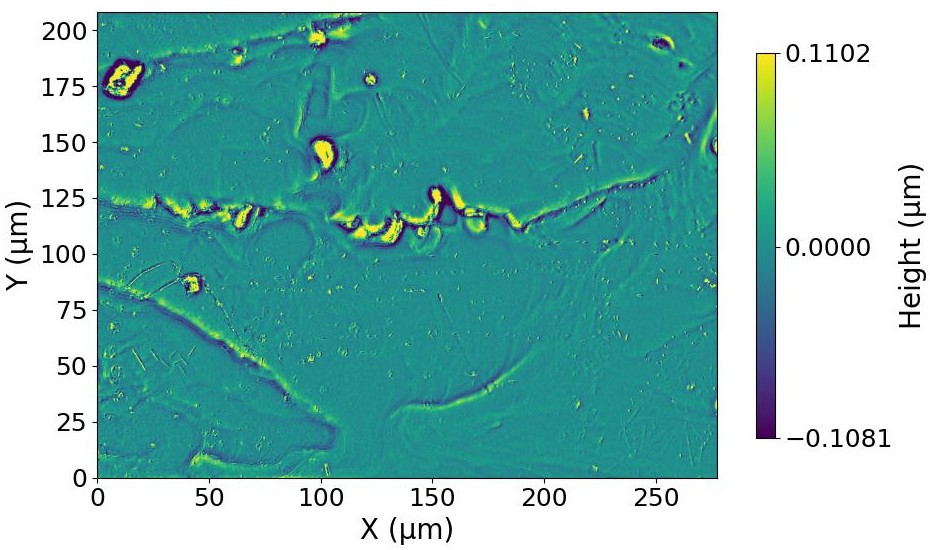}
  \caption{Height map image.}
  \label{fig:NaCl_single_50x_height}
\end{subfigure}\hfill
\caption[NaCl]{50x magnification images of a NaCl sample.}
\label{fig:NaCl_single_50x}
\end{figure}

\begin{figure}[htbp]
\centering
\begin{subfigure}[t]{0.49\textwidth}
  \centering
  \includegraphics[width=0.99\linewidth]{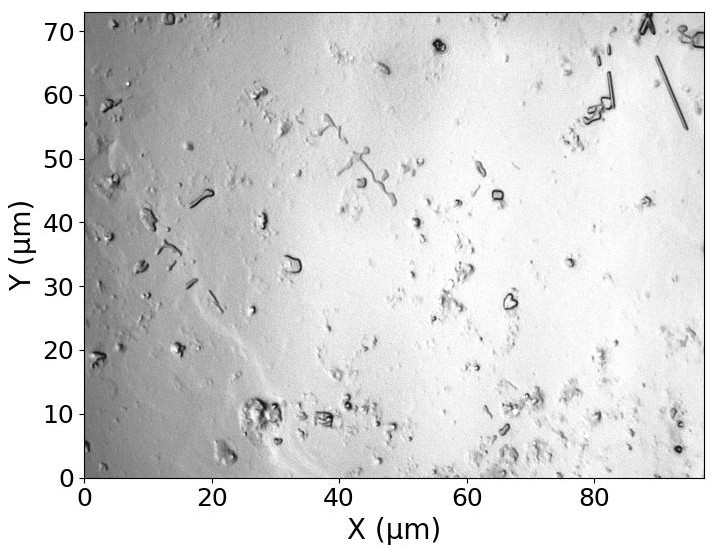}
  \caption{Optical overview image.}
  \label{fig:NaCl_single_150x_refl}
\end{subfigure}
\begin{subfigure}[t]{0.49\textwidth}
  \centering
  \includegraphics[width=1.285\linewidth]{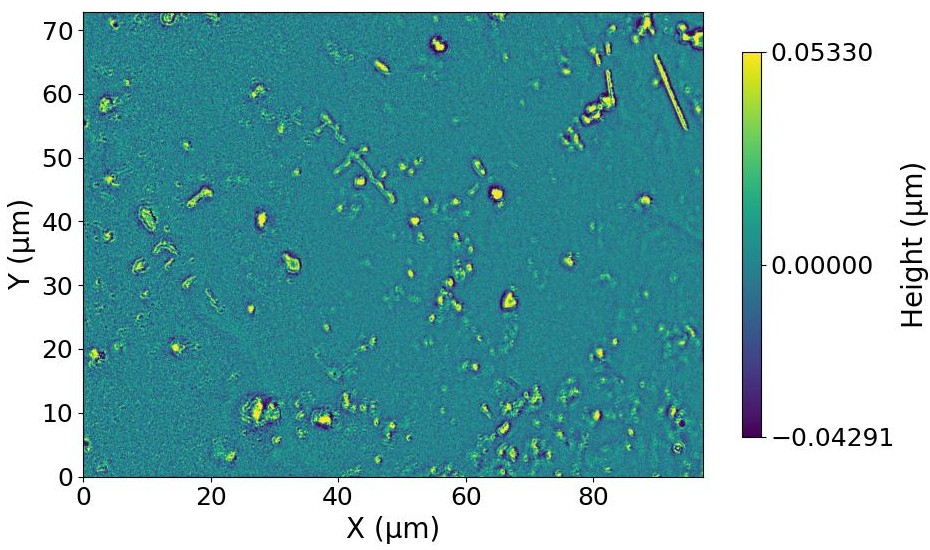}
  \caption{Height map image.}
  \label{fig:NaCl_single_150x_height}
\end{subfigure}\hfill
\caption[NaCl]{150x magnification images of a NaCl sample.}
\label{fig:NaCl_single_150x}
\end{figure}

\begin{figure}[htbp]
   \centering
   \includegraphics[width=0.6\textwidth]{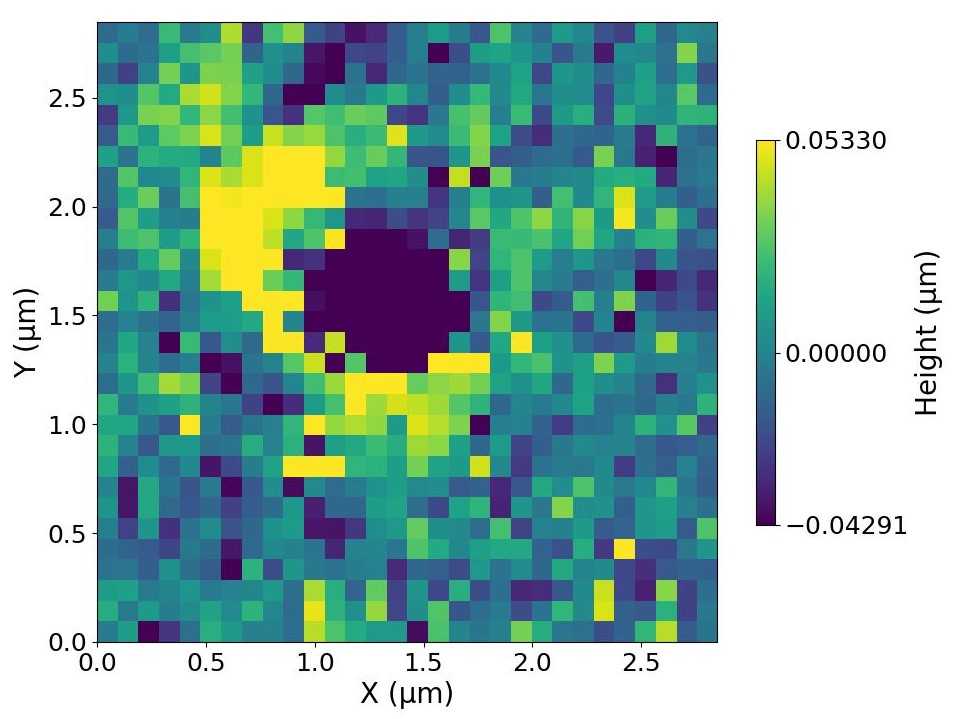}
   \caption{A zoomed in view of a small pit visible in the upper left corner of the 150x image.}
   \label{fig:small_feat}
\end{figure}

We have performed the measurements on a number of samples, and two examples of imaging with magnification 50x and 150x are showcased in Figure  \ref{fig:NaCl_single_50x} and Figure \ref{fig:NaCl_single_150x}, respectively.
The corresponding optical images in those figures are included for overview purposes. 
The images at 50x and 150x magnification are not from the same area of the sample but were selected because they contain numerous small surface features that clearly demonstrate the achievable spatial resolution.
After processing, the resulting height maps exhibit a sufficient spatial resolution to resolve structures with lateral dimensions of approximately 1 µm while maintaining a sub-nanometer vertical resolution.
For illustration of the capabilities of the optical profiler, in Figure \ref{fig:small_feat} we zoom in on a small pit-like feature visible in the upper left corner of Figure \ref{fig:NaCl_single_150x}.
This magnified view demonstrates the level of detail that can be extracted from the processed height data.

The presented data and images are well suited for use of ML-based feature detection algorithms, as the features of interest are well distinguishable from the overall surface.
Moreover, an application of ML is motivated by the large number of individual images that is generated during each measurement and cannot be manually reviewed for identification of potential damage features. 
A suitably trained algorithm could not only perform the identification automatically but also distinguish between natural surface structures and particle-induced features.
Hence, we are currently exploring a range of ML models, including custom developed software and readily available solutions, such as the blob detection algorithms in scikit-image.

\newpage

\newpage

\subsection{Outlook}
\label{sec:outlook}
Now that the development and optimization of the image acquisition chain is in its final stages, we will focus our efforts on further image post-processing and feature enhancement.
Moreover, we will aim to develop ML-based feature identification methods, using our own software tools and the ones provided to us by the group at INFN Milan.
Ultimately, we aim to establish the KEYENCE VK-X300 optical profiler at KIT-IAP as a flexible system for topographical imaging of etched crystals with micrometer-scale spatial and nanometer-scale vertical resolution for searches for dark matter, cosmic rays and other exotic particles.

\newpage

\acknowledgments
We thank Dr. Claudio Galelli of INFN Milan for providing us the NaCl and Olivine samples and for the fruitful discussions, as well as Dr. Keyu Ding of KIT-IAP for the assistance with operating the KEYENCE microscope and with supporting our data analysis efforts.
We also thank Prof. Ulrich A. Glasmacher of the Institute of Earth Sciences, University of Heidelberg for the continuous support for the project and for providing mineral test samples. 
Furthermore, we thank Dr. Christopher J. Kenney of the SLAC National Accelerator Laboratory for providing irradiated samples of mica and silicon for calibration and imaging studies.
We also thank Prof. Yolita Eggeler, Dr. Martin Peterlechner and Dr. Erich Müller of the Laboratory for Electron Microscopy at KIT, for the ongoing support of the project as well as for organizing and performing sub-sample preparation and electron microscopy imaging.
Additionally, we thank Dr. Torsten Scherer of the Institute of Nanotechnology at KIT and Dr. Rafaela Debastiani of the Institute of Nanotechnology at KIT and of the Helmholtz Institute Freiberg for Resource Technology at the Helmholtz-Zentrum Dresden-Rossendorf, for the useful discussions, and for preparation and imaging of samples with nanoCT.
This work was partly carried out with the support of the Karlsruhe Nano Micro Facility (KNMF, www.knmf.kit.edu), a Helmholtz Research Infrastructure at Karlsruhe Institute of Technology (KIT, www.kit.edu). 
The Xradia 810 Ultra (nanoCT) core facility was supported (in part) by the 3DMM2O - Cluster of Excellence (EXC-2082/1390761711).
The presented project is supported in part through the Helmholtz Initiative and Networking Fund (grant agreement no.~W2/W3-118). 
We also gratefully acknowledge the support by the KIT Center Elementary Particle and Astroparticle Physics (KCETA) for this project. 

\clearpage

\section{Investigating cosmic-ray signatures in paleo-detectors with PRImuS}
\label{sec:PRImuS}

Authors: {\it Claudio Galelli, Lorenzo Caccianiga, Lorenzo Apollonio, Paolo Magnani, and Vincent Breton}
\vspace{0.1cm} \\
INFN Milano, UNIMI, GSSI, LPC
\vspace{0.3cm}

\subsection{Introduction}
For the mineral detection community, secondary cosmic rays (CRs) have traditionally been viewed as an irreducible background that mandates the extraction of target samples from deep-underground environments ($>1$ km.w.e.) \cite{baum_mineral_2023}. For the PRImuS (Paleo-astroparticles Reconstructed with the Interactions of MUons in Stone) experiment, the paradigm is swapped: we leverage the high-energy CR-induced nuclear recoil tracks as a primary signal to study the past of our Galaxy, the history of the Earth's atmosphere, and use paleo-detectors as geological clocks.

The core challenge in using natural minerals for rare-event detection is the accurate characterization of the background. By directly measuring the CR-induced track record in minerals with well-constrained geological histories, we could not only open a new window into time-domain astrophysics but hopefully also provide solid experimental validation of the background models used by the broader MDvDM community. We utilize minerals that have a specific exposure and shielding history, where the track-production clock is precisely started, and sometimes stopped, by known geological events. This allows us to move beyond integrated "bulk" measurements and perform differential time-resolved paleo-astroparticle physics.

\subsection{The Simulation Pipeline: From Primary Flux to Lattice Damage}
The simulation\footnote{Freely available at \url{https://github.com/cgalelli/PrimusCode}} framework is structured as a five-step modular pipeline, designed to handle the complex physics of CR propagation and mineral interaction.

\begin{enumerate}
    \item \textbf{Primary flux modeling and propagation in the atmosphere:} the pipeline begins with the selection of a primary cosmic-ray spectrum, which can be a standard, non-altered scenario or a user-defined, modified custom model. The primary CRs are propagated through the Earth's atmosphere using the \texttt{MCEq} (Matrix Cascade Equations) package \cite{fedynitch_2022, Riehn:2017}. This step yields the secondary particle spectra, for muons and neutrons, at the Earth's surface in the requested scenario.
    \item \textbf{Particle-matter interaction in GEANT4:} secondary particles are re-injected into a custom \texttt{GEANT4} \cite{GEANT4:2002zbu} simulation environment, where all the relevant interactions for neutrons and positive and negative muons are computed. Resulting nuclear recoils are compiled in files containing the kinetic energy $E_R$ imparted to a nucleus during a collision, the species of the recoiling particle, the muon's or neutron's remaining energy, and the position in the simulation volume where the interaction occurred.
    \item \textbf{Stopping power and recoil range calculation:} the recoil energy is converted into a physical track length $R$. We utilize the \texttt{SRIM} (Stopping and Range of Ions in Matter) tables to compute the projected range of recoiling ions within the target lattice.
    \item \textbf{Geological convolution and exposure windows:} the instantaneous track production rate is integrated over the geological ``exposure window.'' The pipeline can handle continuous deposition, like the cases of sedimentary scenarios exemplified by the evaporites from the Messinian salinity crisis \cite{caccianiga_2024}, where the software accounts for the varying attenuation as the overburden increases over time.
    \item \textbf{Instrumental response and projection:} finally, the 3D track distribution is projected onto a 2D observation plane to simulate optical microscopy. This step includes:
    \begin{itemize}
        \item Random sampling of track starting points and orientations.
        \item Computation of intersection with a slicing plane.
        \item Modeling of the etching effect, where chemical or plasma processing enlarges the latent track into a visible pit.
        \item Simulated measure of the 2D projection of the pit.
    \end{itemize}
    A visual representation of this step is sketched in figure \ref{fig:projection_sim}.
    
    \begin{figure}[ht!]
     \centering
     \includegraphics[width=0.7\columnwidth]{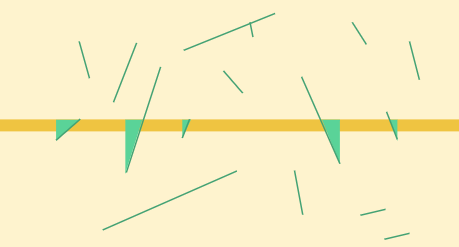}
     \caption{Sketch of the process of the simulation of instrumental effects, etching, and 2D projection for microscope observation. Tracks, in green, intersecting the etching plane, in dark brown, are retained and measured as 2D projections represented as the horizontal sides of the corresponding green triangles.}
     \label{fig:projection_sim}
    \end{figure}
    
\end{enumerate}

\subsection{Target Focus: The Chaîne des Puys (CdP) Volcanic Chronosequence}
The Chaîne des Puys (CdP) in the French Massif Central offers a unique ``paleo-detector array'' consisting of dozens of volcanic centers erupted over the last 100 kyr. The targets of interest for this array are mantle xenoliths (peridotites) brought to the surface by basaltic magma. These xenoliths consist primarily of olivine, a mineral often cited as a possible paleo-detector. Crucially, the eruption process itself acts as a natural ``zeroing'' mechanism. The host basaltic magma reaches temperatures exceeding $1100^\circ$C, which is well above the track-annealing temperature for olivine. This thermal event completely erases any pre-existing tracks accumulated during the millions of years the mineral spent in the upper mantle. Consequently, the track record in a CdP xenolith begins precisely at the moment of eruption and subsequent cooling, providing a "clean" starting point for the CR integration.

\subsubsection{Time-Differential Reconstruction}
The CdP field contains a sequence of eruptions with very well-reconstructed ages. By comparing the track densities in olivine from a sequence of different volcanoes, we can perform a time-differential analysis. Instead of a single measurement, we obtain a series of overlapping exposure windows. By subtracting the integrated record of a younger sample from an older one, we can isolate the CR flux during specific historical intervals, such as the Holocene/Pleistocene transition. A selected list of volcanic events chosen for the simulation is displayed in table \ref{tab:chrono}.

\begin{table}[htbp]
 \centering
 \caption{Chronology of selected volcanic events in the Chaîne des Puys, from \cite{Galelli_2026, Boivin2017}.}
 \label{tab:chrono}
 \begin{tabular}{l c}
 \hline
 \textbf{Volcanic Event} & \textbf{Age (kyr)}\\
 \hline
 Puy de Montcineyre & 7.65$\pm$0.12\\
 Puy de la Vache & 8.64$\pm$0.06\\
 Puy Pariou & 9.5$\pm$0.5\\
 Puy de Dôme & 10.96$\pm$0.15\\
 Puy de Côme & 13.1$\pm$0.7\\
 Puy de Lemptégy II & 30$\pm$4.5\\
 Laschamp Event & 41.4$\pm$1.1\\
 \hline
 \end{tabular}
\end{table}

\subsubsection{Sensitivity to the Laschamp Excursion and Nearby Supernovae}
\begin{figure}[ht!]
 \centering
 \includegraphics[width=0.7\columnwidth]{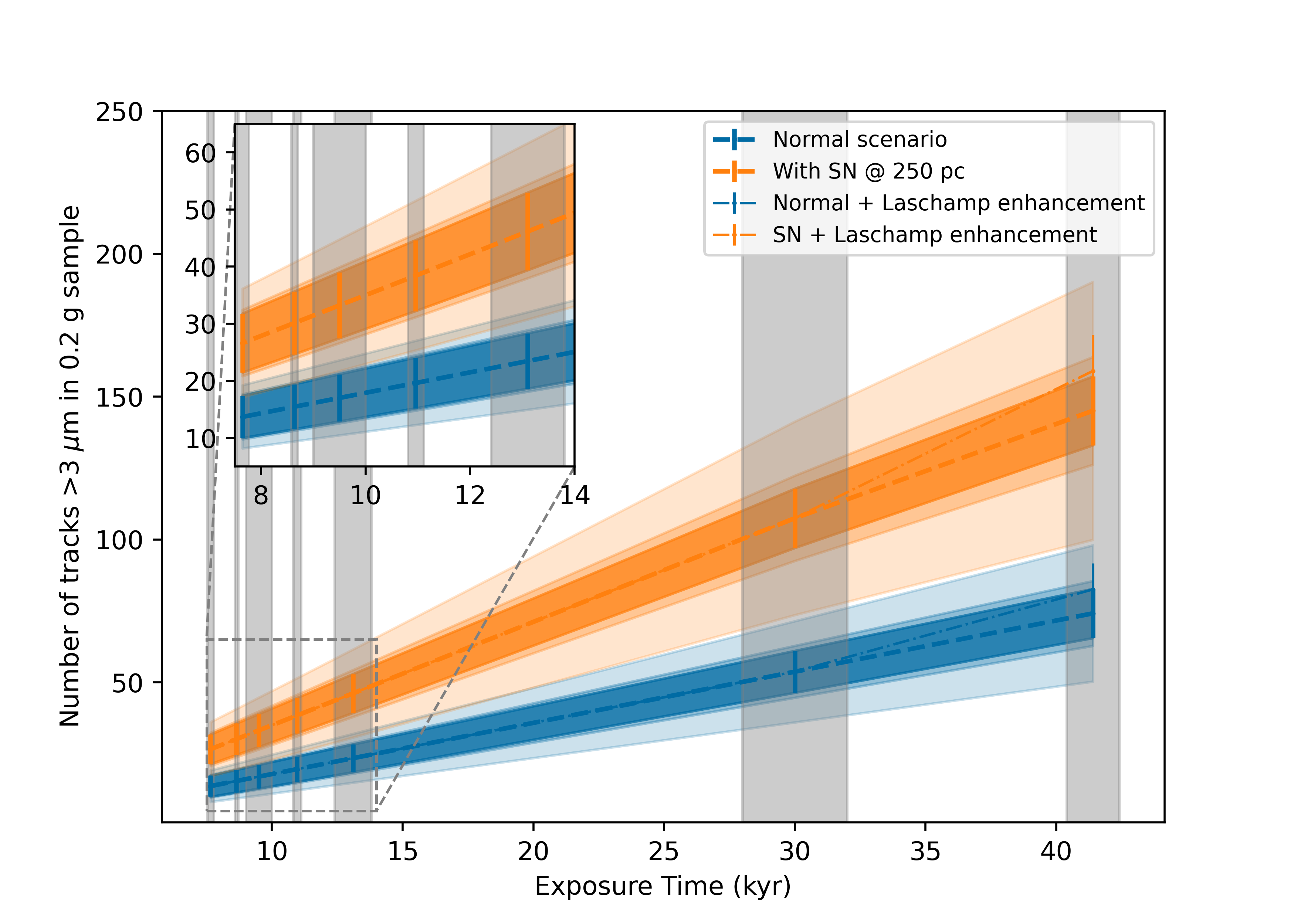}
 \caption{Time evolution of the total integrated number of muon-induced tracks as a function of sample exposure time. The points represent the expected signal for each of the volcanic scenarios in the normal (blue) and SN250 (orange) flux scenarios, with Poissonian, Poissonian + 10\% counting error, and Poissonian + 30\% counting error bands. For each flux, an enhancement due to the lower Earth's magnetic field during the Laschamp event is shown as a dashed line deviating from each flux scenario in the time interval between 30 and 40~kya. The inset shows a zoom in the 7 to 14~kyr range, an interval of time in which eruptions were particularly frequent. Time uncertainties in the eruption dating are represented by the grey vertical bands.}
 \label{fig:time_evolution}
\end{figure}

The 41 kyr window is of particular interest as it encompasses the Laschamp geomagnetic excursion. During this event, the Earth's magnetic dipole intensity dropped to near zero for approximately 500 years, leading to a significant increase in the flux of low-energy cosmic rays reaching the atmosphere \cite{Laj2014}. Our simulations indicate that this increase could be detectable as a ``bump'' in the track-length distribution, depending on experimental precision. Furthermore, the CdP sequence provides sensitivity to the ``local bubble'' supernova history. Specifically, we have modeled the impact of a ``toy-model'' supernova, consistent with the Antlia SNR precursor (dated to $\sim$50 kyr, at a distance of $\sim$250 kpc) \cite{Fesen2021}. The resulting secondary muon flux would leave a distinct increased signature in the olivine tracks, potentially providing the first direct ``kinetic'' confirmation of supernova-enhanced CR fluxes, as shown in figure \ref{fig:time_evolution}, from \cite{Galelli_2026}. Such a result would open the door to the paleo-detector technique as a complement to existing data from cosmogenic isotopes like $^{10}$Be or $^{60}$Fe; more importantly, this technique could peer into the astrophysical past much deeper than isotope-based techniques, which are mostly limited by half-life.

\subsection{PRImuS and the \texttt{OptimusPrimus} Analysis Pipeline}
The experimental setup of the PRImuS experiment is composed of a high-throughput analytical pipeline at INFN Milano. Currently, a pipeline is being implemented for sample preparation, where minerals undergo controlled surface preparation using an Argon-based plasma etching protocol, implemented via a Diener ATTO RF system. This technique should allow for the precise enlargement of latent nuclear tracks while maintaining morphological integrity, a crucial advantage over traditional wet chemical etching. The main instrument for to be used for sample scanning is an automated optical microscopy system by Evident/Olympus. The etched surfaces are imaged using an 8 Mpx color camera looking at a motorized XYZ stage, achieving a lateral resolution of 0.345~$\mu$m/pixel and a vertical precision of O(1~$\mu$m). This setup enables the systematic scanning of large-area samples (up to several cm$^2$) on multiple focus planes. To process the high volume of imaging data required for statistically significant track detection, the software infrastructure for PRImuS includes, in addition to the simulation side, \texttt{OptimusPrimus}, a class for automated feature extraction.

\subsubsection{Deep Learning Segmentation}

The core of \texttt{OptimusPrimus} is a semantic segmentation model based on the U-Net architecture with a \texttt{ResNet34} encoder pretrained on ImageNet. The model is trained to recognize the specific morphology of etched tracks against the background of mineral defects, fractures, and surface roughness. The input images are sliced to $1024 \times 1024$ pixels, and then are segmented into a binary mask where each ``on'' pixel represents a high probability of belonging to a track candidate.

\begin{figure}[ht!]
 \centering
 \includegraphics[width=0.7\columnwidth]{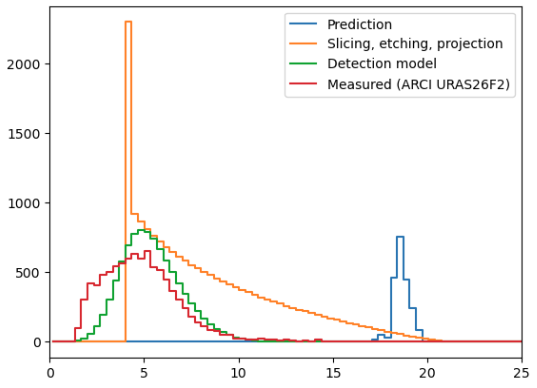}
 \caption{Comparison of the distribution of tracks as segmented and counted by the model in a sample of volcanic obsidian, in red, with the expected fission fragment track spectrum before instrumental effects, in blue, after applying the simulated slicing and etching effects, in orange, and taking into account the model efficiency, in green.}
 \label{fig:projection}
\end{figure}

Once segmented, individual tracks are isolated using contour detection. \texttt{OptimusPrimus} fits an ellipse to each track to extract the major axis ($a$) and minor axis ($b$). The class includes an efficiency correction module that accounts for recall, the probability that a track of length $R$ is detected by the ML model, and precision, the suppression of false positives. By applying these corrections, which are done on a length bin-by-bin basis, we can directly compare the measured 2D ellipse distributions to projected and "measured" theoretical spectra produced by the simulation pipeline. An example of this comparison is shown in figure \ref{fig:projection}.

\subsection{Conclusion}
The integration of complex geological histories with astrophysics, particle simulations, and machine learning has established a possible new observational window into the paleo-universe. Ongoing analysis of CdP xenoliths and neutron-irradiated calibration samples will hopefully provide the first experimental characterization of the CR record in paleo-detectors, paving the way for future rare-event searches.

\subsection*{Acknowledgments}
PRImuS is an INFN experiment funded by the CSN5 Young Scientist Grant 2024. The authors thank Didier Miallier, Thierry Pilleyre, Pierre-Jean Gauthier, Valentin Niess, Denis Andrault, and Emmanuel Gardes for the fruitful discussions in Clermont-Ferrand on xenoliths and olivine samples from the Chaîne des Puys and possible methods of analysis and readout. We also thank the members of the paleo-detector community, especially Alexey Elykov and Emilie LaVoie-Ingram, for their interest and exchange of ideas. 
\clearpage

\section{Reassessing the Directional Signature of the Dark Matter Wind for DMICA}\label{sec:DMICA_directional_signature}

Authors: {\it Shigenobu~Hirose$^1$, Yoichi~Usui$^2$ and Yoji~Kawamura$^1$}
\vspace{0.1cm} \\
$^1$Japan Agency for Marine-Earth Science and Technology (JAMSTEC) \\ $^2$Kanazawa University
\vspace{0.3cm}

\subsection{Introduction}

Ancient mica can record nuclear-recoil damage over geological time scales and therefore provides a possible paleo-detector for dark matter.  In the DMICA concept~\cite{MDDM2025Proceedings}, recoil tracks intersecting a cleavage plane are exposed as etch pits by chemical etching and are counted with depth information using an optical profiler.  The first mica dark-matter search by Snowden-Ifft, Freeman and Price~\cite{snowden-ifft:1995zgn} (SI95) used AFM readout over $0.08\,\mathrm{mm}^2$ of 0.5-Gyr-old mica and found no events in the shallow-depth region of interest, setting an upper limit on the WIMP cross section.  DMICA aims to revisit this idea with much faster optical-profiler readout and with a projected exposure many orders of magnitude larger than the original AFM-scale search.
\par
The large exposure enabled by optical-profiler readout raises a natural question: if a statistically significant excess of DM-like recoil pits is observed, can the same fossil record also reveal the directionality of halo dark matter?  This requires assessing how much of the DM-wind signature survives geological-time averaging, including continental drift of the mica normal and the Galactic motion of the Solar System.  The key angle is $\alpha$, defined as the angle between the mica cleavage-plane normal and the DM-wind direction; $\alpha=0^\circ$ corresponds to the normal being parallel to the wind direction, while $\alpha=90^\circ$ corresponds to the cleavage plane being parallel to the wind direction.  Here we reassess this paleo-directional capability of mica following Snowden-Ifft and Westphal~\cite{SnowdenIfft:1997} (SIW97).

\subsection{Intrinsic angular response and geological-time averaging}

\begin{figure}
    \centering
    \includegraphics[width=0.98\linewidth]{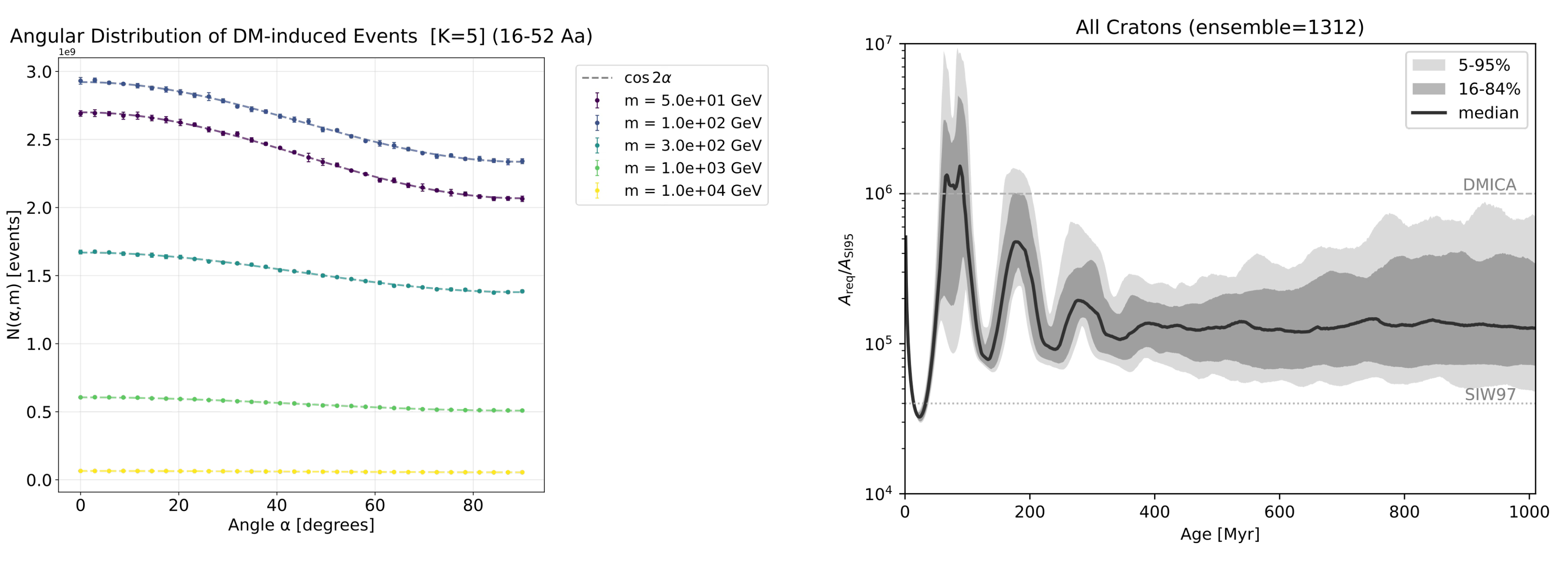}
\caption{
Summary of the paleo-directional benchmark for DMICA.
Left: Monte Carlo pit counts as a function of the angle $\alpha$ for
selected DM masses, with $\cos 2\alpha$ fits.
Right: required exploration area normalized to the SI95 scan area after
geological-time averaging; bands show the 5--95\% and 16--84\% ranges of
the 1312-member ensemble over cratons, rotation models, and Galactic
orbital periods of 185--225 Myr.
}
    \label{fig:DMICA_directional_summary}
\end{figure}

The intrinsic response was estimated with Monte Carlo simulations of DM-induced nuclear recoils in mica.  As shown in Fig.~\ref{fig:DMICA_directional_summary}(left), the pit number is well described by
\begin{equation}
    N(\alpha,m)=c_0(m)+c_2(m)\cos 2\alpha,
    \label{eq:DMICA_angular_response}
\end{equation}
where $m$ is the DM mass.  This form respects the up--down symmetry $N(\alpha,m)=N(\pi-\alpha,m)$ of tracks crossing the cleavage plane and remains smooth at $\alpha=90^\circ$.  The maximum contrast between the most and least favorable orientations is
\begin{equation}
    s_{\rm max}(m)=
    \frac{N(0^\circ,m)-N(90^\circ,m)}{N(0^\circ,m)+N(90^\circ,m)}
    =\frac{c_2(m)}{c_0(m)} .
\end{equation}
We define the intrinsic asymmetry factor as
\begin{equation}
    \omega(m)\equiv \frac{s_{\rm max}(m)}{1-s_{\rm max}(m)/3} .
    \label{eq:DMICA_omega}
\end{equation}
The simulations indicate that $\omega(m)$ is at the $\mathcal{O}(10\%)$ level for light-to-intermediate DM masses and decreases toward heavier masses.
\par
For a mica sample of age $t$, the present-day mica normal $\bm{n}_0$ traces a long history through Earth rotation, revolution, precession, continental drift and Galactic orbital motion.  The observed signal contrast can be written approximately as
\begin{equation}
    s(m,t)\simeq \omega(m)\,\xi(t),
    \label{eq:DMICA_factorization}
\end{equation}
where $\xi(t)$ is the history degradation factor.  Since the angular response depends on $\cos 2\alpha$, we compute
\begin{equation}
    \left\langle\!\left\langle \cos 2\alpha \right\rangle\!\right\rangle(t,\bm{n}_0)
    =\frac{1}{t}\int_0^t \cos 2\alpha(t',\bm{n}_0)\,dt'
\end{equation}
for each possible present-day normal and define
\begin{equation}
    \xi(t)=\frac{1}{2}\left[
    \max_{\bm{n}_0}\left\langle\!\left\langle \cos 2\alpha \right\rangle\!\right\rangle(t,\bm{n}_0)
    -
    \min_{\bm{n}_0}\left\langle\!\left\langle \cos 2\alpha \right\rangle\!\right\rangle(t,\bm{n}_0)
    \right].
    \label{eq:DMICA_xi}
\end{equation}
In practice, fast periodic motions are averaged first.  The remaining slow evolution is described by the declination $\delta$ of the mica normal, mainly controlled by continental drift, and the ecliptic latitude $\beta$ of the DM wind, controlled by Galactic orbital motion.  For each assumed sample age, present-day mica normals are sampled on a Fibonacci grid, their histories are propagated in the $\delta$--$\beta$ plane using paleomagnetic rotation models~\cite{Cao2024Zenodo,Scotese2016Zenodo,EarthByte2024GPlates}, and Eq.~\eqref{eq:DMICA_xi} is evaluated.

\subsection{Paleo-directional reach}

Combining $\omega(m)$ with $\xi(t)$ gives the final directional contrast in the signal-dominated benchmark. The ensemble calculation contains 1312 equally weighted histories, constructed by varying the craton, the rotation model, and the Galactic orbital period over 185--225 Myr.  It shows that $\xi(t)$ decreases for older samples; for a 500-Myr-old sample it is of order $0.04$ in the reference calculation. The right panel of Fig.~\ref{fig:DMICA_directional_summary} shows that the required exploration area is only weakly age-dependent.  This behavior mainly reflects the competition between the decreasing history factor $\xi(t)$ and the increasing accumulated exposure time.  This result should be read as a benchmark for whether mica could retain enough directional information to test the halo-DM-wind origin of a sufficiently large DM-like excess.  The updated estimate is less optimistic than the original benchmark by a factor of a few, but it remains within the large-exposure regime targeted by DMICA.

\subsection{Summary and outlook}

This reassessment confirms that mica has an intrinsic directional response to the DM wind at approximately the ten-percent level, with a stronger effect for lighter DM masses.  Geological-time averaging suppresses the signal through $\xi(t)$, but does not erase it. In a signal-dominated benchmark, the required exploration area remains compatible with the optical-profiler exposure scale envisioned for DMICA.  Thus, if a sufficiently large DM-like excess is observed, ancient mica could retain enough directional information to test whether the excess is consistent with a halo-DM-wind origin.  The natural next step is to move beyond this benchmark and construct a hybrid analysis that combines pit-depth spectra and directional pit counting in a joint likelihood framework.

\acknowledgments
This work was supported by JSPS KAKENHI Grant Number JP25K07350.

\clearpage

\section{Mineral Detection of Cosmic-Ray Boosted Dark Matter}\label{sec:UESTC}

Authors: {\it Jin-Wei Wang}
\vspace{0.1cm} \\
School of Physics, University of Electronic Science and Technology of China, Chengdu 611731, China
\vspace{0.3cm}

\subsection{Introduction}
The existence of dark matter (DM) has been firmly established by a variety of astrophysical and cosmological observations. However, its particle nature remains unknown. Direct detection experiments provide one of the most powerful approaches to probe DM interactions with Standard Model particles. 
Recent experiments such as XENONnT, LZ, and PandaX have achieved remarkable sensitivities to weakly interacting massive particles (WIMPs). However, their sensitivity rapidly deteriorates in the sub-GeV regime because halo DM carries insufficient kinetic energy to produce detectable nuclear recoils.

A robust mechanism to overcome this limitation is cosmic-ray boosted DM (CRDM) \cite{Bringmann:2018cvk}. In this scenario, Galactic cosmic rays (CRs) scatter off halo DM particles and accelerate a small fraction of the DM population to relativistic energies. The resulting boosted component can induce observable nuclear recoils even for very light DM.

Recently, paleo-detectors have emerged as a promising approach for DM detection \cite{snowden-ifft:1995zgn,Baum:2018ekm,drukier:2018pdy}. Instead of operating a detector in real time, paleo-detectors search for permanent damage tracks accumulated in ancient minerals over geological timescales. Owing to their enormous effective exposures of order ${\cal O}(10^5)\ {\rm ton\ yr}$,
they provide unprecedented sensitivity to extremely rare processes.

The combination of CRDM and paleo-detectors is particularly attractive. Since CRDM particles carry much larger kinetic energies than conventional halo DM, they can produce significantly longer recoil tracks, which are efficiently distinguished from the dominant backgrounds.
Motivated by this observation, we perform the first dedicated study of CRDM in paleo-detectors and investigate their sensitivity to DM--proton interactions \cite{Wang:2026you}.

\subsection{CRDM Flux and Track Spectrum}
\label{sec:simulaitons}

To calculate the CRDM flux, we use the local interstellar spectra of cosmic rays obtained from  \texttt{GALPROP} and  \texttt{HELMOD} simulations. The dominant cosmic-ray species included in our analysis are
\{\rm H, He, C, N, O, Ne, Mg, Si, Fe\}, which contribute more than 90\% of the total CRDM flux.
To illustrate the impact of different interaction structures, we consider two benchmark scenarios:

\begin{itemize}
	\item A phenomenological model with a constant DM--proton scattering cross section $\sigma_{\chi p}$;
	\item A simplified model in which DM interacts with quarks through a massive vector mediator.
\end{itemize}

For the vector-mediator model, both elastic scattering and deep inelastic scattering (DIS) processes are included. We find that DIS becomes increasingly important at high energies and significantly modifies the high-energy tail of the CRDM spectrum. 
Neglecting DIS may therefore underestimate the CRDM flux at high energies, particularly in the high-energy region relevant for boosted DM searches.

Once the CRDM flux is obtained, the corresponding recoil spectrum in minerals can be calculated. 
We consider two representative minerals, Gypsum [$\mathrm{Ca(SO_4)\!\cdot\!2(H_2O)}$] and Olivine [$\mathrm{Mg}_{1.6}\mathrm{Fe}^{2+}_{0.4}(\mathrm{SiO}_4)$].

The recoil nuclei lose energy while propagating through the crystal lattice and leave permanent damage tracks. The relation between recoil energy and track length is calculated using the  \texttt{SRIM} package.
Following previous paleo-detector studies, we adopt a benchmark exposure of 100\,{\rm g\,Gyr} and a track-length resolution of approximately 15\,{\rm nm}.

\begin{figure}
    \centering
    \includegraphics[width=0.6\linewidth]{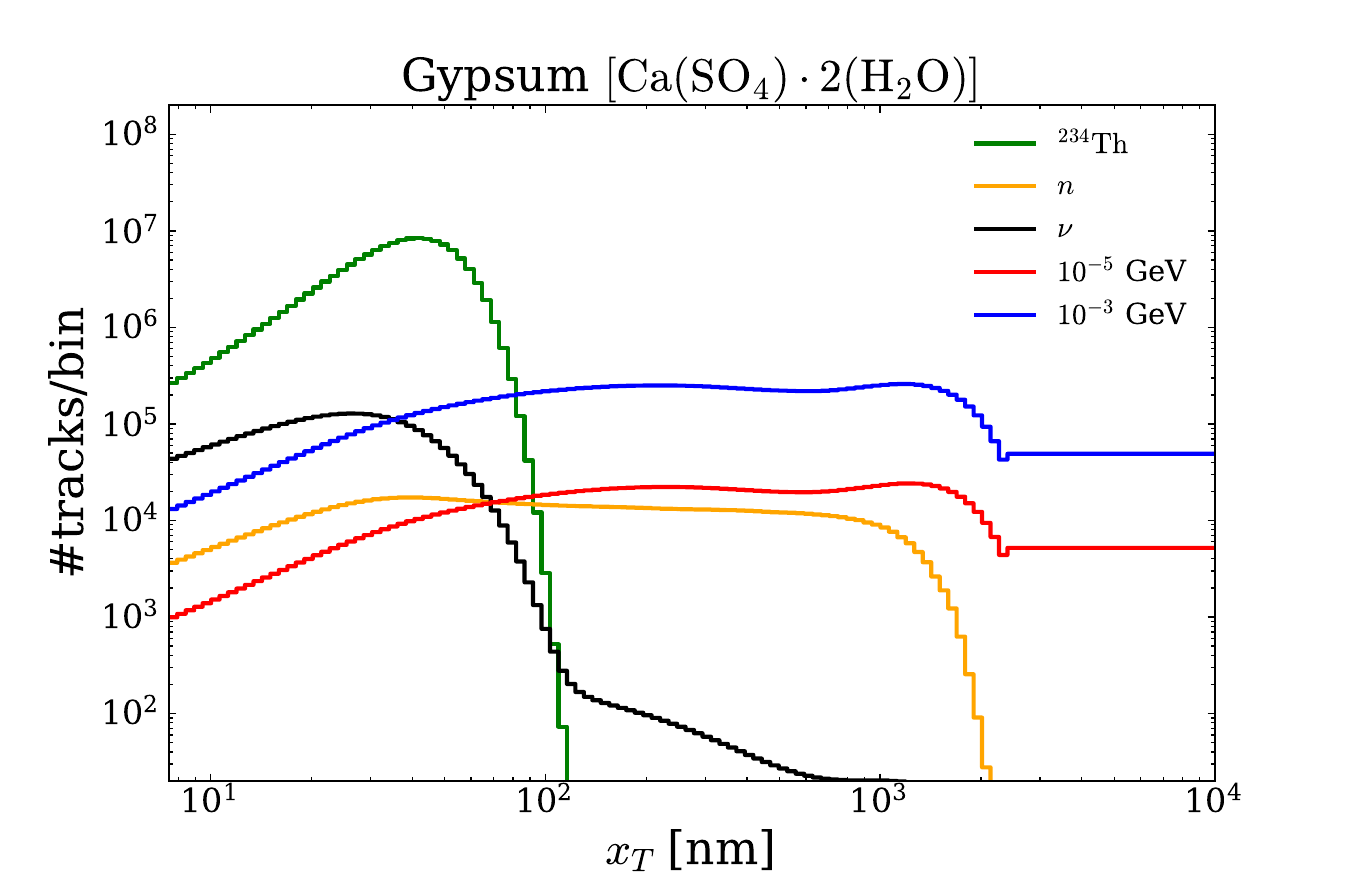}
    \caption{Binned track length distributions in Gypsum [$\mathrm{Ca(SO_4)\!\cdot\!2(H_2O)}$] for CRDM in the vector-mediator model. 
    	The red and blue curves show CRDM signals for $m_\chi=10^{-5}~\mathrm{GeV}$ and $10^{-3}~\mathrm{GeV}$, respectively, while the black, orange, and green curves denote the dominant nuclear recoil backgrounds from neutrinos, radiogenic neutrons, and $^{238}\mathrm{U}\!\rightarrow\!^{234}\mathrm{Th}+\alpha$ decays. 
    	The input parameters are $g_\chi=g_q=0.7$ and $m_V=1~\mathrm{GeV}$.}
    \label{fig:bin_tracks}
\end{figure}

Fig.~\ref{fig:bin_tracks} shows the binned track length spectrum in Gypsum for the vector-mediator model with $g_\chi=g_q=0.7$ and $m_V=1~\mathrm{GeV}$. 
The red and blue curves correspond to $m_\chi=10^{-5}~\mathrm{GeV}$ and $m_\chi=10^{-3}~\mathrm{GeV}$, respectively. The dominant backgrounds arise from solar, atmospheric, and supernova neutrinos (black), radiogenic neutrons (orange), and $^{234}$Th recoils originating from uranium decay chains (green). 
A characteristic feature of CRDM is that it produces substantially longer tracks than all dominant backgrounds. This feature enables efficient background rejection and plays a crucial role in enhancing the sensitivity of paleo-detectors.

\subsection{Projected Sensitivities}
\label{sec:simulation_of_ion_induced_damage}

Projected sensitivities are derived using the public package \texttt{paleoSens}, which employs a profile-likelihood analysis including detector response and background uncertainties.
For comparison, we also derive constraints from the latest XENONnT data. Following the XENONnT analysis, we combine the SR0, SR1a and SR1b datasets and perform a Poisson likelihood analysis of the total event counts.

\begin{figure}
	\centering
	\includegraphics[width=0.6\linewidth]{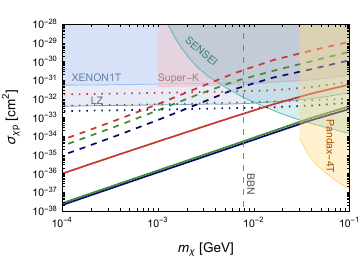}
	\caption{Constraints on the DM--proton scattering cross section as a function of the DM mass. 
		The purple and green curves show the projected sensitivities of paleo-detectors using Gypsum and Olivine, respectively, while the red curves indicate the XENONnT limits. 
		The dotted curves correspond to the constant cross section scenario, whereas the dashed and solid curves denote the vector-mediator model with $m_V=10~\mathrm{MeV}$ and $m_V=1~\mathrm{GeV}$, respectively. Existing constraints from XENON1T \cite{Bringmann:2018cvk}, Super-K \cite{Super-Kamiokande:2022ncz}, LZ \cite{LZ:2025iaw}, PandaX-4T~\cite{PandaX:2023xgl}, SENSEI~\cite{SENSEI:2023zdf}, and BBN~\cite{Giovanetti:2021izc} are also shown for comparison.}
	\label{fig:constraints}
\end{figure}

The resulting constraints are shown in Fig.~\ref{fig:constraints}. The projected sensitivities of paleo-detectors are presented for both Gypsum and Olivine, together with existing constraints from XENONnT, LZ, PandaX-4T, Super-K and SENSEI.
We find that paleo-detectors achieve highly competitive projected sensitivities over a large fraction of the sub-GeV parameter space.
Depending on the underlying interaction model, the sensitivity to the DM–proton scattering cross section can improve by one to two orders of magnitude relative to current XENONnT limits.

The improvement originates from two key advantages. First, paleo-detectors accumulate enormous effective exposures over geological timescales. Second, CRDM-induced recoils generate long tracks that can be efficiently distinguished from conventional short-track backgrounds. 
These results demonstrate that paleo-detectors provide a powerful and complementary approach to probing light DM and highlight their potential for future CRDM searches.

\acknowledgments

The authors wish to thank Alessandro Granelli for the thoughtful discussions at the initial stage of this project. The work of  J.-W.W. was supported by the National Natural Science Foundation of China (NSFC) under Grants
No. 12405119, the Natural Science Foundation of Sichuan Province under Grant No. 2025ZNSFSC0880, and Fundamental Research Funds for the Central Universities (Grant No. Y030242063002070).
\clearpage

\section{Heavy-element paleodetectors for Higgsino dark matter}\label{sec:Higgsino}

Authors: {\it Peter W.~Graham$^{1,2}$, Harikrishnan Ramani$^3$, and Samuel S.~Y.~Wong$^4$}
\vspace{0.1cm} \\
$^1$Leinweber Institute for Theoretical Physics, Department of Physics, Stanford University, Stanford, CA 94305, USA\\
$^2$Kavli Institute for Particle Astrophysics and Cosmology, Department of Physics, Stanford University, Stanford, CA 94305, USA \\
$^3$Department of Physics and Astronomy, University of Delaware and the Bartol Research Institute, Newark, DE 19716, USA\\
$^4$Department of Physics, University of Washington, Seattle, WA 98195, USA\\
\vspace{0.3cm}

This contribution is a summary of Ref.~\cite{Graham:2026ivn} and includes excerpts from that work.

\subsection{Inelastic dark matter}

An electroweak WIMP with nonzero hypercharge generically scatters with nucleons through tree-level $Z$ exchange, with a cross section $\simeq 10^{-39}~\text{cm}^2$.
While such a large elastic cross section has long been ruled out by xenon experiments,
one remaining possibility is the inelastic dark matter scenario~\cite{Tucker-Smith:2001myb}, in which a small mass splitting $\delta$ exists between two nearly degenerate states that only couple to the $Z$ boson off-diagonally. If $\delta$ is larger than the available incoming kinetic energy, the scattering would be kinematically forbidden. A particularly well-motivated realization of inelastic DM is the 1.1~TeV thermal relic Higgsino in the minimal supersymmetric standard model (MSSM), one of the last classic supersymmetric WIMPs standing~\cite{Krall:2017xij}.

For a target nucleus with mass number $A$, the largest $\delta$ that can be scattered is given by
\begin{align} \label{eq:delta_max}
    \delta_{\rm{max},A} = \frac{1}{2} \mu_A v_{\rm max}^2 ~,
\end{align}
where $\mu_A$ is the DM-nucleus reduced mass, and $v_{\rm max}$ is the maximum DM velocity in the Earth frame. For TeV-scale DM, the reduced mass is given by the target nucleus mass, $\mu_A \approx m_A=A m_n$. Hence, the current kinematic reach of xenon experiments is about $\delta \approx 350$~keV, set by the xenon mass ($A_{\rm Xe}=131$). Larger mass splittings would be invisible to both xenon experiments and existing paleodetector proposals, since the heaviest element in existing paleodetector proposals is iron ($A_{\rm Fe}=56$)~\cite{Theodosopoulos:2026ehn}, which is lighter than xenon.

\subsection{Heavy-element paleodetectors}

To overcome this obstacle, we propose a variant of paleodetectors that we term ``heavy-element paleodetectors." Our approach targets inelastic DM by using radiopure minerals that contain heavy elements such as lead ($A_{\rm Pb}=207$).

We identify a geological formation mechanism that, unlike those previously considered in the paleodetector literature, can produce radiopure minerals containing heavy elements: mineral precipitation in deep underground aquifers of geological age containing geothermal brines with high concentrations of Pb~\cite{regenspurg_formation_2016,KharakaHanor2003,Regenspurg2010,Regenspurg2014,McKibben1987SaltonSea,Scheiber2013ScalingInhibitorSoultz}. Geothermal brine can have extremely low uranium concentrations. In the most radiopure cases to our knowledge, samples from geopressured, geothermal aquifers in the Gulf of Mexico basin (at $\sim 5$~km depth) were found to have uranium concentrations as low as $C^{238} \lesssim 3\times 10^{-12}~\text{g}/\text{g}$~\cite{Kraemer_lowU_1981,Kraemer_lowU_1986,geopressure_review_1992}. In fact, calculations show that at an optimal redox potential and pH, uranium concentration could get as low as $C^{238} \lesssim 10^{-13}~\text{g}/\text{g}$~\cite{lowU_calculation_Langmuir_1978,Kraemer_lowU_1981,uranium_solubility_Goodwin_1980}. Precipitates from these radiopure brines are expected to inherit the extremely low uranium concentrations~\cite{Kraemer_lowU_1981}. We term them ``brine precipitates (BPs)."

As a concrete example, we focus on the mineral Laurionite (PbClOH)~\cite{ralph_mindatorg_2025,MindatLaurionite}, a colorless to white crystal, as a promising candidate for heavy-element paleodetectors, though we emphasize that there are many other potential mineral candidates. Laurionite has a resistivity $\rho_r \approx 10^{8}\text{--}10^{10}~\Omega~\text{cm}$~\cite{laurionite_conductivity}, comfortably satisfying $\rho_r \gtrsim 2000~\Omega~\text{cm}$, the condition for materials to record damage tracks~\cite{drukier:2018pdy,GUO2012233}. Laurionite contains Pb with a mass fraction of $w_{\rm Pb}=0.8$. Furthermore, it also contains H, which drastically reduces fast neutron backgrounds~\cite{drukier:2018pdy}. As a chloride, it is soluble in water~\cite{MindatLaurionite}, and it could precipitate from brines with sufficiently high concentrations of Pb and Cl, and with pH between about 6.4 and 10~\cite{regenspurg_formation_2016}. 

\subsection{Ancient, fast dark matter from the LMC}
Conventionally, most analyses assume the Standard Halo Model (SHM)~\cite{drukier:1986tm}, which is a Maxwell-Boltzmann distribution truncated at the Milky Way (MW) escape velocity. However, recent work~\cite{Besla:2019xbx, Smith-Orlik:2023kyl, Donaldson_2022} in galactic dynamics shows that the close pericenter approach of the Large Magellanic Cloud (LMC) to the MW $ \sim 50$~Myr ago~\cite{Patel_2020} in the opposite direction to solar motion may have resulted in an appreciable population of fast, unbound DM in the Earth's frame. The resulting DM velocity tail may have been faster $ \sim 50$~Myr ago than it is today~\cite{Smith-Orlik:2023kyl}.

Paleodetectors are uniquely sensitive to this history. The history of the velocity tail does not necessarily favor the oldest minerals, contrary to usual assumptions, but instead favors those that are just old enough to reach the period with the highest velocity tail. Since there is still significant uncertainty surrounding this astrophysical scenario, we will present our results both for this LMC model for the velocity distribution as well as for the SHM.

\begin{figure}[t]
   \centering
   \includegraphics[width=0.7\textwidth]{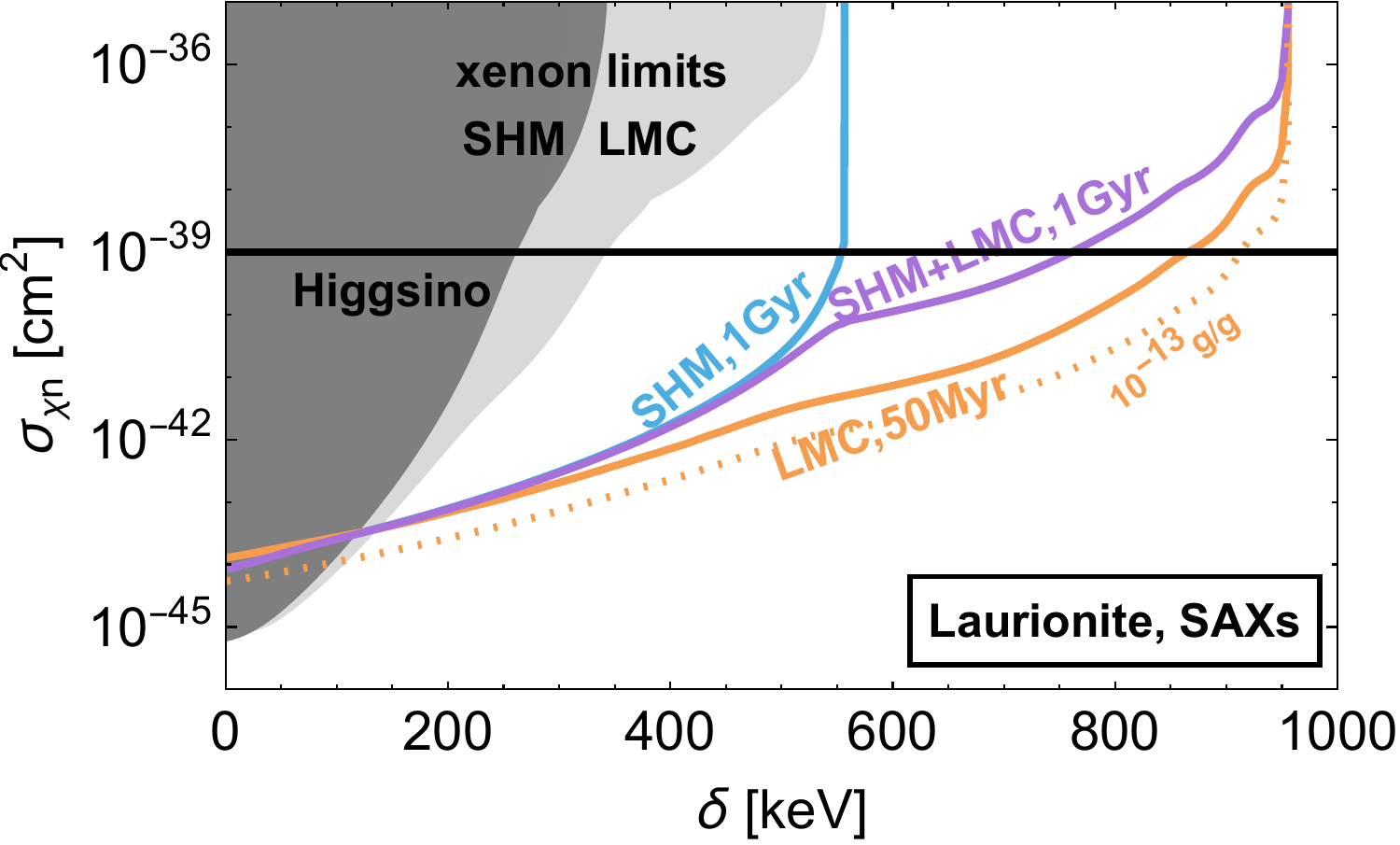}
   \caption{Projected limit on DM-nucleon cross section $\sigma_{\chi n}$ vs mass splitting $\delta$, for a $60~\text{cm}^3$ sample of Laurionite (PbClOH) with resolution $\sigma_x=15$~nm relevant for Small Angle X-ray scattering (SAXs). Results are shown for a Gyr-old sample assuming the Standard Halo Model (SHM) (blue); the SHM followed by the LMC model during the last 50~Myr (purple); and a 50~Myr-old sample in the LMC model (orange). The orange dotted line assumes a theoretically minimal $^{238}\text{U}$ concentration of $C^{238} = 10^{-13}~\text{g}/\text{g}$. The Higgsino cross section and existing limits~\cite{Graham:2024syw} from xenon experiments are also shown.}
   \label{fig:Higgsino_Projection}
\end{figure}

\subsection{Projected limits}
The projected limits are shown in Fig.~\ref{fig:Higgsino_Projection}, where we consider scanning a $60~\text{cm}^3$ sample of Laurionite with Small Angle X-ray scattering (SAXs)~\cite{Schaff:2015}. For the Higgsino benchmark, the reachable mass splitting can be read off by the intersection of a given curve with the horizontal black line denoting the Higgsino target cross section.

Assuming the SHM, a Gyr-old sample can probe $\delta \approx 550$~keV (blue curve). Assuming the LMC model, a 50~Myr-old sample reaches $\delta\approx 870$~keV (orange solid curve). And a Gyr-old mineral subject to the SHM for the first 950~Myr and additionally to the LMC for the last 50~Myr performs worse than its younger counterpart at $\delta \approx 770$~keV (purple curve), due to the extra accumulated backgrounds before the LMC event. We also make an ultimate LMC projection in which a BP sample is found with the theoretically minimal $^{238}\text{U}$ concentration, $C^{238}_{\rm BP,min}=10^{-13}~\text{g}/\text{g}$ (whereas all other scenarios assume the experimentally measured value $C^{238}_{\rm BP}=3\times 10^{-12}~\text{g}/\text{g}$). In this case, we find an ultimate reach of $\delta \approx 920$~keV (orange dotted curve).

\subsection{Larger backgrounds: 2~km depth and radio-impurity}

\begin{figure}[t]
   \centering
   \includegraphics[width=0.7\textwidth]{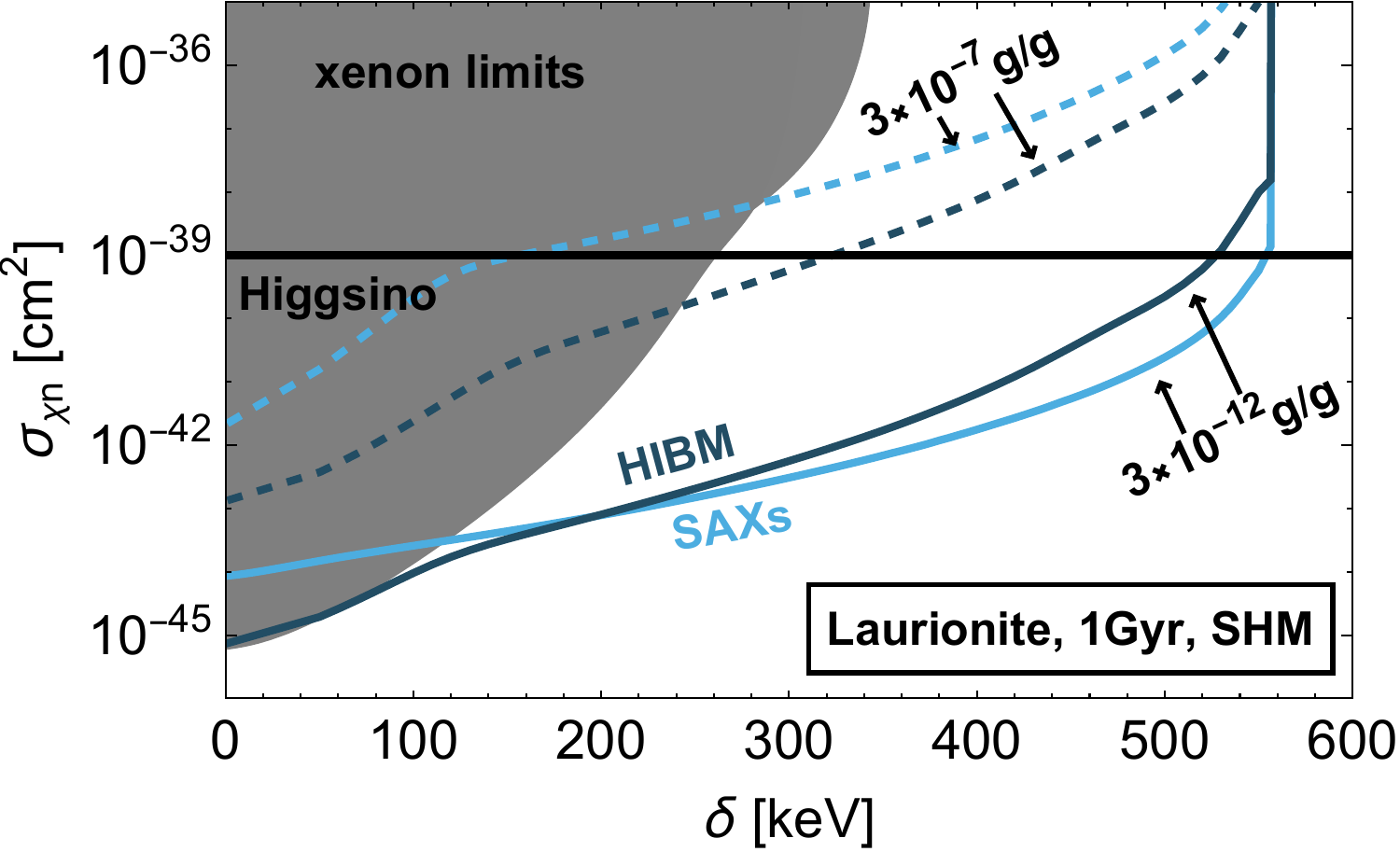}
   \caption{Same as Fig.~\ref{fig:Higgsino_Projection}, but we compare the read-out methods of Small Angle X-ray scattering (SAXs) ($60~\text{cm}^3$ sample, $\sigma_x=15~\text{nm}$) and Helium Ion Beam Microscopy (HIBM) ($6~\text{mm}^3$ sample, $\sigma_x=1~\text{nm}$). Solid curves assume benchmark uranium concentration $C^{238}=3\times 10^{-12}~\text{g}/\text{g}$. Dashed curves show interesting reach even for suboptimal samples with $C^{238}=3\times 10^{-7}~\text{g}/\text{g}$. }
   \label{fig:Higgsino_Projection_Large_BG}
\end{figure}

In Fig.~\ref{fig:Higgsino_Projection_Large_BG}, we compare the reach of the SAXs and the Helium Ion Beam Microscopy (HIBM)~\cite{Hill:2012} read-out methods, accounting for their respective processable sample volumes and resolutions. Excitingly, we find that we can probe unexplored Higgsino parameter space even with a much larger $^{238}\text{U}$ background with the HIBM method. For a 1~Gyr sample, this can be achieved even if the uranium background increases by 5 orders of magnitude to $C^{238}=3\times 10^{-7}~\text{g}/\text{g}$ (dashed curves) relative to the BP benchmark $C^{238}_{\rm BP}=3\times 10^{-12}~\text{g}/\text{g}$. This uranium concentration is comparable to that of a typical mineral in the Earth's crust~\cite{drukier:2018pdy}. We have thus opened up the possibility of using radio-impure minerals as paleodetectors.

Furthermore, being able to tolerate such a large uranium background implies that going to 5~km in depth is no longer necessary. Recall that at 5~km depth, the cosmogenic neutron flux becomes negligible compared to radiogenic neutrons at the level of $C^{238}=10^{-10}~\text{g}/\text{g}$. Therefore, minerals from the deepest existing laboratories at $\sim 2$~km depth are not viable for conventional paleodetectors, since the cosmogenic neutron flux there is $10^4$ times larger than that at 5 km depth~\cite{baum:2018tfw}. Conversely, if we can tolerate $10^4$ times larger radioactivity and still probe new Higgsino parameter space, then existing laboratories at depths of 2~km become viable, opening up vastly more accessible sample sources. 

This large background tolerance is made possible by the large Z-mediated cross section $\simeq 10^{-39}~\text{cm}^2$. In contrast, radiopure samples from $5~\text{km}$ depth are required both for probing the largest mass splittings in inelastic DM and for conventional paleodetector searches. Thus, inelastic DM could serve as an intermediate physics goal for the paleodetector paradigm.

\acknowledgments
We thank Masha Baryakhtar, Emilie LaVoie-Ingram, Patrick Stengel, and Xiuyuan Zhang for helpful discussions. S.~W. is supported in part by the U.S. Department of Energy Office of Science under Award Number DE-SC0024375, and by the Gordon and Betty Moore Foundation through Grant GBMF13898 to the University of 
Washington.
This work was supported in part by NSF Grant No. PHY-2310429, NSF Grant No. PHY-2515007, Simons Investigator Award No. 824870, The University of Delaware Research Foundation and the John Templeton Foundation Award No. 63595.
\clearpage

\section{(Minicharged) magnetic monopole with mineral detectors}
\label{sec:Kamada}
Authors: {\it Ayuki Kamada}
\vspace{0.1cm} \\
University of Warsaw
\vspace{0.3cm}

\subsection{Introduction}

Thanks to a long exposure, mineral detectors are sensitive to even very rare (low flux) relics of the Universe, which do not necessarily account for the whole of dark matter.
Magnetic monopole is one of such well-motivated relics, as well as topological defect~\cite{yin:2025wuv} and Q-ball~\cite{kamada:2025mji}.
Monopole is introduced to make Maxwell-equations symmetric under dual transformation: $\vec{E} \to \vec{B}$, $\vec{B} \to - \vec{E}$, $\rho_e \to \rho_m$, $\vec{J}_e \to \vec{J}_m$, $\rho_m \to - \rho_e$, $\vec{J}_m \to - \vec{J}_e$.
The lightest monopole is stable as the lightest charged particle.
Monopole can appear in well-motivated high-energy theories as a topological defect, when a simple group is spontaneously broken into its subgroup including U(1).

\subsection{Monopole interaction}
Classical interaction of monopole is described by classical electrodynamics: monopole generates classical electromagnetic field, which in turn exerts force on charged particles (and vice versa).
On the other hand, quantum interaction is rather uncertain.
Single-valuedness of electron wave function leads to Dirac quantization of magnetic charge in units of $g_{\rm D} = 68.5 e$.
This large charge makes unreliable usual Feynman-diagrammatic computation in quantum electrodynamics (QED).
More specifically, excitation and ionization of atoms in the presence of magnetic field generated by monopole can be computed reliably (Bethe-Bloch formula for heavy ion with the replacement of $Z e \to g v / c$)~\cite{Derkaoui:1998uv}.
But QED processes such as photonuclear interaction, pair production and (though subdominant for a large mass) bremsstrahlung may need special care, especially when a process involves large-virtuality photons~\cite{Wick:2000yc}.

\subsection{Relic monopoles}
Monopoles are accelerated by Galactic and inter-galactic magnetic fields in the late/present Universe and thus have a large energy, depending on its mass~\cite{Perri:2025qpg}.
Semi-/ultra-relativistic monopoles do not cluster in our Galaxy and thus one needs to estimate the flux based on global (cosmological) mass density
\begin{eqnarray}
F_{\rm mon} = 6 \times 10^{-13}  {\rm /cm^2 / s / sr} \times \left( \frac{\rho_{\rm mon}}{\rho_{\rm dm, glo} = 2 \times 10^{-6}  {\rm GeV} / c^2 {\rm / cm^3}} \right) \left( \frac{v}{c} \right) \left( \frac{10^{16}  {\rm GeV}}{m_{\rm mon}} \right)
\end {eqnarray}
rather than local (around the solar system) mass density
\begin{eqnarray}
F = 6 \times 10^{-13}  {\rm /cm^2 / s / sr} \times \left( \frac{\rho}{\rho_{\rm dm, loc} = 0.3 \, {\rm GeV} / c^2 {\rm / cm^3}} \right) \left( \frac{v}{233 \, {\rm km/s}} \right) \left( \frac{10^{16} {\rm GeV}}{m} \right) \,.
\end{eqnarray}
To my best knowledge, in a series of old studies using mica, only \cite{Fleischer:1969mki} considers semi-/ultra-relativistic monopole.
They focus on double or more charged monopole, since single-charged monopole does not leave etchable tracks.

\subsection{Minicharged monopoles}

Once we introduce another (dark) electrodynamics in addition to our (visible) electrodynamics, visible magnetic charge does not need to be a multiple of $g_{\rm D}$.
Dark electrodynamics appears in a particle-physics model called dark sector: one example is mirror world, which has a copy of standard-model particles and dynamics; it is introduced to restore spatial parity symmetry, which is broken in weak interaction.
Dark photon may mix with visible photon (mixing parameter $\epsilon$), meaning that electron feels dark electromagnetic field in addition to visible electromagnetic field and dark monopole generates visible magnetic field in addition to dark magnetic field.
On the other hand, as it is, this does not lead to a force between electron and dark monopole, since the forces mediated by visible photon and by dark photon cancel with each other. (This is why mixing does not violate single-valuedness of electron wave function.)
The situation changes once dark electrodynamics is spontaneously broken and dark magnetic flux is confined into cosmic string.
Then, electron located far away from the string feels only visible electromagnetic field generated by dark monopole, namely, dark monopole looks like visible monopole whose magnetic charge is $\epsilon$.

\subsection{Relic network}

In this setup, dark monopole and anti-monopole are connected by a cosmic string.
Evolution of this system depends on dark charge of dark monopole $N$ in units of that of cosmic string~\cite{Kibble:2015twa}.
When $N=1$ (``dumbbell''), dark monopole and anti-monopole attract each other with a linear potential and thus annihilate very quickly.
When $N=2$ (``necklace''), dark monopole is connected and attracted to two anti-monopoles and thus do not annihilate very quickly.
On the other hand, evolution leaves only one cosmic string per Hubble volume.
Dark monopoles do not exist everywhere, but only along the string.
Encounter between dark monopoles and the Earth (or mineral detectors) is unlikely.
When $N\geq3$ (``network''), even after evolution, dark monopoles exist every where.
This is the case of our interest.

One caveat on this network is energy density of strings.
Since the distance between strings is comparable with that between dark monopoles, energy density of strings is related to that of dark monopoles:
\begin{eqnarray}
\rho_{\rm str} \sim 0.1 \rho_{\rm dm, glo} \times \left( \frac{\mu}{\rm MeV^2} \right) \left( \frac{\rho_{\rm mon}}{\rho_{\rm dm, glo}} \right)^{2/3} \left( \frac{v}{c} \right) \left( \frac{10^{16}  {\rm GeV}}{m_{\rm mon}} \right)^{2/3}
\end {eqnarray}
where $\mu$ is a string tension.
Not to spoil success of $\Lambda$CDM model in large-scale structure of the Universe, string tension and thus a dark photon mass needs to be around or below $100 \, {\rm MeV}$.
Such a dark photon with mixing parameter $\epsilon = 10^{- 8} \text{--} 10^{- 6}$ is still allowed by collider and beam-dump experiments~\cite{Alekhin:2015byh}.

\acknowledgments

A. K. thanks Daniele Perri for the collaboration, and Shigenobu Hirose for encouraging us to work on this study.

\clearpage

\newcommand{\sub}{\rm{sub}}
\newcommand{\rmmin}{\rm{min}}
\newcommand{\rmmax}{\rm{max}}
\newcommand{\obs}{\rm{obs}}
\newcommand{\NFW}{\rm{NFW}}
\newcommand{\solar}{\rm{solar}}
\newcommand{\ddedect}{d_{\rm{detect}}}

\section{Searching for the truly ``Dark" subhalos with Paleo-detectors}\label{sec:MIT_XZ}

Authors: Xiuyuan Zhang$^{1}$, Lina Necib$^{1}$, Denis Erkal$^{2}$

\noindent Affiliations: 

\noindent $^{1}$MIT Kavli Institute, Cambridge, MA, USA \\
$^{2}$University of Surrey, Guildford, UK

\subsection{Introduction}

One of the distinctive predictions of $\Lambda$CDM is the abundant population of small-scale dark matter (DM) substructures, or subhalos, within larger host halos~\cite{10.1093/mnras/183.3.341, 2008Natur.454..735D, 2008MNRAS.391.1685S, Klypin_2011}. Depending on the cutoff in the matter power spectrum, the smallest halos are expected to contain little or no visible matter, making them effectively inaccessible to electromagnetic observations. Nevertheless, these dark subhalos provide important probes of the nature of DM and of small-scale structure formation, as their properties are closely tied to the primordial power spectrum~\cite{Zentner_2003}. In particular, their internal density profiles inform the mass--concentration relation, $c(M)$~\cite{Correa_2015}, while their abundance tests the low-mass end of the subhalo mass function~\cite{2008MNRAS.391.1685S}. Together, these observables can help distinguish between different DM models.

Paleo-detectors have been proposed as an alternative approach to direct detection, using ancient minerals to record nuclear recoil tracks over geological timescales~\cite{PhysRev.133.A1443, doi:10.1126/science.149.3682.383, annurev:/content/journals/10.1146/annurev.ns.15.120165.000245, GUO2012233, Bramante_2022}. Time series of paleo-detectors can also probe time-dependent DM substructure, including dark disks and subhalos with masses of $10^4$--$10^8\,M_{\odot}$~\cite{baum:2021chx}.

In this study\footnote{A more detailed discussion can be found in Ref.~\cite{zhang2025darknesscrustsearchingtruly}.}, we build on this time-series paleo-detector framework to explore the detectability of subhalos over a wider mass range, focusing on the dependence of the expected sensitivity on the mass--concentration relation and DM model parameters.

\subsection{Subhalo Encounter Rate}

To estimate the impact of low-mass dark subhalos on DM direct-detection experiments, we first characterize the expected encounter rate between such subhalos and the Earth.

We define an effective cross section, $\sigma(v_0,M_{\rm sub})$, within which a subhalo of mass $M_{\rm sub}$ and initial velocity $v_0$ can intersect the Earth and produce an observable effect in direct-detection experiments. For a subhalo population with velocity distribution $f(v_0)$ and number density $n$, the differential incoming rate for subhalos with velocities in the interval $[v_0,v_0+dv_0]$ is given by
$f(v_0)\sigma(v_0,M_{\rm sub})v_0 n\,dv_0$. The differential encounter rate per unit subhalo mass is therefore
\begin{equation}\label{eq6}
    \frac{dR_{\rm sub}}{dM_{\rm sub}}
    =
    \int \frac{dn_{\rm sub}}{dM_{\rm sub}}\,
    f(v_0)\,v_0\,\sigma(v_0,M_{\rm sub})\,dv_0 ,
\end{equation}
where $dn_{\rm sub}/dM_{\rm sub}$ denotes the local differential number density of dark subhalos.

We assume that the velocity distribution of dark subhalos follows the Standard Halo Model~\cite{LEWIN199687, RevModPhys.85.1561}. For the local differential number density, $dn_{\rm sub}/dM_{\rm sub}$, we adopt the parametrization given in Ref.~\cite{10.1093/mnras/stw1957}, which was shown to be broadly consistent with the local subhalo abundance inferred from the public Via Lactea II simulation catalogs~\cite{2008Natur.454..735D} for subhalos within $50~{\rm kpc}$ of the Galactic center.

Integrating over the mass range $10^{-6}M_{\odot}$--$10^{5}M_{\odot}$, we obtain a total encounter rate
\begin{equation} \label{eq:rate}
\int_{10^{-6}M_{\odot}}^{10^5 M_{\odot}}
\frac{dR_{\rm sub}}{dM_{\rm sub}}\,dM_{\rm sub}
\sim 10^{-8}~{\rm yr}^{-1}.
\end{equation}

\subsection{Results} \label{sec:experimental studies}

Although the yearly encounter rate with low-mass subhalos is only $\sim 10^{-8}~{\rm yr}^{-1}$, such rare events may be probed with the long integration times offered by paleo-detectors~\cite{PhysRev.133.A1443, doi:10.1126/science.149.3682.383, annurev:/content/journals/10.1146/annurev.ns.15.120165.000245, GUO2012233}. If a subhalo encounter occurred during the exposure history of a mineral sample, the resulting time-dependent enhancement in the recoil track density could be used to search for the flyby and constrain the subhalo properties.

\begin{figure}[t]
\begin{center}
  \includegraphics[width=0.95\linewidth]{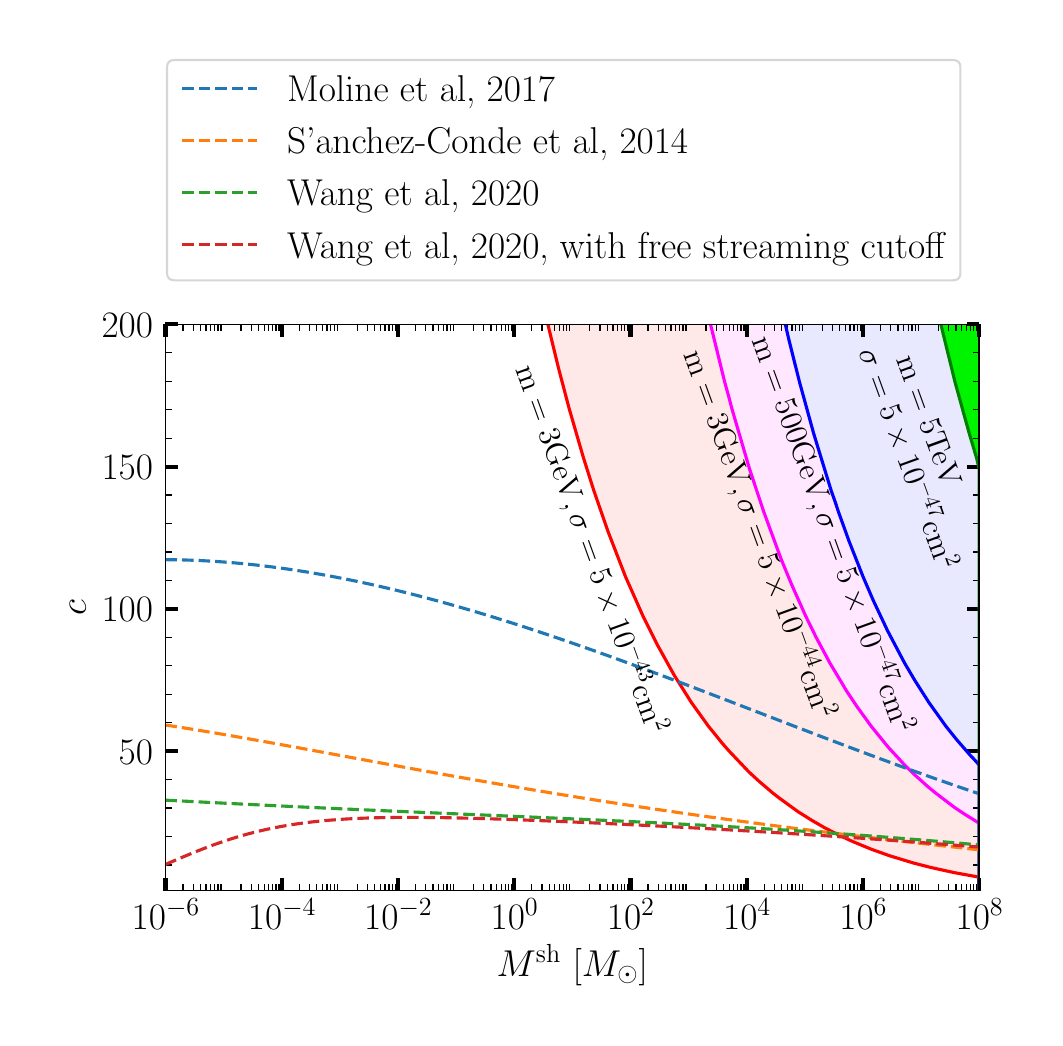}
 \caption{Projected constraints on the subhalo mass--concentration plane for representative DM models. Dashed curves show analytic mass--concentration relations from previous studies, while shaded regions indicate the regions accessible for different DM masses and cross sections. The accessible region grows for lighter DM masses and larger cross sections, suggesting that light DM models are better suited for testing existing analytic $c(M)$ predictions.}
  \label{fig:CM}
 \end{center}
 \end{figure}

In Fig.~\ref{fig:CM}, we show the projected sensitivity to a subhalo encounter in the mass--concentration plane. Dashed curves indicate analytic mass--concentration relations from previous studies, while the shaded regions correspond to four representative choices of the DM mass and scattering cross section. Three of these benchmark points remain allowed by current constraints, whereas the case $m_\chi=500~{\rm GeV}$ and $\sigma=5\times10^{-47}~{\rm cm}^2$ is already excluded by the latest LZ result~\citep{mount2017luxzeplinlztechnicaldesign, aalbers2024darkmattersearchresults}. We include this excluded benchmark only to illustrate how the sensitivity region changes with DM mass and cross section. Overall, lighter DM candidates with larger scattering cross sections yield broader sensitivity regions and can probe a meaningful portion of existing analytic $c(M)$ relations.

\begin{figure}[t]
\begin{center}
  \includegraphics[width=0.95\linewidth]{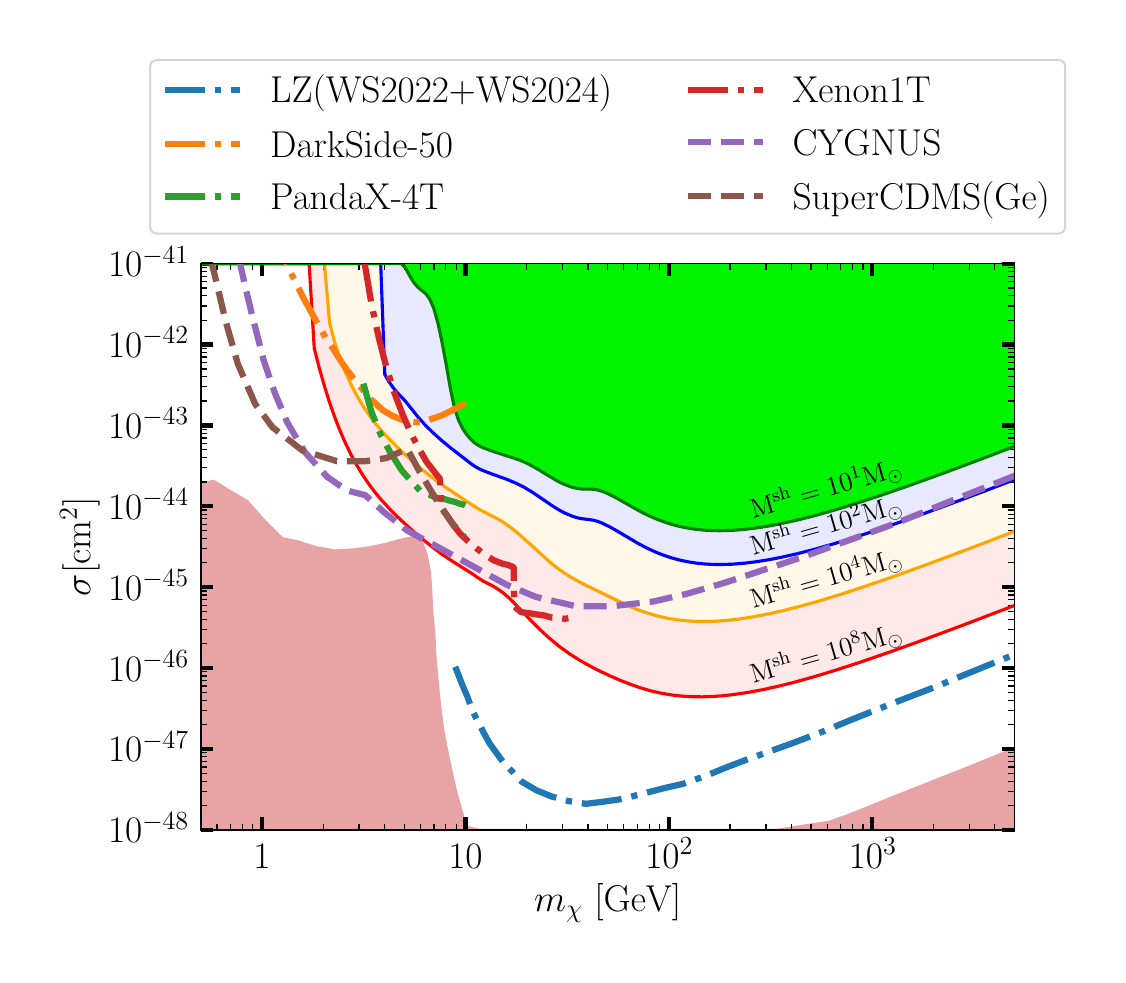}
 \caption{DM mass and cross section required to detect subhalo flyby events for different assumed subhalo masses. The latest LZ bound~\citep{mount2017luxzeplinlztechnicaldesign, aalbers2024darkmattersearchresults} is shown as a dash-dotted line, projected sensitivities for CYGNUS~\citep{vahsen2020cygnusfeasibilitynuclearrecoil} and SuperCDMS(Ge)~\citep{agnese:2016cpb} are shown as dashed curves, and the neutrino fog~\cite{Monroe_2007, Strigari_2009} is marked by the dark red region.}
  \label{fig:MS}
 \end{center}
 \end{figure}

To examine how the detectable subhalo mass depends on the DM mass and cross section, we next fix the mass--concentration relation to the form proposed by Moline et al.~\cite{Molin__2017} and invert the result of Fig.~\ref{fig:CM}. The corresponding sensitivity regions are shown in Fig.~\ref{fig:MS} for subhalo masses of $10^8M_{\odot}$, $10^4M_{\odot}$, $10^2M_{\odot}$, and $10M_{\odot}$, shown in red, yellow, blue, and green, respectively. The latest LZ constraint~\citep{mount2017luxzeplinlztechnicaldesign, aalbers2024darkmattersearchresults} is shown as a dash-dotted line, while projected sensitivities for CYGNUS~\citep{vahsen2020cygnusfeasibilitynuclearrecoil} and SuperCDMS(Ge)~\citep{agnese:2016cpb} are shown as dashed curves. The neutrino fog~\cite{Monroe_2007, Strigari_2009} is indicated by the dark red region.

For the adopted $c(M)$ relation and subhalo mass range, Fig.~\ref{fig:MS} shows that detectable flyby events are primarily associated with light DM candidates, $m_\chi \lesssim 10~{\rm GeV}$. At larger DM masses, the required parameter space is largely excluded by current LZ constraints~\citep{mount2017luxzeplinlztechnicaldesign, aalbers2024darkmattersearchresults}. In the range $2~{\rm GeV}\lesssim m_\chi \lesssim 10~{\rm GeV}$, paleo-detector searches for subhalo flybys could probe parameter space beyond current bounds and complementary to future experiments such as CYGNUS~\citep{vahsen2020cygnusfeasibilitynuclearrecoil} and SuperCDMS(Ge)~\citep{agnese:2016cpb}. More concentrated substructures, such as axion miniclusters or dark compact objects, could potentially extend the detectable region beyond the LZ bound, although their distinct mass functions would require a separate analysis. For spin-dependent interactions, the sensitivity is expected to be qualitatively similar but more strongly dependent on the chemical composition of the target mineral~\cite{drukier:2018pdy}. In particular, Ref.~\cite{drukier:2018pdy} noted that suitable target materials may be difficult to identify because the relevant isotopes require an unpaired neutron and are rare in nature.

In this work, we estimated the encounter rate of low-mass subhalos with Earth and showed how a flyby event recorded in Paleo-detectors could constrain subhalo properties, including the mass-concentration relation. For $\Lambda$CDM, we find an encounter rate of $\sim 10^{-8}{\rm yr}^{-1}$ for conventional direct detection experiments over the mass range $10^{-6}M_{\odot}$-$10^5M_{\odot}$. Given a sufficiently massive flyby, Paleo-detectors could probe subhalos from $\sim 10^5M_{\odot}$ down to $\sim 100M_{\odot}$ for light DM with $m_\chi\simeq 3{\rm GeV}$ and $\sigma_{\chi N}\simeq 5\times 10^{-43}{\rm cm}^2$, with the reach depending on the assumed concentration model. Future work should refine the treatment of tidal disruption, redshift evolution, track formation, and alternative DM scenarios such as atomic DM or strongly interacting subcomponents. Overall, Paleo-detectors provide a complementary probe of both DM particle properties and Milky Way substructure over geological timescales.

\acknowledgments

We would like to thank  He Feng, Marianne Moore, Sandip Roy, and Tracy Slayter for helpful conversations.
XZ is partially supported by DOE award DE-SC0024112.
LN is supported by the Sloan Fellowship, the NSF CAREER award 2337864, NSF award 2307788, and by
the NSF award PHY2019786 (The NSF AI Institute
for Artificial Intelligence and Fundamental Interactions,
\url{http://iaifi.org/}). DE acknowledges the support of the Australian Research Council through project number DP220102254.
\clearpage

%
\section{Cosmic Wall Scattering and Recoil Spectra in Paleo Detectors}
\label{sec:CosmicWalls}

Authors: {\it Wen~Yin$^{1}$} \\
\vspace{0.1cm}
$^1$Tokyo~Metropolitan~University,~Department~of~Physics
\vspace{0.3cm}

\subsection{Direct Detection of Cosmic Walls Using Ancient Minerals}

Paleo detectors, namely ancient minerals that retain fossilized damage tracks from nuclear recoils, provide a unique time-integrated probe of rare cosmological events.\footnote{This proceeding is based on Ref.\,\cite{yin:2025wuv}.}
In this contribution, I discuss the direct detection of cosmic walls, including relativistic bubble walls from late-time first-order phase transitions and domain-wall-like configurations, through wall-induced nuclear recoils.

Cosmic walls arise in many extensions of the Standard Model. They have been discussed in connection with cosmic birefringence~\cite{Takahashi:2020tqv}, gravitational waves~\cite{Kitajima:2023cek}, and possible time variations of fundamental constants that may be relevant for cosmological tensions~\cite{Sekiguchi:2020teg}. 
Recently, it was shown that when an axion-like particle (ALP) starts to oscillate due to a potential induced by hidden Yang--Mills dynamics, a phase-transition-like phenomenon can be triggered~\cite{Sugeno:2025kwx}. 
Alternatively, if a nonlinear ALP-like fluid undergoes a nonlinear transition, an enormous number of domain walls can be produced~\cite{Narita:2025jeg,Miyazaki:2025tvq}. Such late-time walls are also natural targets of this proposal. 
Late-time walls are usually constrained indirectly through their gravitational effects on the cosmic microwave background~\cite{Zeldovich:1974uw}. By contrast, direct detection of a wall-crossing event is much less explored, although related ideas have been considered for non-scaling domain walls, for example in Ref.\,\cite{GNOME:2023rpz}.

If a cosmic wall passed through the Earth during the geological lifetime of a mineral sample, it could have induced correlated nuclear recoils. Such recoils would leave damage tracks in minerals such as muscovite, olivine, or apatite. Since a wall-crossing event is expected to occur at most $\mathcal{O}(0.1-1)$ times over the relevant cosmological history, passive detectors with very long exposure times are essential. Paleo detectors are therefore particularly well suited for this purpose.

\subsection{Wall-Induced Transition Probability}

I consider a scalar field $\phi$ associated with the wall and its effective coupling to a nucleon or nucleus $N$. The wall background can induce the transition
\begin{equation}
  N \to N+\phi ,
\end{equation}
where the outgoing nucleus obtains a recoil momentum. The transition probability is computed in the wall background by treating the wall profile as a space-time dependent external field. 

The relevant observables are the final recoil momentum in the Earth frame, $p_f^{\rm Earth}$, and the recoil angle, $\theta_{\rm Earth}$, measured with respect to the wall velocity. I therefore evaluate the differential transition probability
\begin{equation}
  \partial_{\log_{10} p_f^{\rm Earth}}
  \partial_{\sin\theta_{\rm Earth}}
  P_{N\to N\phi}.
\end{equation}
This quantity determines both the recoil momentum spectrum and the angular distribution of the damage tracks.

\begin{figure}[htbp]
  \centering

  \begin{subfigure}[b]{0.30\linewidth}
    \centering
    \includegraphics[width=\linewidth]{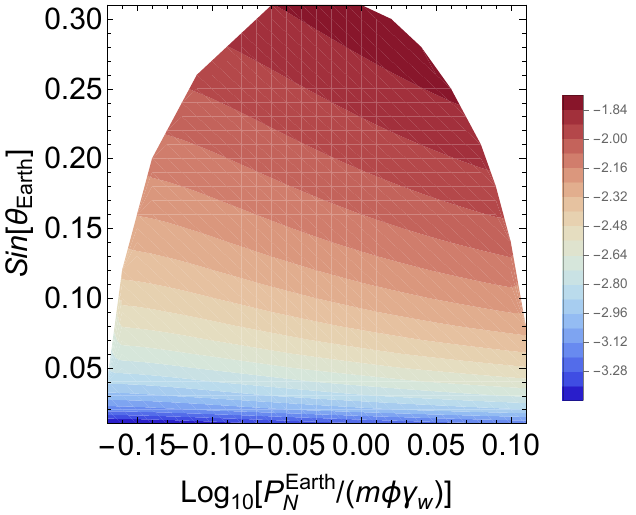}
    \caption{$\gamma_w=1.5$}
    \label{fig:bubble-mphi-1em2-p0-1p5}
  \end{subfigure}
  \hfill
  \begin{subfigure}[b]{0.29\linewidth}
    \centering
    \includegraphics[width=\linewidth]{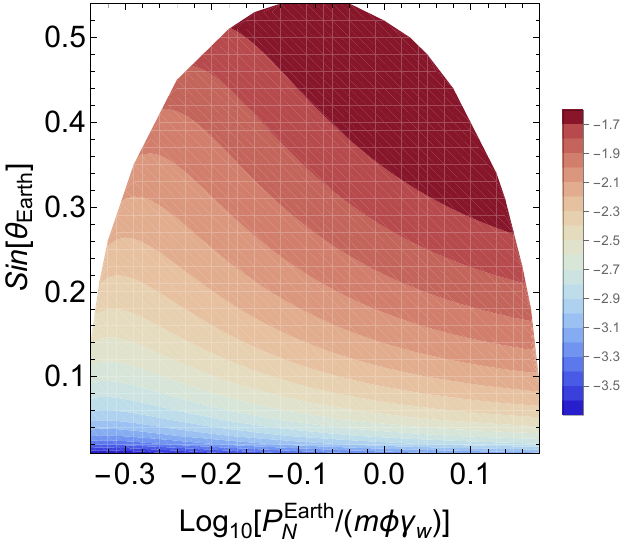}
    \caption{$\gamma_w=1.7$}
    \label{fig:bubble-mphi-1em2-p0-1p7}
  \end{subfigure}
  \hfill
  \begin{subfigure}[b]{0.30\linewidth}
    \centering
    \includegraphics[width=\linewidth]{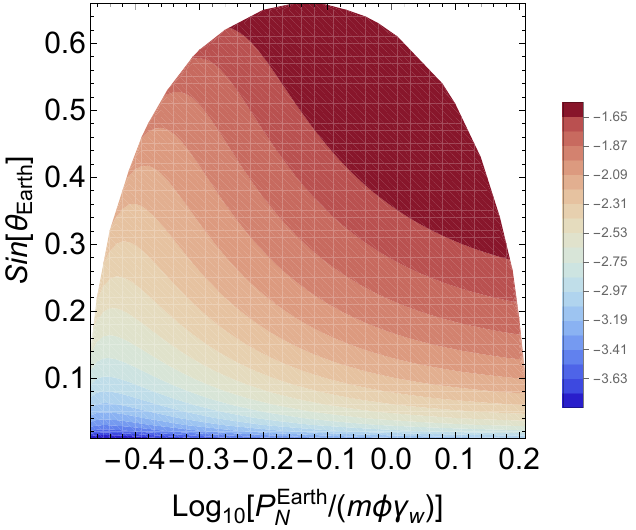}
    \caption{$\gamma_w=1.9$}
    \label{fig:bubble-mphi-1em2-p0-1p9}
  \end{subfigure}

  \vspace{2mm}

  \begin{subfigure}[b]{0.30\linewidth}
    \centering
    \includegraphics[width=\linewidth]{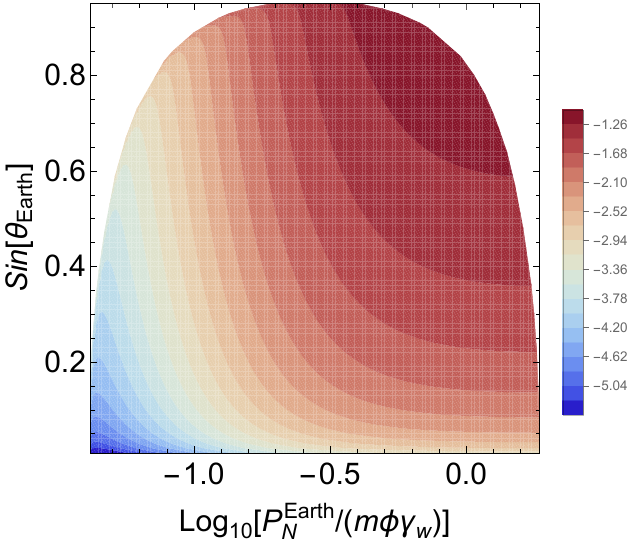}
    \caption{$\gamma_w=5$}
    \label{fig:bubble-mphi-1em2-p0-5}
  \end{subfigure}
  \hfill
  \begin{subfigure}[b]{0.30\linewidth}
    \centering
    \includegraphics[width=\linewidth]{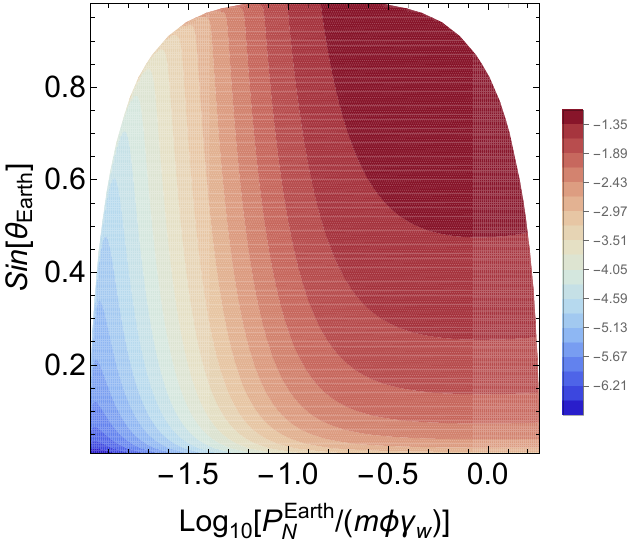}
    \caption{$\gamma_w=10$}
    \label{fig:bubble-mphi-1em2-p0-10}
  \end{subfigure}
  \hfill
  \begin{subfigure}[b]{0.30\linewidth}
    \centering
    \includegraphics[width=\linewidth]{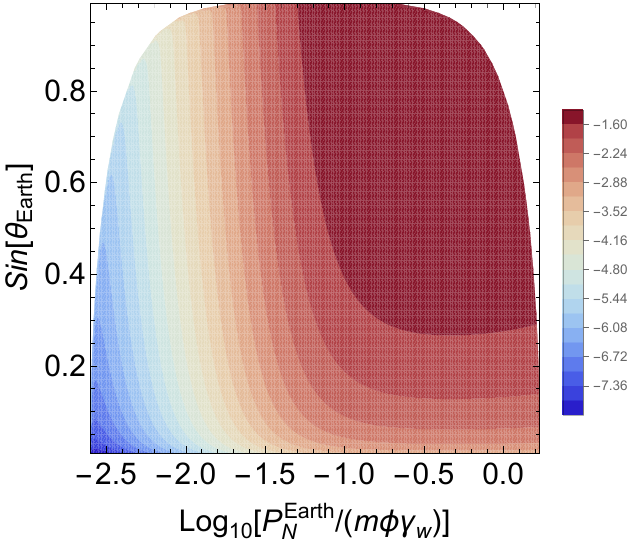}
    \caption{$\gamma_w=20$}
    \label{fig:bubble-mphi-1em2-p0-20}
  \end{subfigure}

  \caption{
  Contour plots of
  $\log_{10}\!\left[
 \frac{\partial^2 P_{N\to N\phi}}
     {\partial \log_{10} p_f^{\rm Earth}\,
      \partial \sin\theta_{\rm Earth}}
  \right]$
  in the bubble-wall scenario for $m_{\phi}=10^{-2}m_N$.
  Representative wall Lorentz factors are shown.
  }
  \label{fig:all-p0-0.01}
\end{figure}

Figures \ref{fig:all-p0-0.01} show the differential transition probability in the bubble-wall scenario for $10^{-2}m_N$, respectively. The displayed values of the wall Lorentz factor range from the mildly relativistic regime to $\gamma_w=20$. The horizontal and vertical axes correspond to $\log_{10}p_f^{\rm Earth}$ and $\sin\theta_{\rm Earth}$. For most of the parameter region, $\theta_{\rm Earth}\lesssim45^\circ$, which implies that the recoil tracks tend to be nearly aligned with the direction of the wall motion. This directional correlation is a characteristic signal of a wall-crossing event.

The main observable in paleo detectors is a set of damage tracks. The wall interpretation predicts two important features. First, the tracks should be globally correlated, since a cosmic wall crossing is a coherent event on the scale of the Earth. Second, the recoil directions should exhibit anisotropy, whose pattern is related to the properties of the wall, as illustrated in Figures~\ref{fig:all-p0-0.01}. 
Future paleo-detector searches with larger mineral exposures, better spatial resolution, and multiple geographically separated samples would substantially improve the sensitivity by incorporating the recoil spectra obtained in this work.

\subsection*{Acknowledgements}
This work is supported by JSPS KAKENHI Grant Nos. 22K14029 (W.Y.), 23K22486 (W.Y.), and  26K00695 (W.Y.). W.Y. is also supported by Selective Research Fund and Incentive Research Fund from Tokyo Metropolitan University. 
\clearpage

%

\section{Directional Dark Matter Detection and Beyond: Diamond as a Platform for Mineral Detectors}
\label{sec:UMD}
{\it Daniel Ang, Chinmay Bharathulwar, Priyanshu Bhattacharya, Mason Camp, Anson Cook, Gavishta Liyanage, Maximilian Shen, Jiashen Tang, Ronald Walsworth}
\\Quantum Technology Center, University of Maryland
\vspace{0.3cm}\\

\subsection{Introduction}

Diamond is a monoatomic, wide-bandgap crystal that is both highly radiopure and 
hosts a versatile room-temperature quantum sensor in the nitrogen-vacancy (NV) center. These properties make it an attractive medium in which to detect and characterize nuclear-recoil damage tracks left by rare particle interactions. In particular, next-generation Weakly Interacting Massive Particle (WIMP) dark matter (DM) detectors will soon encounter the ``neutrino fog,'' where solar-neutrino backgrounds obscure a DM signal~\cite{ohareNewDefinitionNeutrino2021}. Directional detection can overcome this limit by exploiting the differing angular distributions of neutrinos and WIMPs, rejecting backgrounds through imaging of nuclear-recoil damage tracks paired with conventional diamond-based real-time event detection~\cite{rajendran_method_2017,marshall_directional_2021,ebadi_directional_2022}. Natural diamonds extend this reach into the past: with Gyr-scale exposure ages and intrinsic radiopurity, they are candidate paleodetectors in which optically stable color centers could record an integrated DM/neutrino dose over geological time. More broadly, diamond is an ideal pathfinder mineral for the emerging program of mineral-based detection~\cite{baum_mineral_2023,drukier:2018pdy}. Although less naturally abundant than other mineral candidates such as quartz or olivine, it uniquely combines high-purity synthetic and well-characterized natural samples, a deep knowledge base of defects, color centers, and computational modeling, and the capability to host a highly versatile and developed NV quantum sensors that aid the study of track formation and morphology. Together, these features allow the readout and simulation techniques the field requires to be developed and experimentally validated in a single, well-understood material before being applied to other minerals.

Building on our contributions to previous MD$\nu$DMs~\cite{Ang2024,Ang2025}, we report recent progress at the University of Maryland (UMD) Quantum Technology Center (QTC) on (i) multi-scale reconstruction of individual nuclear-recoil tracks, now reported in full in Ref.~\cite{ang_multiscale_2026}; (ii) a maturing simulation-and-inference toolchain for predicting observed damage track morphologies in mineral detection experiments; and (iii) a new natural-diamond paleodetection direction based on H3 centers.
\subsection{Multi-scale reconstruction of single-ion tracks}
\label{sec:multiscale}
We recently demonstrated multi-scale readout of individual sub-MeV nuclear-recoil tracks in nitrogen-rich diamond~\cite{ang_multiscale_2026}. Single \SI{800}{keV} carbon ions implanted at the Sandia National Laboratories microbeam facility produce spatially localized ensembles of NV centers after annealing, detected by confocal photoluminescence with a mean of \num{17.5(0.4)} NVs per ion across 222 sites. We then developed an experimentally calibrated forward model using SIIMPL~\cite{janson2003hydrogen} to model vacancy production during implantation followed by kinetic Monte Carlo modeling of the annealing process which produces nitrogen-vacancy (NV) centers~\cite{mitchell2023spparks}. We found that applying a correction for second-nearest-neighbor defect reactions (discovered from first principles calculation) brings the predicted yield from $\sim$24 into agreement with the measured value ($\sim$16.6). Crucially, head--tail directional information is partially retained through annealing and is recoverable by a simulation-based inference (SBI)~\cite{tejero2020sbi} machine-learning (ML) classifier with $\sim$95\% efficiency (correctly assigning head--tail direction for 95\% of tracks) at a false-positive rate below 5\%. The ML algorithm maintains similar directional reconstruction performance on simulated 100 keV tracks. Finally, spin-coherence measurements ($T_2\approx\SI{1.47}{\micro s}$) indicate that nanoscale, magnetic-gradient NV imaging of the resulting tracks is feasible~\cite{zhang_selective_2017,ebadi_directional_2022}. Full details are given in Ref.~\cite{ang_multiscale_2026}.

Alongside this single-ion work, we are advancing several complementary efforts to develop novel methods for imaging damage tracks in diamond, including a light-sheet quantum diamond microscope (LS-QDM) for rapid, widefield 3D NV-based imaging of the lattice strain imprinted by a recoil damage track~\cite{marshall_high-precision_2022,Ang2025}. With the first generation LS-QDM at the QTC, we measured a light-sheet waist of $\sim$\SI{6.5}{\micro m} in mm-scale diamond samples and demonstrated volumetric strain mapping in both isotopically purified ($>$99\% $^{12}$C) and natural-abundance (1.1\% $^{13}$C) synthetic diamonds. Work is underway to refine the performance of the device and analyze the resulting 3D images of growth-induced strain structures.

\subsection{Toward end-to-end event reconstruction in mineral detectors}
\label{sec:pipeline}

\begin{figure}[ht]
   \centering
   \includegraphics[width=\textwidth]{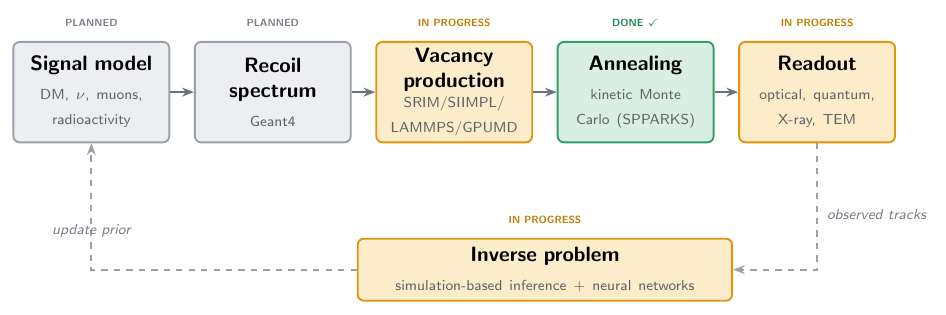}
   \caption{End-to-end simulation-and-inference pipeline for mineral track detection. A signal model and recoil spectrum (future work) feed vacancy production (binary-collision / molecular dynamics), defect annealing (kinetic Monte Carlo), and readout (optical, quantum, X-ray, TEM); the inverse problem is solved by simulation-based inference. Each stage can be applied to any candidate mineral.}
   \label{fig:pipeline}
\end{figure}

A central goal is an end-to-end simulation pipeline that maps a signal model (e.g., dark matter, solar/atmospheric neutrinos, muons, and radiogenic backgrounds) to the damage tracks produced in mineral detectors and subsequently observed by diverse readout methods (Fig.~\ref{fig:pipeline}). Such a forward simulation pipeline can then be utilized to address the inverse problem, which provides the ability to analyze and interpret experimental data. At UMD, work has focused on developing forward simulations for vacancy production (using software packages such as SRIM, SIIMPL, and GPUMD), annealing (SPPARKS), and developing an SBI-based, ML-driven framework to solve the inverse problem. Other stages of the pipeline will be developed in tandem with other members of the MDDM collaboration, which UMD recently joined.

\subsection{Forward modeling at scale with machine-learned molecular dynamics}
\begin{figure}[ht]
   \centering
   \includegraphics[width=1.0\textwidth]{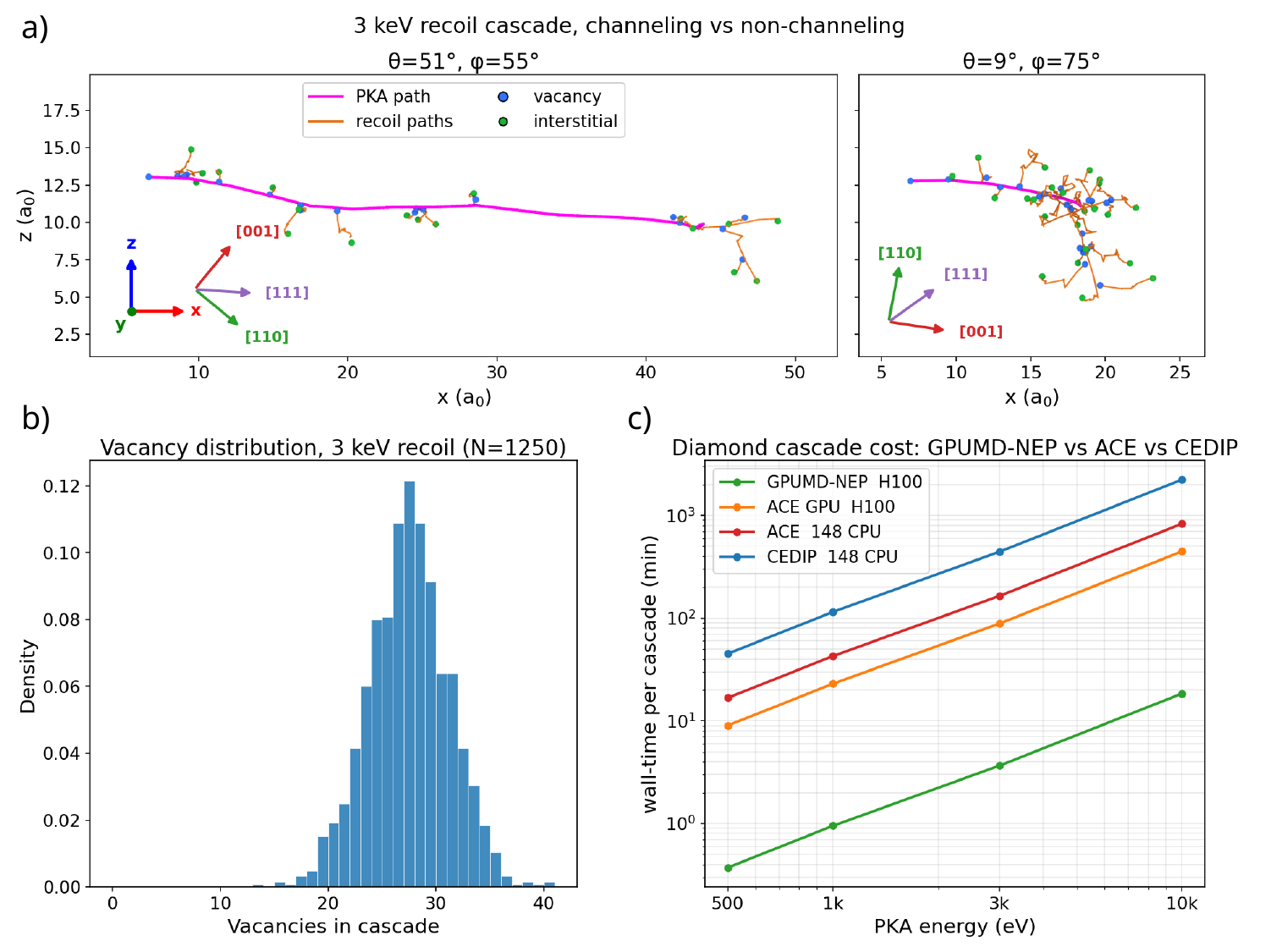}
   \caption{GPU-accelerated-MD (GPUMD) simulation of nuclear-recoil damage cascades in diamond. (a) Representative vacancy distribution for a 3~keV recoil, with the crystal axes indicated; typical behaviors of both channeling and non-channeling cascades are shown. (b) Distribution of the number of vacancies per cascade over N=1250 runs. (c) Wall-time calculation cost for 0.5--10 keV cascades, showing how GPUMD enables simulating cascades beyond the reach of conventional MD. }
   \label{fig:cascade}
\end{figure}

Fully atomistic molecular dynamics simulation tools (MD) capture channeling, recombination, phonon transport, and other damage cascade phenomena that typical binary-collision codes such as Stopping and Range of Ions in Matter (SRIM) cannot. However, conventional MD methods are prohibitively slow for high-energy cascade work in diamond, requiring thousands of CPU hours per $\sim$10 keV track with typical codes such as LAMMPS~\cite{lammps2022}. We have transitioned to machine-learning-driven MD using the GPUMD package~\cite{Fan2022GPUMD} with a neuroevolution potential (NEP)~\cite{Fan2021nep} trained on first-principles density functional theory (DFT) data~\cite{rowe_carbon_2020} and supplemented with electronic stopping~\cite{Haiek2022espnn} and close-range theory~\cite{nordlund_2025} to enable accurate modeling of cascade dynamics. The most recent version of the potential successfully reproduces DFT steady-state properties as benchmarked via elastic constants (Table~\ref{tab:elastic}). Test runs show a $\sim$100$\times$ speedup that should enable recoil cascades up to $\sim$\SI{40}{keV}. Figure~\ref{fig:cascade} shows some preliminary test cascades and the resulting vacancy statistics. This approach generalizes to any mineral, requiring only suitable DFT training data for less-studied materials.

\begin{table}[ht]
\centering
\begin{tabular}{ccc}
\hline
Constant & NEP (this work) & optB88-vdW DFT \\
\hline
$C_{11}$ (GPa)& 1039 & 1061 \\
$C_{12}$ (GPa)& 127  & 126  \\
$C_{44}$ (GPa)& 557  & 559  \\
\hline
\end{tabular}
\caption{Elastic constants of diamond: NEP potential trained at UMD vs. DFT-based values (computed with the optB88-vdW functional) from the JARVIS-DFT database~\cite{choudharyJointAutomatedRepository2020}.}
\label{tab:elastic}
\end{table}

\begin{figure}[htbp]
   \centering
   \includegraphics[width=0.7\textwidth]{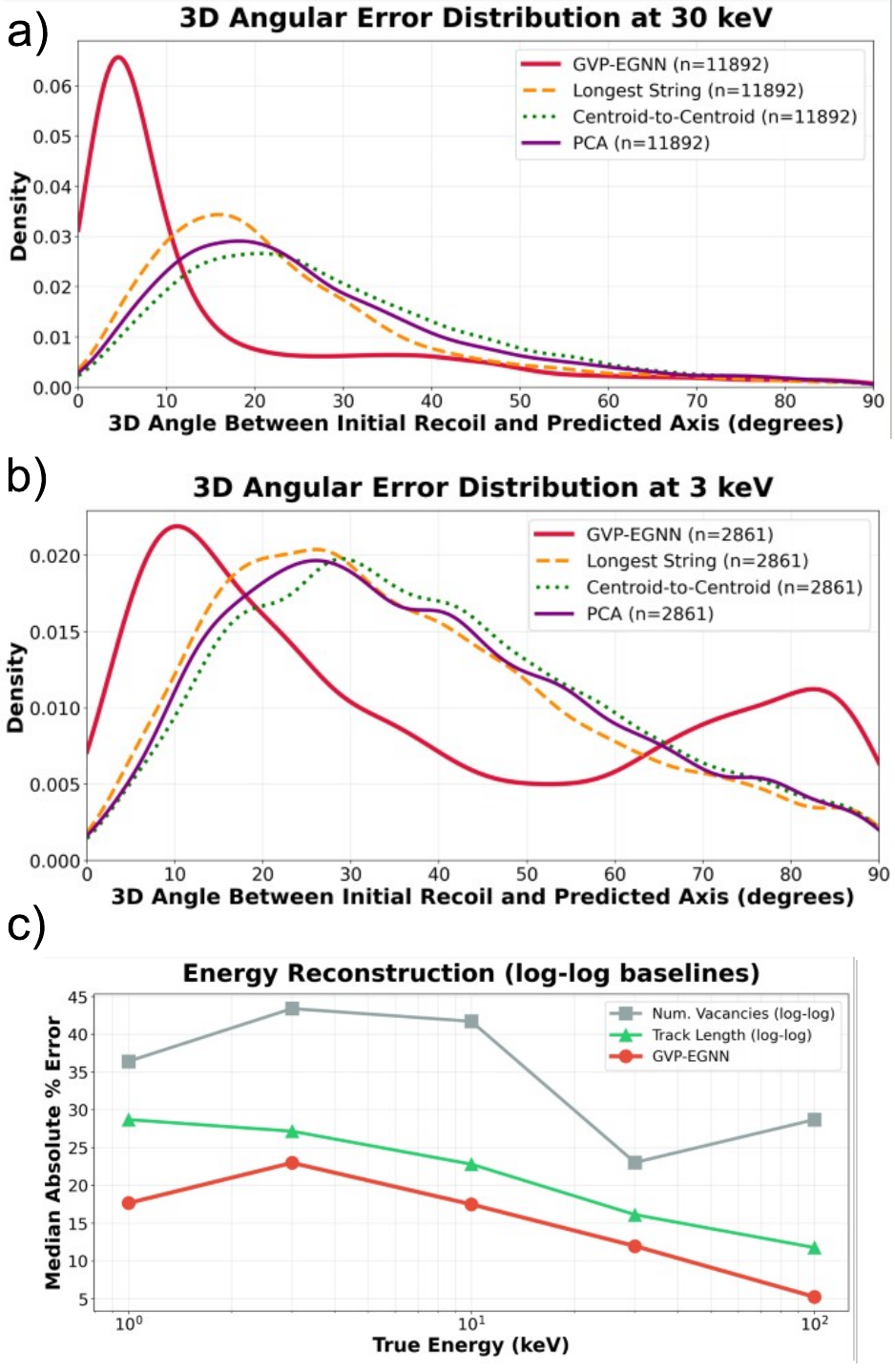}
   \caption{Simulation-based-inference reconstruction of recoil direction and energy from 3D vacancy distributions, using SRIM-generated tracks. (a, b) Distribution of the 3D angle between the true recoil axis and the predicted axis, for the GVP-EGNN model compared to three classical summary-feature baselines (longest string, centroid-to-centroid, principal component analysis), at (a) 30~keV and (b) 3~keV recoil energy. GVP-EGNN is sharply peaked at small angles in both cases, though at 3~keV it exhibits a secondary peak near $80^\circ$ that is still under investigation. (c) Median absolute percent error in energy reconstruction as a function of true recoil energy, comparing GVP-EGNN to two classical baselines (number of vacancies, track length). GVP-EGNN outperforms both across the full energy range, with the largest relative gains at low energy.}
   \label{fig:sbi}
\end{figure}

\subsection{Solving the inverse problem with simulation-based inference}
To address the inverse problem, we use simulation-based inference (SBI)~\cite{tejero2020sbi}, training a neural posterior estimator on $\sim$200k SRIM-generated 3D tracks. Whereas Ref.~\cite{ang_multiscale_2026} addressed only one-dimensional head--tail classification, our most recent model uses a geometric graph neural network (GVP-EGNN)~\cite{jing_gvp_2021} to encode track information and is capable of inferring both the energy and the recoil direction in full 3D. As shown in Fig.~\ref{fig:sbi}, this substantially outperforms conventional summary features (principal-component axis, longest string, centroid-to-centroid, vacancy count, track length), with the largest gains at low recoil energy where simple descriptors fail. Efforts are ongoing to refine the model to incorporate diamond crystal structure using SIIMPL~\cite{siimpl} and the effects of finite track imaging resolution.

\subsection{Natural diamond as a paleodetector}
\label{sec:paleo}
While most mineral detection efforts at UMD have focused on synthetic diamond, natural diamond is also an attractive paleodetection target. Diamonds typically reside in the mantle for $\sim$1--3\,Gyr before kimberlite eruptions carry them to the surface, giving exposure times comparable to the oldest paleo-mineral candidates~\cite{gurney2010,smit_shirey_2019_ages}. The lattice is also intrinsically radiopure, with sub-ppb uranium abundances confined to inclusions rather than the lattice~\cite{kramers_lead_1979}, while synthetic diamond is widely available for method development. The main challenges are the comparatively uranium-rich host kimberlite, which generates a radiogenic background~\cite{kresten1974}, and deformation-induced defects which could mimic or obscure genuine recoil tracks.

The readout for paleodetection also differs from synthetic-diamond directional detection. NV readout requires isolated substitutional nitrogen (type Ib) in order to form NVs, which is rare in nature: over 95\% of natural diamonds are type Ia, with nitrogen aggregated into A-centers (2N) and B-centers (4N)~\cite{breeding_shigley_2009}. We therefore propose the H3 (N--V--N) center as the natural-diamond-compatible readout target. The H3 center is an optically-active point defect that is formed when radiation-induced vacancies migrate during annealing and are trapped at A-centers~\cite{crossfield_1974}; mantle temperatures readily drive this migration, and the resulting H3 centers are stable to temperatures exceeding 2000$^\circ$C~\cite{breeding_green_2018}. The calibrated track-formation framework of Sec.~\ref{sec:pipeline} and Ref.~\cite{ang_multiscale_2026} can be extended to incorporate these constraints and predict H3 yields under geological thermal histories. At high ($\gtrsim$$10^3$ ppm) nitrogen concentrations, however, H3 photoluminescence is quenched by nearby aggregated nitrogen via a non-radiative energy-transfer process~\cite{crossfield_1974,collins_2001}, thus likely requiring filtering of candidate samples via FTIR screening. H3 centers are otherwise bright and photostable \cite{pant_2022}, rendering them promising for light-sheet and super-resolution imaging being developed for color center readout; e.g., stimulated-emission-depletion imaging of single H3 centers has already been demonstrated~\cite{kolesov_2018,laporte_psaltis_2016}. H3 centers thus merit experimental characterization in natural diamond to assess their suitability for paleodetection. As part of a preliminary study, we obtained three natural diamond samples from De Beers containing H3 centers and plan to examine them with optical microscopy.

\acknowledgments
This work was supported by, or in part by, the Argonne National Laboratory under Award No. 2F60042; the DOE Fusion program under Award No. DESC0021654; the U.S. Army Research Laboratory under Contract Nos. W911NF1920181 and W911NF2420143; the LPS/QTC Jumping Electron Fellowship through award H9823022C0029; and the University of Maryland Quantum Technology Center.

\clearpage

\section{Projected Sensitivity of Paleo-Detectors to WIMP-Nucleus Effective Interactions}\label{sec:UT_Austin}

Authors: {\it Dionysios P. Theodosopoulos}
\vspace{0.1cm} \\
Weinberg Institute for Theoretical Physics, Department of Physics,
The University of Texas at Austin, Austin, TX 78712, USA
\vspace{0.3cm}

\subsection{Introduction}

A wide range of cosmological and astrophysical observations indicate that non-baryonic cold dark matter (DM) constitutes most of the matter in the Universe, yet its particle nature remains unknown. Weakly Interacting Massive Particles (WIMPs) are among the most well-motivated DM candidates and are the target of extensive direct-detection searches.

Paleo-detectors offer a complementary approach to direct DM detection by exploiting geological timescales to achieve enormous effective exposures. Instead of operating large detector volumes, as in conventional direct-detection experiments, paleo-detectors exploit ancient minerals that have remained deep underground for up to $\sim 1~\mathrm{Gyr}$, preserving damage tracks produced by nuclear recoils from rare scattering events.

In this contribution, we summarize the results of Ref.~\cite{Theodosopoulos:2026ehn}, which investigated the projected sensitivity of paleo-detectors to WIMP--nucleus interactions within the framework of a non-relativistic effective field theory (NREFT)~\cite{Fan:2010gt,Fitzpatrick:2012ix}. The study considered the complete set of NREFT operators for elastic scattering as well as inelastic scattering into heavier WIMP states~\cite{Tucker-Smith:2001myb}, using the recoil track length in the target mineral as a proxy for the nuclear recoil energy.

The results show that paleo-detectors can achieve sensitivities competitive with conventional direct-detection experiments for $\mathcal{O}(\mathrm{TeV/c^2})$ DM masses, while providing sensitivity improvements of several orders of magnitude for $\mathcal{O}(\mathrm{GeV/c^2})$ DM masses.

\subsection{Non-Relativistic Effective Field Theory}
\label{sec:NREFT}

We describe WIMP--nucleon interactions using the non-relativistic effective field theory (NREFT) framework developed in Refs.~\cite{Fan:2010gt,Fitzpatrick:2012ix}. Assuming Galilean invariance and momentum conservation, all elastic interactions can be constructed from the Hermitian building blocks
\begin{equation}
 i\frac{\vec q}{m_N}, \qquad
 \vec v^{\,\perp}=\vec v+\frac{\vec q}{2\mu_N}, \qquad
 \vec S_\chi,\qquad
 \vec S_N ,
\end{equation}
where $\vec q$ is the momentum transfer, $\vec v$ the WIMP--nucleon relative velocity, $\mu_N$ the reduced mass of the WIMP-nucleon system, and $\vec S_\chi$ and $\vec S_N$ the WIMP and nucleon spin operators, respectively.
The most general interaction Lagrangian can be written as
\begin{equation}
\mathcal{L}_{\rm int}
=
\sum_{N=n,p}\sum_i c_i^{(N)}\,\mathcal O_i\,
\chi^+\chi^-N^+N^- ,
\end{equation}
where $\mathcal O_i$ are the dimensionless NREFT operators listed in Table~\ref{tab:NREFT_operators}~\cite{Fitzpatrick:2012ix,Anand:2013yka}. Throughout this work we assume isoscalar couplings, $c_i^p=c_i^n$, and elastic nucleon scattering ($N^+=N^-$). The operators $\mathcal O_1$ and $\mathcal O_4$ correspond to the conventional spin-independent (SI) and spin-dependent (SD) interactions, respectively.
\begin{table}[t]
\centering
\small
\setlength{\tabcolsep}{6pt}   
\renewcommand{\arraystretch}{0.9} 
\captionsetup{justification=raggedright,singlelinecheck=false}
\begin{tabular}{l l}
\hline\hline
\multicolumn{2}{c}{NREFT operators relevant for elastic WIMP--nucleon scattering} \\
\hline
$\mathcal{O}_1 = 1_\chi 1_N$ 
& $\mathcal{O}_9 = i\, \vec{S}_\chi \cdot ( \vec{S}_N \times \vec{q}/m_N )$ \\

$\mathcal{O}_3 = i\, \vec{S}_N \cdot ( \vec{q}/m_N \times \vec{v}^{\perp} )$ 
& $\mathcal{O}_{10} = i\, \vec{S}_N \cdot \vec{q}/m_N$ \\

$\mathcal{O}_4 = \vec{S}_\chi \cdot \vec{S}_N$ 
& $\mathcal{O}_{11} = i\, \vec{S}_\chi \cdot \vec{q}/m_N$ \\

$\mathcal{O}_5 = i\, \vec{S}_\chi \cdot ( \vec{q}/m_N \times \vec{v}^{\perp} )$ 
& $\mathcal{O}_{12} = \vec{S}_\chi \cdot ( \vec{S}_N \times \vec{v}^{\perp} )$ \\

$\mathcal{O}_6 = ( \vec{S}_\chi \cdot \vec{q}/m_N )
                 ( \vec{S}_N \cdot \vec{q}/m_N )$
& $\mathcal{O}_{13} = i ( \vec{S}_\chi \cdot \vec{v}^{\perp} )
                       ( \vec{S}_N \cdot \vec{q}/m_N )$ \\

$\mathcal{O}_7 = \vec{S}_N \cdot \vec{v}^{\perp}$ 
& $\mathcal{O}_{14} = i ( \vec{S}_\chi \cdot \vec{q}/m_N )
                       ( \vec{S}_N \cdot \vec{v}^{\perp} )$ \\

$\mathcal{O}_8 = \vec{S}_\chi \cdot \vec{v}^{\perp}$ 
& $\mathcal{O}_{15} = - ( \vec{S}_\chi \cdot \vec{q}/m_N )
                       [ ( \vec{S}_N \times \vec{v}^{\perp} )
                       \cdot \vec{q}/m_N ]$ \\
\hline\hline
\end{tabular}
\caption{Operators defining the NREFT of WIMP--nucleon elastic interactions~\cite{Fan:2010gt,Fitzpatrick:2012ix}. The operators $\mathcal{O}_{1}$ and $\mathcal{O}_{4}$ correspond to canonical SI and SD interactions, respectively. Operator $\mathcal{O}_{2}$ is quadratic in $\vec{v}^{\perp}$ and is therefore not considered here~\cite{Anand:2013yka}.}
\label{tab:NREFT_operators}
\end{table}

In addition to elastic DM scattering ($\chi^-\equiv\chi^+$), we also consider inelastic scattering, where the incoming and outgoing dark matter states differ by a mass splitting
$\delta_m = m_{\chi,\mathrm{out}}-m_{\chi,\mathrm{in}}>0$
\cite{Tucker-Smith:2001myb}. The modified kinematics can be incorporated through the replacement~\cite{Barello:2014uda}
\begin{equation}
\vec v^{\,\perp}
\rightarrow
\vec v_{\rm inel}^{\,\perp}
=
\vec v^{\,\perp}
+
\frac{\delta_m}{|\vec q|^2}\vec q ,
\end{equation}
from which the corresponding inelastic operators are obtained. In this contribution, we summarize the projected paleo-detector sensitivity to the operators of Table~\ref{tab:NREFT_operators} for both elastic and inelastic WIMP--nucleon scattering.

\subsection{Dark Matter Signals and Backgrounds in Paleo-Detectors}
\label{DMpaleo}

In paleo-detectors, WIMP--nucleus scattering produces nuclear recoils that leave damage tracks in ancient minerals. Since the track length $x_T$ is determined by the stopping power of the recoiling nucleus, it serves as a proxy for the recoil energy $E_R$. For a mineral composed of multiple nuclear species (Z,A), the total differential track-production rate (per unit exposure, and per unit track length) is
\begin{equation}
    \frac{dR}{dx_T}=\sum_{(Z,A)}\xi_{(Z,A)}\left(\frac{dR}{dE_{R}}\right)_{(Z,A)}\left(\frac{dE_R}{dx_T}\right)_{(Z,A)}~,\label{dRdx}
\end{equation}
where $\xi_{(Z,A)}$ is the mass fraction of each nuclear species, $(dR/dE_R)_{(Z,A)}$ denotes the recoil energy spectrum for a given nuclear species, as defined in Refs~\cite{Fitzpatrick:2012ix,Anand:2013yka}, and $(dE_R/dx_T)_{(Z,A)}$ is the stopping power calculated using \texttt{SRIM}.

We consider four benchmark target minerals: gypsum, halite, olivine, and muscovite. These minerals are commonly found in marine evaporites and ultrabasic rocks and are characterized by low ${}^{238}$U concentrations, reducing radiogenic backgrounds. Detailed compositions and event-rate predictions for all NREFT operators can be found in Ref.~\cite{Theodosopoulos:2026ehn}.

We adopt two benchmark track-readout scenarios. The high-resolution (HR) scenario assumes a $10\,\mathrm{mg}$ sample and a spatial resolution of $1\,\mathrm{nm}$, providing optimal sensitivity to low-mass WIMPs. The high-exposure (HE) scenario assumes a $100\,\mathrm{g}$ sample and a resolution of $15\,\mathrm{nm}$, offering greater sensitivity to heavier WIMPs.

The dominant backgrounds are astrophysical neutrinos and radiogenic processes. Although similar to those encountered in conventional direct-detection experiments, their relative importance is modified by the enormous effective exposures achievable over geological timescales. A detailed discussion of signal and background modeling is provided in Refs.~\cite{drukier:2018pdy,Theodosopoulos:2026ehn}.

\subsection{Projected Sensitivity of Paleo-Detectors}
\label{sec:sensitivity}

In this section, we summarize the projected sensitivity of paleo-detectors to WIMP--nucleon interactions, following the results originally obtained in Ref.~\cite{Theodosopoulos:2026ehn}. We employ a profile-likelihood–ratio approach, which allows us to project the sensitivity of paleo-detectors to a DM signal using the full spectral information (DM signals plus background).

\begin{figure}
    \captionsetup{justification=raggedright,singlelinecheck=false}
    \centering
    \begin{subfigure}{0.32\textwidth}
    \includegraphics[width=\linewidth]{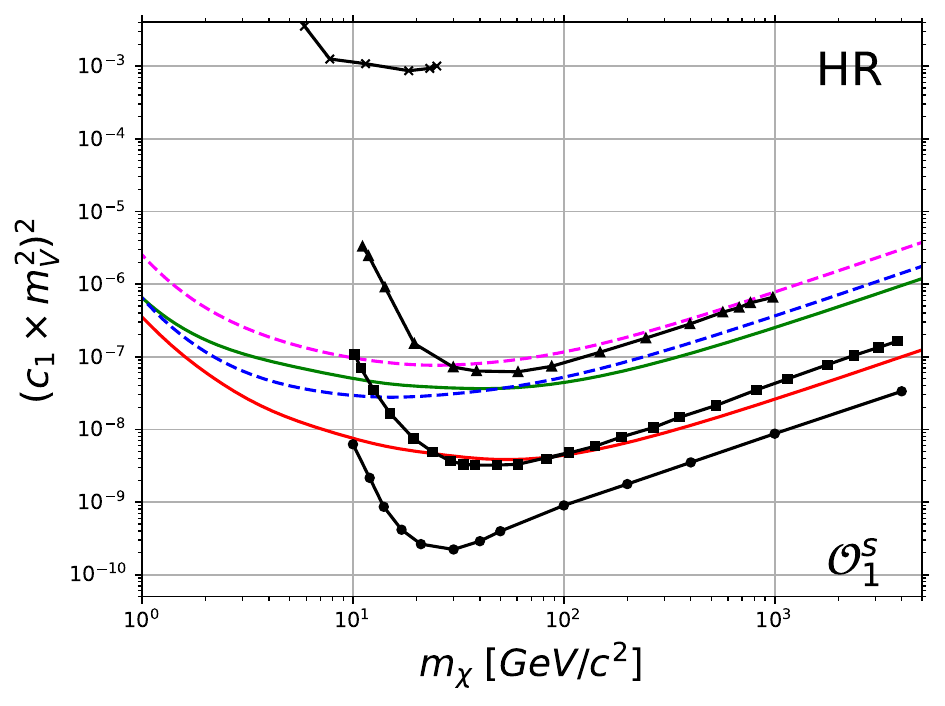}
    \end{subfigure}
    \begin{subfigure}{0.32\textwidth}
    \includegraphics[width=\linewidth]{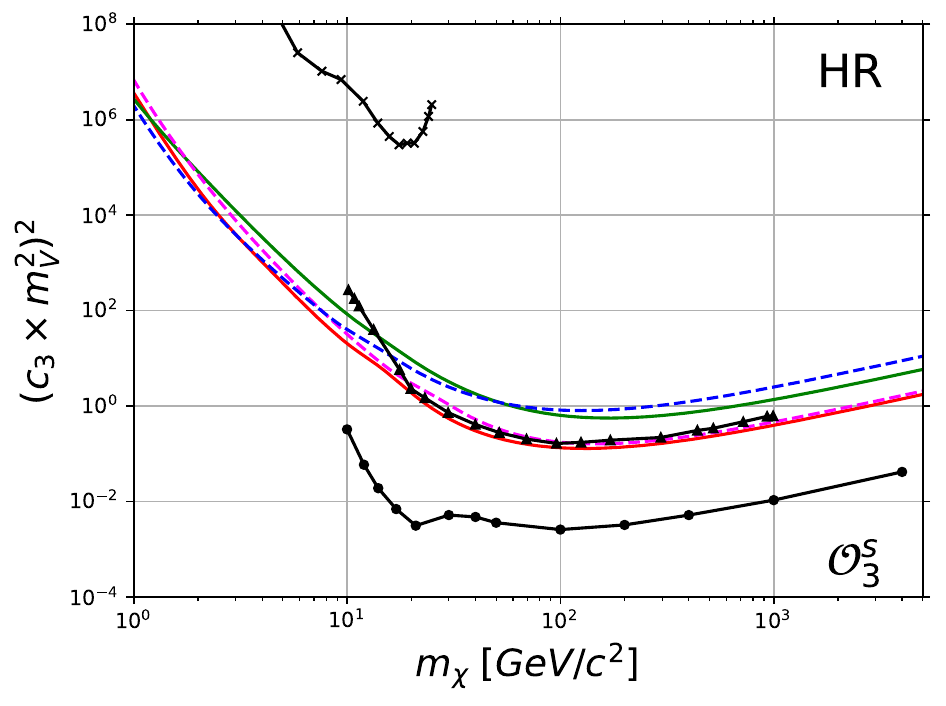}
    \end{subfigure}
    \begin{subfigure}{0.32\textwidth}
    \includegraphics[width=\linewidth]{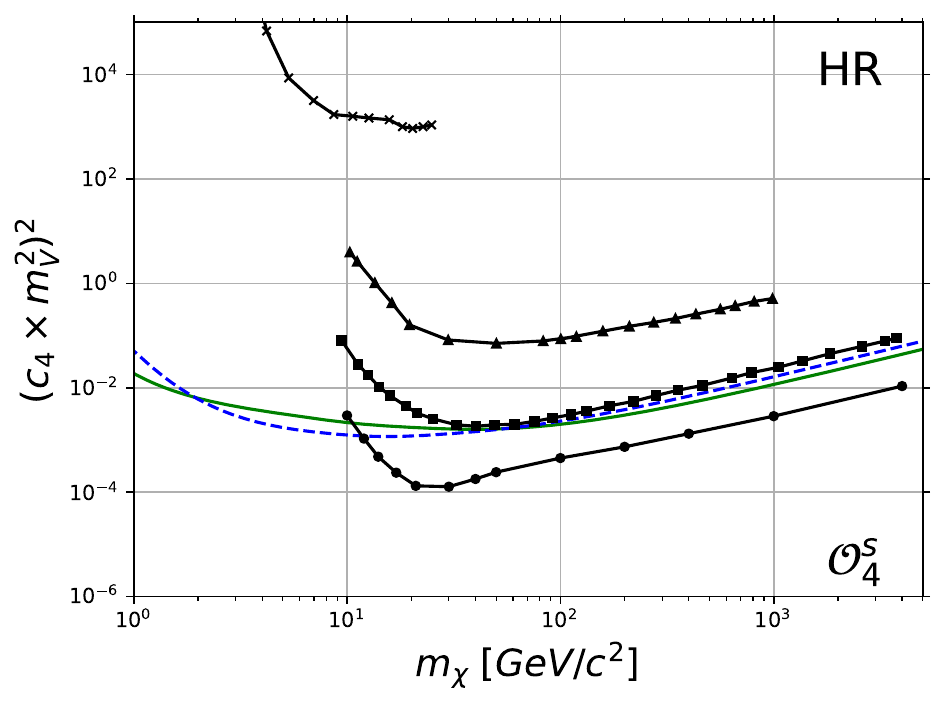}
    \end{subfigure}

    \vspace{0.1ex}

    \begin{subfigure}{0.32\textwidth}
    \includegraphics[width=\linewidth]{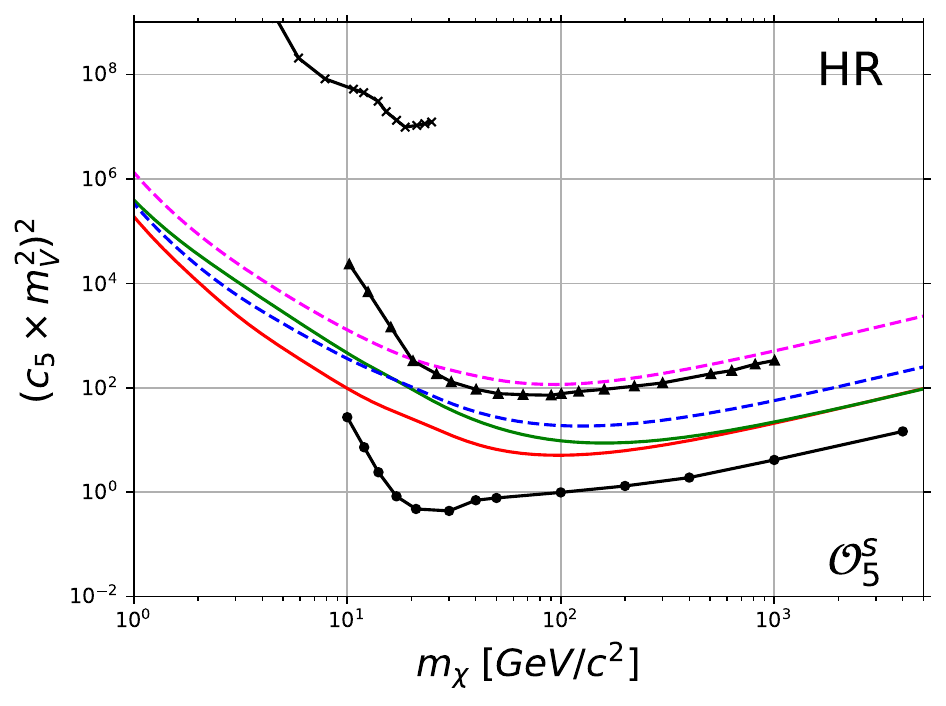}
    \end{subfigure}
    \begin{subfigure}{0.32\textwidth}
    \includegraphics[width=\linewidth]{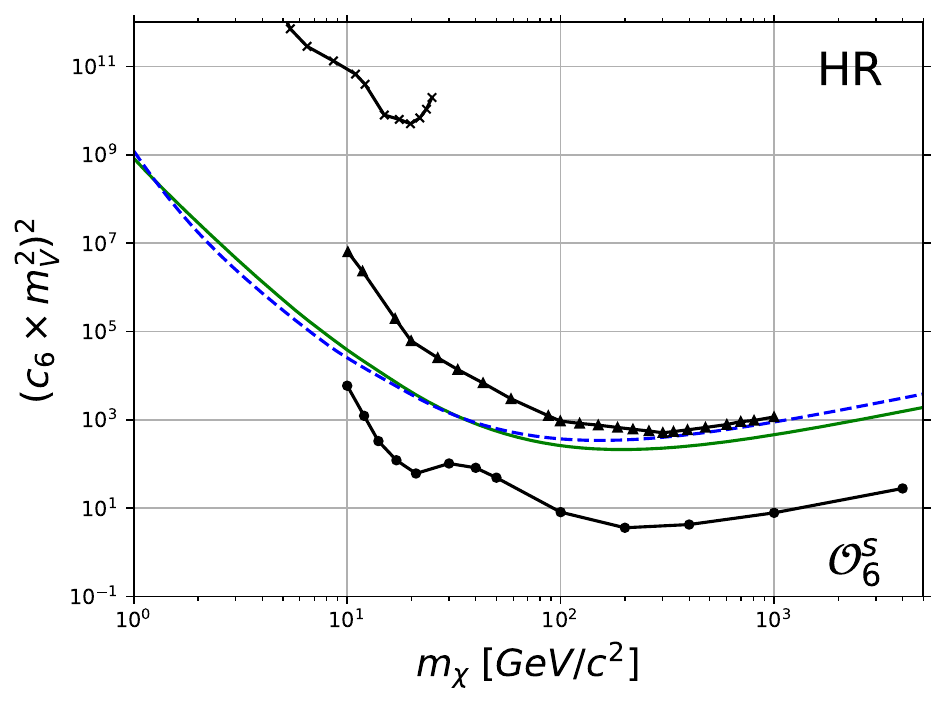}
    \end{subfigure}
    \begin{subfigure}{0.32\textwidth}
    \includegraphics[width=\linewidth]{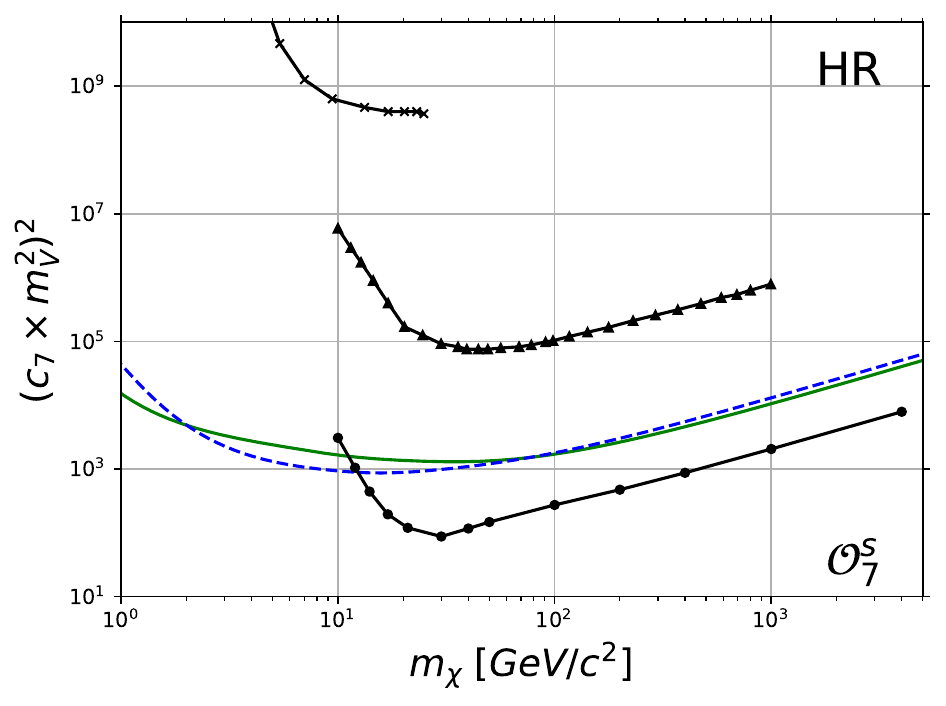}
    \end{subfigure}

    \vspace{0.1ex}

    \begin{subfigure}{0.32\textwidth}
    \includegraphics[width=\linewidth]{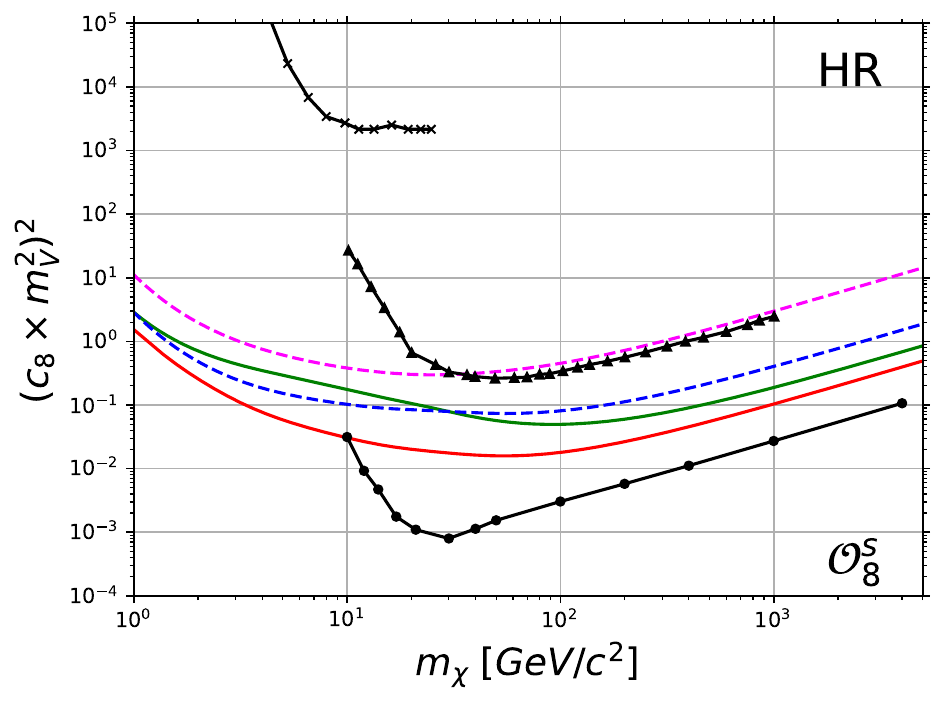}
    \end{subfigure}
    \begin{subfigure}{0.32\textwidth}
    \includegraphics[width=\linewidth]{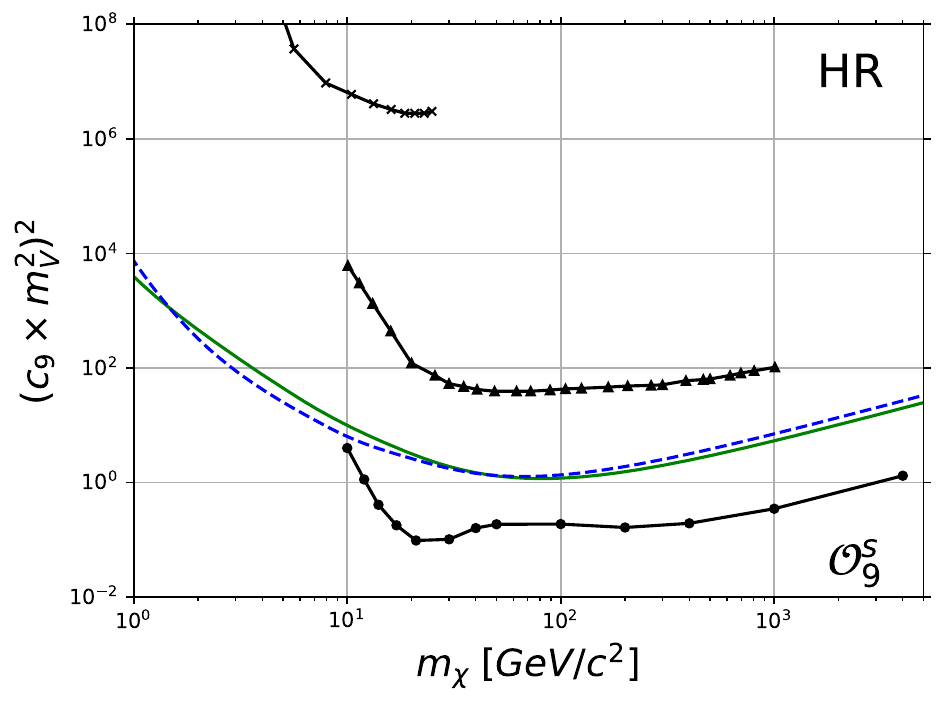}
    \end{subfigure}
    \begin{subfigure}{0.32\textwidth}
    \includegraphics[width=\linewidth]{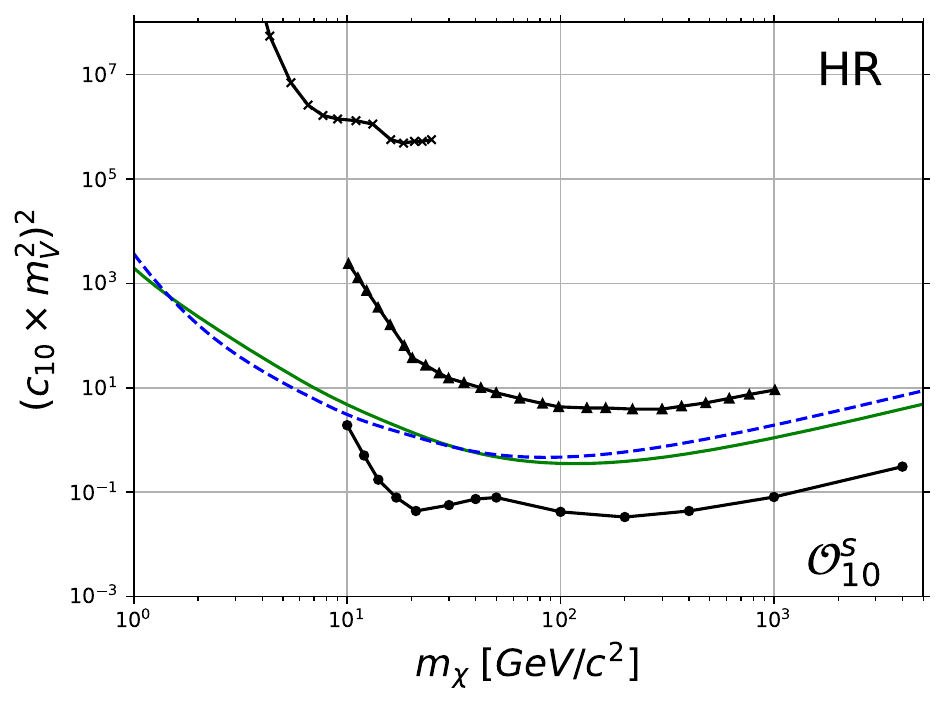}
    \end{subfigure}

    \vspace{0.1ex}

    \begin{subfigure}{0.32\textwidth}
    \includegraphics[width=\linewidth]{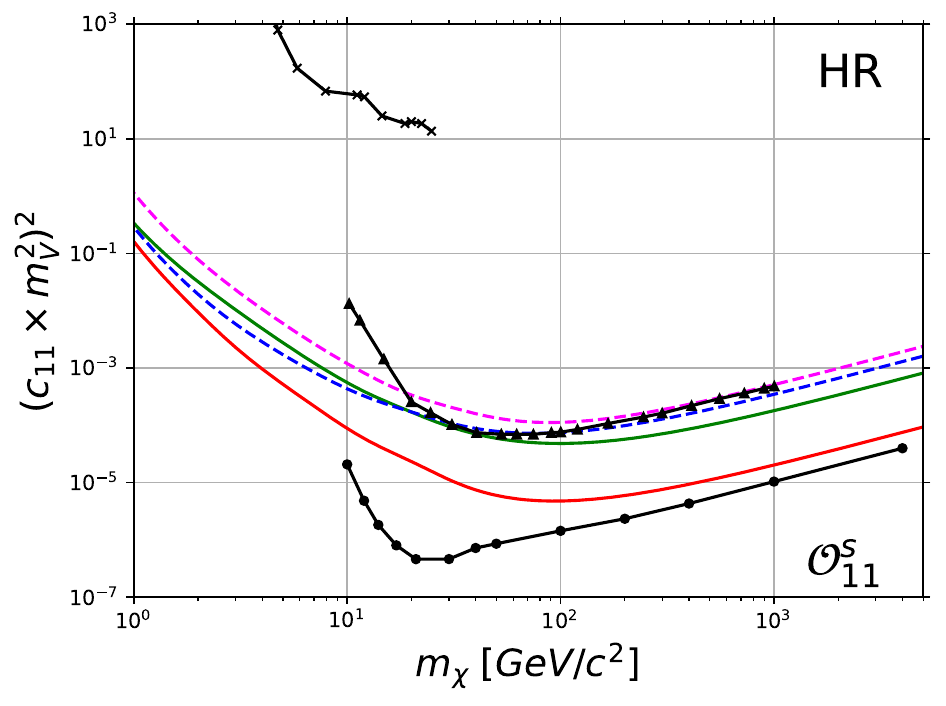}
    \end{subfigure}
    \begin{subfigure}{0.32\textwidth}
    \includegraphics[width=\linewidth]{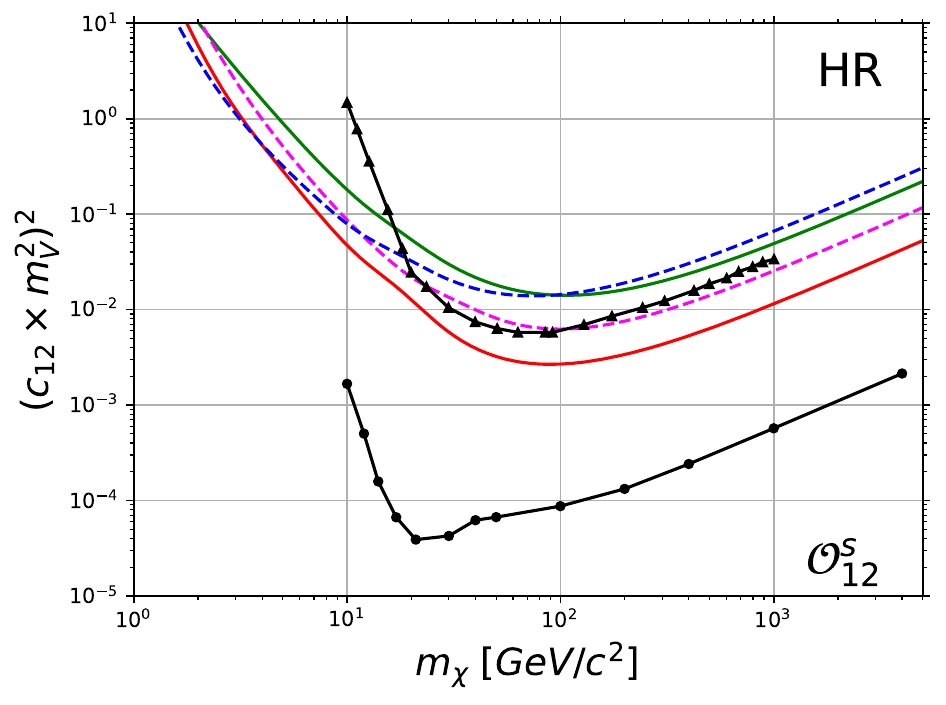}
    \end{subfigure}
    \begin{subfigure}{0.32\textwidth}
    \includegraphics[width=\linewidth]{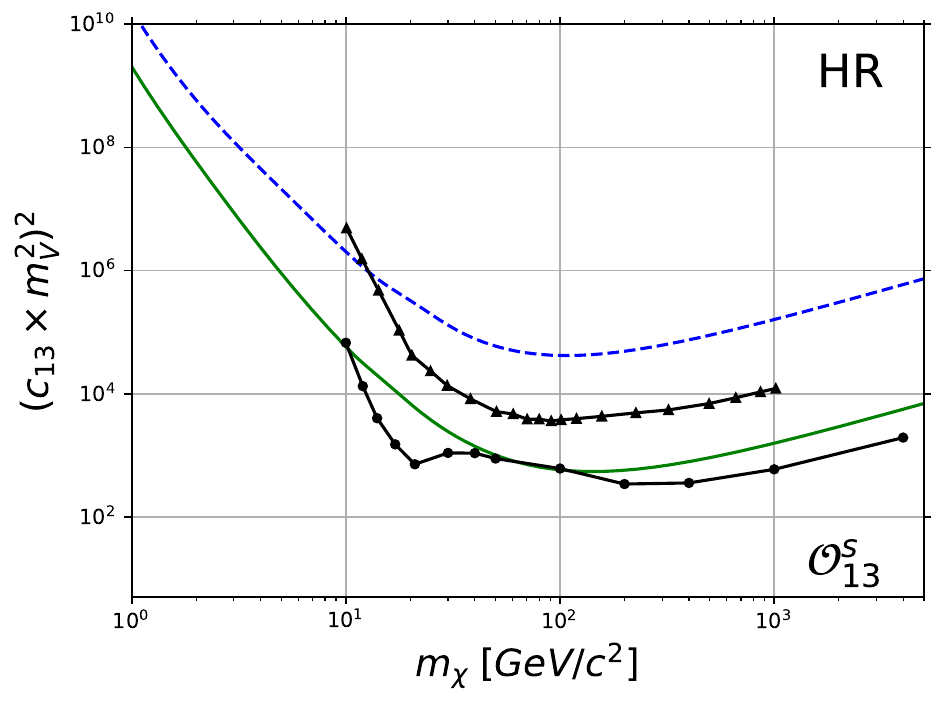}
    \end{subfigure}

    \vspace{0.1ex}

    \begin{subfigure}{0.32\textwidth}
    \includegraphics[width=\linewidth]{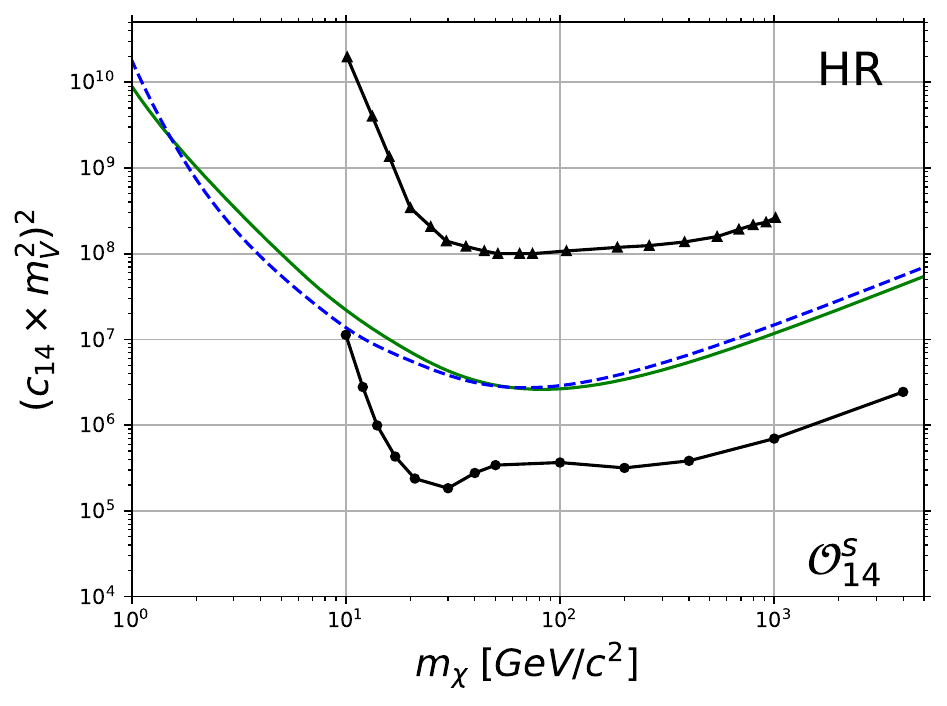}
    \end{subfigure}
    \begin{subfigure}{0.32\textwidth}
    \includegraphics[width=\linewidth]{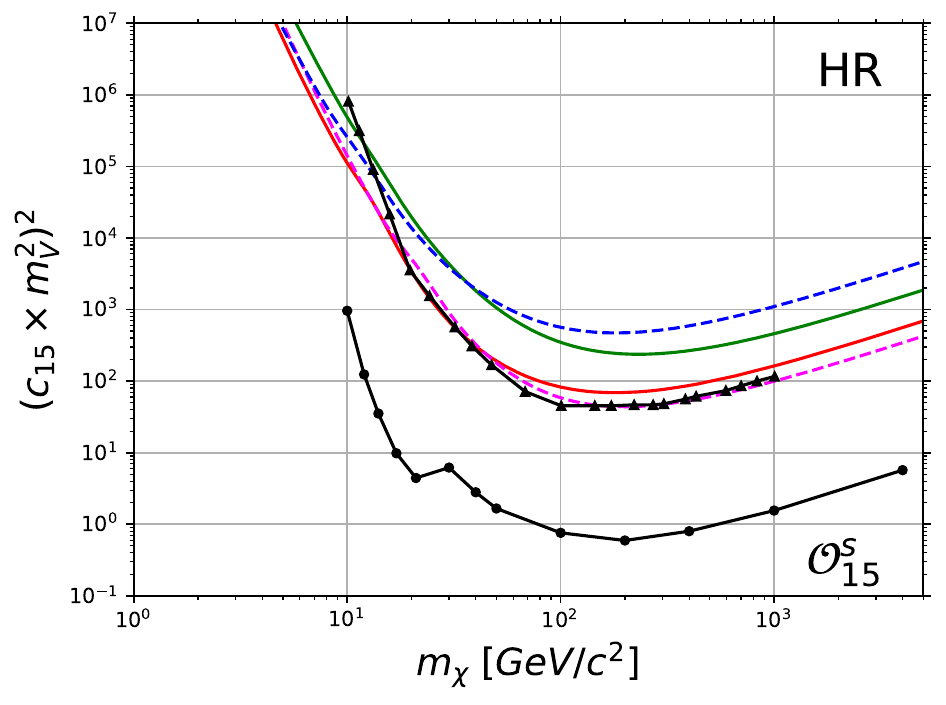}
    \end{subfigure}\hspace{4.55em}
    \begin{subfigure}{0.21\textwidth}
    \includegraphics[width=\linewidth]{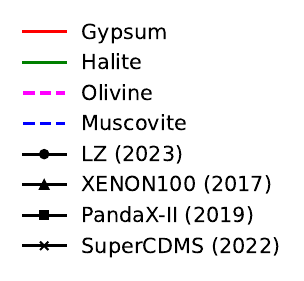} 
    \end{subfigure}
    \caption{Projected 90\% confidence level upper limits on the dimensionless isoscalar WIMP--nucleon NREFT coupling constants for elastic scattering, in the HR scenario. The solid (dashed) lines indicate minerals with $C^{238}=10^{-11}$ g/g ($C^{238}=10^{-10}$ g/g). Black lines show the NREFT results from conventional DD experiments: the 90\% confidence level upper limits from XENON100 \cite{XENON:2017fdd}, LUX--ZEPLIN \cite{LZ:2023lvz}, PandaX--II \cite{PandaX-II:2018woa}, as well as the 95\% Bayesian credible region of the two-dimensional marginalized posterior distribution from SuperCDMS \cite{SuperCDMS:2022crd}. Figures adapted from Ref.~{\cite{Theodosopoulos:2026ehn}}.}
    \label{fig:HR}
\end{figure}

\begin{figure}
    \captionsetup{justification=raggedright,singlelinecheck=false}
    \centering
    \begin{subfigure}{0.32\textwidth}
    \includegraphics[width=\linewidth]{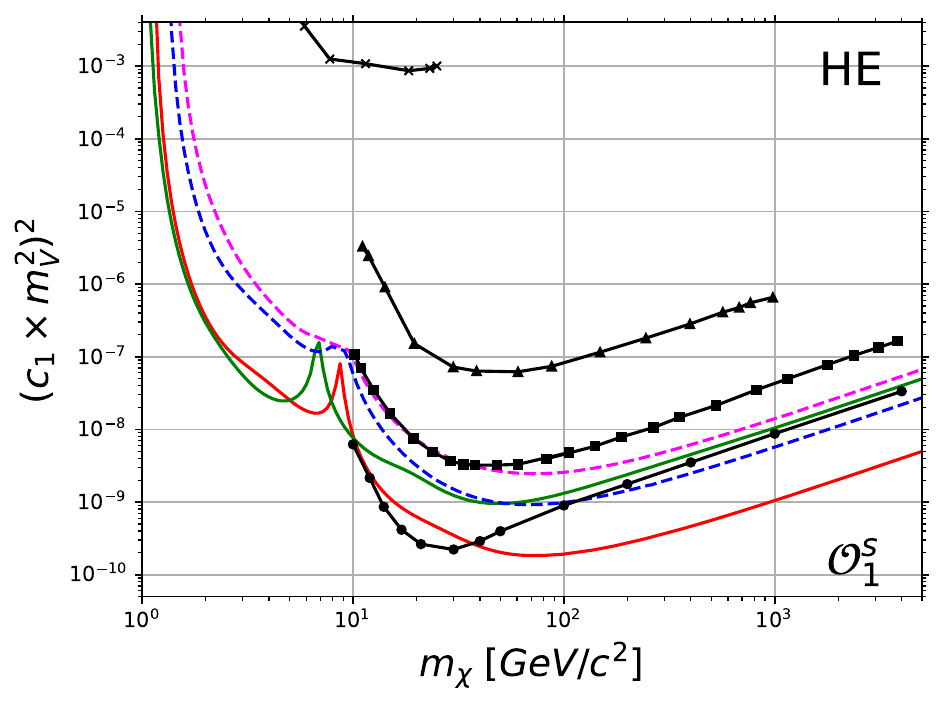}
    \end{subfigure}
    \begin{subfigure}{0.32\textwidth}
    \includegraphics[width=\linewidth]{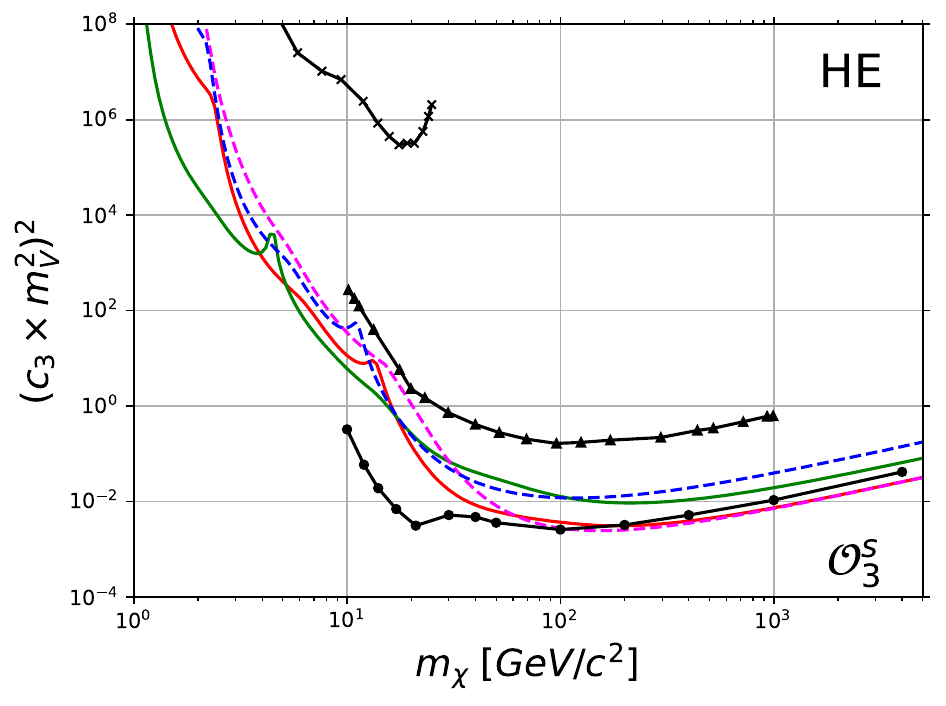}
    \end{subfigure}
    \begin{subfigure}{0.32\textwidth}
    \includegraphics[width=\linewidth]{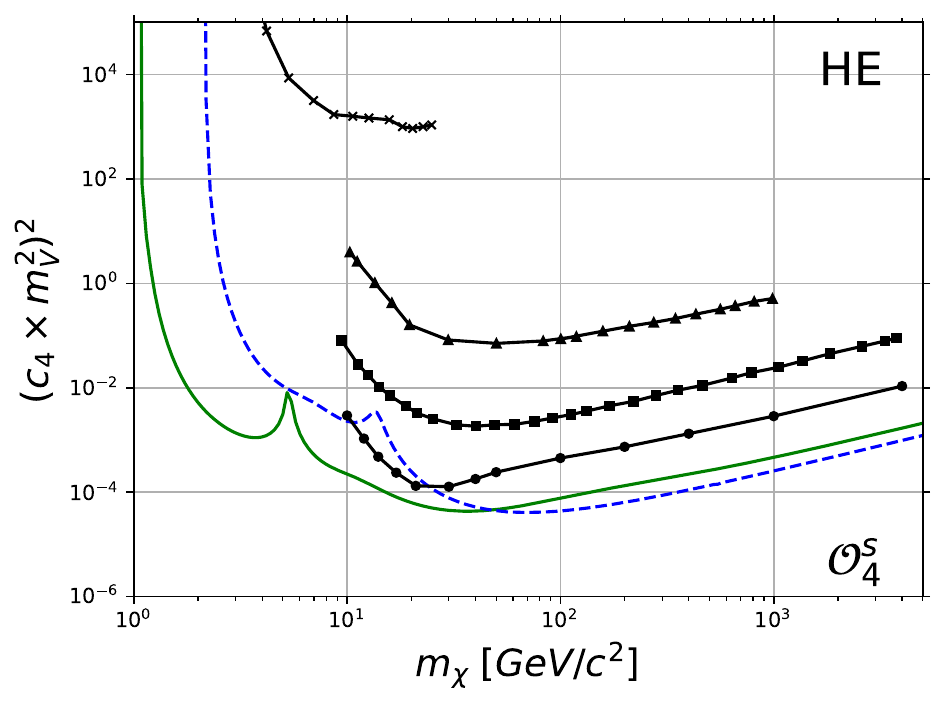}
    \end{subfigure}

    \vspace{0.1ex}

    \begin{subfigure}{0.32\textwidth}
    \includegraphics[width=\linewidth]{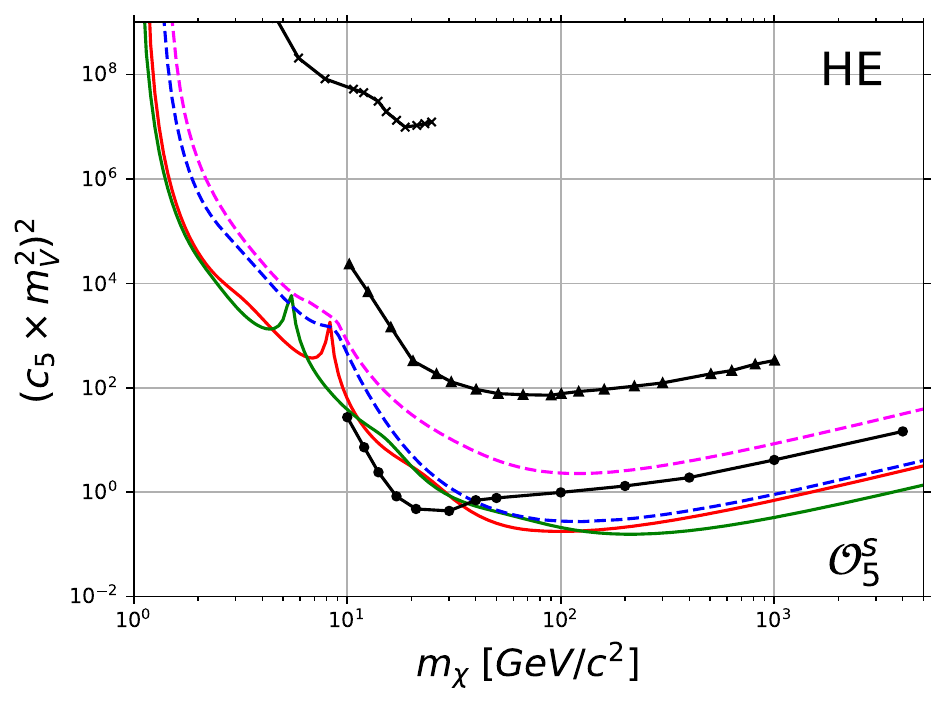}
    \end{subfigure}
    \begin{subfigure}{0.32\textwidth}
    \includegraphics[width=\linewidth]{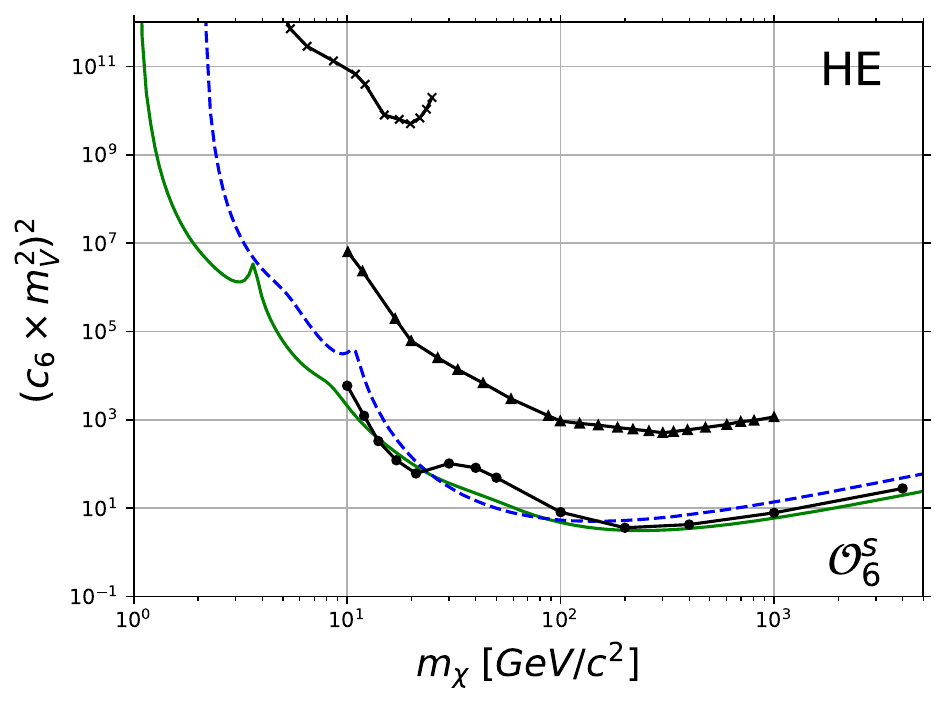}
    \end{subfigure}
    \begin{subfigure}{0.32\textwidth}
    \includegraphics[width=\linewidth]{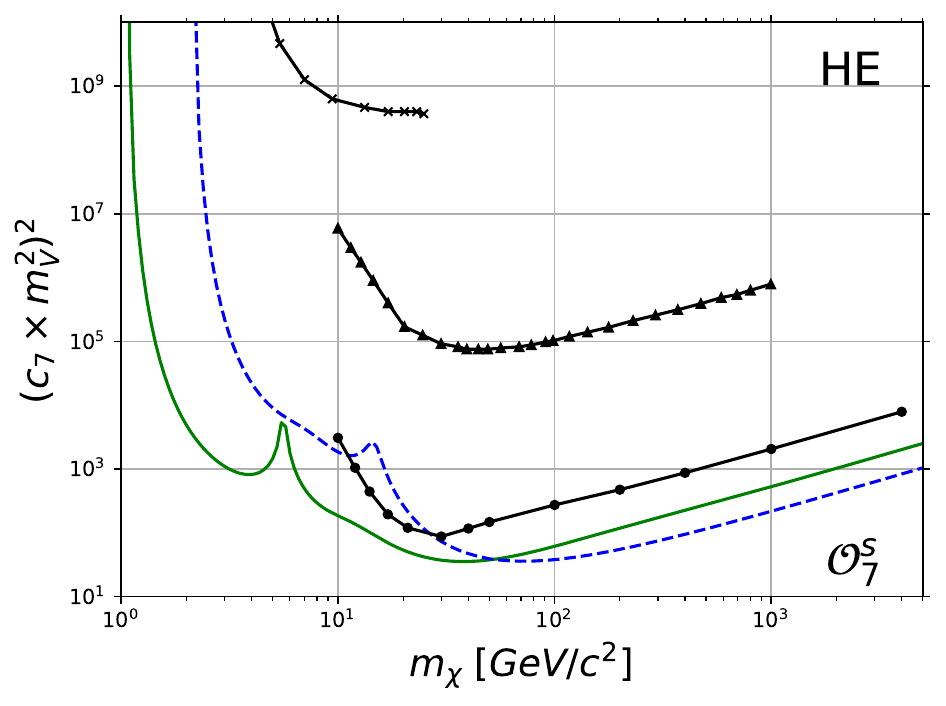}
    \end{subfigure}

    \vspace{0.1ex}

    \begin{subfigure}{0.32\textwidth}
    \includegraphics[width=\linewidth]{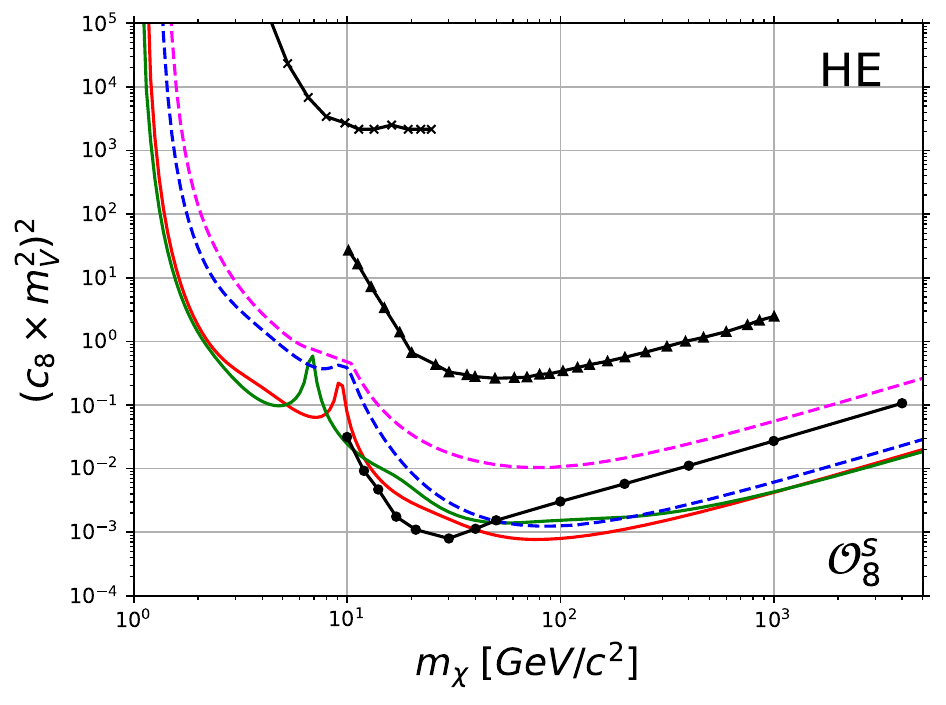}
    \end{subfigure}
    \begin{subfigure}{0.32\textwidth}
    \includegraphics[width=\linewidth]{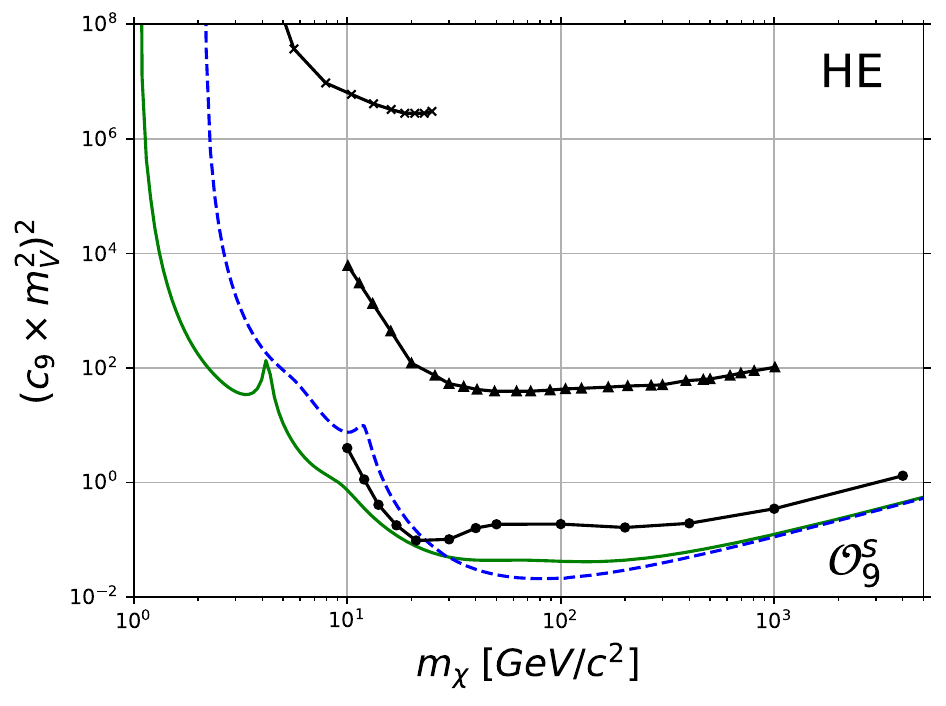}
    \end{subfigure}
    \begin{subfigure}{0.32\textwidth}
    \includegraphics[width=\linewidth]{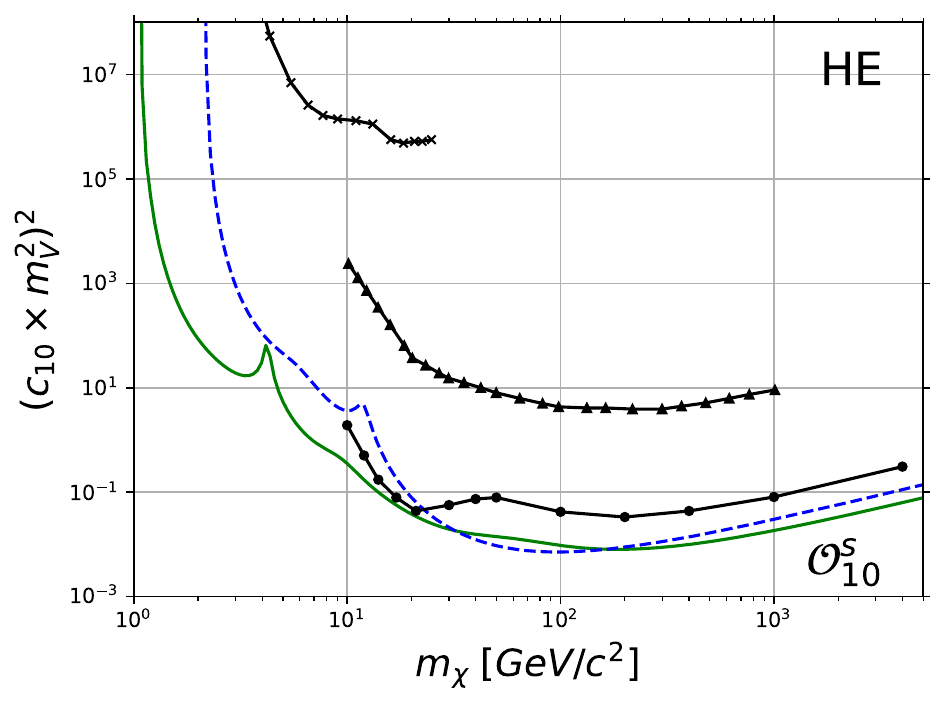}
    \end{subfigure}

    \vspace{0.1ex}

    \begin{subfigure}{0.32\textwidth}
    \includegraphics[width=\linewidth]{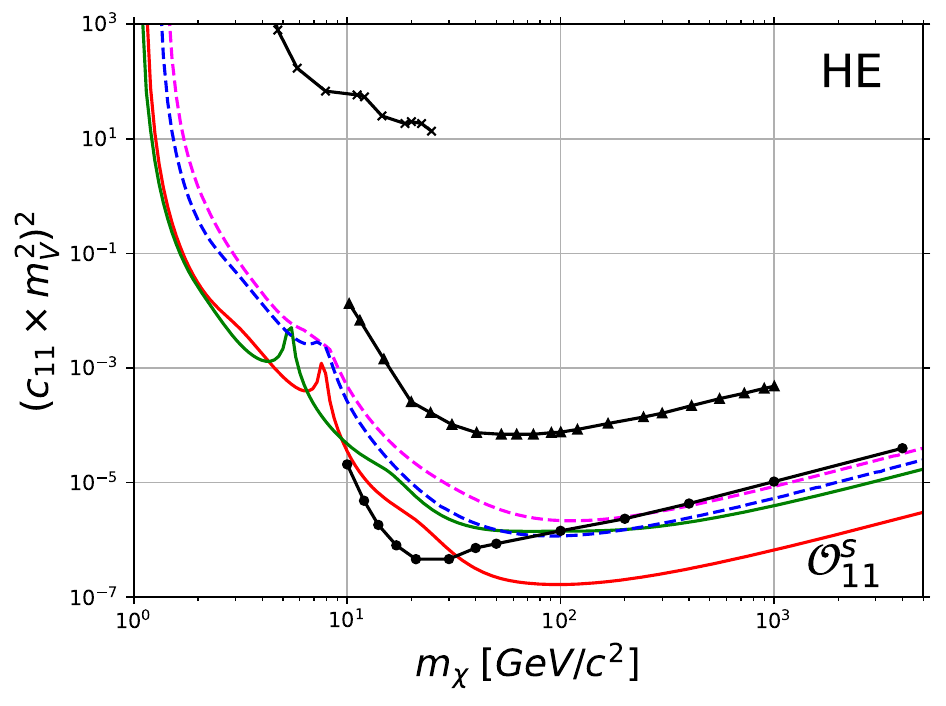}
    \end{subfigure}
    \begin{subfigure}{0.32\textwidth}
    \includegraphics[width=\linewidth]{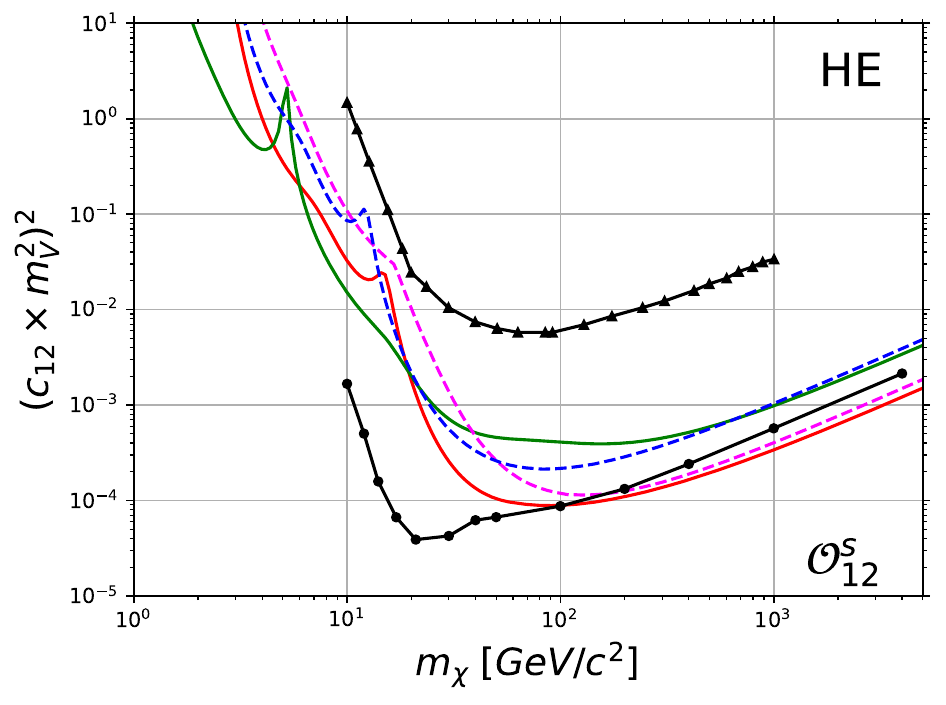}
    \end{subfigure}
    \begin{subfigure}{0.32\textwidth}
    \includegraphics[width=\linewidth]{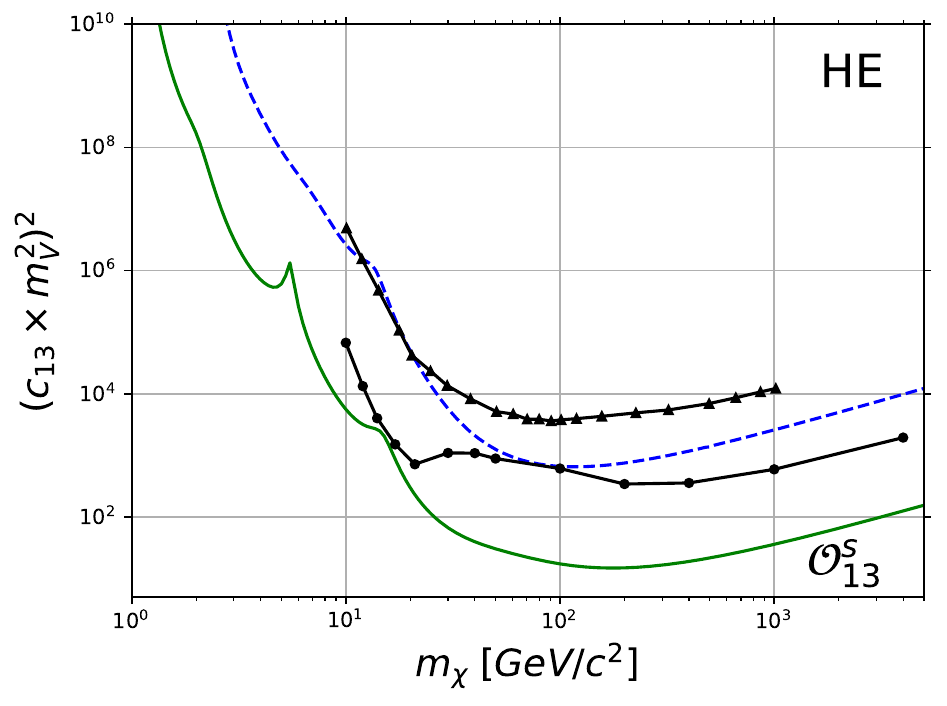}
    \end{subfigure}

    \vspace{0.1ex}

    \begin{subfigure}{0.32\textwidth}
    \includegraphics[width=\linewidth]{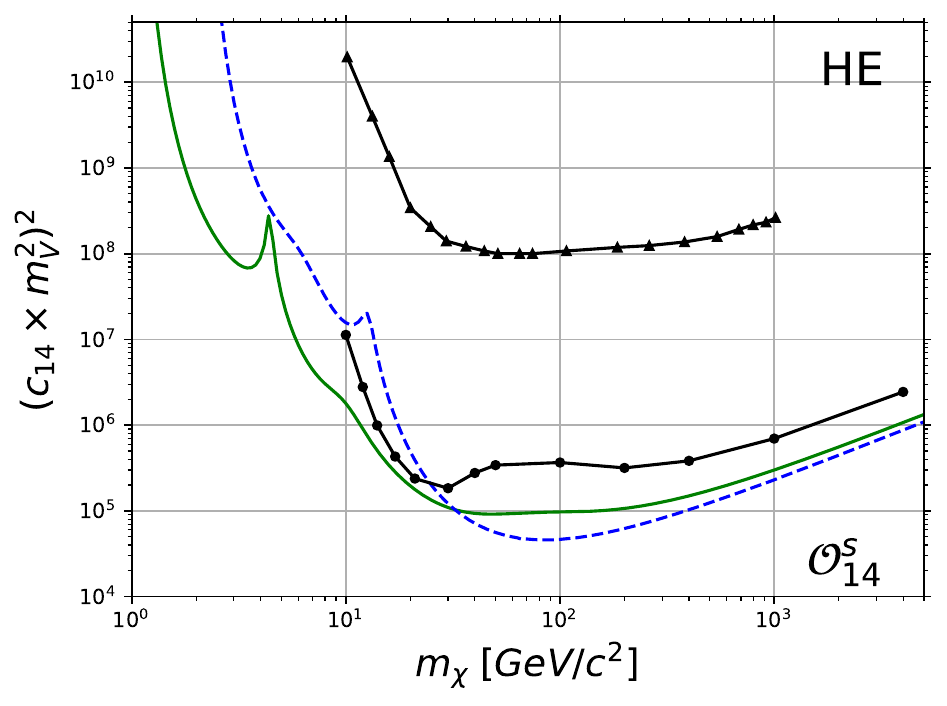}
    \end{subfigure}
    \begin{subfigure}{0.32\textwidth}
    \includegraphics[width=\linewidth]{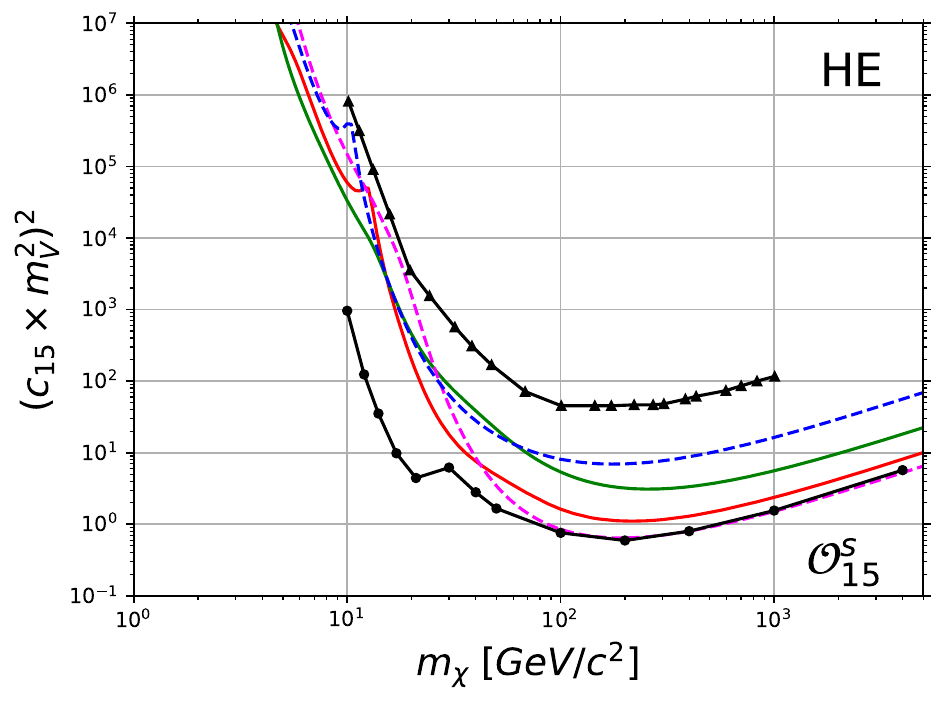}
    \end{subfigure}\hspace{4.55em}
    \begin{subfigure}{0.21\textwidth}
    \includegraphics[width=\linewidth]{Figures/UT_Austin/legend.pdf} 
    \end{subfigure}
    \caption{Projected 90\% confidence level upper limits on the dimensionless isoscalar WIMP--nucleon NREFT coupling constants for elastic scattering, in the HE scenario. The solid (dashed) lines indicate minerals with $C^{238}=10^{-11}$ g/g ($C^{238}=10^{-10}$ g/g). Black lines present the NREFT results from conventional DD experiments: the 90\% confidence level upper limits from XENON100 \cite{XENON:2017fdd}, LUX--ZEPLIN \cite{LZ:2023lvz}, PandaX--II \cite{PandaX-II:2018woa}, as well as the 95\% Bayesian credible region of the two-dimensional marginalized posterior distribution from SuperCDMS \cite{SuperCDMS:2022crd}. Figures adapted from Ref.~{\cite{Theodosopoulos:2026ehn}}.}
    \label{fig:HE}
\end{figure}

Projected $90\%$ confidence-level exclusion limits on the isoscalar NREFT coupling constants as a function of the DM mass $m_\chi$ are presented in Figs.~\ref{fig:HR} and \ref{fig:HE} for elastic scattering in the HR and HE scenarios. The results are compared with existing constraints from XENON100~\cite{XENON:2017fdd}, PandaX-II~\cite{PandaX-II:2018woa}, LUX--ZEPLIN~\cite{LZ:2023lvz}, and SuperCDMS~\cite{SuperCDMS:2022crd}.

The HR scenario provides superior paleo-detector sensitivity to that of conventional direct detection (DD) experiments for light DM, $m_\chi \lesssim 10~\mathrm{GeV}/c^2$, where solar-neutrino backgrounds dominate and excellent track-length resolution is required to distinguish signal from background. For heavier DM, $m_\chi \gtrsim 10~\mathrm{GeV}/c^2$, sensitivity is primarily enhanced by large exposure, making paleo-detectors one more potential powerful way for direct DM detection in addition to other DD experiments.

\begin{figure}
    \captionsetup{justification=raggedright,singlelinecheck=false}
    \centering
    \begin{subfigure}{0.32\textwidth}
    \includegraphics[width=\linewidth]{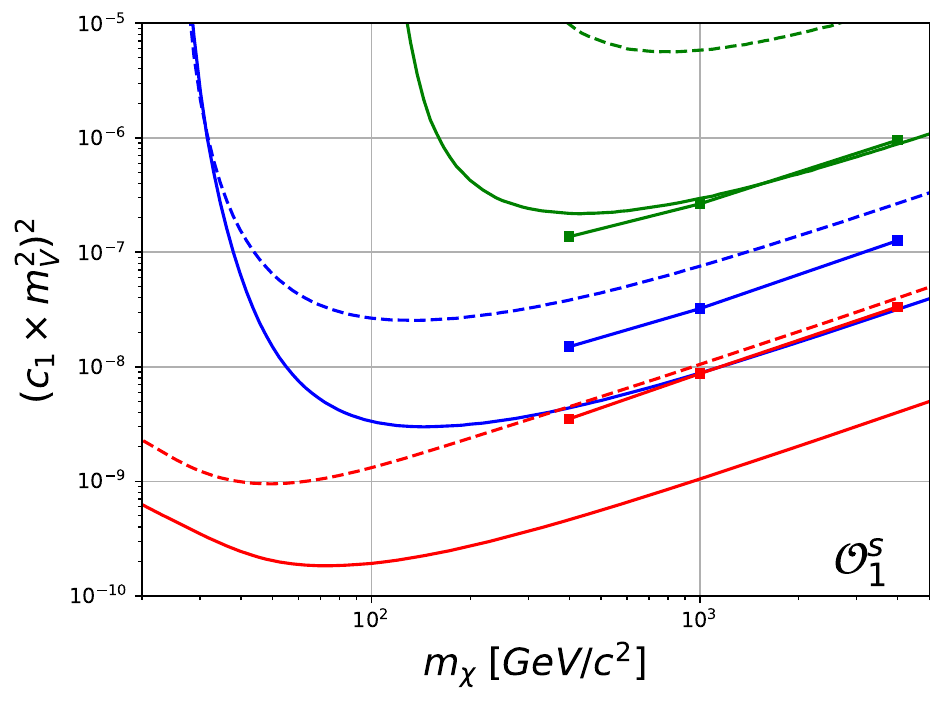}
    \end{subfigure}
    \begin{subfigure}{0.32\textwidth}
    \includegraphics[width=\linewidth]{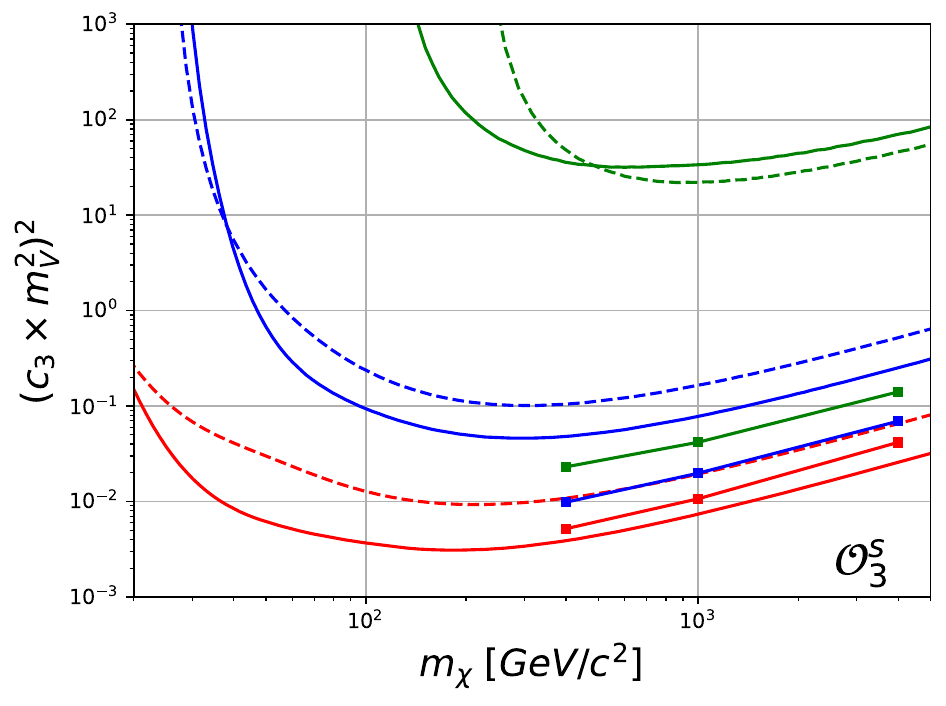}
    \end{subfigure}
    \begin{subfigure}{0.32\textwidth}
    \includegraphics[width=\linewidth]{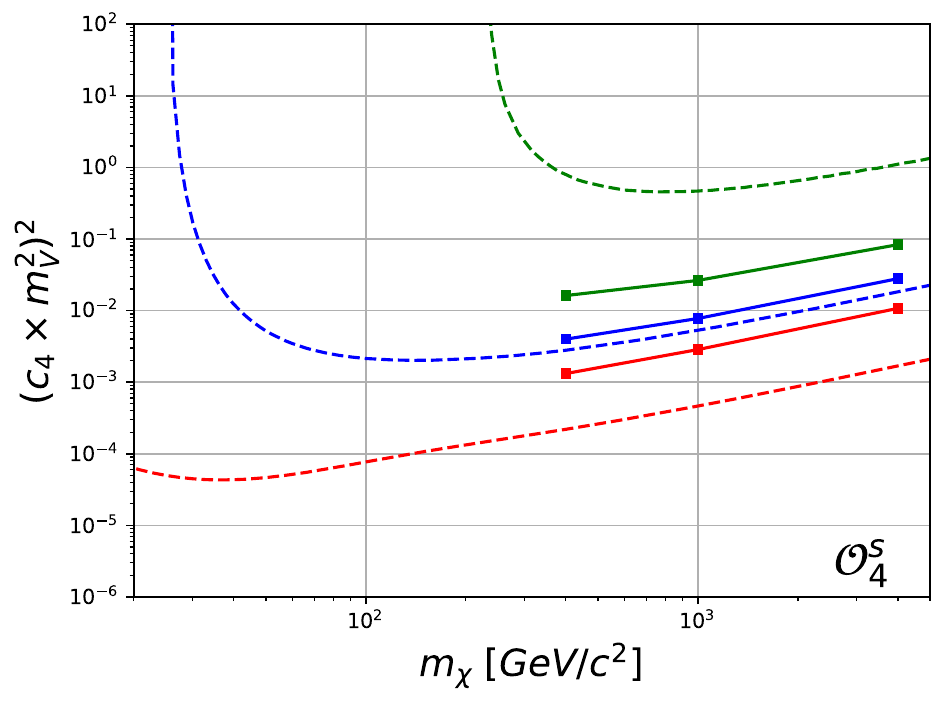}
    \end{subfigure}

    \vspace{0.1ex}

    \begin{subfigure}{0.32\textwidth}
    \includegraphics[width=\linewidth]{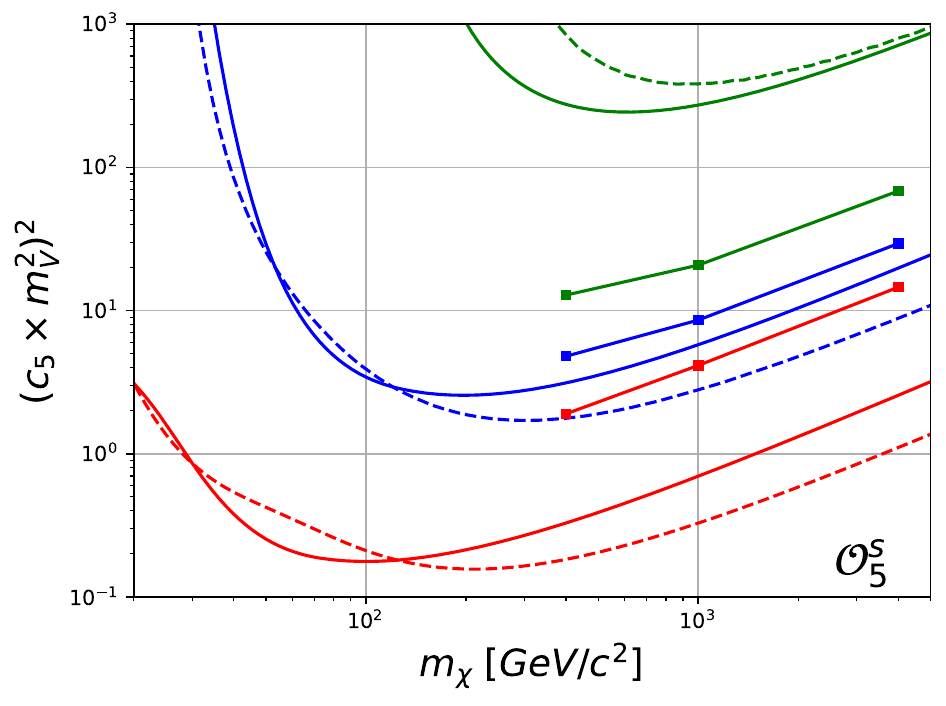}
    \end{subfigure}
    \begin{subfigure}{0.32\textwidth}
    \includegraphics[width=\linewidth]{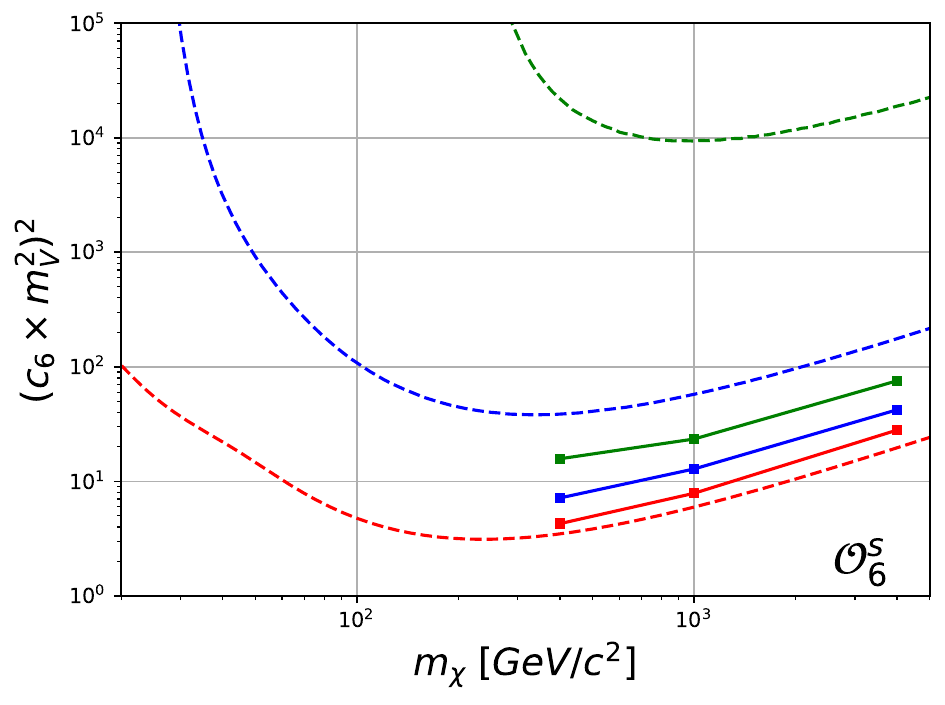}
    \end{subfigure}
    \begin{subfigure}{0.32\textwidth}
    \includegraphics[width=\linewidth]{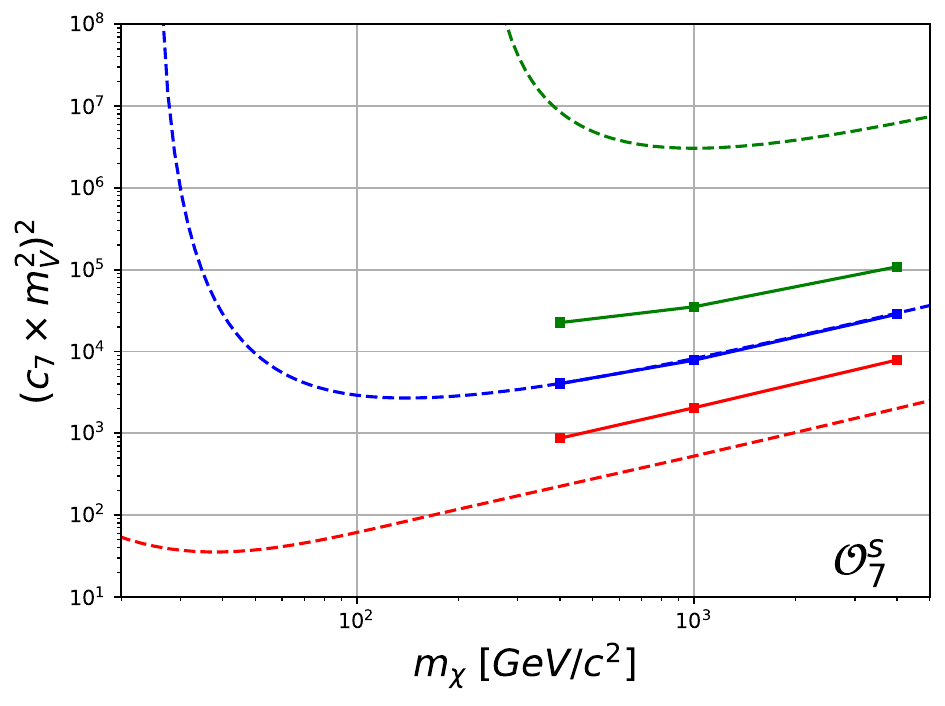}
    \end{subfigure}

    \vspace{0.1ex}

    \begin{subfigure}{0.32\textwidth}
    \includegraphics[width=\linewidth]{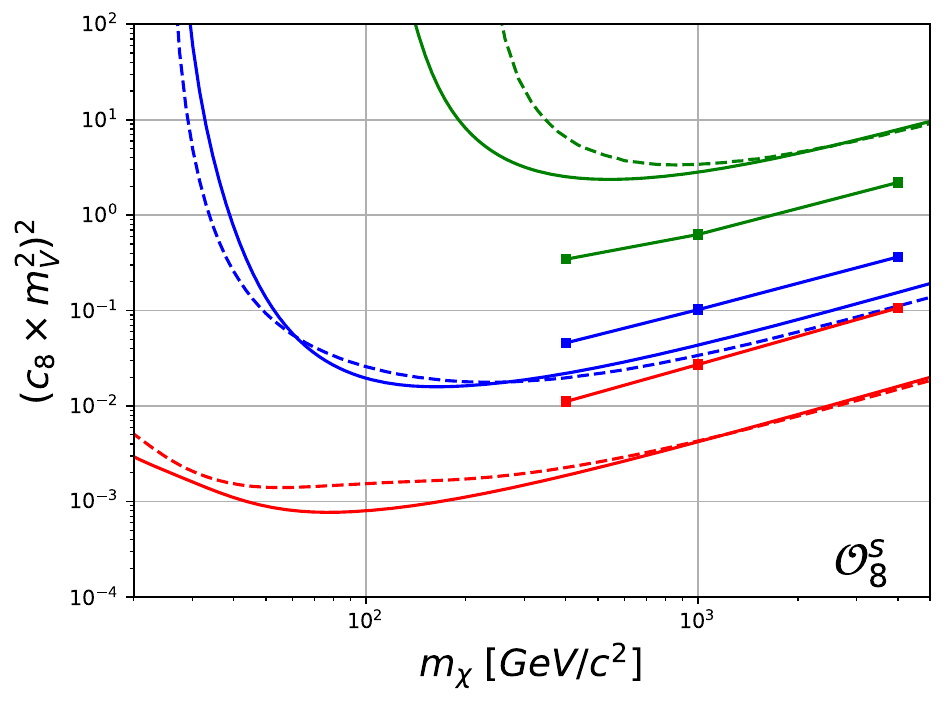}
    \end{subfigure}
    \begin{subfigure}{0.32\textwidth}
    \includegraphics[width=\linewidth]{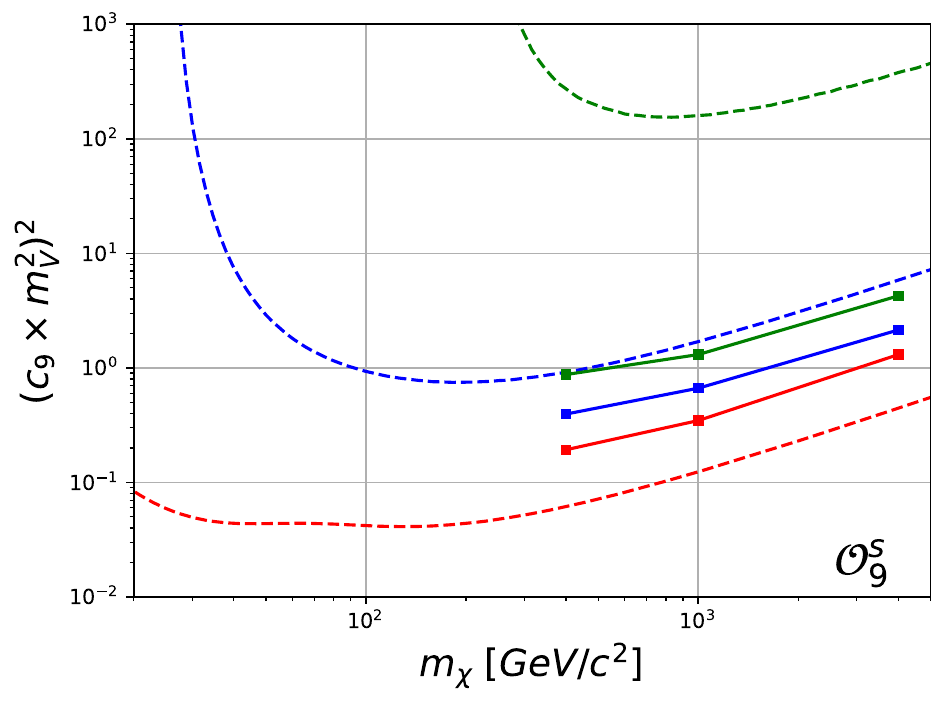}
    \end{subfigure}
    \begin{subfigure}{0.32\textwidth}
    \includegraphics[width=\linewidth]{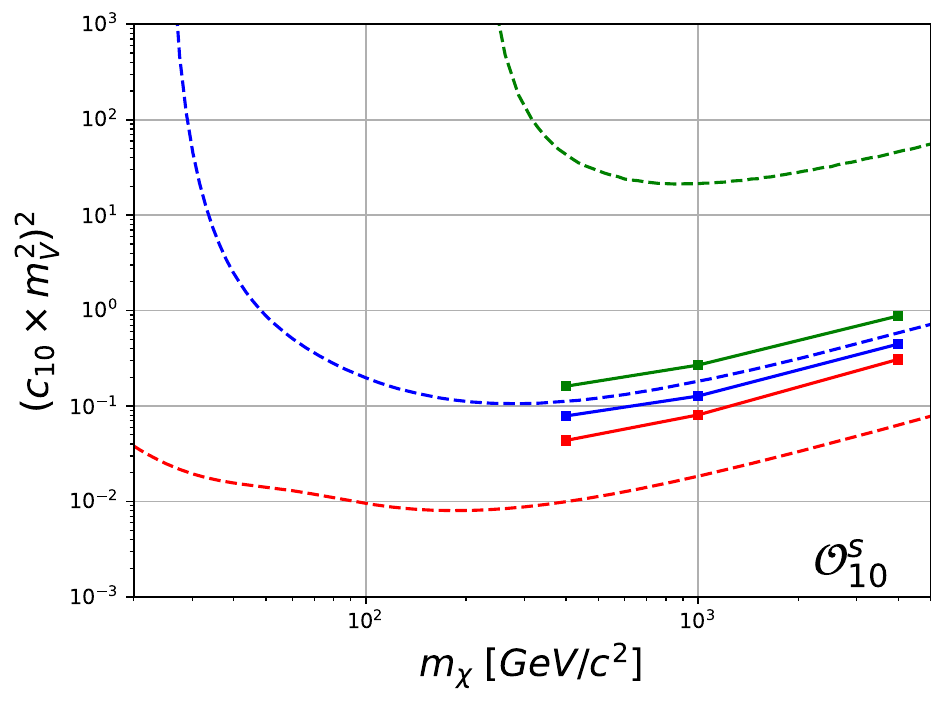}
    \end{subfigure}

    \vspace{0.1ex}

    \begin{subfigure}{0.32\textwidth}
    \includegraphics[width=\linewidth]{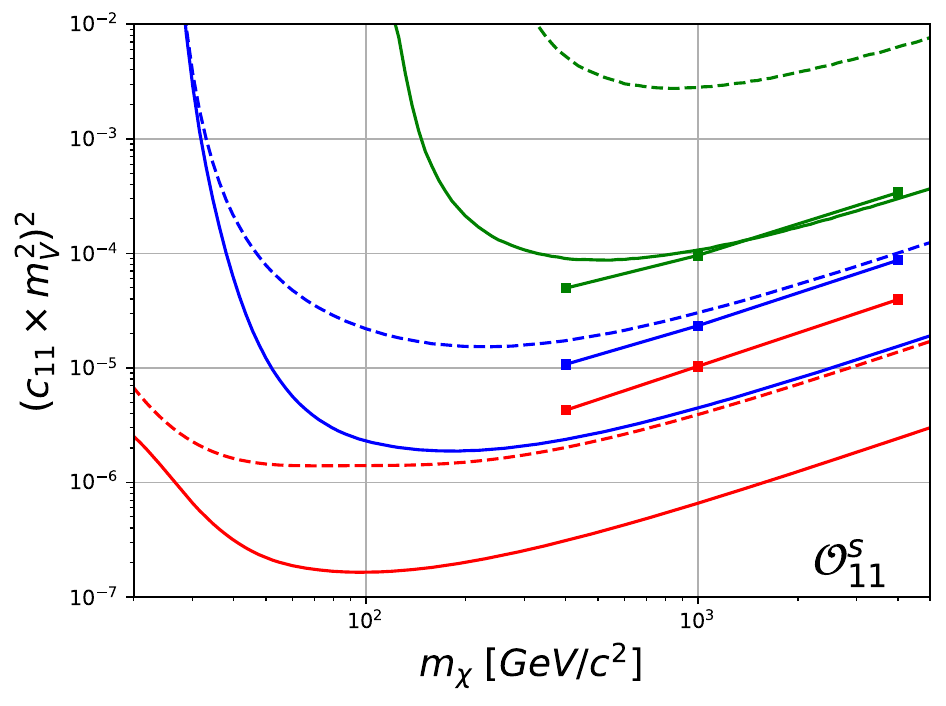}
    \end{subfigure}
    \begin{subfigure}{0.32\textwidth}
    \includegraphics[width=\linewidth]{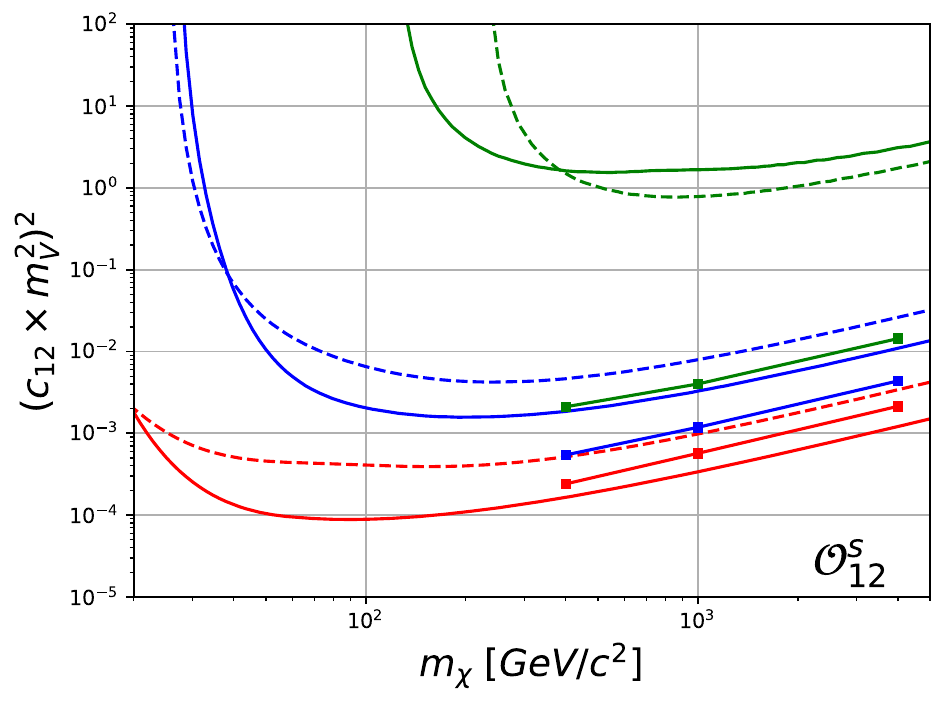}
    \end{subfigure}
    \begin{subfigure}{0.32\textwidth}
    \includegraphics[width=\linewidth]{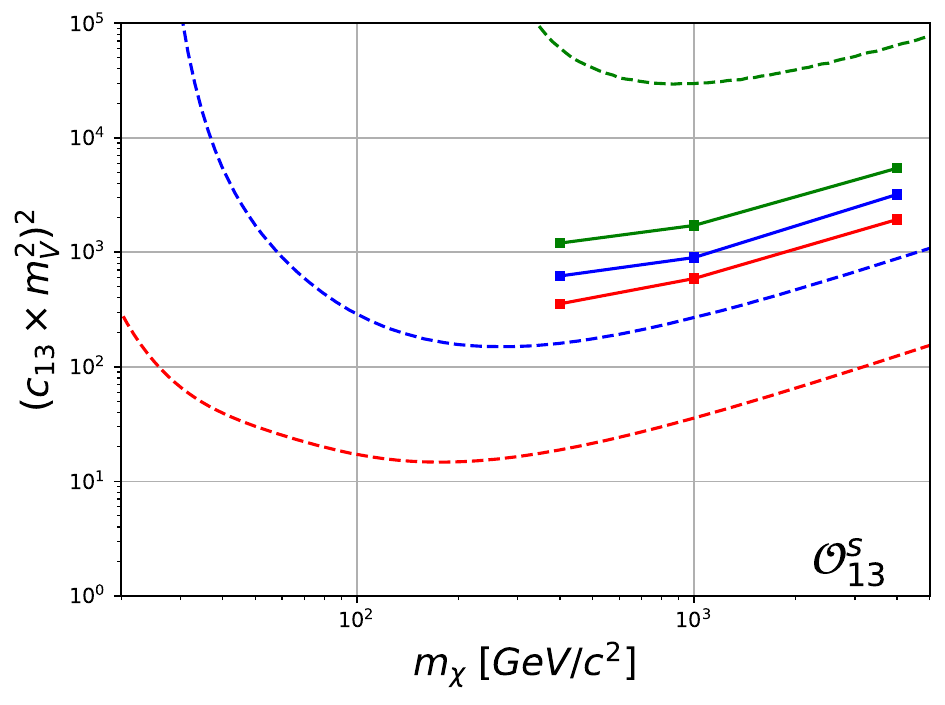}
    \end{subfigure}

    \vspace{0.1ex}

    \begin{subfigure}{0.32\textwidth}
    \includegraphics[width=\linewidth]{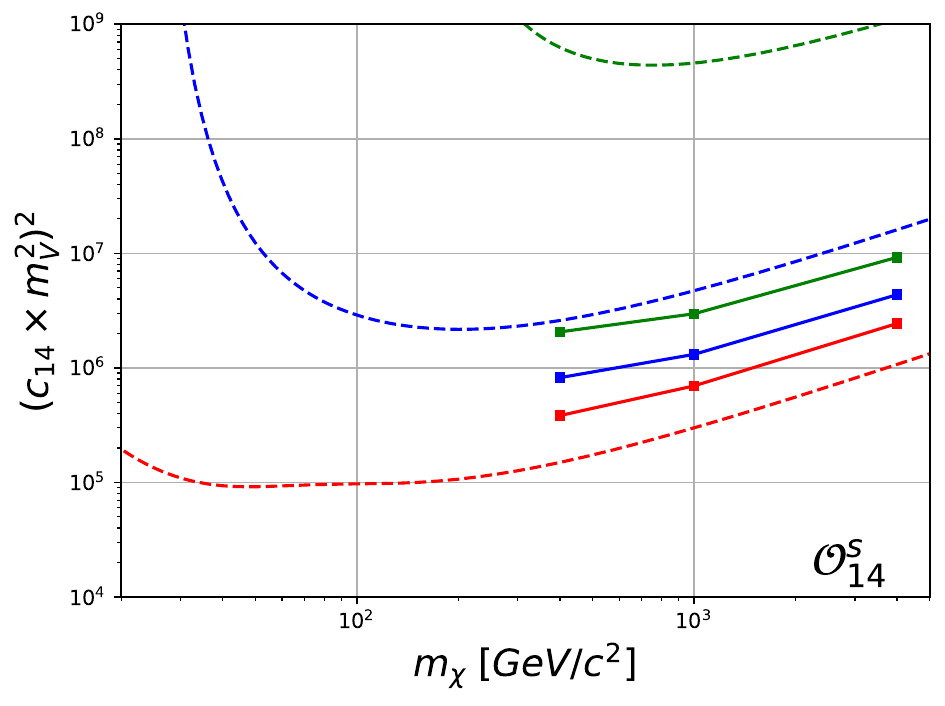}
    \end{subfigure}
    \begin{subfigure}{0.32\textwidth}
    \includegraphics[width=\linewidth]{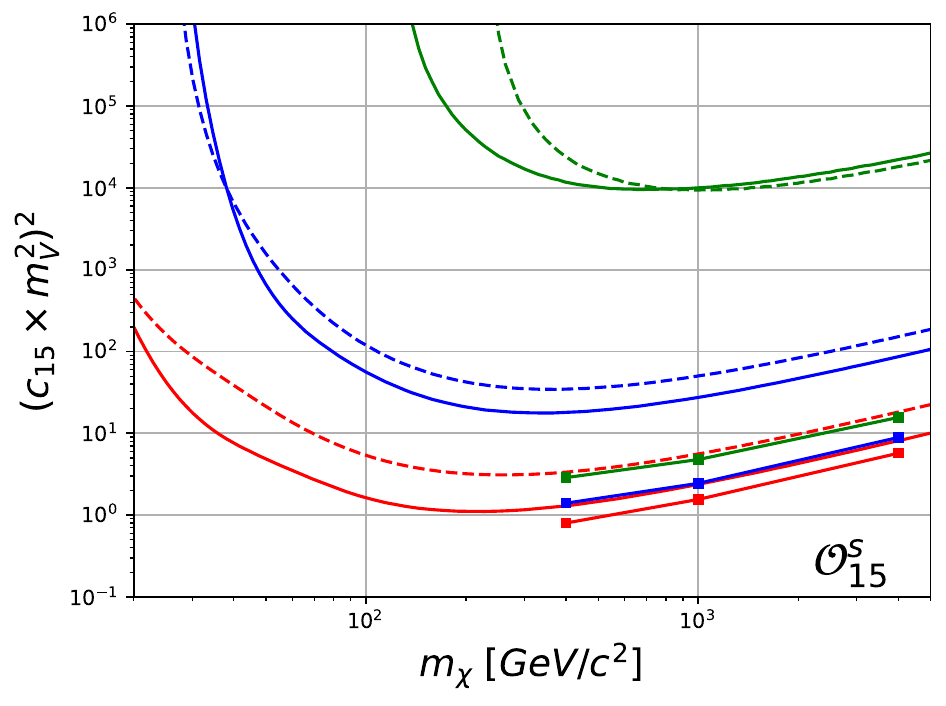}
    \end{subfigure}\hspace{4.55em}
    \begin{subfigure}{0.21\textwidth}
    \includegraphics[width=\linewidth]{Figures/UT_Austin/legend.pdf} 
    \end{subfigure}
    \caption{Projected 90\% confidence–level upper limits on the dimensionless isoscalar WIMP–nucleon NREFT coupling constants for inelastic scattering, in the HE scenario. Colored lines correspond to different mass splittings $\delta_{m}$, while square markers show the NREFT limits from the LUX–ZEPLIN experiment \cite{LZ:2023lvz}. The target minerals are gypsum (solid lines) and halite (dashed lines). Figures adapted from Ref.~{\cite{Theodosopoulos:2026ehn}}.}
    \label{fig:inelastic}
\end{figure}

We also consider inelastic scattering with mass splittings up to $\delta_m=100~\mathrm{keV}/c^2$. Figure~\ref{fig:inelastic} presents the projected limits for gypsum and halite in the HE scenario. Increasing the mass splitting suppresses the scattering rate and shifts sensitivity toward larger DM masses. For $\delta_m=50~\mathrm{keV}/c^2$, paleo-detectors are projected to improve upon LUX--ZEPLIN limits for several operators and remain sensitive to DM masses below those accessible to conventional experiments. For larger splittings, $\delta_m \sim 100~\mathrm{keV}/c^2$, the signal becomes strongly suppressed and sensitivity is progressively lost. 

In summary, paleo-detectors are projected to provide leading sensitivity to elastic WIMP--nucleon interactions at low masses ($1-10\ \mathrm{GeV}/c^2$) and competitive or superior sensitivity at higher masses ($>10\ \mathrm{GeV}/c^2$) through their large effective exposures. For inelastic scattering, they can improve upon existing limits for mass splittings up to approximately $50~\mathrm{keV}/c^2$, while sensitivity rapidly deteriorates for larger splittings. Sensitivity projections for additional target minerals are presented in the Appendix of Ref.~\cite{Theodosopoulos:2026ehn}.

\acknowledgments

D. T. acknowledges support from
the University of Texas at Austin and thanks Katherine Freese, Chris Kelso, and Patrick Stengel for
valuable discussions and collaboration. 
D. T. also acknowledges
support by the U.S. Department of Energy, Office of
Science, Office of High Energy Physics program under
Award No. DE-SC-0022021.
\clearpage

\section{Searching for Heavy Composite Dark Matter with Muscovite Mica and WIMP-like Dark Matter with Olivine}\label{sec:Queens}

Authors: {\it Levente~Balogh$^{1,2}$, Yilda~Boukhtouchen$^{1,3}$, Joseph~Bramante$^{1,3}$, Andrew~Buchanan$^{1,3}$, Audrey Fung$^{1,3,7}$, Kevin Gao$^{1,3}$, Alexander~Hayes$^{1,3}$, Matthew~Leybourne$^{1,4}$, Thalles Lucas$^{1,2}$, Jennika~McIntosh$^{1,3}$, Anupam~Ray$^{1,3,6\, \textcolor{red}{*}}$, Aaron~Shugar$^{5}$, and Aaron Vincent$^{1,3}$;}
\vspace{0.1cm} \\
\vspace{0.1cm} \\
{$^1$ Arthur B. McDonald Canadian Astroparticle Physics Research Institute, 64 Bader Lane,
Queen’s University, Kingston, Ontario, Canada}\vspace{0.1cm} \\
{$^2$ Department of Mechanical and Materials Engineering, Queen’s University, Kingston, Ontario, Canada}\vspace{0.1cm} \\
{$^3$ Department of Physics, Engineering Physics, and Astronomy, Queen’s University, Kingston, Ontario, Canada}\vspace{0.1cm} \\
{$^4$ Department of Geological Sciences and Geological Engineering, Queen’s University, Kingston, Ontario, Canada}\vspace{0.1cm} \\
{$^5$ Department of Art History and Art Conservation, Queen’s University, Kingston, Ontario, Canada}\vspace{0.1cm} \\
{$^6$ Perimeter Institute for Theoretical Physics, Waterloo, Ontario, N2J 2W9, Canada}\\
{$^7$Asia Pacific Center for Theoretical Physics, Postech, Pohang 37673, Korea}
\\
\subsection{Introduction}
Dark matter, which constitutes a substantial fraction of the Universe, remains one of the most profound mysteries in modern physics, with its fundamental nature still unknown. While its gravitational effects are well established across multiple astrophysical and cosmological scales, any non-gravitational interactions with Standard Model particles have yet to be observed. Among the many proposed dark matter candidates, composite dark matter, namely bound states of dark-sector constituents formed through nucleosynthesis-like processes, phase transitions, or dissipative collapse, represents a particularly rich and relatively underexplored region of theoretical parameter space. Just as visible matter organizes into atoms, nuclei, and macroscopic structures through the interplay of attractive forces and binding energies, analogous dynamics in the dark sector can give rise to composite states spanning a wide range of masses and spatial scales (see~\cite{Bramante:2026wzh} for a recent review).

Recently, we have been focusing on heavy composite dark matter candidates and show that muscovite mica can serve as an excellent paleo-detector to search for their interactions. Heavy dark matter, especially those with masses above the Planck scale $(\sim 10^{19}~\rm{GeV})$, can not be probed by conventional direct detection experiments as the expected dark matter flux through a typical meter-scale detector falls below one particle per year.  Paleodetectors offer a promising alternative by exploiting geological exposure times which are often over billions of years.  Among potential mineral targets, muscovite mica is particularly attractive because it possesses low intrinsic radioactivity, excellent cleavage properties, and a demonstrated ability to preserve damage features over geological timescales. Furthermore, mica can be obtained in large, high-quality sheets, enabling macroscopic detector areas.

The central idea explored in a recent article (see Ref.~\cite{Boukhtouchen:2026rfz}) is that heavy composite dark matter deposits substantial energy while traversing  mica sample. If the local energy deposition is fast enough and exceeds the melting threshold of the mineral, a  melt region is produced along the trajectory. Such melt-track damage can remain preserved over billion-year timescales, providing a unique geological record of dark matter interactions. The key challenge, however, is how to reliably detect and characterize such damages, and we identify a novel readout technique that enables detection of these melt-tracks with high precision. \vspace*{-0.75 cm}\\
\subsection{Composite Dark Matter Interactions and Melt-Track Formation}
\label{sec2}
In the analysis presented in Ref.~\cite{Boukhtouchen:2026rfz}, we focus on energy deposition by heavy composite dark matter, considering both opaque composites and diffuse composites. These two limiting cases encompass a broad range of possible composite dark matter realizations and capture the dominant mechanisms by which energy is transferred to the target material.

In the opaque limit, the composite behaves as a dense object whose constituents are effectively shielded from incoming nuclei. Consequently, target nuclei interact primarily with the geometric surface of the composite, and the energy deposition is determined by the object's size, velocity, and geometric cross section. In contrast, in the diffuse limit, the composite is sufficiently dilute that target nuclei can penetrate its interior and scatter directly from its constituents. In this case, the energy deposition depends not only on the overall size of the composite but also on its internal structure and constituent density.

Although the microscopic interactions differ between these two regimes, both can be described within a common framework based on the rate of energy deposited per unit path length in the target material. This deposited energy ultimately determines whether the local temperature exceeds the melting threshold required for melt-track formation in mica. A detailed derivation of the corresponding energy-deposition rates and their dependence on the composite properties can be found in~\cite{Boukhtouchen:2026rfz}. For reference, in the following, we briefly outline the theoretical melt-track radius calculation in the opaque (geometric) limit.

In the opaque limit, a macroscopic dark matter state interacts by displacing all target nuclei within its geometric cross-sectional area $ \pi R_{D}^2$
where $R_{D}$ is the radius of the dark matter composite. The total kinetic energy deposited per unit path length is therefore given by $dE_{\rm dep}/dx =\rho v^2_D \pi R_{D}^2$ where $\rho$ denotes the density of the mica and $v_D \approx 220$~km/s denotes typical velocity of the dark matter composite. Of this deposited energy, only a fraction contributes to local heating and subsequent melt-track formation; we parameterize this efficiency by $\eta = 0.7$.

Now, to form a contiguous melt track of radius $R_{\rm melt}$, the deposited thermal energy must raise the temperature of the enclosed cylindrical volume to the melting point of the mineral and supply the latent heat of fusion. This requires an energy per unit length: $dE_{\rm req}/dx \approx \rho \left(C_p \Delta T_{\rm melt} + H_f \right)\pi R_{\rm melt}^2$ where $C_p$ is the mineral specific heat capacity, $\Delta T_{\rm melt}$ is the temperature increment to the melting point and $H_f$ is the specific latent heat for melting the mica. Equating the effective heat deposited, $\eta dE_{\rm dep}/dx $ to $dE_{\rm req}/dx$
yields the condition for melt-track formation, which is independent of the target density
\begin{equation}
R_{\rm melt} = R_D \sqrt{\frac{\eta v_D^2}{C_p \Delta T_{\rm melt} + H_f}} \approx 220 R_D\,.
\end{equation}
Clearly, for a typical composite radius of $\mathcal{O} (1)\mu m$, which is of primary interest here, we have to look for $\mathcal{O} (100)\mu m$-scale damage in the mica sample. Here, it is also worth noting that Ref.~\cite{Boukhtouchen:2026rfz} validated this theoretical melt-track radius using SRIM/TRIM simulations~\cite{zieglersrim2010}, finding excellent agreement in the small-composite-radius regime, while deviations become increasingly significant for larger composite sizes $(R_D \geq 1 \mu m)$. The primary origin of this deviation is that, for larger composites, the deposited energy becomes sufficiently large that SRIM's underlying assumption of an unperturbed crystalline lattice breaks down, causing the cascade-based picture to systematically underestimate the melt radius relative to the geometric expectation.
\subsection{Experimental Readout of Melt-Track in Mica with X-ray Fluorescence Technique}
\label{sec3}
A major task in this work is now the identification of micron-scale damage features in mica samples. To this end, we develop a novel and practical readout technique capable of efficiently surveying large mica areas~\cite{Boukhtouchen:2026rfz}. The method is essentially based on X-ray fluorescence transmission imaging in combination with a copper (Cu) backing sheet.

\begin{figure}[!t]
\centering
\includegraphics[width=0.9\linewidth]{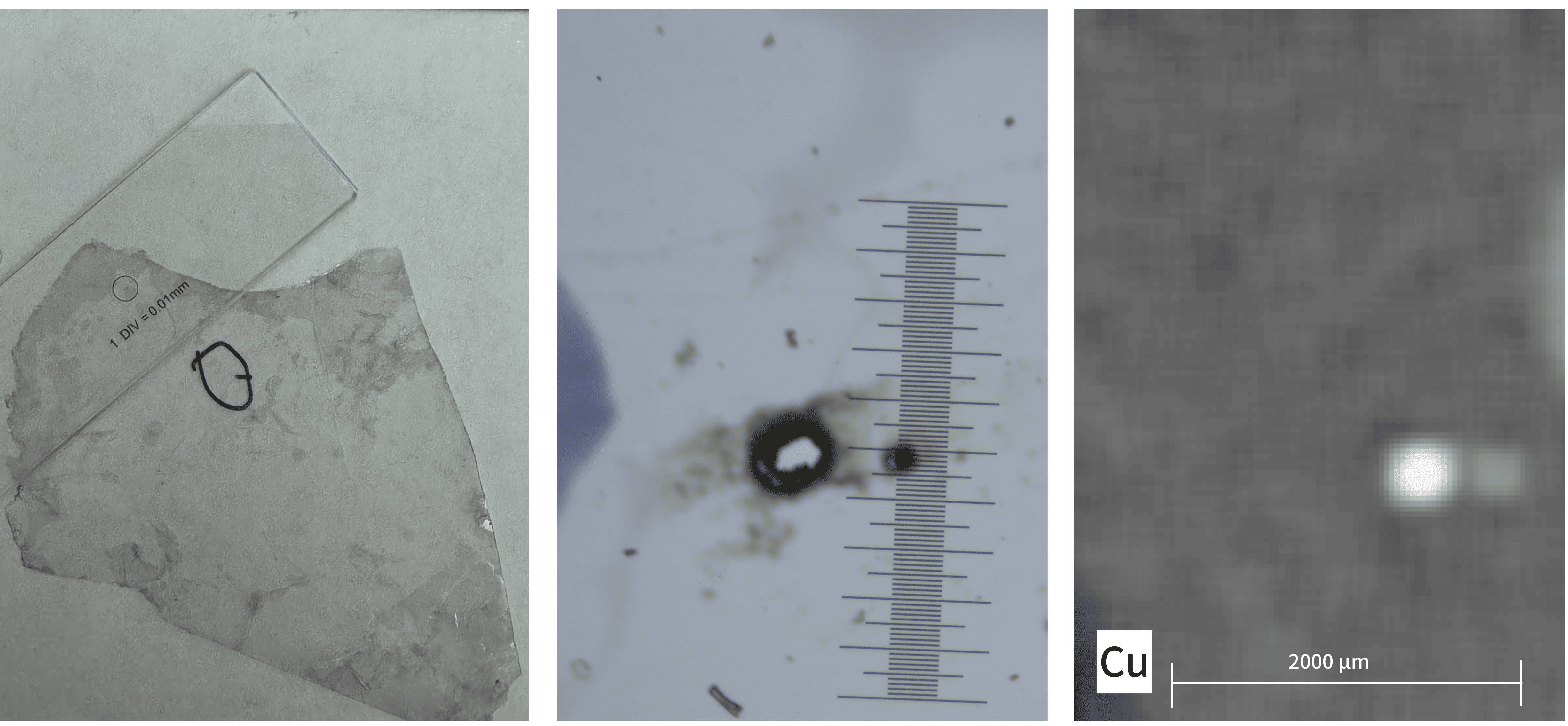}
\caption{From left to right, three views of the same calibration sample.
\textit{Left:} wide-field photograph with the two laser-ablated calibration spots circled; 
\textit{Middle:} optical microscopy of the marked region, showing the laser-ablated melt tracks of
$\sim 150$ and $\sim 50~\mu$m diameter; \textit{Right:} X-ray fluorescence map, in which each laser-induced feature appears as a localized enhancement in the characteristic Cu X-ray emission. See Ref.~\cite{Boukhtouchen:2026rfz} for more details.}
\label{fig:xraycalibration}
\end{figure}

To calibrate the technique, two artificial damage features ($\sim$ 50 and 150 $\mu m$) were produced via laser ablation. The mica sample was then scanned over an area of 8 $\times$ 4.5 mm with a step size of 30 $\mu m$ and a total scan time of 18 minutes. Both laser-melted regions are clearly identified in the fluorescence map as localized enhancements in count rate relative to the intact mica background (see Fig.~\ref{fig:xraycalibration}), demonstrating successful calibration of the method.
\subsection{Results: Sensitivity on Heavy Dark Matter Interactions}
\label{sec4}
Building upon the energy-deposition models of Section~\ref{sec2} and the XRF readout calibration of Section~\ref{sec3}, we can combine the relevant geological, astrophysical, and detector inputs to determine the projected dark matter sensitivity. A detailed discussion of the geological selection criteria adopted in this analysis, including sample selection, long-term track preservation in muscovite mica,  can be found in Ref.~\cite{Boukhtouchen:2026rfz}.

For our sensitivity estimates, it is important to account for the attenuation of the dark matter flux as it traverses the Earth's overburden before reaching the mica sample. Since the amount of attenuation depends on the column density encountered along the dark matter trajectory, an accurate description of the Earth's internal density structure is required.

\begin{figure}[!t]
\centering
\includegraphics[width=0.48\textwidth]{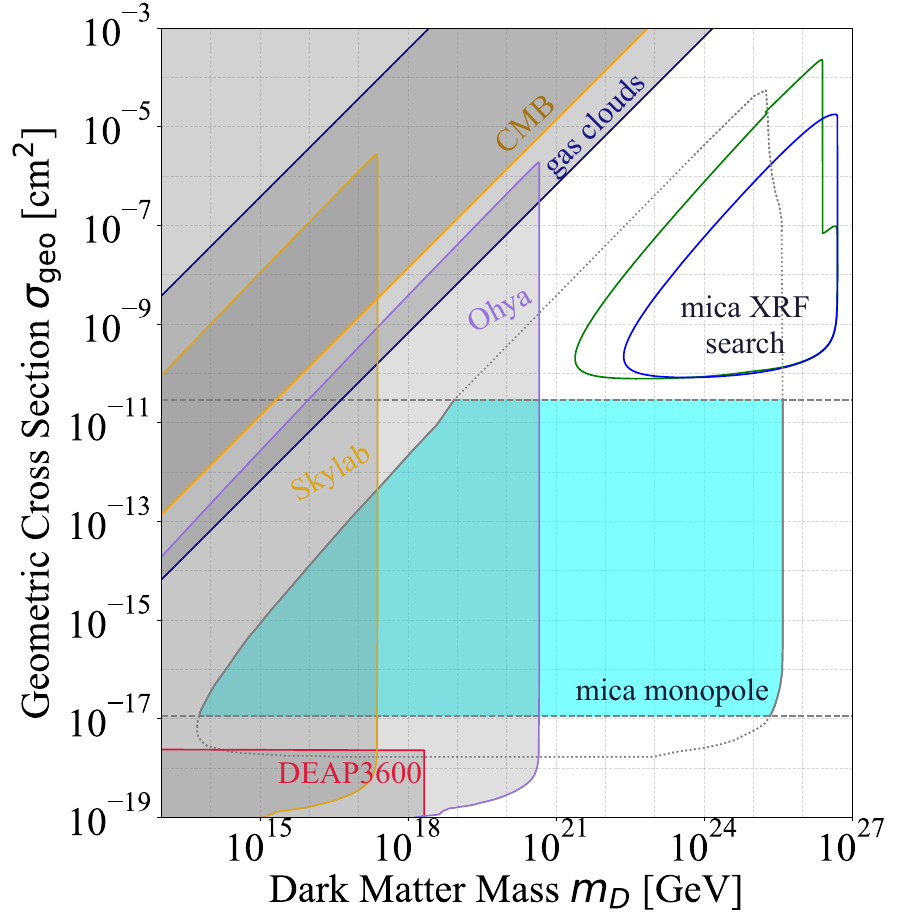}
\hspace{0.1 cm}
\includegraphics[width=0.48\textwidth]{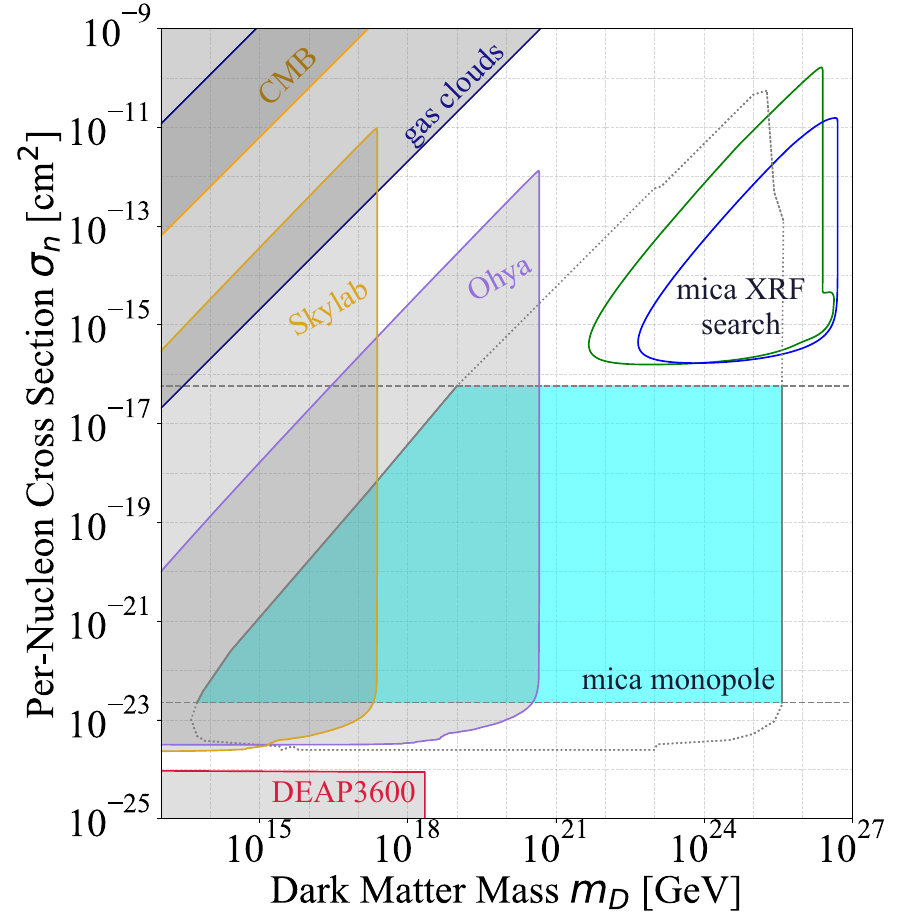}
\caption{Projected 90\% C.L. sensitivity to heavy composite dark matter interactions, assuming a benchmark exposure of 1 m$^2$ $\times$ 10$^9$ year of cratonic muscovite. (Left) is for the opaque geometric limit of energy deposition and (right) is for diffuse loosely bound interaction regimes discussed in Section~\ref{sec2}. Existing astrophysical/terrestrial limits (shaded) are shown for comparison (See Ref.~\cite{Boukhtouchen:2026rfz} for more details).}
\label{fig:bounds}
\end{figure}

The other key effect is the orientation of the mica sample relative to the incoming dark matter flux. Since the expected signal depends on the path traversed by dark matter, the effective path length can vary significantly with the sample’s orientation and geographic location. Moreover, the Earth’s rotation continuously changes the direction of the incoming dark matter wind, resulting in a time-dependent overburden. Properly accounting for the sample orientation is therefore essential for accurately determining the surviving dark matter flux and the resulting experimental sensitivity. In Fig.~\ref{fig:bounds}, we present results for two limiting configurations: one with the sample oriented parallel to the Earth’s surface (green) and another with the sample oriented perpendicular to it (blue). The true sensitivity for any given sample orientation is expected to lie between these two extremes. Here, it is worth-noting that one might wonder whether variations in the dark matter density profile over billion year timescales could affect this sensitivity; however, the Standard Halo Model provides a sufficiently robust description for our purposes.

\subsection{WIMP-like dark matter searches}
\begin{figure}
    \centering
    \includegraphics[width=0.7\linewidth]{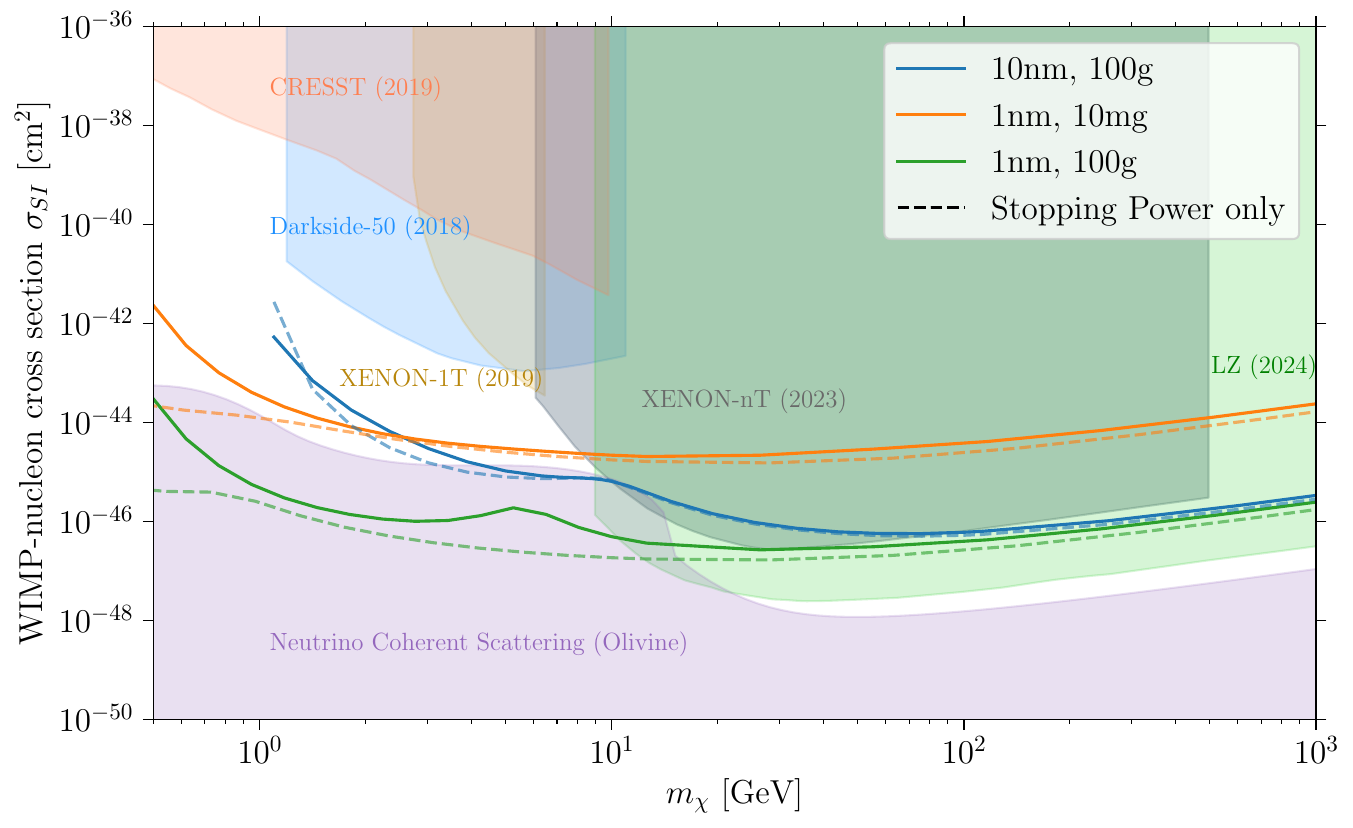}
    \caption{Projected sensitivities to WIMP-like dark matter using an olivine target, under different exposure and resolution assumptions. Dashed lines show previous expectation using stopping power only to compute track length; solid lines show full results using Monte Carlo-informed TLD. Figure from \cite{Fung:2025cub}.}
    \label{fig:wimplimits}
\end{figure}

A second major direction in paleodetection is the theoretical and experimental validation of predicted track length distributions (TLDs) from background and new physics signals. Ref.~\cite{Fung:2025cub} used Monte Carlo simulations to properly characterize the the expected TLD given nuclear recoil energy $E_R$ in olivine. As previous studies had used a one-to-one relationship between $E_R$ and track length, this more rigorous treatment showed that the smearing produced by stochastic paths of the primary knock-on atom in the mineral leads to slightly worsened sensitivities, as can be seen in Fig.~\ref{fig:wimplimits}. This is especially true at low recoil energies, where many collisions will fail to produce tracks at all. Also worth noting is that in all cases, the signal in these low-recoil energy searches falls well below the expected background, which mainly comes from U/Th decay chain neutrons. This means that accurate and precise modeling of background is essential for a likelihood-based reconstruction of any beyond the SM signal.

Current theory directions at Queen's include in-depth calculations of reconstruction techniques such as small-angle X-ray scattering (SAXS). One important obstacle to producing reliable prediction is exact knowledge of the expected track shape at low recoil energies--these are being investigated using molecular dynamics simulations.

Ongoing experiments at the Reactor Materials Testing Laboratory (RMTL) at Queen's University aim to compare tracks produced in olivine and galena from an incident 3-5 MeV proton beam with predicted distributions from simulations. Preliminary tests have been made, and an upgraded thermal dissipation/sample holding device has been designed.

\clearpage

\section{Fluorescent Microscopy to Detect Nuclear Recoil-Induced Damage}\label{sec:Hedges_Talk}

Authors: {\it Samuel C. Hedges and Patrick Huber}
\vspace{0.1cm} \\Center for Neutrino Physics, Virginia Tech, Blacksburg, VA, USA
\vspace{0.3cm}

\subsection{Imaging Nuclear Recoil Damage in
LiF with Light Sheet Microscopy}
In certain materials, vacancies created by nuclear recoil damage can form optically-active color centers~\cite{cogswell:2021qlq}. 
The threshold for producing this damage is material-dependent, typically on the order of tens of eV, allowing this detection channel to be sensitive to low energy nuclear recoils. 
The induced damage to the underlying crystal lattice can persist at room temperature over long timescales, and it is difficult for low energy electromagnetic interactions from gamma rays or electrons to produce similar defects~\cite{araujo2025nuclear}. 
The generated color centers can potentially be read out non-destructively through fluorescence microscopy.

\subsubsection{Color center detection in lithium fluoride}
There is much existing effort observing fluorescent damage tracks in lithium fluoride (LiF) \cite{PICCININI2024107140,bilski:2024ghu}. 
Color-center production in LiF is well studied, and the produced color centers are excitable with visible wavelengths of excitation light. 
LiF also boasts an \textit{in-situ} calibration capability produced via thermal neutron capture on lithium-6. 
The capture process has a large cross section and produces back-to-back monoenergetic alphas and tritons. 

Two common readout techniques for fluorescent damage tracks are widefield microscopy \cite{bilski:2024ghu} and confocal microscopy~\cite{PICCININI2024107140}. Widefield microscopy can rapidly read out damage, but is restricted to shallow depths, and positional resolution is degraded by out-of-focus fluorescence. Confocal microscopy can probe further depths with improved positional resolution, but is limited by slow read-out times. 

Recently, Selective-Plane Illumination Microscopy (SPIM) has demonstrated the readout of color centers in bulk LiF~\cite{araujo2025nuclear,vladimirov:2023}. SPIM shares the rapid readout of widefield microscopy, but the excitation light predominantly interacts only in the illumination plane, allowing it to probe deep into samples while retaining good position resolution.

\subsubsection{Preliminary studies from the VT benchtop mesoSPIM}
A customized version of the Benchtop mesoSPIM~\cite{vladimirov:2023} light sheet microscope has recently been commissioned at Virginia Tech. This instrument is dedicated to observing nuclear recoil damage, and features an upgraded single-photon counting camera. With the VT mesoSPIM, hundreds of mm${}^3$ of LiF were scanned over a few days, and data sets with different proportions of fast-to-thermal neutron exposure were collected. Fast elastic scatters and thermal neutron captures have been observed in the data, which is currently under analysis.

We have also conducted preliminary studies of bleaching of color centers in LiF. These studies suggest that while some bleaching occurs, excitation laser power can be increased to further reduce readout time. Additional bleaching studies are underway.

\subsubsection{Simulating fluorescent damage tracks}
\begin{figure}
\centering
\begin{subfigure}{0.48\textwidth}
  \includegraphics[width=\linewidth]{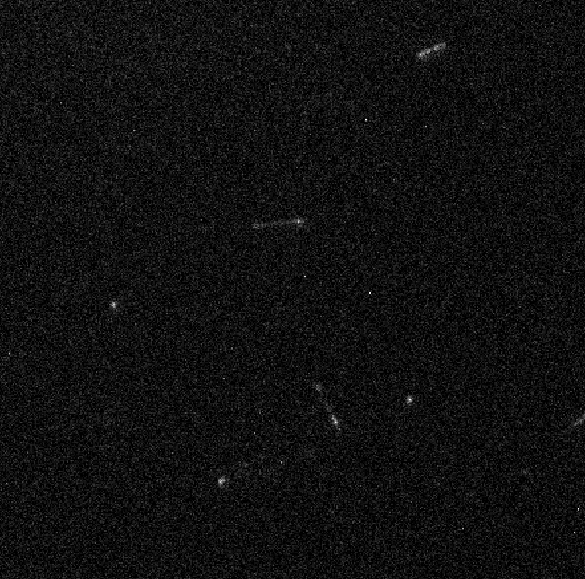}
  \caption{Data from the VT mesoSPIM.}
\end{subfigure}
\hfill
\begin{subfigure}{0.48\textwidth}
  \includegraphics[width=\linewidth]{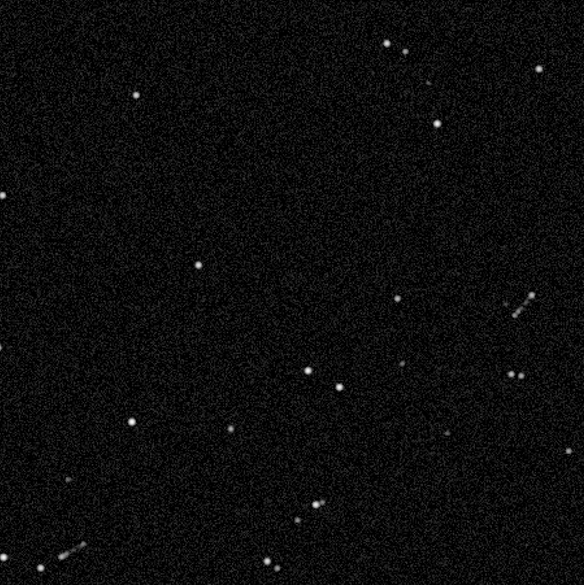}
  \caption{Simulated data.}
\end{subfigure}
\caption{Comparison of simulated and observed images of neutron-irradiated LiF.}
\label{Fig:mesoSPIM_sim_and_data}
\end{figure}

A custom simulation framework has been developed, in which particle interactions in LiF are modeled with Geant4~\cite{GEANT4:2002zbu}, while specific damage tracks are sampled from pre-computed libraries generated with TRIM~\cite{zieglersrim2010}. The simulated data incorporate measured optical and noise parameters specific to the VT mesoSPIM, and the framework produces simulated images visually similar to those acquired with the mesoSPIM. While further tuning of the simulation is required, the goal is to use simulated data to test analysis code for bias, and to generate training sets for machine-learning recoil track identification. An example simulated microscope image of LiF tracks along with acquired data (prior to post-processing) can be found in Fig.~\ref{Fig:mesoSPIM_sim_and_data}.

\subsection{Calorimetric readout of
paleo-detectors}
The majority of existing paleodetector studies focus on the length of produced damage tracks, which is proportional to the deposited energy. While SPIM allows for new searches for Dark Matter enabled by its rapid readout time, there are fundamental limits to its position sensitivity, dependent on the wavelength of excitation light. For sub-micron scale damage tracks, obtaining useful information about energy depositions from track length will be marginal. 

However, the number of produced color centers (i.e. the brightness of the fluorescence integrated across the track) is also proportional to the deposited energy. SPIM is well-suited to utilize this quantity as an energy estimator. We contrast analyses using the number of vacancies versus track length for a Dark Matter sensitivity analysis, and show an additional benefit by combining these two quantities to identify recoiling particle species. 

Simulations of tens of thousands of damage tracks were completed across a broad range of energies using TRIM. TRIM can produce damage tracks using two different modes: \textbf{quick mode}, which employs the Kinchin-Pease statistical approximation of the number of vacancies produced at each primary knock-on atom (PKA) interaction site (secondaries are not tracked), and \textbf{full-cascade mode}, in which both the PKA and secondaries are tracked as the PKA deposits energy. Full-cascade mode is necessary for low energy depositions, where the concept of a track length becomes nebulous as a result of the small number of produced vacancies. It has been posited there is a bug in the full-cascade mode output of TRIM~\cite{AGARWAL202111}; our findings indicate for LiF the full and quick modes differ by approximately a factor of two or less.

\begin{figure}
    \centering
    \includegraphics[width=0.7\linewidth]{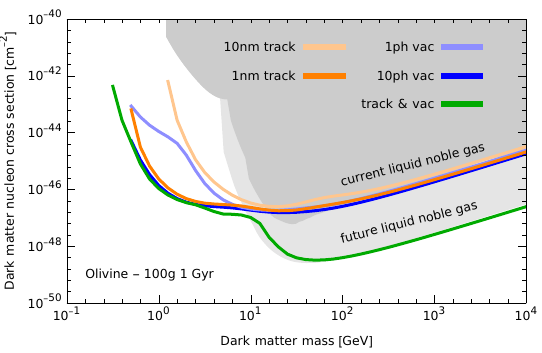}
    \caption{Projected $90\%$ C.L. sensitivity limits on the spin-independent
      dark matter--nucleon cross section as a function of dark matter
      mass, for $100\,\mathrm{g}$ of $1\,\mathrm{Gyr}$ old
      olivine. The green curve shows the combined analysis using the
      joint (track length, vacancy count) likelihood; the orange and
      blue curves show the track-only and vacancy-only envelopes. The dark-gray and light-gray
      shaded regions indicate the current and projected reach of
      liquid noble gas direct-detection experiments. All results use
      full cascade simulations. Figure and caption from Ref.~\cite{hedges2026calorimetricapproachpaleodetectiondark}.}
    \label{fig:dm}
\end{figure}

A combined analysis of both track length and number of vacancies provides additional information on the identity of the PKA. As neutron and WIMP dark matter cross sections are expected to scale differently based on the nuclear species, the combined analysis improves background rejection, probing additional dark matter phase space. For additional details on this analysis, see Ref.~\cite{hedges2026calorimetricapproachpaleodetectiondark}. The key result is that the combined sensitivity is comparable to future large liquid-noble gas detectors for $m_\chi>10\,$GeV and far exceeds it for $m_\chi<10$\,GeV as shown in Fig.~\ref{fig:dm}.

\acknowledgments

This material is based upon work supported by the Department of Energy National Nuclear Security Administration through Defense Nuclear Nonproliferation's Enabling Capabilities in Technology Consortium under Award Number DE-NA0004197. It was also supported by the
U.S. Department of Energy Office of Science under award number DE-SC0020262. It was also supported by a National
Science Foundation Growing Convergence Research award number 2428507.

\clearpage

\section{The Bedretto Underground Laboratory for Fundamental Physics: Developing~a~Future~Underground~Microscopy~Facility}\label{sec:UZH_Bedretto}

Authors: {\it Luisa~H\"otzsch\,$^1$, Florian J\"org\,$^1$, Laura Baudis\,$^1$, Alexey Elykov\,$^2$, Patrick Huber\,$^3$, Patrick Stengel\,$^4$}

\vspace{0.1cm}
\noindent $^1$Universit\"at Z\"urich, Switzerland

\noindent $^2$Karlsruhe Institute of Technology, Germany

\noindent $^3$Virginia Tech, United States of America

\noindent $^4$Jo\v{z}ef Stefan Institute, Slovenia

\vspace{0.3cm}

Rare event searches such as direct dark matter detection or measurements of the coherent elastic neutrino nucleus scattering (CE$\nu$NS) process require strong minimization of any source of interfering backgrounds. Therefore, such experiments are located in background suppressed environments provided by underground laboratories. The large rock overburden largely attenuates the flux of atmospheric muons and other cosmic rays.

We present the studies for a new underground laboratory site situated in the Bedretto tunnel in Ticino, Switzerland, that would constitute one of the world's most background free environments, and discuss its potential as an underground microscopy facility for the detection of neutrinos and dark matter in lithium fluoride.

\subsection{Background characterization of BedrettoLab Physics site}

The Bedretto Underground Laboratory for Geosciences and Geoenergies (BedrettoLab), operated by the Eidgenössische Technische Hochschule (ETH) Z\"urich, is located inside the Bedretto tunnel in the southern Swiss Alps below the Gotthard massif\,\cite{bedretto_charaterization}. At a distance of 3.5\,km from the tunnel entrance (TM3500), the tunnel features a maximum vertical overburden of about 1\,500\,meters, making it a very promising candidate site for a future low background physics laboratory. 

\begin{figure}[ht]
    \centering
    \includegraphics[height=0.175\paperheight]{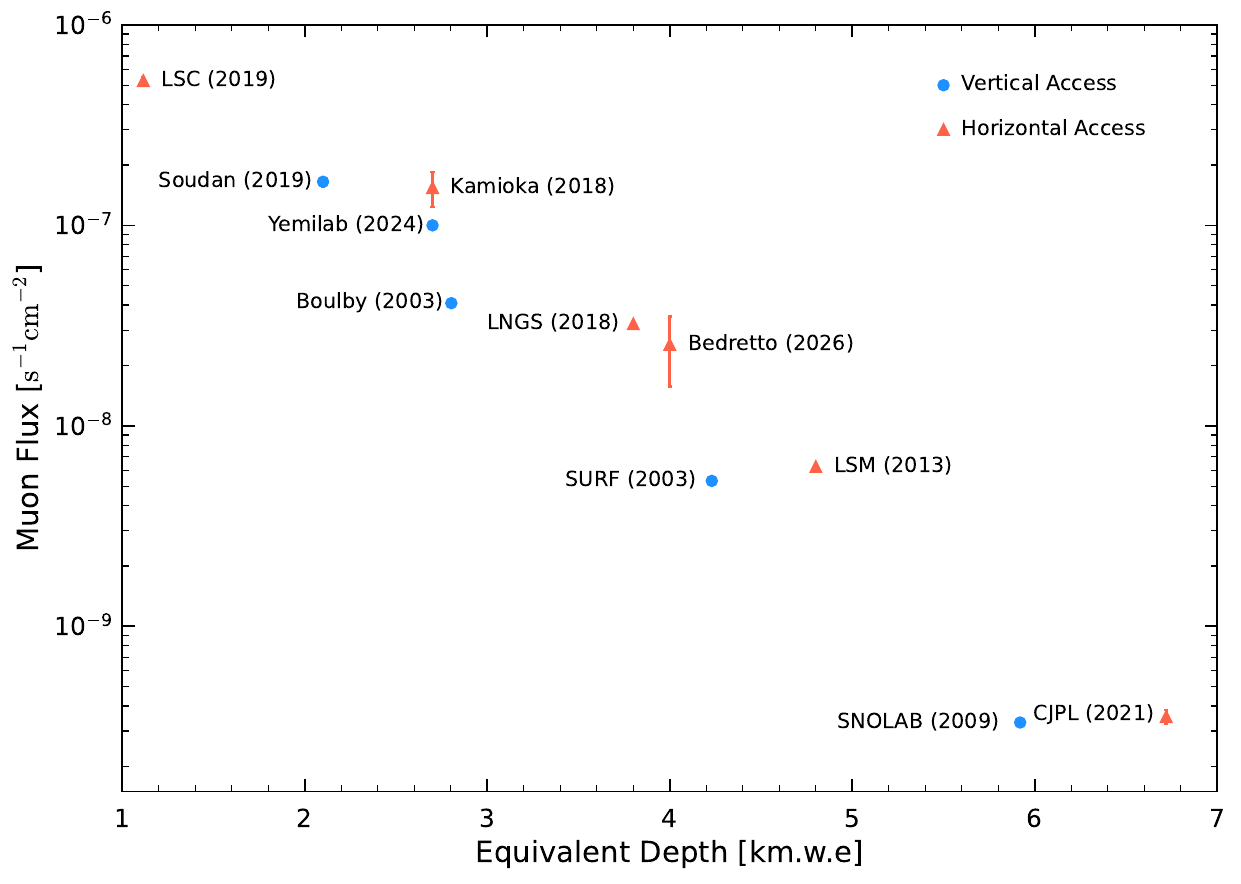}%
    \includegraphics[height=0.175\paperheight]{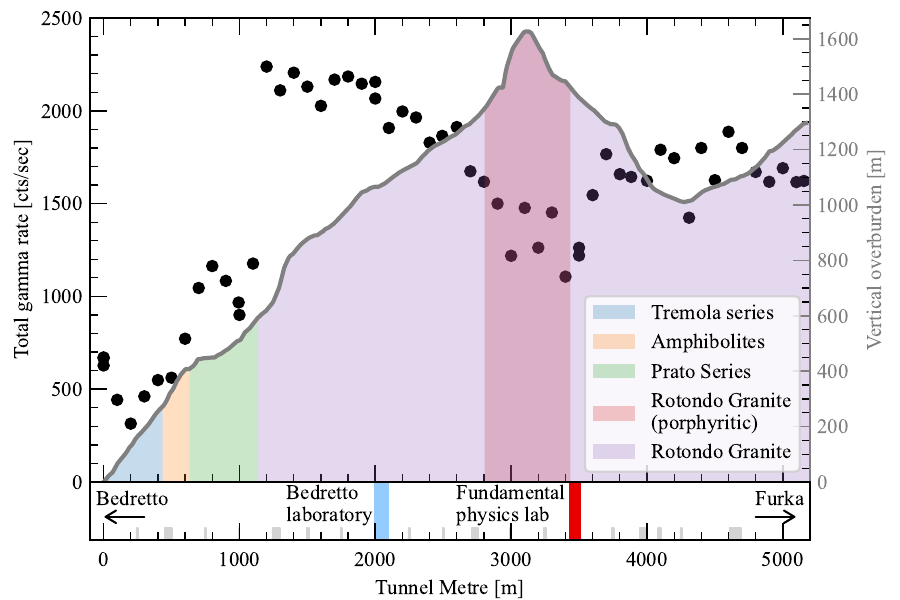}
    \caption{\textbf{Left:} Comparison of the muon flux at different deep underground laboratories across the world. At an equivalent depth of about 4\,km.w.e, the envisioned BedrettoLab Physics would be the second deepest lab in Europe. \textbf{Right:} Measurement of the gamma ray flux along the Bedretto tunnel (black points) together with the elevation profile (gray). The foreseen location of the new BedrettoLab Physics at TM3500 is marked in red.
    Figures taken from\,\cite{Penning:2025mek}.}
    \label{fig:UZH_Bedretto_muon_gamma_flux}
\end{figure}

In 2025, we conducted a comprehensive background characterization campaign in the bare rock cavern at the TM3500 site\,\cite{Penning:2025mek}. The main focus of this campaign was a measurement of the muon flux, which was found to be $\Phi_\mu = (2.54~\pm~0.97)~\times 10^{-8}~\mathrm{s^{-1}} \mathrm{cm^{-2}}$, corresponding to a reduction by a factor of $10^6$ compared to the surface flux. As shown in Figure\,\ref{fig:UZH_Bedretto_muon_gamma_flux} (left), this would make a future laboratory at this site (BedrettoLab Physics) the second deepest European underground laboratory, outperforming established research infrastructures such as the Laboratori Nazionali del Gran Sasso (LNGS), Italy in terms of the expected muon background flux.

Further radiative background sources have been assessed including the neutron and gamma flux of $\Phi_\mathrm{n} = (5.56 \pm 0.26)\times10^{-5}~\mathrm{s}^{-1} \mathrm{cm}^{-2}$ and $\Phi_\gamma = (5.67\pm0.37)\,\mathrm{s}^{-1} \mathrm{cm}^{-2}$ respectively (see Figure\,\ref{fig:UZH_Bedretto_muon_gamma_flux}, right). The radon in air concentration was found to fluctuate between 1.5 and 3\,kBq/cm$^3$. These measurements were carried out in the bare rock cavern, providing guidance for a future outfitting of the cavern. In particular, they show that background levels comparable to other deep underground laboratories can be achieved using standard methods like concrete wall cladding and fresh air ventilation. Furthermore, non-radiative background components such as vibrations and electromagnetic interference, which are relevant for gravitational wave searches, were found at exceptionally low levels\,\cite{Penning:2025mek}.

The path towards a permanent underground facility for physics in the Bedretto tunnel follows a multi staged approach. First science experiments would become possible within a container setup. A full underground laboratory infrastructure is foreseen to be established via dedicated cavern space excavation at the TM3500 site. 
Based on the engineering experience in the tunnel from the existing geology laboratory space, such excavations can be performed using the commonly utilized drill and blast method, allowing for their completion within about a year or less.

\subsection{Underground microscopy with mesoSPIM at BedrettoLab Physics}

\begin{figure}
    \centering
    \includegraphics[width=0.5\linewidth]{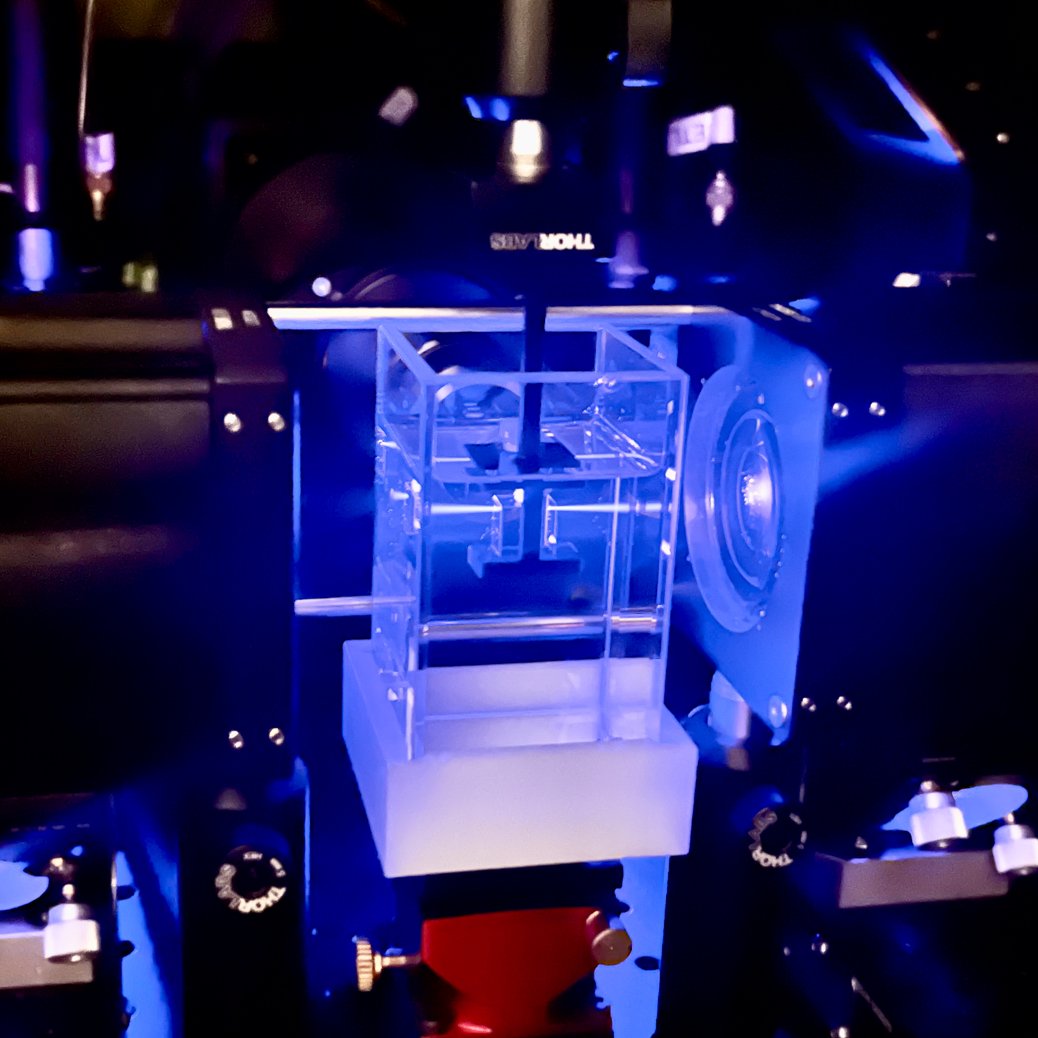}
    \caption{Picture of a LiF crystal being imaged in the benchtop mesoSPIM at the center for microscopy and image analysis (ZMB) at UZH.}
    \label{fig:uzhmesospim}
\end{figure}

A promising imaging technique for color centers from nuclear recoil damage features in crystals is fluorescence light-sheet microscopy, specifically mesoscale Selective Plane Illumination Microscopy (mesoSPIM)~\cite{vladimirov:2023}.
Originally developed at the Center for Microscopy and Imaging Analysis (ZMB) at the University of Zurich (UZH), these devices have been successfully used to image color center tracks produced by thermal-neutron captures, fast neutrons, and gamma rays in artificial lithium fluoride (LiF) crystals~\cite{araujo2025nuclear}.
A photograph of the imaging process of a 1\,cm$^3$ LiF crystal at UZH is shown in Figure\,\ref{fig:uzhmesospim}.

Aiming for the detection of dark matter and a measurement of CE$\nu$NS from reactor neutrinos via nuclear recoil detection in LiF crystals, we propose to deploy customized mesoSPIM devices at the future Bedretto Underground Laboratory for Fundamental Physics.
This would establish the world's first underground microscopy facility, providing the necessary shielding from cosmic ray backgrounds to enable competitive rare event searches.
In the envisioned BedrettoLab Physics facilities at TM3500, mesoSPIM devices could be installed and operated with additional shielding from environmental radioactivity in low-background, clean-room conditions.
In a direct dark matter search, such an underground mesoSPIM device could permanently house a LiF crystal, allowing for regular scanning of the crystal without above-ground exposure to cosmic rays, and thereby enabling both integrated and diurnal and annual signal modulation searches.
Additionally, underground mesoSPIM devices could be used to image LiF crystals that have been exposed to the antineutrino flux from nuclear reactors. This would enable a measurement of the CE$\nu$NS process on Li and F for the first time.

We will conduct an initial characterization of backgrounds for LiF crystals at BedrettoLab Physics by exposing them on site and measuring them afterwards using the existing mesoSPIM devices at UZH.
Additionally, in the first stage of a container-based underground infrastructure at Bedretto, the use of a smaller, portable mesoSPIM device is also feasible.
Finally, an initial deployment of LiF crystals at the Leibstadt Nuclear Power Plant (KKL) in Switzerland is planned, followed by imaging of the crystals at the UZH facilities.
These measurements will allow for the validation of background models in LiF at the reactor site, and the development of shielding requirements during transport from the reactor to a potential future microscopy facility at BedrettoLab Physics.
The close geographical proximity of Bedretto to both UZH and KKL (about 2 and 3 hours by car, respectively) is a key advantage for these efforts, minimizing the cosmic-ray background exposure during transport.

Together with the detailed background characterization campaign\,\cite{Penning:2025mek} and the foreseen multi-stage development of the TM3500 site, these efforts %
strongly motivate the 
realization of an underground microscopy facility for rare-event searches at the Bedretto Underground Laboratory for Fundamental Physics.

\clearpage

\section{Crystal Defect Creation by Nuclear Recoils in Cryogenic Calorimeters}\label{sec:Kluck}

Authors: {\it Holger~Kluck and Jens~Burkhart}
\vspace{0.1cm} \\
MBI - Marietta-Blau-Institut für Teilchenphysik der Österreichischen Akademie der Wissenschaften, 1010 Wien, Austria
\vspace{0.3cm}

\subsection{Introduction}

Mineral detectors look for signals of new and interesting physics by studying defects in the crystal lattice of natural minerals caused by recoiling nuclei. As the measurement of the defects happens sometimes Gyr after the searched-for interaction, this can be understood as the extreme case of a ``long-lived'' signal. In contrast, cryogenic calorimeters are used to study the temperature increase caused by the nuclear recoil in the laboratory—a very ``short-lived'' signal. For cryogenic calorimeters, crystal defects are a source of systematic uncertainty, as their creation energy is missing from the measured thermal signal. Although the detection methods are different, nuclear recoils and crystal defect creation are of common interest for mineral detectors and cryogenic calorimeters alike.

In this contribution, we will discuss crystal defects from the perspective of cryogenic calorimetry based on our work for the \textsc{Eloise} \cite{Kluck2023}/\textsc{Incidence} projects in the framework of the \textsc{Cresst} experiment, searching for galactic Dark Matter particles \cite{Angloher:2025fzw}; the \textsc{Nucleus} experiment, aiming for the precision measurement of Coherent Elastic Neutrino-Nucleus Scattering (CE\textnu{}NS) \cite{NUCLEUS:2026pnv}; and the \textsc{Crab} experiment, establishing monoenergetic nuclear recoils caused by thermal neutron capture as a novel calibration standard \cite{Abele2025a}, hereafter called the ``\textsc{Crab} peaks''. All three experiments operate \ce{Al_2O_3}- and \ce{CaWO_4}-based calorimeters with detection thresholds down to the $\mathcal{O}(\qty{10}{\eV})$ level. Hence, our energy range of interest is firmly in the sub-keV domain, where solid-state effects become relevant.

After briefly introducing rare-event searches with cryogenic calorimeters in \cref{sec:kluck:cryogenicCalorimetery}, we will present our simulations of crystal defect creation by nuclear recoils using the Molecular Dynamics (MD) program \texttt{LAMMPS} (\cref{sec:kluck:MDSim}). The possibility of using measurements of \textsc{Crab} peaks to cross-check MD simulations will be reviewed in \cref{sec:kluck:nuclRecoil}. In \cref{sec:kluck:precisionMeasurement}, we give an outlook on the possibility of measuring crystal defect creation in situ with cryogenic calorimeters. Finally, we conclude in \cref{sec:kluck:conclusion}.

\subsection{Nuclear Recoils in Cryogenic Calorimeters}
\label{sec:kluck:cryogenicCalorimetery}
A monocrystal operated at $\mathcal{O}(\qty{10}{\milli\kelvin})$ acts as a cryogenic calorimeter (see, e.g., \cite{Enss2005}). At this base temperature $T$, the atoms are at their lattice sites and the potential energy $E_\mathrm{pot}$ of the lattice is at a relative minimum. Depending on the interaction being searched for, energy $E$ can be transferred to an atom in various ways, e.g., via elastic scattering induced by a Dark Matter particle, CE\textnu{}NS, or the capture of a thermal neutron. If $E$ exceeds the threshold displacement energy $E_\mathrm{dis}$, which is specific to the given atom and crystal, the scattered atom leaves its lattice site as a \emph{primary knock-on atom} (PKA) with a kinetic energy $E_\mathrm{kin}$, leaving behind a phononically and electronically excited crystal and a distorted lattice with an increased potential energy $E'_\mathrm{pot}=E_\mathrm{pot}+E_\mathrm{def}$. Consequently, the kinetic energy of the PKA, $E_\mathrm{kin}=E-E_\mathrm{exc}-E_\mathrm{def}$, is reduced by the excitation energy $E_\mathrm{exc}$ and the \emph{defect creation energy} $E_\mathrm{def}$. Via scattering, the PKA dissipates energy and contributes to the excitation of the crystal. If $E_\mathrm{kin}>E_\mathrm{dis}$, the PKA may cause secondary displacements and start a nuclear recoil cascade. After some time, the crystal reaches a new thermal equilibrium with $T'=T+(E_\mathrm{kin}+E_\mathrm{exc})/k_\mathrm{B}$. For a more detailed description of radiation-induced crystal damage, see, e.g., \cite{Nordlund2018}.

The temperature increase $\Delta T = T'-T$ caused by the recoil (cascade) can be measured in various ways. For example, \textsc{Cresst}, \textsc{Crab}, and \textsc{Nucleus} use a tungsten \emph{transition edge sensor} (TES) \cite{Proebst1995,Ferger1996,Meier2000}: a thin tungsten film evaporated onto the crystal surface. By choosing a suitable base temperature, the tungsten film is in the transition region between its superconducting and normal-conducting phases. Hence, $\Delta T$ increases the TES resistance relative to a reference resistor; the corresponding change in electric current can be measured after amplification, e.g., with a \textsc{Squid}. As the crystal is coupled to a heat sink, a dedicated Ohmic heater element on the crystal surface can be used in a feedback loop to drive the crystal back to its base temperature, preparing the setup for the next measurement. The detector is calibrated by measuring its response $\Delta T$ for a known energy, provided, e.g., by the electron capture (EC) decay of \ce{^{55}Fe} to \ce{^{55}Mn}, which features K\textsubscript{\textalpha,1} and K\textsubscript{\textalpha,2} X-ray lines with an average energy of \qty{5.9}{\keV}. Once anchored at a known absolute energy, the energy scale is linearised by measuring the detector response to injected artificial heat pulses of known relative energy.

Crystal defects act as a systematic uncertainty in the measured $\Delta T$ and hence the reconstructed energy. If the displaced atoms end up as interstitials, the measured temperature increase does not correspond to the full signal energy $E$: $\Delta T = (E-E_\mathrm{def})/k_\mathrm{B}$. A complete correspondence $\Delta T = E/k_\mathrm{B}$ is only given if no crystal defects were created or if all defects anneal on a timescale shorter than the measurement duration. However, at mK temperatures, annealing is strongly suppressed with respect to room temperature. Therefore, a reliable understanding of the extent and frequency of crystal defect creation is crucial for the operation of cryogenic calorimeters as precision detectors for low-energy signals such as DM scattering, CE\textnu{}NS, or \textsc{Crab} peaks.

\subsection{Molecular Dynamics Simulations of Crystal Defects}
\label{sec:kluck:MDSim}
As shown by S.~Sassi et al., MD calculations are a suitable tool for studying the creation of crystal defects in the target crystals of cryogenic DM detectors, e.g., in \ce{Al_2O_3} for PKAs with $E \leq \qty{200}{\eV}$ \cite{Sassi2022}. The calculation operates on the unit cell (or multiples thereof) of the crystal under study. For each atom in the cell, the forces exerted on it by the other atoms are derived, e.g., from a parameterised interatomic potential, and the atom's position and velocity are updated accordingly over a given time step. For a more detailed description of the MD method, see, e.g., \cite{Haile1997}.

We extended the original work of S.~Sassi et al. for \ce{Al_2O_3} in two ways: first, to higher energies, to be relevant for CE\textnu{}NS experiments at nuclear reactors, with equidistant steps of \qty{20}{\eV} up to \qty{540}{\eV}, and one dedicated MD calculation at \qty{1144}{\eV} for the study of the \textsc{Crab} peak in \ce{Al_2O_3} (see \cref{sec:kluck:nuclRecoil}); second, following the work of K.~Nordlund et al. \cite{Nordlund2024}, we considered the quantum-mechanical zero-point vibrations.

Using crystal lattice data from \cite{Lewis1982}, we implemented a supercell with periodic boundary conditions in the open-source MD code \texttt{LAMMPS} \cite{lammps2022}, large enough to contain all the nuclear recoil cascades for the energies under study: \num{\approx 12e3} atoms ($9 \times 9 \times 5$ times the unit cell) for energies \qty{\leq 540}{\eV}, and \num{\approx 145e3} atoms for the \qty{1144}{\eV} simulation. The interatomic potential of P. Vashishta et al. \cite{Vashishta2008,Vashishta2009} was used, and the zero-point vibrations were taken into account via a quantum thermal bath \cite{Barrat2011,Dammak2009}. For each value of $E$, a random atom from the innermost unit cell of the supercell was selected \num{1000} times, and its direction was sampled isotropically. To dissipate the excess lattice vibration energy, the supercell was kept at \qty{40}{\milli\kelvin} for most calculations. We calculated $E_\mathrm{def}$ as the difference between the potential energy of the supercell before the PKA started and \qty{8}{\ps} afterwards, when the cell had reached equilibrium again.

\begin{figure}
	\begin{subfigure}{0.5\textwidth}
		\includegraphics[width=\linewidth]{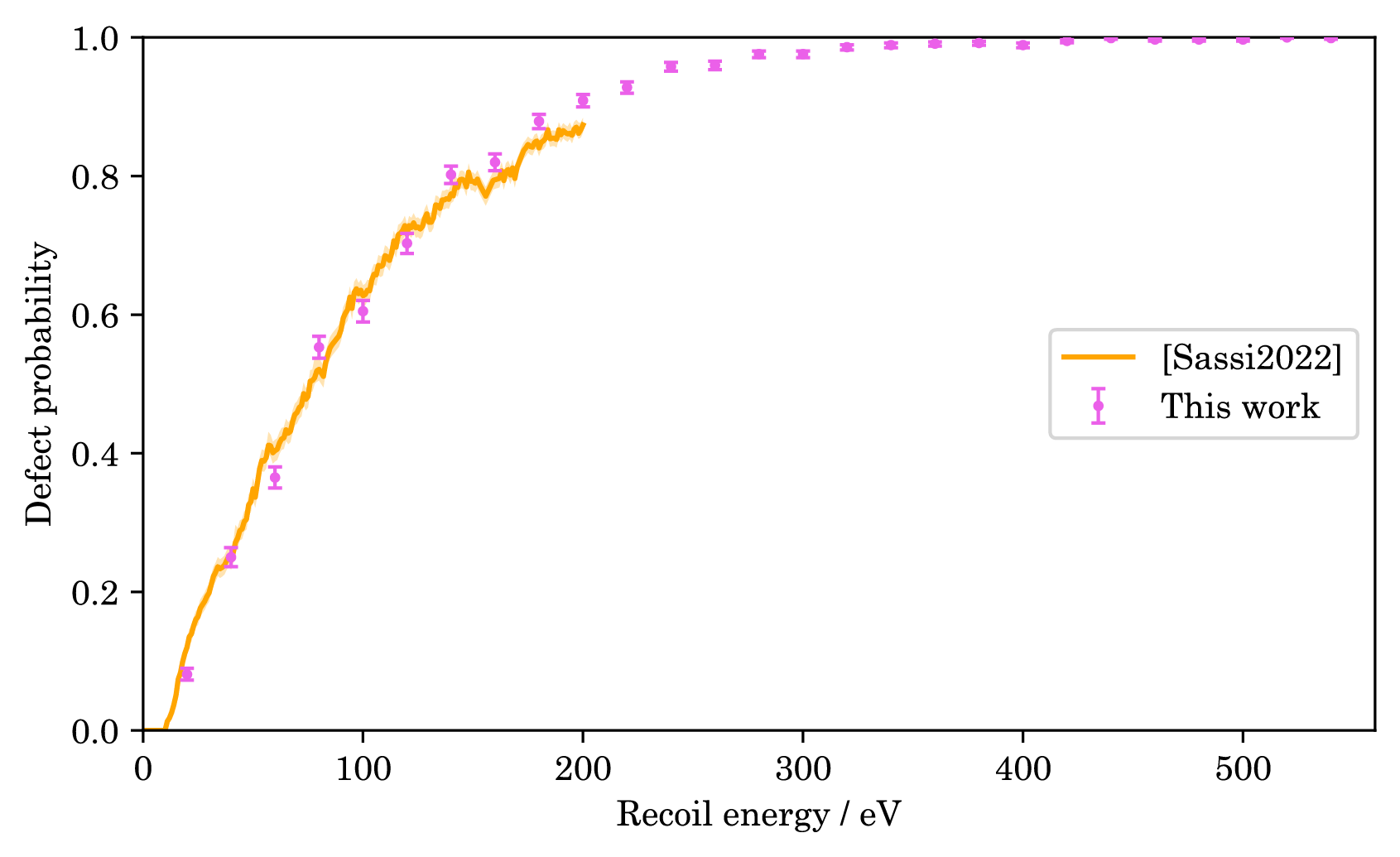}
	\end{subfigure}~
	\begin{subfigure}{0.5\textwidth}
		\includegraphics[width=\linewidth]{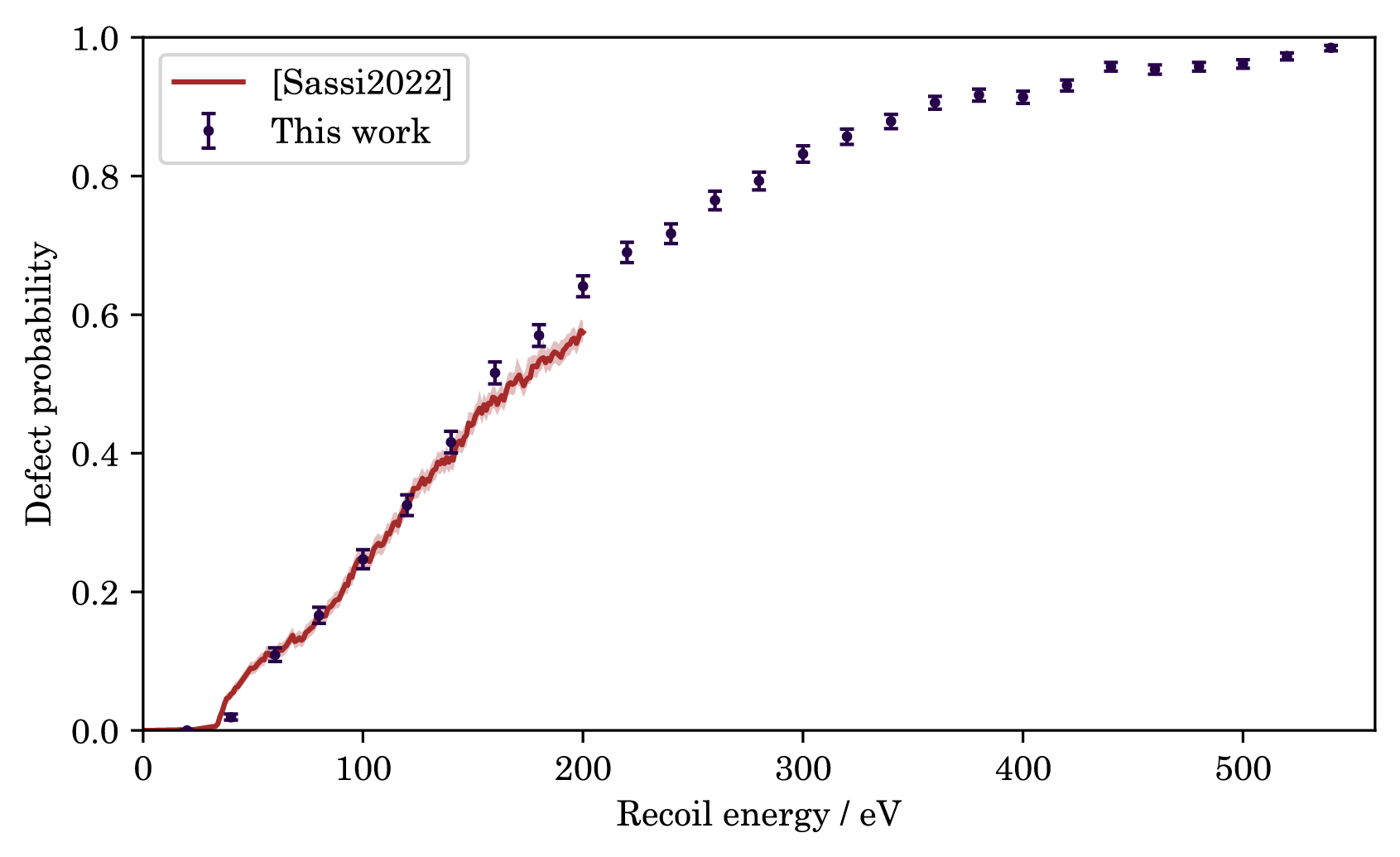}
	\end{subfigure}\\
	\caption{Simulated probabilities of creating at least one crystal defect in \ce{Al_2O_3} as a function of the kinetic energy of the recoiling \emph{primary knock-on atom} (PKA): \emph{left} for an Al-PKA, \emph{right} for an O-PKA. Figures taken from \cite{Burkhart2026}.} \label{fig:kluck:prob}
\end{figure}

\begin{figure}
	\begin{subfigure}{0.5\textwidth}
		\includegraphics[width=\linewidth]{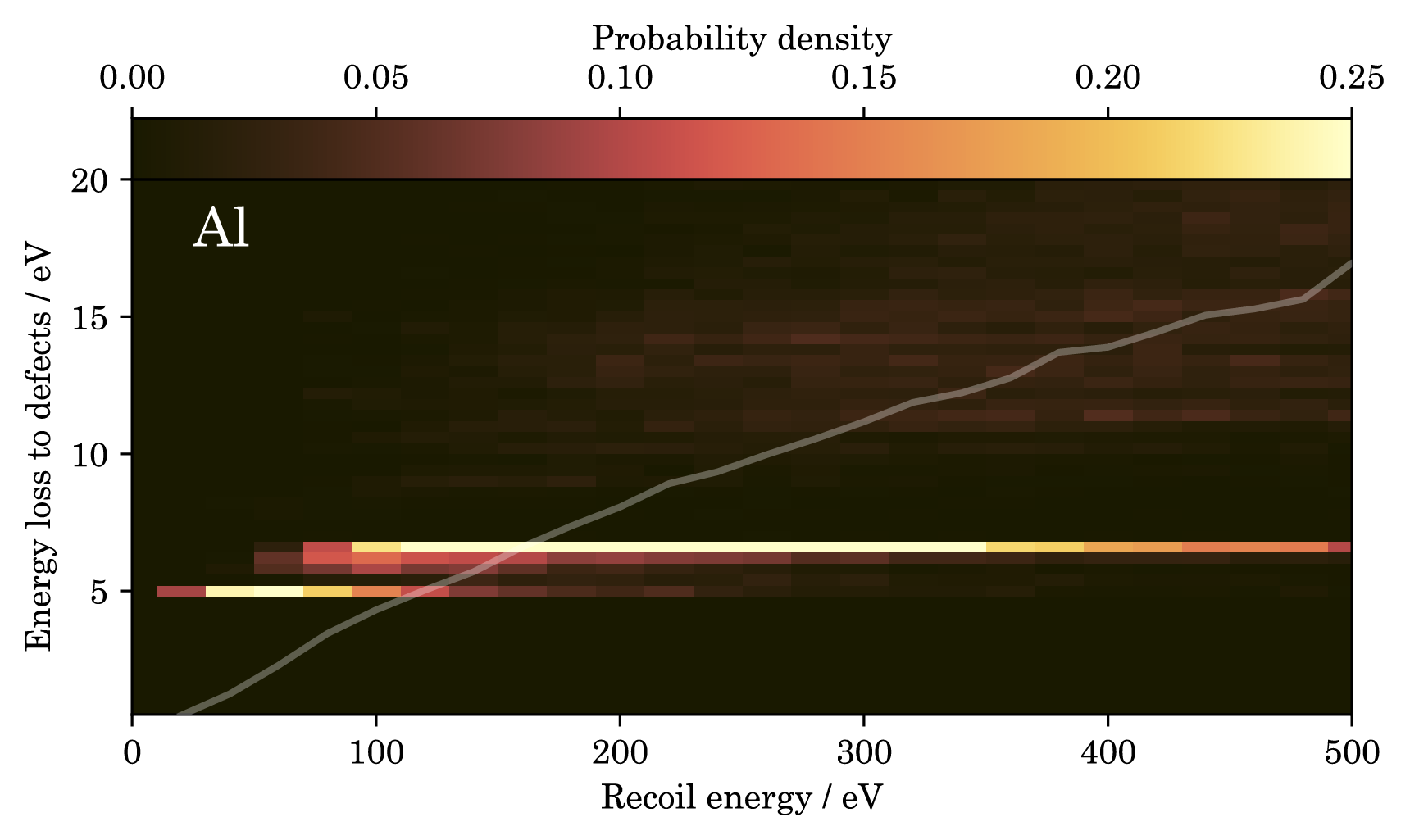}
	\end{subfigure}~
	\begin{subfigure}{0.5\textwidth}
		\includegraphics[width=\linewidth]{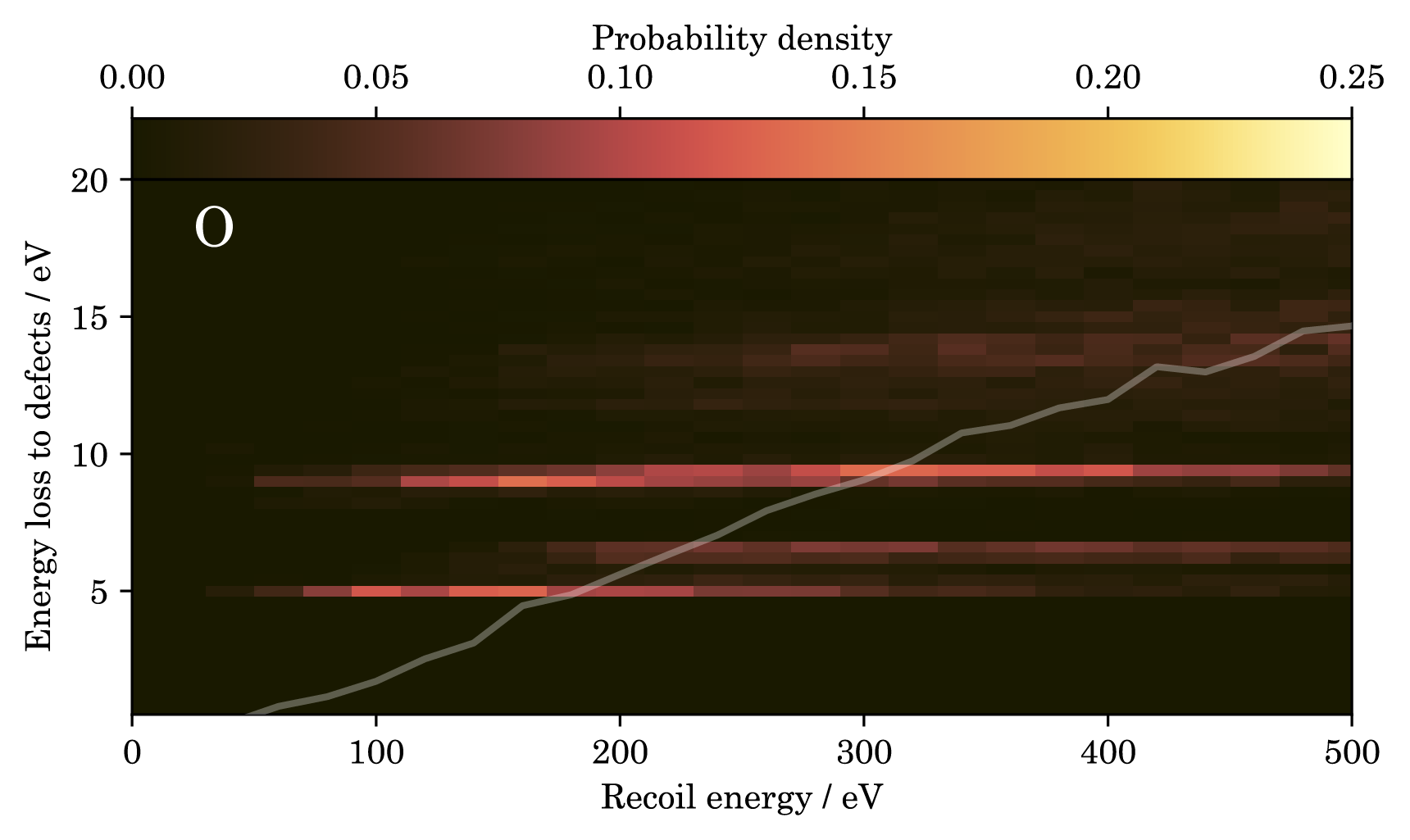}
	\end{subfigure}\\
	\begin{subfigure}{0.5\textwidth}
		\includegraphics[width=\linewidth]{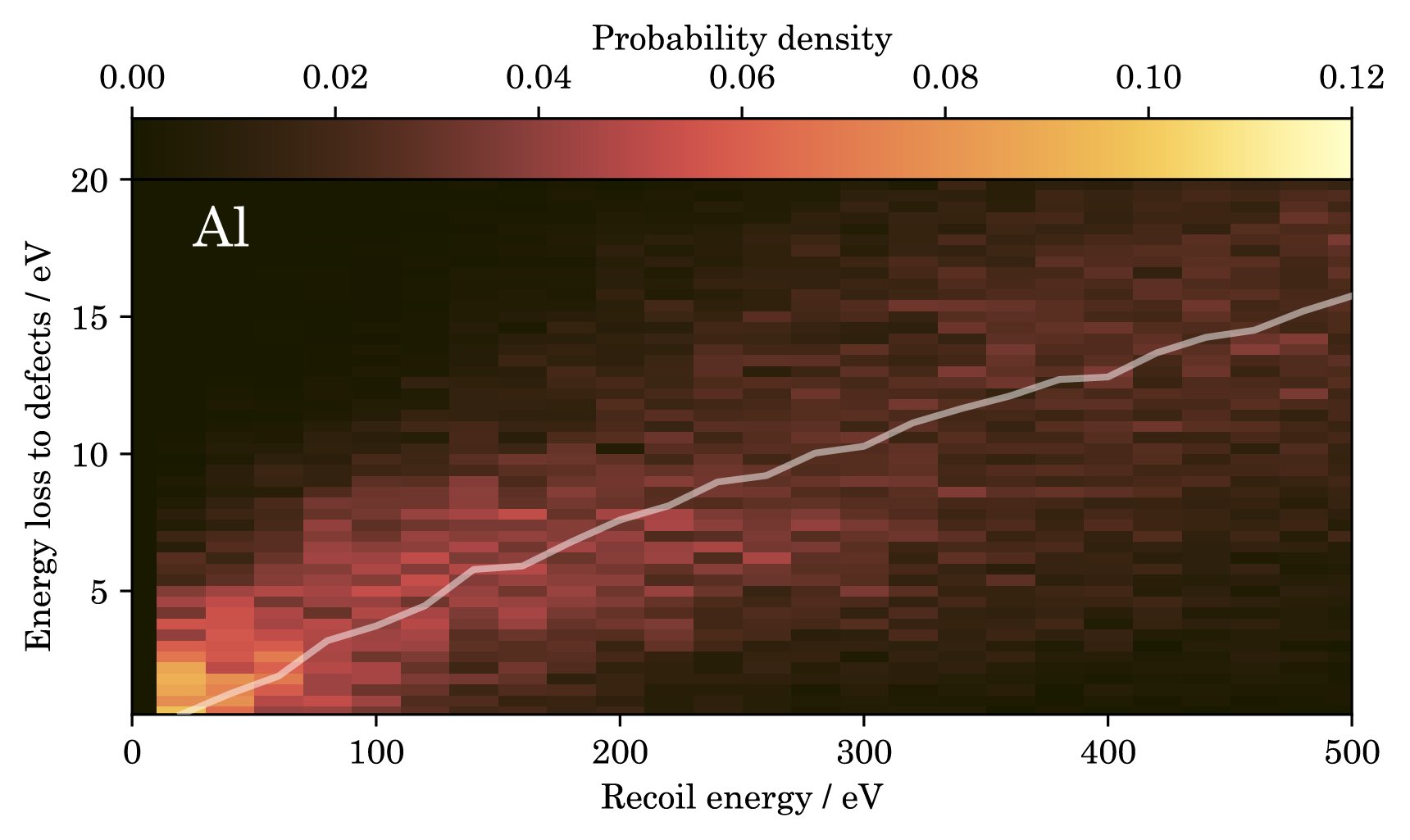}
	\end{subfigure}~
	\begin{subfigure}{0.5\textwidth}
		\includegraphics[width=\linewidth]{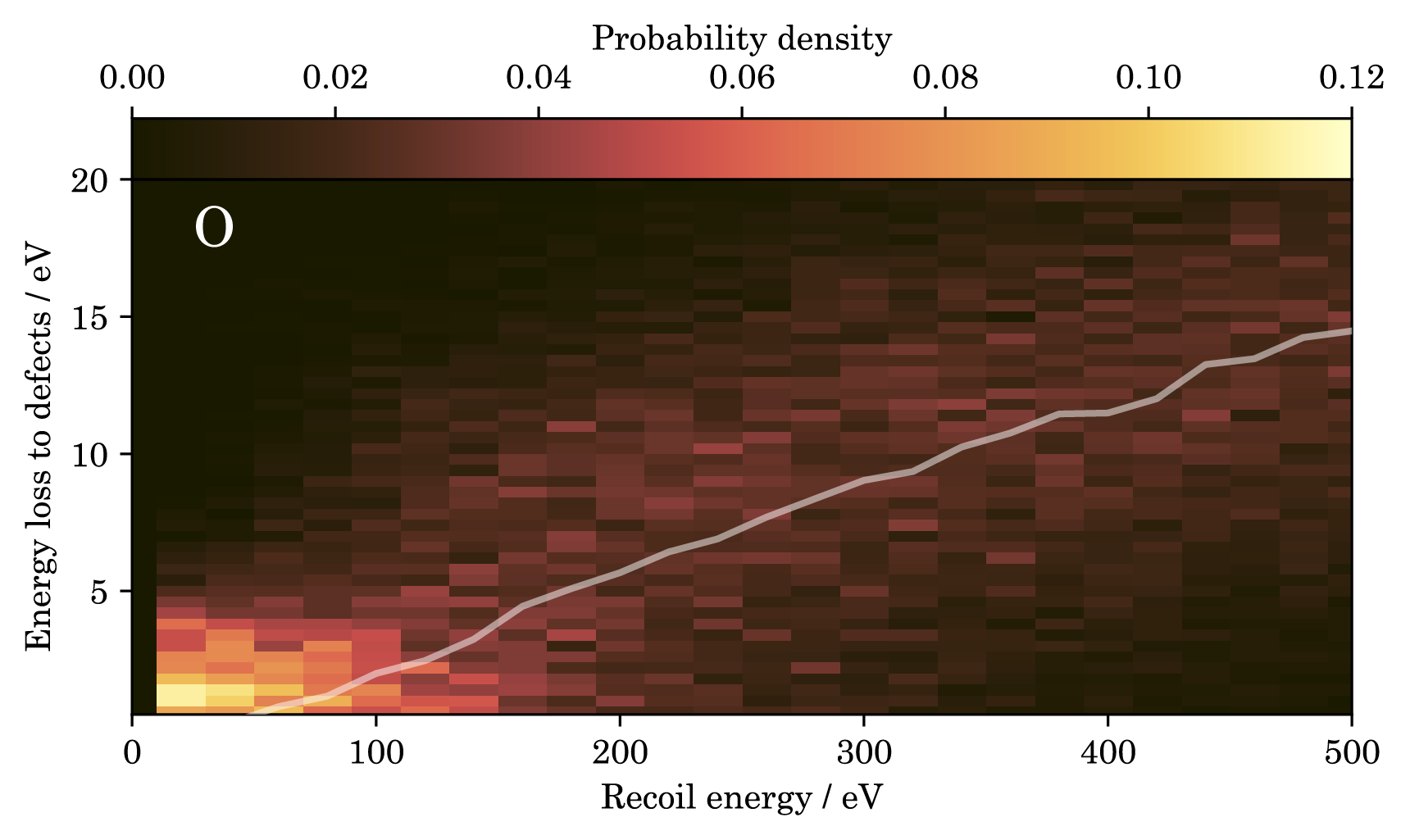}
	\end{subfigure}\\
	\caption{Simulated energy losses due to crystal defect creation in \ce{Al_2O_3} as a function of the energy of the recoiling \emph{primary knock-on atom} (PKA): for an Al-PKA in the \emph{left column}, for an O-PKA in the \emph{right column}; without considering quantum-mechanical zero-point vibrations in the \emph{top row}, and with them in the \emph{bottom row}. The \emph{white} curves are the average energy loss per recoil-energy bin. Figures taken from \cite{Burkhart2026}.}
	\label{fig:kluck:edef}
\end{figure}

\begin{figure}
	\centering
	\begin{subcaptiongroup}
		\begin{overpic}[width=\textwidth]{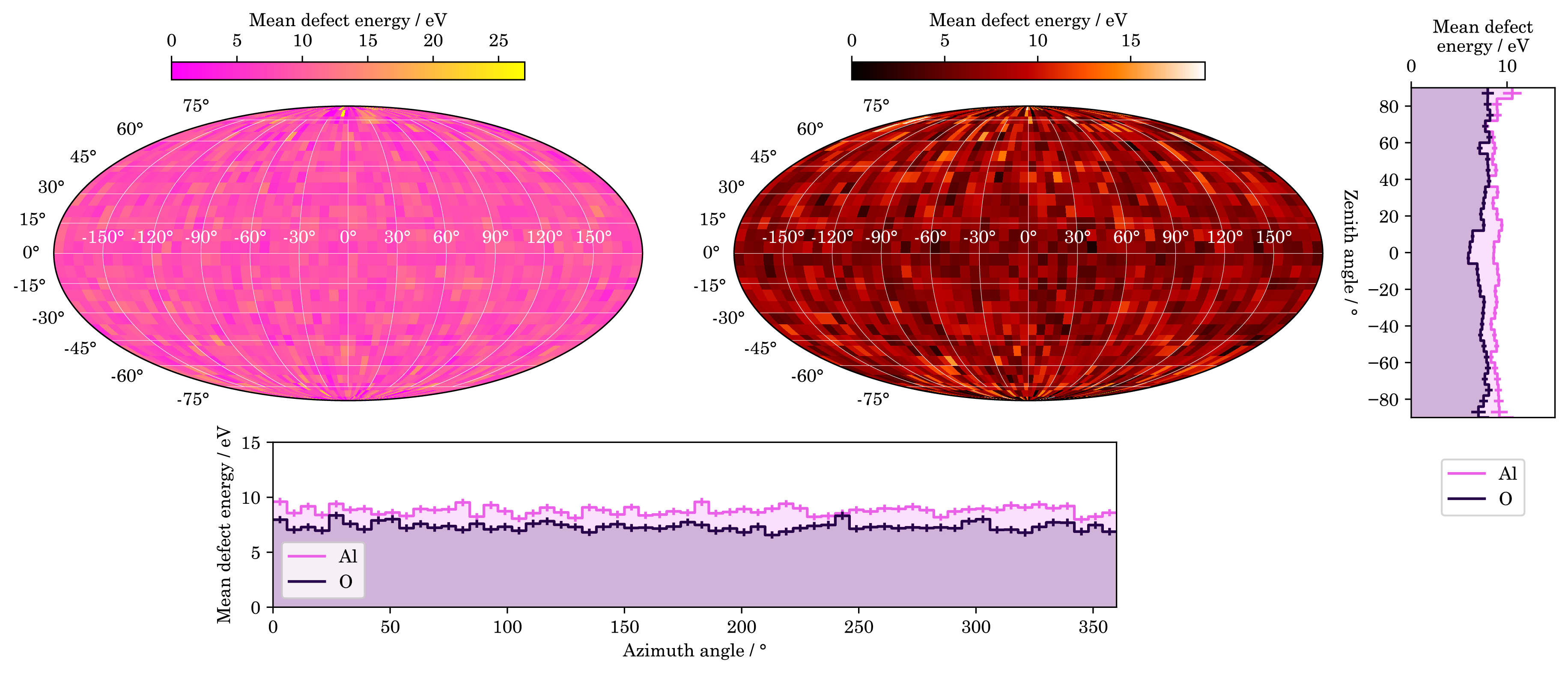}
			\phantomcaption
			\label{fig:kluck:dir:Al}
			\put(3,35){\captiontext*{}}
			\phantomcaption
			\label{fig:kluck:dir:O}
			\put(45,35){\captiontext*{}}
			\phantomcaption
			\label{fig:kluck:dir:azimuth}
			\put(9,13){\captiontext*{}}
			\phantomcaption
			\label{fig:kluck:dir:zenith}
			\put(85,40){\captiontext*{}}
		\end{overpic}
	 \end{subcaptiongroup}
	 \caption{Directional dependence of the energy loss due to defect creation inside \ce{Al_2O_3}, marginalised over all simulated energies: Mollweide projections for (\protect\subref{fig:kluck:dir:Al}) Al-PKAs and (\protect\subref{fig:kluck:dir:O}) O-PKAs, and marginal distributions over the (\protect\subref{fig:kluck:dir:azimuth}) azimuth and (\protect\subref{fig:kluck:dir:zenith}) zenith directions. Figures taken from J. Burkhart \cite{Burkhart2026}.} \label{fig:kluck:dir}
\end{figure}

\Cref{fig:kluck:prob} shows the probability that at least one crystal defect is created as a function of the recoil energy $E$ for the two different atomic species in \ce{Al_2O_3}. As expected, the probability increases significantly over the extended energy range $(\qty{200}{\eV},\qty{540}{\eV}\rbrack$ accessible with CE\textnu{}NS, but missing in the work of S.~Sassi et al. Especially for O-PKAs, defect creation is nearly certain for $E \gtrsim \qty{300}{\eV}$.
The distribution of the corresponding energy loss $E_\mathrm{def}$ due to crystal defect creation is shown in \cref{fig:kluck:edef}: on average, around \qty{3}{\percent} of the kinetic recoil energy is lost. Furthermore, we see the importance of considering quantum-mechanical zero-point vibrations, resulting in a significant smearing of $E_\mathrm{def}$. We also found an indication of a slight directional dependence of $E_\mathrm{def}$ in the case of O-PKAs; see \cref{fig:kluck:dir}: in the equatorial plane, i.e., for a zenith angle of \qty{0}{\degree}, less energy is lost than otherwise (\cref{fig:kluck:dir:zenith}).

\subsection{Possibility of Experimental Verification}
\label{sec:kluck:nuclRecoil}
A cross-check of the accuracy of the MD calculations could be possible by comparing them to a measured \textsc{Crab} peak: monoenergetic nuclear recoils of precisely known energy. Originally proposed in Refs.~\cite{Thulliez2021,Fuss2022}, the method is based on the radiative capture of a thermal neutron by a nucleus $X$ inside the crystal under study, creating an excited daughter nucleus: ${}^A_Z X(\mathrm{n}, \mathrm{\gamma})^{A+1}_Z X$. As the average thermal neutron energy (\qty{25}{\milli\eV}) can be neglected, the excitation energy is completely given by the neutron separation energy, which is known from the literature. If $X$ de-excites via the emission of a single \textgamma{} ray, the recoil energy $E$ of $X$ can be calculated precisely via two-body kinematics. In a suitably small target crystal, the emitted \textgamma{} ray escapes and the only deposited energy is the recoil energy, creating a monoenergetic peak in the spectrum. Hence, a comparison of the observed $E$ with the prediction can reveal the impact of solid-state effects such as crystal defects.

This method is applicable to any nuclide with a suitable capture cross section, \textgamma{} branching ratio, and natural abundance; see \cite[table 3]{Abele2025a}. The most favourable target nuclide in these terms is \ce{^{182}W}, producing a recoiling \ce{^{183}W} nucleus with $E=\qty{112}{\eV}$. MD calculations by G.~Soum-Sidikov et al. predicted the peak on average at $E-\qty{8.9}{\eV} \approx \qty{103.1}{\eV}$ \cite{SoumSidikov2024}, i.e., a relative shift of \qty{\approx 8}{\percent}. The peak was first measured by the \textsc{Crab} and \textsc{Nucleus} collaborations at \qty{106.7(1.9:2.0)}{\eV} in a \ce{CaWO_4}-based cryogenic calorimeter of \qty{0.75}{\gram} with a baseline resolution of $\sigma = \qty{6.0}{\eV}$; a moderated \ce{^{252}Cf} source delivered \qty{\approx 0.25}{\per\s} thermal neutrons \cite{Abele2023}. This was independently confirmed by the \textsc{Cresst} collaboration, observing the peak in three different detectors at energies \qty{106.2+-1.8}{\eV}, \qty{107.8+-1.3}{\eV}, and \qty{113.3+-1.4}{\eV} with \textsigma{} values of \qty{7.92}{\eV}, \qty{4.95}{\eV}, and \qty{4.83}{\eV}, respectively \cite{Angloher2023c}. The four observations feature a spread of at most \qty{\approx 4}{\percent} relative to their average. These remarkably small deviations show the feasibility of using a keV source together with the applied linearisation scheme (see \cref{sec:kluck:cryogenicCalorimetery}) for the calibration of an energy range of interest that is an order of magnitude less energetic than the source. Furthermore, they indicate that the precision of the reconstructed energy is at a level comparable to the expected impact of crystal defects, raising the prospect of a direct measurement of $E_\mathrm{def}$ with a cryogenic calorimeter.

In \ce{Al_2O_3}, the recoil peak due to \ce{^{27}Al}(n, \textgamma)\ce{^{28}Al} with $E=\qty{1144}{\eV}$ was first observed by \textsc{Cresst} at \qty{1113.6+-6.5}{\eV} \cite{Angloher2025a}. Here, our MD calculations (see \cref{sec:kluck:MDSim}) predicted $E_\mathrm{def}=\qty{31.9+-8.5}{\eV}$ \cite{Angloher2025a}, i.e., $E-E_\mathrm{def}=\qty{1112.1}{\eV}$, which is tantalisingly close to the observed value. However, a detailed statistical investigation showed that the uncertainty in the energy scale of the detectors has to be further reduced before an in situ measurement of $E_\mathrm{def}$ with a cryogenic calorimeter can be claimed: for a statistical significance of $>3 \sigma$, an uncertainty of \qty{<0.7}{\percent} is required \cite{Angloher2025a}. The peak itself was confirmed by \textsc{Crab} \cite{Abele2025b}.

\subsection{Outlook to Precision Measurements of Nuclear Recoils}
\label{sec:kluck:precisionMeasurement}
One possibility for improving the detection precision of \textsc{Crab} peaks is an increase in the thermal neutron flux. For this purpose, the \textsc{Crab} collaboration established an experimental site at the \textsc{Triga}-Mark II nuclear research reactor of TU Wien in Vienna, Austria \cite{Abele2025a}. A dedicated beam line delivers a thermal neutron flux of \qty{121+-17}{\per\square\cm\per\s} to the calorimeter inside a cryostat, increasing the statistics of the measured nuclear recoils. Assuming a detector resolution of \qty{5}{\eV}, which is feasible, cf.\ \cref{sec:kluck:nuclRecoil}, the impact of crystal defect creation on the \textsc{Crab} peak should be measurable, as shown in \cite[fig.~21]{Abele2025a} for the example of \ce{CaWO_4}.

In addition, the \textsc{Crab} setup at TU Wien features an array of \ce{BaF_2} detectors to measure the escaped \textgamma{} ray from nuclear de-excitation in coincidence with the nuclear recoil, improving the signal-to-background ratio \cite{Abele2025a}. Furthermore, the \textsc{Crab} and \textsc{Nucleus} collaborations aim for an improved energy calibration of the calorimeter by using two newly developed calibration devices. First, an X-ray fluorescence (XRF) source provides several fluorescence lines between \qtyrange{0.6}{6}{\keV} \cite{Abele2025} of known energy. This is especially promising for a precise measurement of the Al \textsc{Crab} peak in \ce{Al_2O_3}, as this peak is bracketed by the $\mathrm{K_\alpha}$ lines of Cu and Al at energies of \qty{927.7}{\eV} and \qty{1486.6}{\eV}, respectively. A first calibration of an \ce{Al_2O_3} calorimeter with the XRF source has already indicated a non-trivial detector response that requires further investigation \cite{Abele2025b}. Independently, an LED light pulser, emitting photons at \qty{255}{\nm}, acts as a second calibration device, providing peaks of freely tunable energy. As the amplitude and width of the peaks are governed by Poisson statistics, the energy scale of the LED pulser can be calibrated in situ \cite{DelCastello2024}.

\subsection{Conclusion}
\label{sec:kluck:conclusion}
Cryogenic calorimeters based on monocrystals are an established tool for searching for nuclear recoils caused by rare interactions of potential new physics beyond the Standard Model of particle physics. For example, calorimeters based on \ce{Al_2O_3} and \ce{CaWO_4} with detection thresholds of $\mathcal{O}(\qty{10}{\eV})$ are used by \textsc{Cresst} to search for Dark Matter particles and will be used by \textsc{Nucleus} to search for deviations from Standard Model CE\textnu{}NS. Crystal defects caused by nuclear recoils are a potential systematic uncertainty, as the defect creation energy is not visible in the measured thermal signal; MD simulations predict an impact on the $\mathcal{O}(\qty{10}{\eV})$ scale. Therefore, a reliable understanding of crystal defect creation is crucial, based on both simulation and measurement.

Studies of the energy calibration of the detectors with monoenergetic nuclear recoil peaks of known energy, as established by the \textsc{Crab} collaboration, also indicate an accuracy at the $\mathcal{O}(\qty{10}{\eV})$ scale. Hence, feasible improvements in detector calibration methods, as currently being developed by all three experiments, raise the possibility of measuring defect creation in situ with cryogenic calorimeters, providing crucial data to verify the MD simulations of low-energy nuclear recoils. A verified simulation, in turn, provides a reliable understanding of the detector physics, which is crucial for any search for rare events of new physics.

\acknowledgments
This contribution would not be possible without the work of the \textsc{Cresst}, \textsc{Crab}, and \textsc{Nucleus} collaborations on which this review is based. We thank Dominik Fuchs for reviewing the manuscript and providing useful comments. J.~Burkhart and H.~Kluck were funded by the Austrian Science Fund (FWF) through project ``\textsc{Eloise}'' \href{http://dx.doi.org/10.55776/P34778}{DOI:10.55776/P34778}. Collaboration with Veronika~Palušov\'a and Miroslav~Macko from the Institute of Experimental and Applied Physics (IEAP) of the Czech Technical University (CTU) in Prague was supported by Austria's Agency for Education and Internationalisation (OeAD) through project CZ 13/2023 ``\textsc{Incidence}''.

\clearpage

\section{First Principles Insights into Defect Formation Processes in Dark Matter Detection Materials}\label{sec:VTQM}

Authors: {\it Jordan Chapman\textsuperscript{1} and Vsevolod Ivanov\textsuperscript{1,2,3}}
\vspace{0.1cm} \\
\textsuperscript{1} Virginia Tech National Security Institute, Blacksburg, Virginia 24060, USA 

\noindent{\textsuperscript{2} Department of Physics, Virginia Tech, Blacksburg, Virginia 24061, USA}

\noindent{\textsuperscript{3} Virginia Tech Center for Quantum Information Science and Engineering, Blacksburg, Virginia 24061, USA}

\vspace{0.3cm}

\subsection{Introduction}

Point defects in solid-state materials can be either advantageous or detrimental to the effectiveness of dark matter detectors, dependent on the detection scheme. For instance, interactions between neutral particles and crystal lattices can lead to the formation of stable, optically active point defects called color centers. In this case, single color centers hold the potential to describe rare events, such as dark matter or neutrino interactions with crystal lattices, in macroscopic volumes of detector materials with high resolution. We have previously determined the optical properties of color centers in lithium fluoride (LiF) using first principles calculations \cite{GuerreroPerez2025}, which were used to identify the defects formed from neutron and gamma radiation via fluorescence microscopy \cite{araujo2025nuclear}. Conversely, metastable defects in solid-state phonon and charge detectors have been investigated as a source of low-energy excess backgrounds that obscure the signatures of dark matter and neutrino interactions. While the source of such defects is uncertain, low-energy excesses have been observed universally across detector materials, backgrounds, and detection systems \cite{Baxter2025}. 

Density functional theory (DFT) calculations can offer insights into the interactions between lattice point defects and solid-state phenomena like excitons and free charge carriers. We have characterized the dynamics of defect transformations in two prominent dark matter detector materials, LiF and germanium (Ge) \cite{Germanium_EDELWEISS}, using a nudged elastic band (NEB) method \cite{NEB_solidstate} coupled with the Vienna Ab Initio Software Package (VASP) \cite{VASP1,VASP2}. Our results illustrate the complex nature of the interactions between point defect formation and lattice phenomena, a full understanding of which can inform improved dark matter detection systems. 

\subsection{Exciton-Assisted \textit{F} Center Formation in LiF}
\label{sec:LiF}

 We have performed NEB calculations to estimate the energy barriers associated with Frenkel pair formation, i.e., ejection of a fluorine (F) atom from its lattice site to form a fluorine vacancy (V\textsubscript{F}) and a fluorine interstitial (F\textsubscript{i}), a prominent step in the formation of optically active \textit{F} centers. The excited state forces were calculated along the minimum energy path (MEP) to explore the coupling of the LiF lattice with excitons \cite{delgrande2026}.

\begin{figure}
   \centering
   \includegraphics[width=1.0\textwidth]{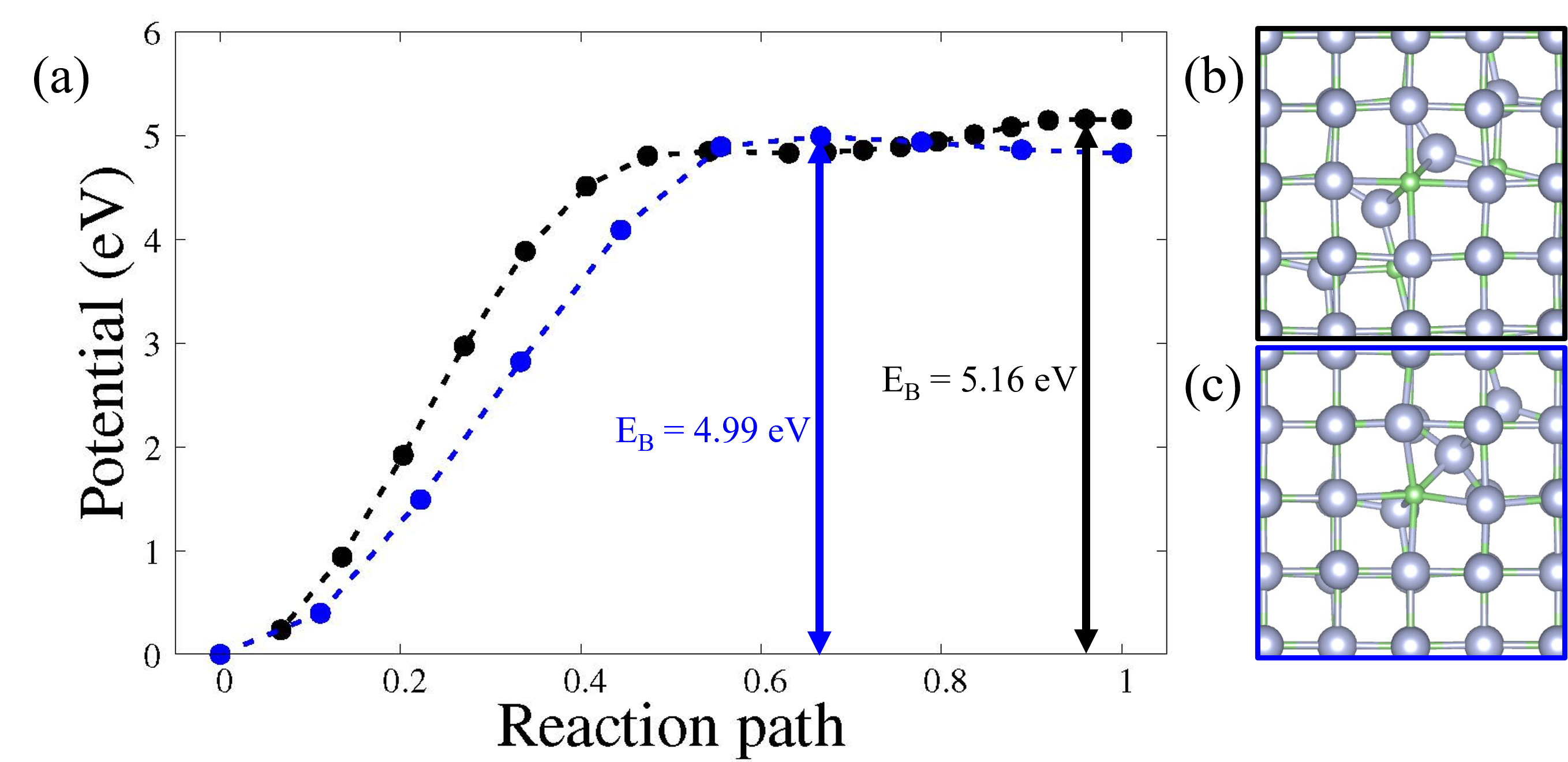}
   \caption{(a) MEPs of S-type (black) and V-type Frenkel pairs. Frenkel pair geometries of (b) S-type and (c) V-type Frenkel pairs.}
   \label{fig:FrenkelPairNEB}
\end{figure}

 We have investigated the Frenkel pair formation of two F\textsubscript{i} geometries: a split (D) configuration and a void (V) configuration. MEPs associated with the Frenkel pair formation of each geometry are shown in Figure\,\ref{fig:FrenkelPairNEB}. Our NEB results indicate that the V-type Frenkel pair has a lower formation energy of 4.83 eV compared to that of the S-type Frenkel pair at 5.14 eV. We also find that both formation mechanisms have activation barriers only slightly larger than the formation energies of their respective stable end configurations; however, the V-type Frenkel pair is more stable, lying approximately 0.16 eV below the saddle point, compared to 0.02 eV for the S-type Frenkel pair.

\begin{figure}
   \centering
   \includegraphics[width=0.7\textwidth]{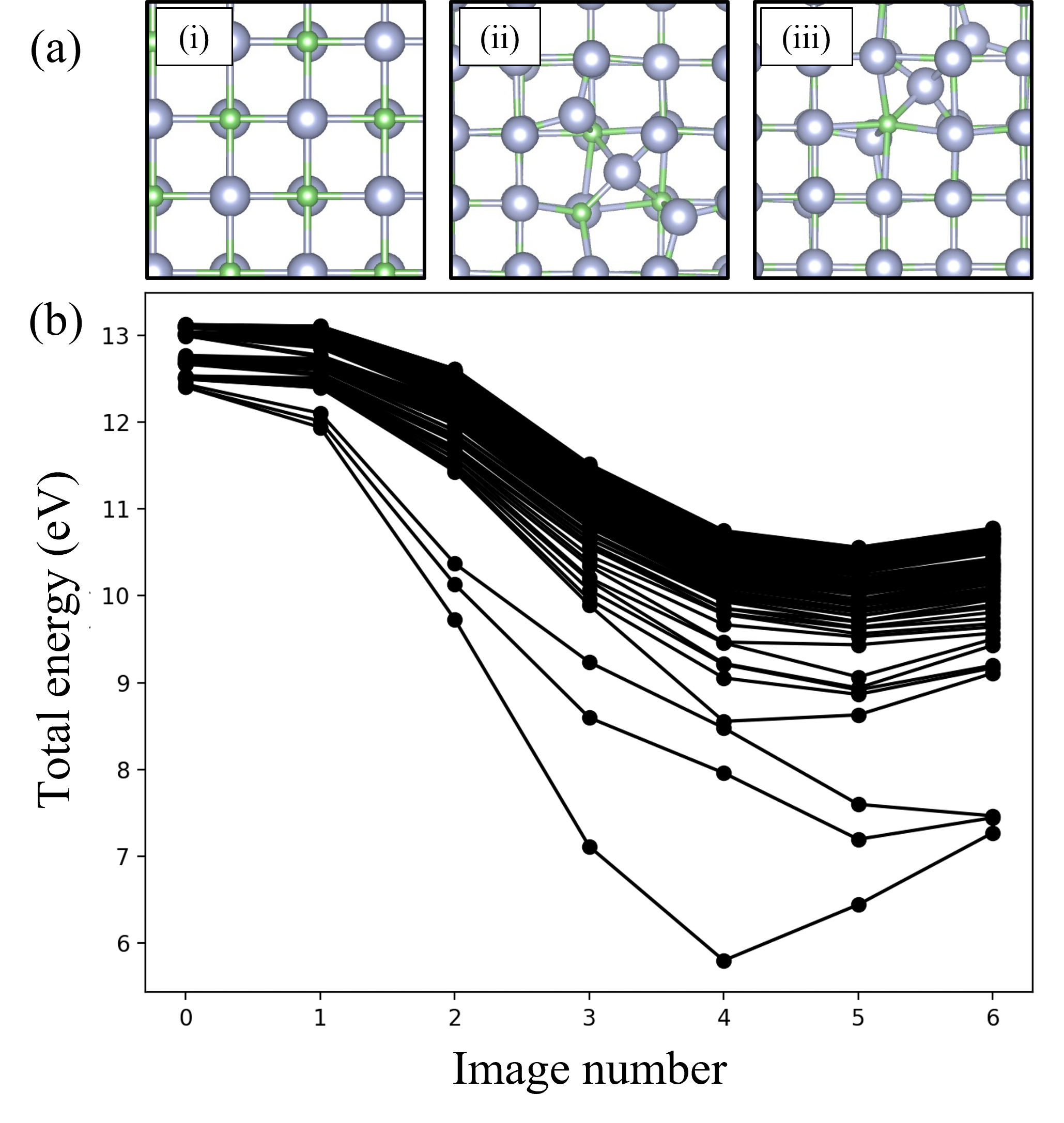}
   \caption{(a) Ionic geometries of (i) pristine LiF, (ii) ground state saddle point, and (iii) V-type Frenkel pair. (b) Total energy including self-consistent field (SCF) and exciton energy of the defect system along its MEP.}
   \label{fig:ExcitonicNEB}
\end{figure}

Excited state force calculations reveal that the excitonic forces strongly couple to the LiF lattice, as shown in Figure\,\ref{fig:ExcitonicNEB}. Strong exciton energies near the pristine LiF geometry are found to alter the potential energy surface (PES) such that the transition state of the V-type Frenkel pair lies near an energy minimum when it is no longer in its ground state electronic configuration. Excited state forces along the ground-state MEP support the claim that excitons play an important role in \textit{F} center formation \cite{LiF_exciton}; our calculations show that relaxation of the pristine lattice at an excited state may be sufficient to bypass the large energy barrier of Frenkel pair formation at the ground state.

\subsection{Metastable Intrinsic Defects in Germanium} \label{sec:Ge}

Defects in Ge are being explored as a partial explanation for low-energy excesses that limit the detection threshold of phonon dark matter detectors. We have screened for metastable intrinsic defects in Ge and elucidated the MEPs that connect their stable geometries using NEB calculations. Briefly, defect screening was done by combining self-vacancies (V\textsubscript{Ge}) with self-interstitial atoms (Ge\textsubscript{i}) at various lattice sites separated by up to approximately 7 \AA.

\begin{figure}
   \centering
   \includegraphics[width=.8\textwidth]{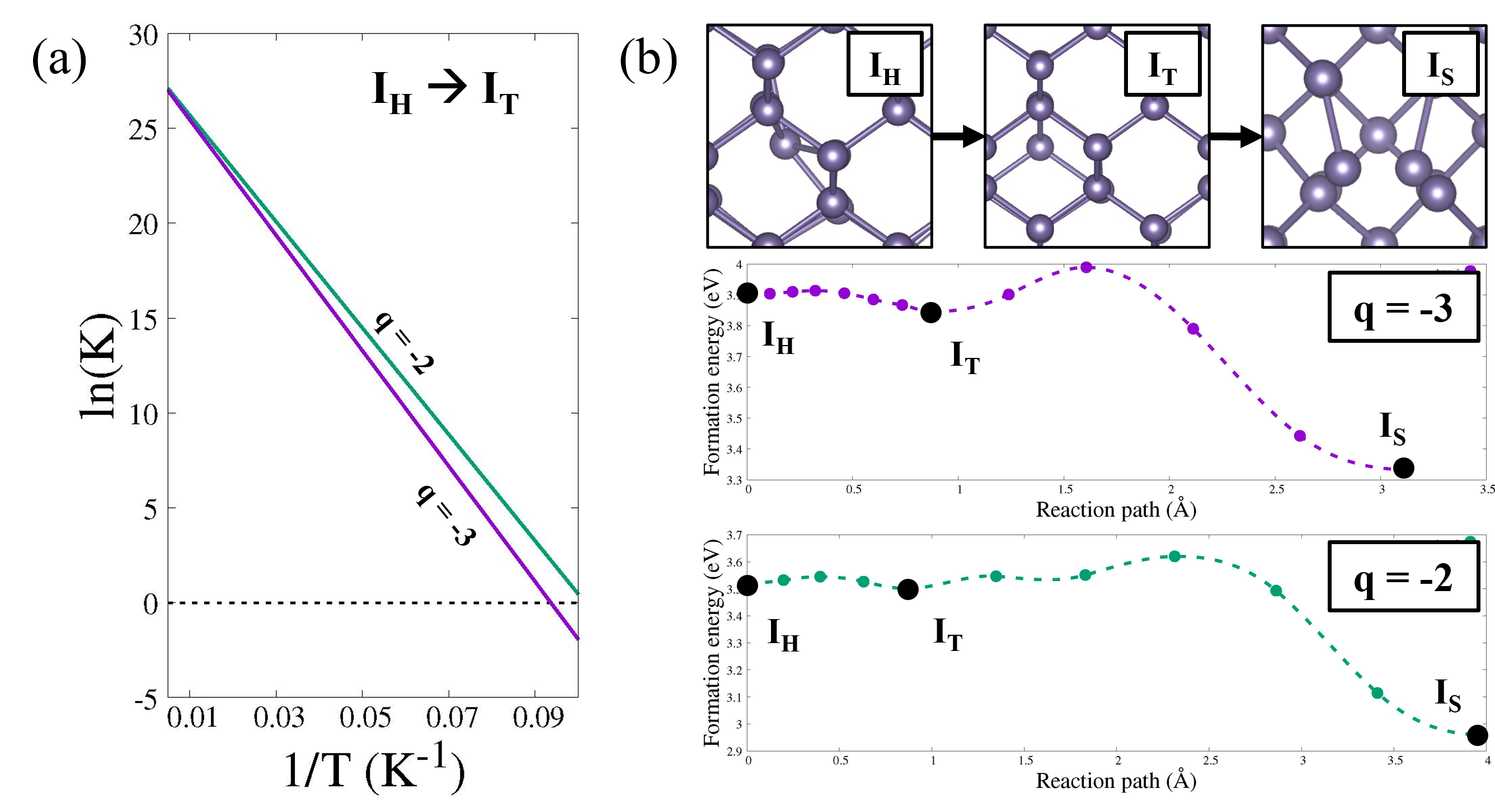}
   \caption{(a) Linearized rate laws for the transition from I\textsubscript{H} to I\textsubscript{T}. (b) MEPs connected stable interstitial geometries and their visualizations. }
   \label{fig:MetastableGeInterstitial}
\end{figure}

Our screening indicates that the lowest-energy metastable defect is the single Ge\textsubscript{i} defect in its hexagonal (I\textsubscript{H}) geometry with a formation energy of 3.51 eV at the -2 charge state. At the same charge state, I\textsubscript{H} has only a slightly larger formation energy than does its tetragonal defect geometry (I\textsubscript{T}) and is separated by an energy barrier of about 0.2 eV. The linearized rate laws and MEPs of the metastable I\textsubscript{H}-to-I\textsubscript{T} conversion are shown in Figure\,\ref{fig:MetastableGeInterstitial}. At a normal temperature of phonon detector operate (10 mK), we find that the conversion from I\textsubscript{H} to I\textsubscript{T} is not possible with only thermal energy available in the lattice.

\begin{figure}
   \centering
   \includegraphics[width=.8\textwidth]{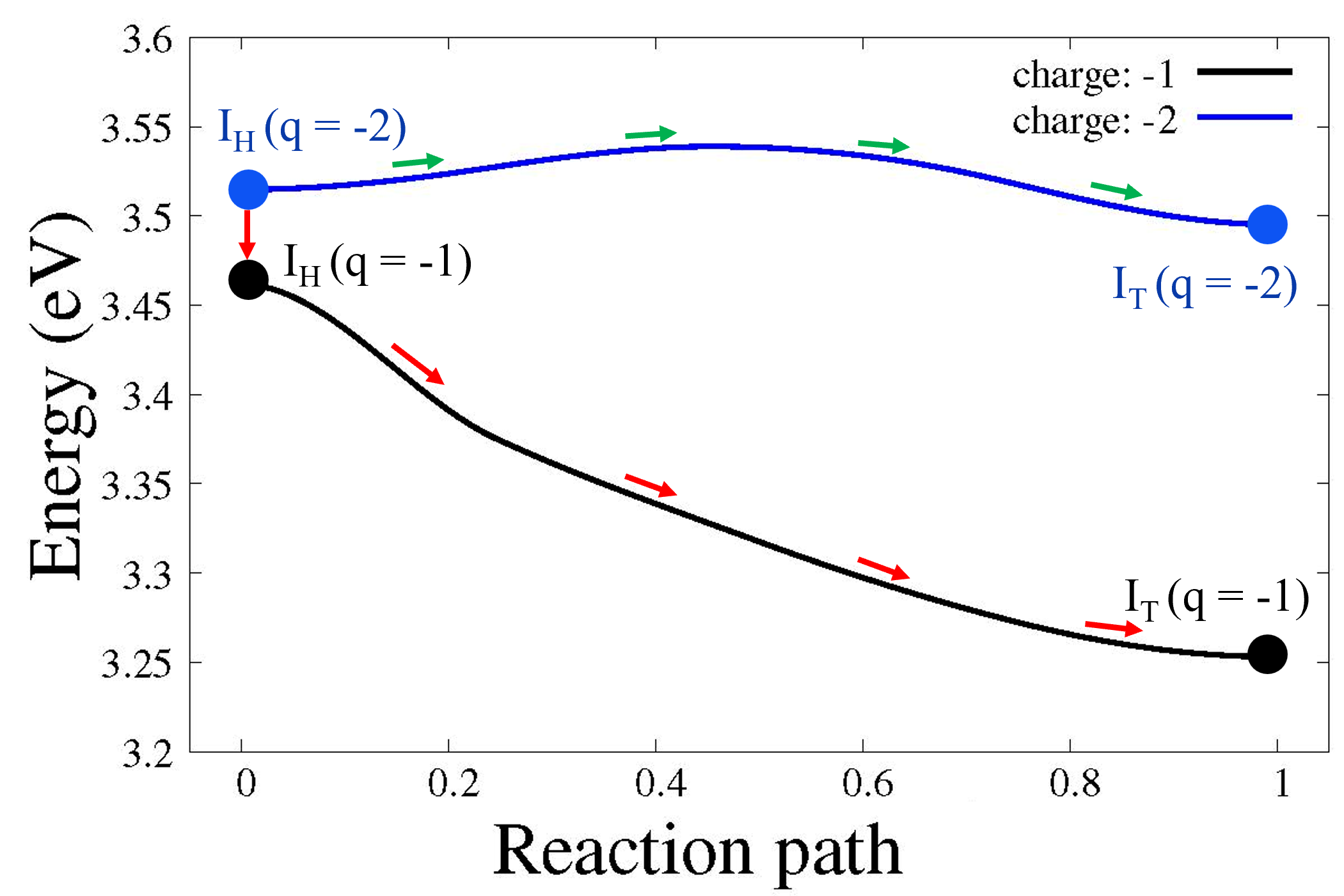}
   \caption{Proposed mechanisms of I\textsubscript{H} collapse to I\textsubscript{T} geometry. The green arrows represent a barrier hopping mechanism, whereas the red arrows represent the recombination of the I\textsubscript{H} defect with a hole, causing loss of stability and relaxation to I\textsubscript{T} configuration.}
   \label{fig:DefectCarrierRecombination}
\end{figure}

Interestingly, the I\textsubscript{H} defect does not stably exist at charge states greater than -2, as predicted with DFT calculations. Instead, we find that if ionized to a +1 charge state, the I\textsubscript{H} geometry lies near a saddle point on the potential energy surface that will spontaneously relax the defect toward its I\textsubscript{T} geometry, as shown in Figure\,\ref{fig:DefectCarrierRecombination}. Thus, we have presented a mechanism by which defects may relax and release sub-1 eV amounts of energy into the lattice that can disperse to trigger other defect relaxations. 

\acknowledgments

This work has been supported through the U.S. National Science Foundation Growing Convergence Research Grant OIA-2428507, titled ``Collaborative Research: GCR: Mineral Detection of Dark Matter'' and the DARPA QuSeN program. The authors acknowledge Advanced Research Computing at Virginia Tech (arc.vt.edu) for providing computational resources and technical support.

\clearpage

\section{Ultra-heavy Exotic Particle Search with the Olivine in Meteorites}\label{sec:Toho_NAKA}

Authors: {\it Tatsuhiro Naka$^{1}$, Naoki Mizutani$^{1}$, Shota Futamura$^{2}$, Takenori Kato$^{2}$, Minako Hashiguchi$^{2}$,  Issei Saikyo$^{1}$, Yohei Igami$^{3}$ }
\vspace{0.1cm} \\
$^{1}$Toho University, $^{2}$Nagoya University, $^{3}$Kyoto University  
\vspace{0.3cm}

\subsection{Introduction}
Paleo detectors have the potential to establish a new methodology for cosmic-ray physics by combining geological timescales with the excellent tracking capability of ancient minerals. In recent years, remarkable progress in observational astronomy, astroparticle phenomenology, and numerical relativity has further enhanced the scientific motivation for paleo-detector research.

Our group is exploring the application of paleo detectors to a wide range of new physics searches beyond dark matter and neutrinos. In particular, extraterrestrial minerals, such as meteorites, are attractive targets because they preserve records of cosmic rays that cannot be observed in underground experiments on Earth.

Studies using extraterrestrial minerals were actively pursued during the 1970s and 1980s. In particular, minerals returned by the Apollo missions were extensively analyzed. One of the major research topics was the study of galactic cosmic-ray elements (GCREs), where geological exposure times enabled detailed measurements of the elemental abundances of nuclei heavier than iron~\cite{Douglas1971}.

Another intriguing result was the observation of a track longer than 1 mm in pyroxene extracted from a lunar sample. Possible explanations included ultra-heavy GCRE nuclei with atomic numbers greater than 80 or exotic particles such as magnetic monopoles~\cite{Price1971}.

More recently, the OLYMPIYA project investigated etched tracks in olivine crystals extracted from pallasite meteorites. In that study, three candidate events with an estimated atomic number of approximately (Z $\sim$ 120) were reported~\cite{olympiya}. If confirmed, these events could correspond to previously unknown nuclear species, such as nuclei located within the predicted island of stability, which have not yet been synthesized in accelerator experiments. These results illustrate the unique potential of paleo detectors to search for exotic particles and rare astrophysical phenomena.

For both GCRE studies and exotic-particle searches, neutron star mergers (NSMs) are particularly important because they are considered one of the primary production sites of heavy nuclei~\cite{NSM_rprocess}. Although the NSM rate in the Milky Way is low, approximately one event every 0.1 Myr, a mineral with an age of 1 Gyr could have experienced several tens of nearby NSMs within a distance of about 1 kpc. Consequently, paleo detectors may provide access to rare astrophysical events and new physics that are beyond the reach of conventional artificial detectors.

\subsection{Etched Track Detection Ability for Olivine} \label{Tracking}
Latent tracks in extraterrestrial minerals can be directly observed at high spatial resolution using electron microscopy. However, such observations are generally limited to tracks located near the mineral surface because most of the observed tracks originate from relatively low-energy cosmic-ray particles, such as those produced by solar energetic particle events.

On the other hand, extremely rare events cannot be efficiently searched for using electron microscopy or other high-resolution imaging techniques because of their limited observation area. Therefore, optical microscopy following chemical etching is indispensable for wide-area searches.

Olivine, one of the most abundant minerals in meteorites and other extraterrestrial materials, can be etched using an etchant based on ethylenediaminetetraacetic acid (EDTA). Under typical etching conditions, tracks become observable after treatment at approximately 100 ${}^\circ$C for more than one hour.

\begin{figure}
   \centering
   \includegraphics[width=0.6\textwidth]{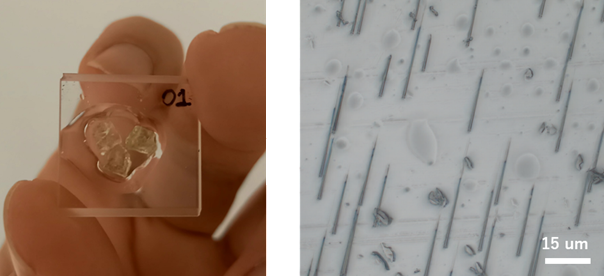}
   \caption{($\it{left}$) Image of olivine sample when etching treatment and optical microscope observation ($\it{right}$) Etched tracks of Xe ion with 0.5-2 GeV treated by EDTA base etchant of 14 hours at 108 $^{\circ}\mathrm{C}$.}
   \label{fig:olivine}
\end{figure}

Figure\,\ref{fig:olivine} shows examples of etched tracks produced by Xe ions with energies of 0.5–2 $GeV$ observed using an optical microscope. To evaluate the track formation characteristics of olivine, we performed calibration experiments using energetic heavy-ion beams at several accelerator facilities in Japan. The irradiation conditions are summarized in Table\,\ref{table:ion}.

\begin{table}[h]
  \centering
\caption{Summary of ion beam test to evaluate detectability as etched tracks}
 \label{table:ion}
  \scalebox{0.8}{
\begin{tabular}{lccc
}
   Ion & Effective Irradiated Energy [GeV] & $S_{e}$, $S_{n}$ [MeV/mg/cm$^{2}$] & Detectability \\ \hline
   Xe & 16 &  14.9, 0.0049  & No\\
   Xe &  10 &20.4, 0.0075 &  $\surd$ \\ 
  Xe &  7.5 &25.9, 0.01  &  $\surd$  \\ 
Xe & 4 & 36.8, 0.02 &  $\surd$    \\ 
Xe & 0.1 & 56.1, 0.4  &  $\surd$    \\ 
Au & 0.35 & 87.2, 0.4  &  $\surd$ \\ \hline
\end{tabular}
}

\end{table}

The calibration results indicate that no etched tracks were observed for 16 GeV Xe ions, whereas clear track formation was confirmed for 10 GeV Xe ions. These results suggest that the threshold stopping power for etched-track formation in olivine is approximately 15–20 MeV/mg/cm$^{2}$. This threshold provides an important benchmark for evaluating the sensitivity of olivine to various astrophysical particles and exotic heavy particles.

For such energetic heavy ions, the dominant energy-loss mechanism in olivine is electronic stopping. The resulting dense ionization along the ion trajectory produces a transient temperature rise, known as a thermal spike, which locally melts the crystal lattice. During the subsequent rapid cooling process, the crystal structure is not completely restored near the ion trajectory, leaving an amorphous damage core. The size of this amorphous region depends strongly on the stopping power of the incident ion and therefore determines the etching characteristics of the track.

As a consequence, the etching velocity can be used to estimate the atomic number of the incident ion. Previous studies ~\cite{olympiya} reported a calibration between the etching velocity and the atomic number of energetic heavy ions. In the present study, we are performing an independent cross-check of this charge-identification method using accelerator-produced heavy ions. Our preliminary results show etching velocities of approximately (20 $\pm$ 5) $\mu$m/h for Au ions and (15 $\pm$ 5) $\mu$m/h for Xe ions near their stopping regions. These values are consistent with the results reported by the OLYMPIYA project. This investigation is currently ongoing.

In contrast, the track formation mechanism for low-velocity heavy particles with velocities below the Bohr velocity remains poorly understood. In this velocity regime, the dominant energy-loss mechanism follows the Lindhard–Scharf model ~\cite{LS}, where lattice defects are produced through cascades of atomic collisions rather than dense electronic excitation. However, systematic studies of latent-track formation and subsequent etched-track formation in olivine are still scarce. Clarifying the response of olivine in this low-velocity region is therefore one of the major objectives of our ongoing research.

\subsection{Ultra-Heavy Galactic Cosmic-ray Elements} 
Recent advances in numerical-relativity simulations of neutron star mergers (NSMs), together with compositional analyses of extraterrestrial materials, suggest that the rapid neutron-capture process (r-process) occurring in NSMs is one of the primary production mechanisms of ultra-heavy elements, including transuranium nuclei. However, the occurrence rate of NSMs in the Milky Way is estimated to be only about one event every (10$^{5}$) years. Therefore, paleo detectors with geological exposure ages on the order of 1 Gyr provide a unique opportunity to investigate the long-term record of these rare astrophysical events. In particular, etched-track analysis has the potential to determine both the atomic number and the energy of incident ultra-heavy GCREs.

\begin{figure}
   \centering
   \includegraphics[width=0.6\textwidth]{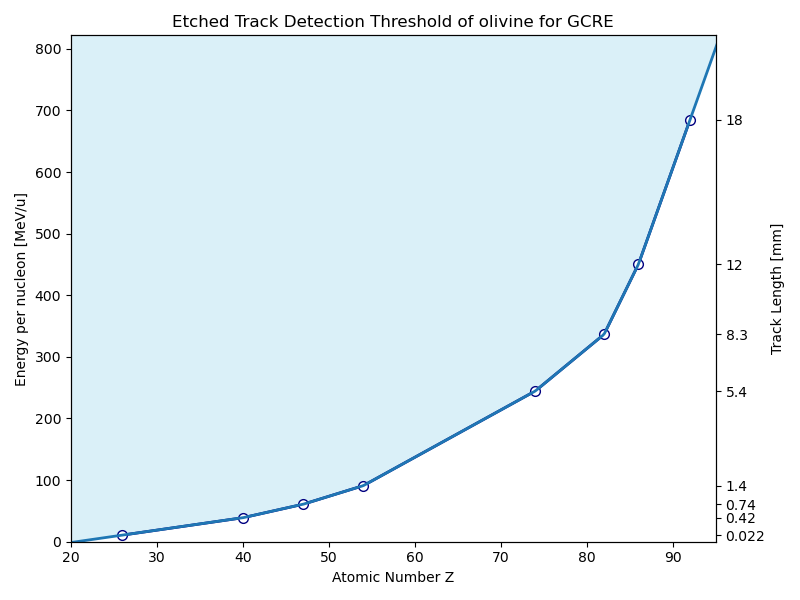}
   \caption{Detectable etched track upper threshold of energy and track length with respect to atomic number for olivine. Blue shadow region indicates undetectable region as etched tracks.  }
   \label{fig:GCRE_olivine}
\end{figure}

Based on the track-formation threshold discussed in Sec\,\ref{Tracking}, the detectable energy range for GCREs as a function of atomic number is shown in Figure\,\ref{fig:GCRE_olivine}. Galactic cosmic-ray nuclei with typical energies of the order of 1–10 GeV/u can readily penetrate meteorites. As they lose energy while traversing the mineral, they eventually stop in olivine with residual energies of approximately 100 MeV/u, where etched-track formation becomes possible.

Since the etching velocity in the stopping region exhibits a strong correlation with the atomic number of the incident ion, the atomic-number distribution of GCREs can be reconstructed from etched-track measurements. This capability provides a powerful method for investigating the composition of ultra-heavy cosmic rays over geological timescales and may offer new insights into r-process nucleosynthesis associated with neutron star mergers.

\subsection{ Strange Quark Matter from Neutron Star Mergers}

Neutron star mergers (NSMs) are also potential sources of exotic particles. For example, if the interiors of neutron stars contain strange quark matter, fragments of strange quark matter (SQM) may be ejected during the merger process~\cite{SQM_NSM}.

\begin{figure}
   \centering
   \includegraphics[width=0.6\textwidth]{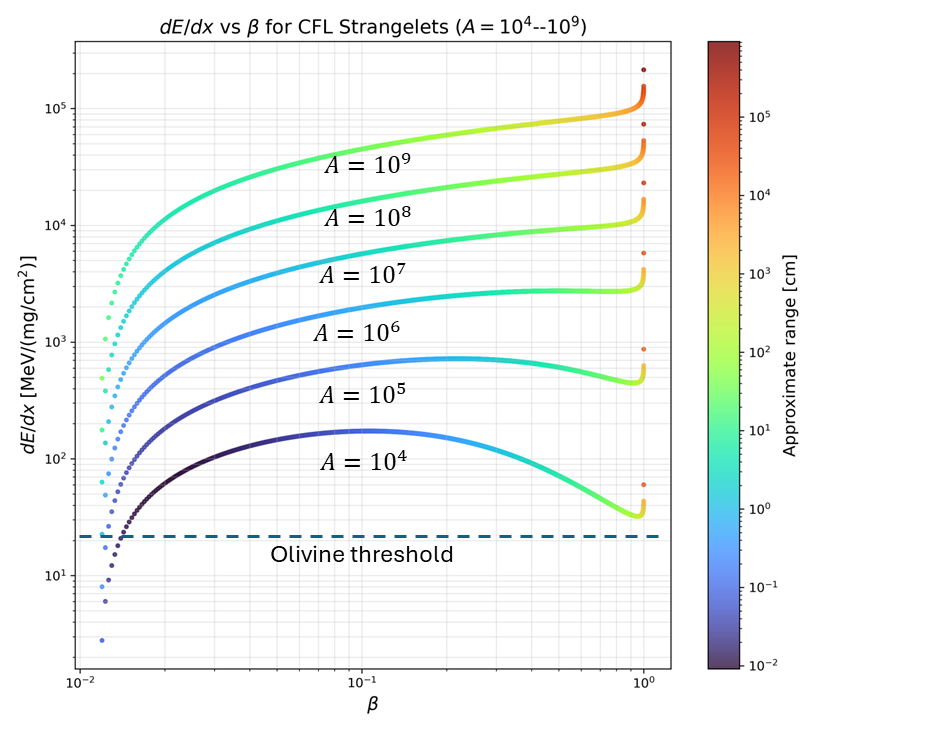}
   \caption{Energy loss of Strange Quark Matter (SQM) with respect to velocity and track length as color for each baryon number case.  }
   \label{fig:SQM_dedx}
\end{figure}

Assuming that SQM exists in a stable phase, such as the Color–Flavor Locked (CFL) phase, stable strangelets with baryon numbers larger than (10$^{4}$) are expected. In the CFL model, the finite mass of the strange quark results in a net positive surface charge~\cite{CFL}. Consequently, the electric charge of SQM is approximately given by
$Z \simeq 0.32A^{2/3}$, where (A) is the baryon number.
SQM fragments can be accelerated by the strong electromagnetic fields associated with NSMs. The maximum attainable momentum is approximately $pc \sim 10^{10} {\rm\,GeV}$.
Consequently, SQM with masses below approximately (10$^{11}\ {\rm GeV}/c^{2}$) can be accelerated to relativistic velocities. In this regime, the dominant energy-loss mechanism in matter is electronic stopping described by the Bethe–Bloch equation.
Such relativistic SQM particles are unlikely to survive to the Earth's surface because of atmospheric energy loss. In contrast, they can readily penetrate meteorites in space and eventually stop within minerals such as olivine.
Figure\,\ref{fig:SQM_dedx} shows the calculated stopping power as a function of velocity for several baryon numbers. 
The range is estimated by assuming a constant (dE/dx) at the initial ($\beta$), using Barkas's effective charge formula.
The calculation indicates that absolutely stable SQM with baryon numbers greater than 10$^{4}$ would produce long etched tracks in olivine, making paleo detectors based on extraterrestrial minerals a promising approach for searching for SQM originating from neutron star mergers.

\subsection{ Automated Optical Scanning System}

Etched-track analysis is indispensable for the search for extremely low-flux cosmic-ray events. Since a large number of tracks must be analyzed over wide areas, an automated optical microscope system is essential. For this purpose, we have developed an automated scanning microscope system, named the QTS.

The QTS is equipped with an automated XYZ stage that enables fully automated scanning over large areas. Tomographic images are acquired at multiple focal planes, and candidate tracks are identified using dedicated image-processing algorithms.
The system has recently been demonstrated using muscovite mica samples. To improve the scanning efficiency, an automatic surface-recognition algorithm was developed by utilizing the high density of $\alpha$-recoil tracks naturally present on the mica surface. This procedure enables reliable autofocus and efficient large-area scanning.

\begin{figure}
   \centering
   \includegraphics[width=0.6\textwidth]{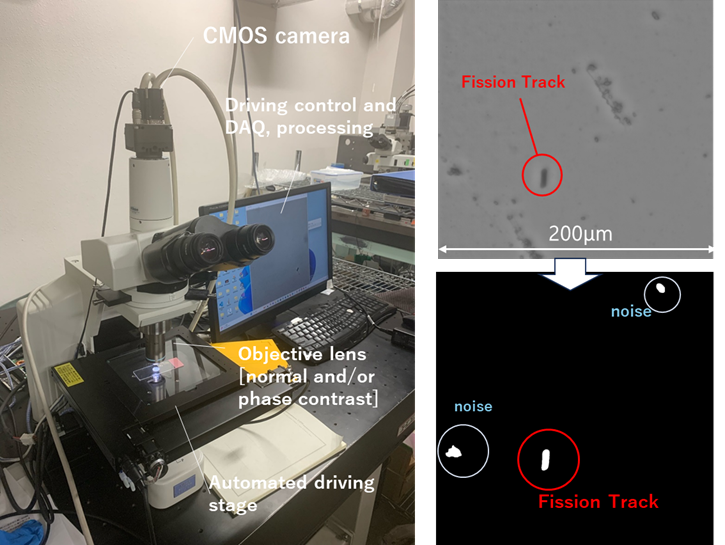}
   \caption{($\it{left}$) Automated optical microscope system (QTS) ($\it{right}$) (a) example image of optical microscope for the mucscovite mica with a fission track. (b) Final residual image for current selection system. Finally, noise was rejected by eye check.  }
   \label{fig:QTS_FT}
\end{figure}

As an initial demonstration, we developed an automated selection algorithm for fission tracks (FTs) produced by the spontaneous fission of $^{238}$U and applied it to the age determination of muscovite mica. Although the geological age of muscovite is commonly determined by the K–Ar dating method, direct dating based on etched fission tracks is important because it provides information on the track-retention (closure) age, allowing comparison with the K–Ar age.
Figure\,\ref{fig:QTS_FT} shows examples of FT candidates selected automatically by the image-processing algorithm. The selected candidates were subsequently verified by visual inspection to eliminate false-positive events. The measured FT density was\,$ (5.2 \pm 0.2)\times10^{5}\ {\rm cm^{-3}}$, while the $^{238}$U concentration determined by ICP–MS was\,$ (4.2 \pm 0.3)\times10^{13}\ {\rm cm^{-3}}$.
Using the standard fission-track dating equation, the preliminary age of the sample was estimated to be (1.6 $\pm$ 0.2)$\times10^{8}$ years. This value is approximately one-third of the age obtained by the conventional K–Ar method. Although the origin of this discrepancy is still under investigation, the present study successfully demonstrates the feasibility of automated fission-track dating using the QTS system.

\acknowledgments

This work was supported by JSPS KAKENHI Grant Number JP25K01016 and carried out  by the Interdisciplinary Research Strategy Projects of the Institute for Space-Earth Environmental Research (ISEE), Nagoya University.

\clearpage


\section{Research Updates from the University of Michigan -- MDvDM 2026}
\label{sec:Michigan}

Authors: {\it Emilie LaVoie-Ingram\,$^1$, Hannah Ross\,$^1$, Joshua Spitz\,$^1$, Andrew Calabrese-Day\,$^1$, Audrey Wu\,$^1$, Pranav Parvathaneni\,$^1$, Thomas Haddock\,$^1$, Zhexian Zhang\,$^1$, Kai Sun\,$^2$}

\vspace{0.1cm}
\noindent $^1$Department of Physics, University of Michigan, Ann Arbor, MI 48103 USA

\noindent $^2$Department of Materials Science \& Engineering, University of Michigan, Ann Arbor, MI 48103 USA

\subsection{Overview}
The current focus of the paleo-detection program at the University of Michigan is the detection of solar, supernova, and atmospheric neutrinos, and WIMP dark matter using ancient samples of quartz [$\mathrm{SiO_2}$], olivine [$\mathrm{Mg_{1.8}Fe_{0.2}SiO_4}$], and other minerals. When an incident particle interacts with an atomic nucleus in a crystal lattice, the resulting recoil can produce a track-like defect in the crystal which can be preserved for upwards of a billion years. These defects, known as nuclear recoil damage tracks, are only a few nanometers wide with lengths dependent on the energy of the incident particle and the mass of the recoiling nucleus ~\cite{jordan:2020gxx, baum:2019fqm, drukier:2018pdy}. These paleo-detectors provide the opportunity to detect WIMP dark matter, atmospheric and astrophysical neutrinos, and can open new avenues for studying the history of the Earth and solar system ~\cite{jordan:2020gxx, baum:2019fqm, drukier:2018pdy}. 

Since these tracks must be distinguished from mineral features accumulated over geologic timescales, the sensitivity of paleo-detection strongly depends on the detailed characterization of nuclear recoil damage track morphology and existing mineral defect backgrounds, in particular those created by cosmogenic and radiogenic sources. In preparation for the search for natural tracks induced by astrophysical particle interactions, experimental work has advanced in studying track characteristics in target mineral candidates. These studies require efforts in mineral collection and preparation, ion implantation, imaging strategies, and the development of track detection algorithms. Complimentary efforts in molecular dynamics simulations and radiogenic/cosmogenic background simulations are also underway.

\subsection{Track Morphology Study through Ion Irradiation}

Paleo-detection requires detailed characterization of nuclear recoil damage track-like features in target minerals. To improve our understanding of damage track formation in a candidate paleo-detector mineral, a sample of natural olivine with low Fe occupancy ($\mathrm{Mg_{1.8}Fe_{0.2}SiO_4}$) was irradiated with 15~MeV Au$^{+5}$ ions at the Michigan Ion Beam Laboratory. Lamellae were removed from the irradiated mineral perpendicular to the ion trajectory and imaged with Scanning Transmission Electron Microscopy (STEM); track width and geometry were analyzed along the ion trajectory using the STEM images, with depths related to energies via TRIM simulations \cite{zieglersrim2010}. Motivating this study was an interest in determining the suitability of olivine to form nuclear recoil damage tracks, and in understanding how track morphology changes with the energy of the recoiling nucleus. Additionally, the experiment provided an opportunity to develop a mineral imaging and analysis pipeline using high-resolution STEM, applicable for multiple species, energies, and minerals. This technique could also potentially be used in concert with higher throughput imaging methods for application with naturally occurring recoil tracks. A more detailed description of the experimental methodology and analysis in this section is provided in Ref.~\cite{TrackWidthPaper}.

\begin{figure}[h]
    \centering
    \includegraphics[width=0.3\textwidth]{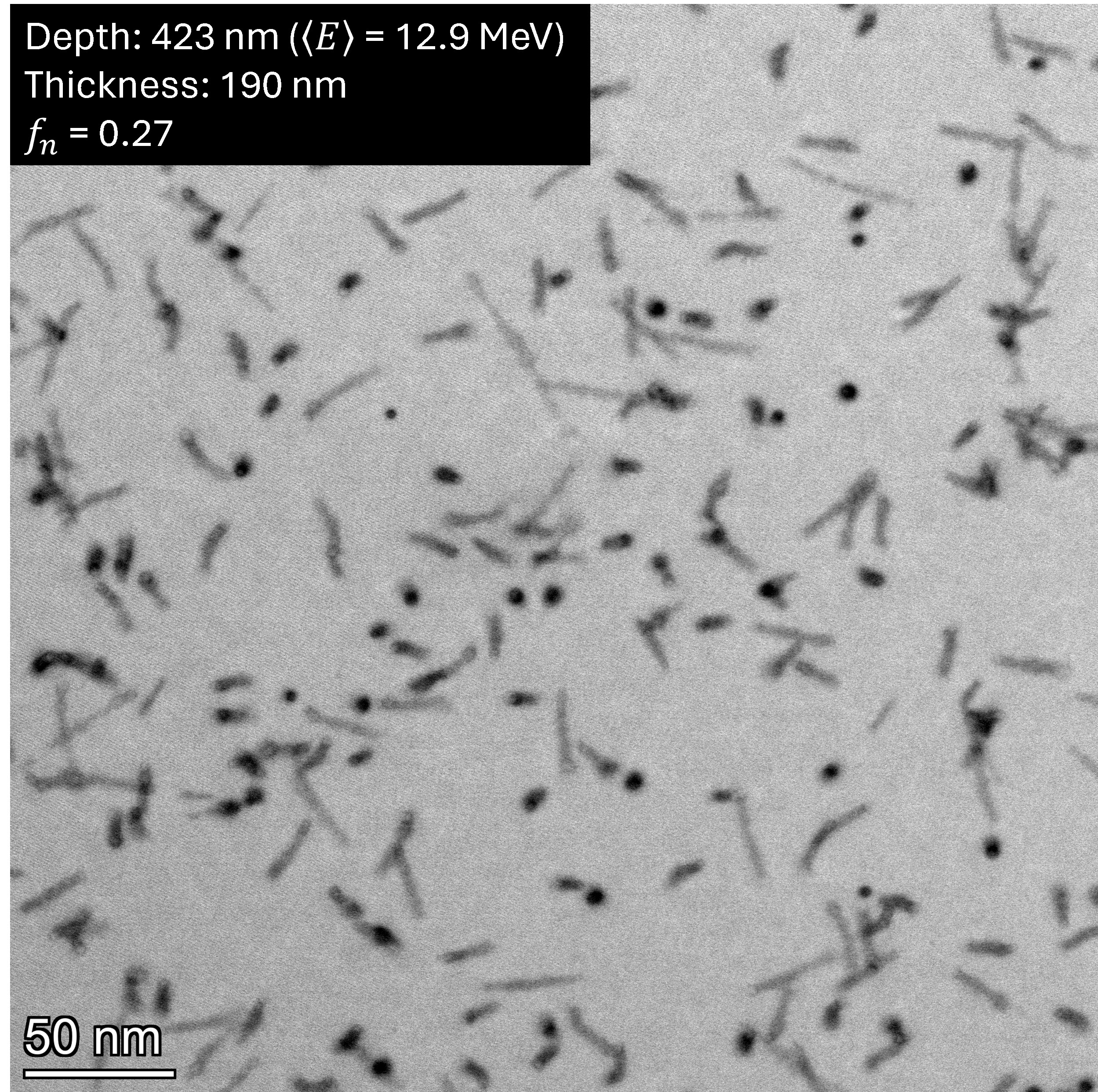}
    \hspace{0.01\textwidth}
    \includegraphics[width=0.3\textwidth]{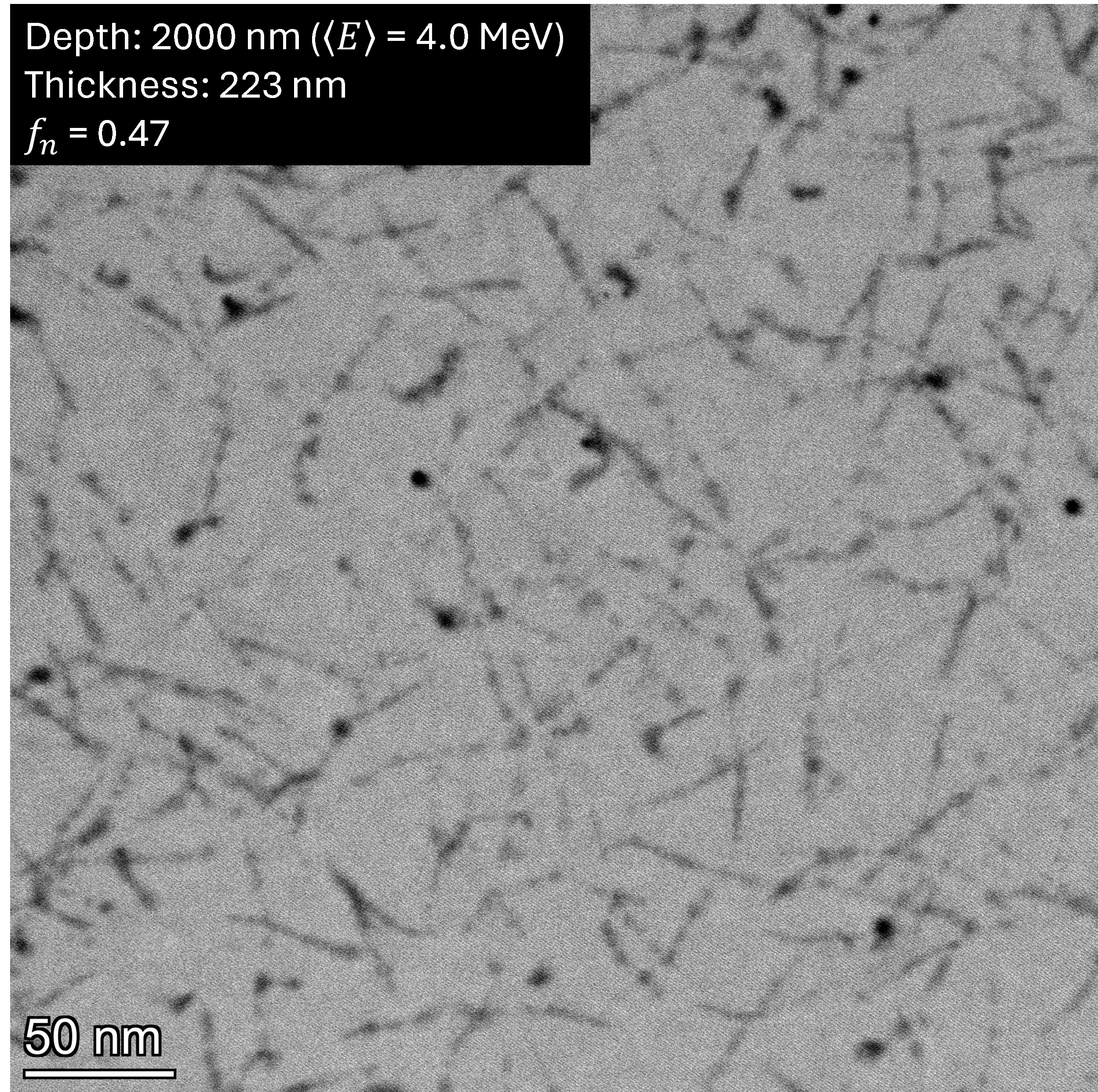}
    \hspace{0.01\textwidth}
    \includegraphics[width=0.3\textwidth]{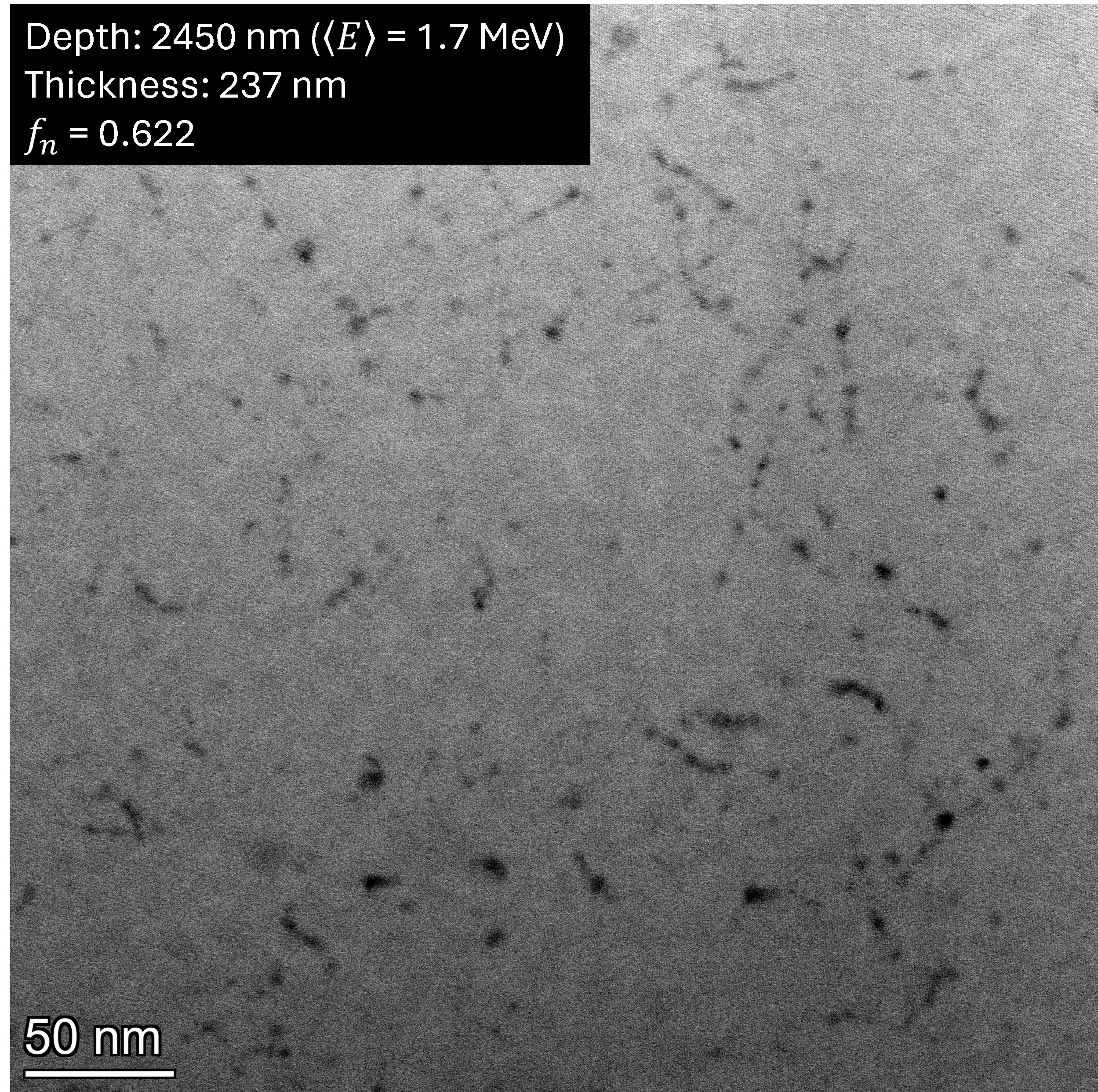}
    \caption{Adapted from Fig. 5 of Ref.~\cite{TrackWidthPaper}. STEM-BF images of Au$^{+5}$-irradiated olivine at target depths of 423, 2000, and 2450~nm. Each image is shown with corresponding target depth, median kinetic energy of ions surviving at depth from TRIM simulations, sample thickness, and nuclear stopping power fraction $f_n$ according to SRIM. These three regions are demonstrative of the change in track continuity across their depth, with $\sim$2000~nm corresponding to the SRIM-predicted transition from electronic to nuclear stopping.}
    \label{fig:threeDepths}
\end{figure}

\begin{figure}[h]
    \centering
    \includegraphics[width=1\textwidth]{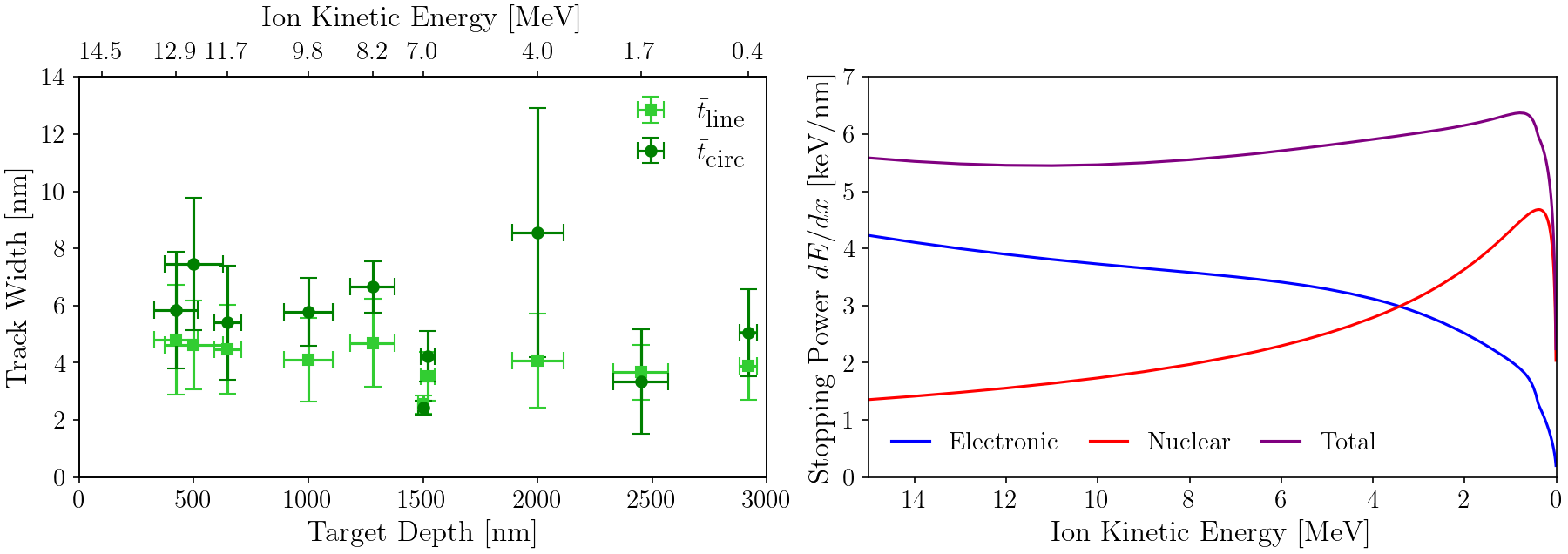}
    \caption{Figure 7 from \cite{TrackWidthPaper} Left: measured widths of linear (light green) and circular (dark green) Au$^{+5}$ tracks in olivine at various target depths. The upper horizontal axis shows the median ion kinetic energy for surviving ions in TRIM simulations at corresponding target depths on the lower horizontal axis. Right: energy loss $dE/dx$ as a function of Au kinetic energy from SRIM, showing electronic (blue), nuclear (red), and total (purple) contributions.}
    \label{fig:depthsEnergiesSrim}
\end{figure}

Fig.~\ref{fig:threeDepths} shows three STEM images collected at depths of 423, 2000, and 2450~nm, corresponding to median ion energies of 12.9, 4.0, and 0.4~MeV using SRIM. At the 423~nm depth, near the irradiated surface, ion tracks appear continuous. At intermediate depths, the tracks become increasingly irregular, before transitioning near the end of the ion range to primarily discontinuous regions of damage. These observations correspond qualitatively with the energies predicted by SRIM to be at the transition from an electronic stopping power dominated regime to a nuclear stopping power dominated regime, shown to be between 3-4~MeV in the right panel of Fig.~\ref{fig:depthsEnergiesSrim} \cite{zieglersrim2010}. In electronically dominated stopping power regions, ions lose energy primarily through interactions with the electrons in the crystal, producing continuous damage along the ion trajectory ~\cite{rymzhanovVelocityEffectSwift2023, amekuraLatentIonTracks2024}. As the ion energy gets lower, the contribution from nuclear stopping power increases. The primary mechanism for damage becomes interactions with individual atomic nuclei; these discontinuous tracks can manifest as strings of point defects \cite{zieglersrim2010, priceObservationsChargedParticleTracks1962, mauretteTrackFormationMechanisms1970, seitzAcceleratorIrradiationsMinerals1970}. This agreement with the observed evolution in track appearance suggests that the formation of defects is influenced not only by the incident ion species and structure of the crystal, but also by the mechanism through which energy is deposited into the crystal lattice.

Track widths measured from the STEM images are shown in the left panel of Fig.~\ref{fig:depthsEnergiesSrim} as a function of depth into the irradiated sample, corresponding to median ion energies predicted by SRIM. Across the ion trajectory, the average measured width remains relatively constant, ranging from 3-8~nanometers, despite the changes in morphology. These dimensions are consistent with the formation of well-defined nanometer-scale damage tracks over a broad range of ion energies, indicating that the track-forming process remains robust across a wide range of MeV-scale ion energies (irradiation depths). These results support olivine as an attractive paleo-detector candidate due to its robust track formation over a broad range of deposited energies. 

Building on these initial irradiation studies, future work will progressively approach the conditions expected in an astrophysical particle interaction. Ion irradiation studies will be conducted in quartz, olivine, and other paleo-detector candidates like Muscovite mica using ions that are already present in the crystal lattices of those minerals. These ion tracks will be a closer proxy to the nuclear recoils expected from neutrino and dark matter interactions. We will study a range of energies with ion irradiation, from a few keV up to several MeV. Finally, irradiation in both synthetic, single-crystal and natural, polycrystalline minerals will be conducted to determine the effects of sample structure on the formation of a track (discussed further in Section 17.3).

\subsection{Progress Towards Natural Track Detection and Imaging Strategies}

The next challenge in paleo-detection is identifying features of interest in naturally occurring minerals, where variability within geological samples, crystal imperfections, and the low density of recoil events introduce additional complexity. Successful paleo-detection requires not only sufficient spatial resolution to resolve nanometer-scale damage tracks, but also careful characterization of the host mineral and use of imaging techniques that are capable of scanning large ($\gtrsim$1 milligram) masses.

It is necessary to account for numerous additional factors when selecting natural minerals as paleo-detector candidates. Geological processes such as annealing and serpentinization may modify or erase damage tracks, while grain boundaries, inclusions, defects, and other naturally occurring variations within mineral samples complicate their identification with various microscopy techniques. These features may also have yet-unpredictable effects on a recoiling nucleus, potentially creating differing damage effects depending on the crystalline or amorphous structures encountered by the recoiling nucleus, or inclusions causing nuclei to scatter in a new direction. Natural minerals may also be sensitive to damage during imaging, especially with electron microscopy which can amorphize samples, erasing tracks. Careful characterization of candidate minerals is therefore necessary to distinguish particle-induced damage from naturally occurring features. 

Electron microscopy techniques provide sufficient spatial resolution to directly image individual damage tracks, as demonstrated by the irradiation studies presented in the previous section. However, techniques like TEM and STEM are fundamentally limited in throughput, with only picogram- to nanogram-scale sample volumes accessible within practical imaging times. TEM is a technique that the University of Michigan will continue to use for high-resolution, sparse sampling of natural tracks in paleo-detectors, but competitive atmospheric neutrino and dark matter measurements will likely require throughput on the order of milligrams to grams, motivating the identification of imaging techniques that can bridge the gap between spatial resolution and volumetric throughput. X-ray microscopy offers a promising path towards this goal. Diffraction-based x-ray microscopy and x-ray tomography techniques such as ptychography can image significantly larger sample volumes while nearing the necessary nanometer-scale spatial resolution, making them attractive candidates for future paleo-detector readout methods. An initial proof-of-concept study conducted at beamline 6-2c at the Stanford Linear Accelerator Center demonstrated the ability to resolve larger, 100nm artificial features using transmission x-ray microscopy. Future studies in ptychography and ptychographic-laminography will be conducted at Argonne National Laboratory's Advanced Photon source, where we will attempt to identify the presence of ion tracks with radii of 3-8~nm. Concurrently, we are characterizing the ability of the Zeiss X-Radia 810 Ultra recently acquired by the Michigan Center for Materials Characterization to identify regions of interest in our target minerals. These approaches will help establish a multi-modal experimental pipeline for detecting recoil damage tracks preserved in ancient geological samples.

\subsection{Simulation Progress}
Simulation is an important tool in the study of paleo-detector candidates, both from the impacts of mineral properties on the formation of defects, and backgrounds induced in natural geological samples. 

LAMMPS is a molecular dynamics code capable of simulating the kinetics of track formation \cite{lammps2022}. The University of Michigan paleo-detector group is focused on simulating ion irradiation in various paleo-detectors with LAMMPS, and we are in the process of validating our code with experimental irradiation data. Measurements of track width and length, number of vacancies, and degree of amorphization can be calculated with the results of molecular dynamics simulations, and can be compared to data collected with transmission electron microscopy. As well as validating simulations, the UM group is working to simulate on the order of several nanoseconds, exceeding the typical timescale of picoseconds that molecular dynamics codes are built for. We are collaborating with different institutions to approach this problem, ultimately with the goal of simulating the full relaxation of a crystal lattice after the initial creation of a track. This will be a necessary step when comparing simulation to real data. 

One of the most prominent background signals in paleo-detectors are tracks induced by muons produced in our atmosphere and fast neutrons produced in surrounding rocks from muon spallation. Cosmic rays interact with the gas molecules in Earth's atmosphere and create a shower of particles, the most problematic being muons and fast neutrons, that can reach the surface and at depth to induce defects. These tracks are identical to our tracks of interest, thus it's essential that we carefully predict the number of background tracks induced in our samples given the location, depth, overburden composition, and age of the mineral. The flux of muons and fast neutrons from this process quickly attenuates with depth underground, and in the ideal case we can obtain natural samples from around 5 kilometers deep, where the background becomes negligible \cite{drukier:2018pdy}. Unfortunately, this will be difficult to do since active mining extends at most to 3 or 4 kilometers. 
Knowing that we will inevitably have a large cosmogenic background in our samples, the group at UM is in the process of simulating and validating this process using a workflow of several software: CRY (Cosmic-ray Shower Library) \cite{hagmann2012cry}, MUTE (MUon iTensity codE) \cite{Woodley:2024eln}, Geant4 \cite{GEANT4:2002zbu}, and SRIM (Stopping and Range of Ions in Matter) \cite{zieglersrim2010}. Our workflow generates a flux of muons from the atmosphere, propagates them through a given depth and rock composition, and then models the resulting flux as a particle source in Geant4. The induced recoils are captured within a target crystal volume, modeled accurately in a host rock, in an underground lab, or in basement storage. The number of recoils per energy is then converted into number of recoils per track length, using the stopping power of the mineral calculated per recoil species and energy from SRIM. Similar to molecular dynamics, this cosmogenic background code is in the process of being validated with experimentally collected data at various underground laboratories, from one of which we have natural quartz samples.

\subsection{Conclusion}
Progress at the University of Michigan has primarily focused on testing various imaging methodologies that balance both resolution and mass throughput, studying the morphology of track formation with ion irradiation, obtaining ancient mineral samples from various sites around the world, and validating simulations of both track formation and backgrounds induced in natural samples. Numerous upcoming irradiation experiments with both synthetic, lab-grown minerals and ancient, polycrystalline minerals will help us study the impacts of pre-existing defects, lattice strain, and grain boundaries on track formation. In addition to irradiation, the aid of molecular dynamics simulations will ultimately help us understand the impacts of deep, geological environments on track formation in natural samples, such as temperature and pressure changes induced by tectonic plate movement.

\acknowledgments
This research is supported by the National Science Foundation (Grant \#EAR-2050374) and the Gordon and Betty Moore Foundation (Grant \#12234).

\clearpage

\section{Surface Characterization of Mineral
Samples Relevant for Particle Detection}\label{sec:UNF}

Authors: {\it Chris Kelso, Gregory Wurtz, Rabeya Rabu, and Andre Peterson}
\vspace{0.1cm} \\
University of North Florida
\vspace{0.3cm}

\subsection{Mineral Sample Handling}
As part of a wider collaboration effort at UNF, Virginia Tech (VT), the University of Michigan (UM)  and SLAC, new procedures have been developed for preparing, handling, and analyzing mineral samples. For the imaging techniques investigated, detailed sample preparation procedures have been developed, including macro- and micro-cutting (FIB-based), polishing, and embedding. Procedures and a centralized database has also been developed, implemented, and utilized for sharing samples across all institutions. VT, UM, UNF, and SLAC have developed strategies for meta-analysis (non-imaging) of mineral samples, including dating, composition analysis, and radiogenic estimation. Using these strategies, a number of ancient mineral samples (olivine, quartz, Jack Hills quartz, halite) from various deep underground locations, which are suitable candidates as mineral detectors have been chemically analyzed, cataloged, stored, and transferred among the institutions in the collaboration.

As part of these efforts, we have developed the ability at UNF to create lamella using the Focused-Ion Beam (FIB) on our mineral samples.  These lamella will be available for use in the wider collaboration/community as well as for our work here at UNF where we will utilize them for Electron Backscatter Diffraction (EBSD) and transmission EBSD (TKD) analyses.
    \begin{figure}[htbp]
    \centering
    \includegraphics[width=0.45\textwidth]{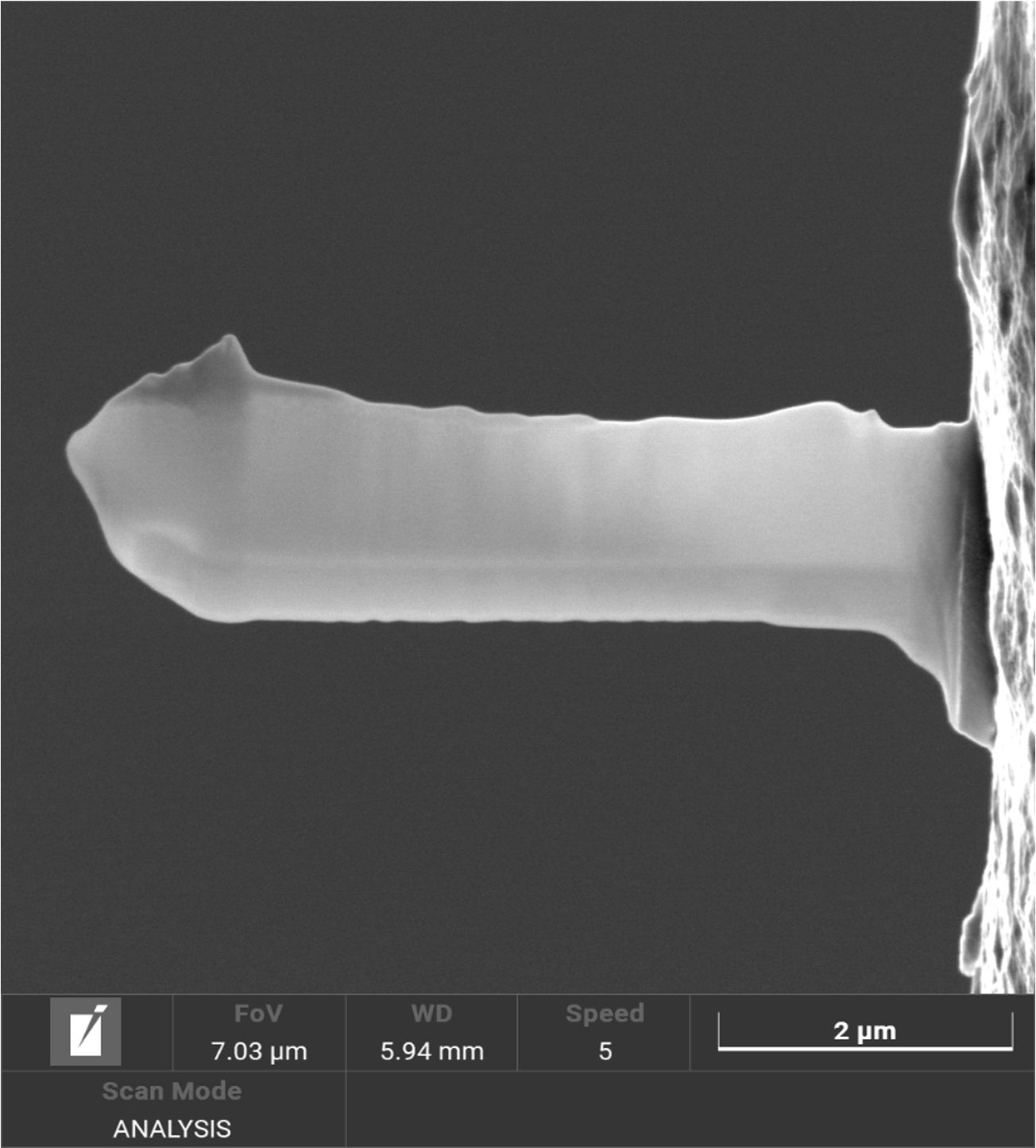}
    \caption{\scriptsize Lamella created at UNF from a UM, \ce{Au^5+} irradiated quartz (\ce{SiO_2}) sample.}
    \label{fig:UNF_lamella}
    \end{figure}

\subsection{AFM}
\label{sec:AFM}

\begin{figure}[htbp]
    \centering
    \includegraphics[width=.95\textwidth]{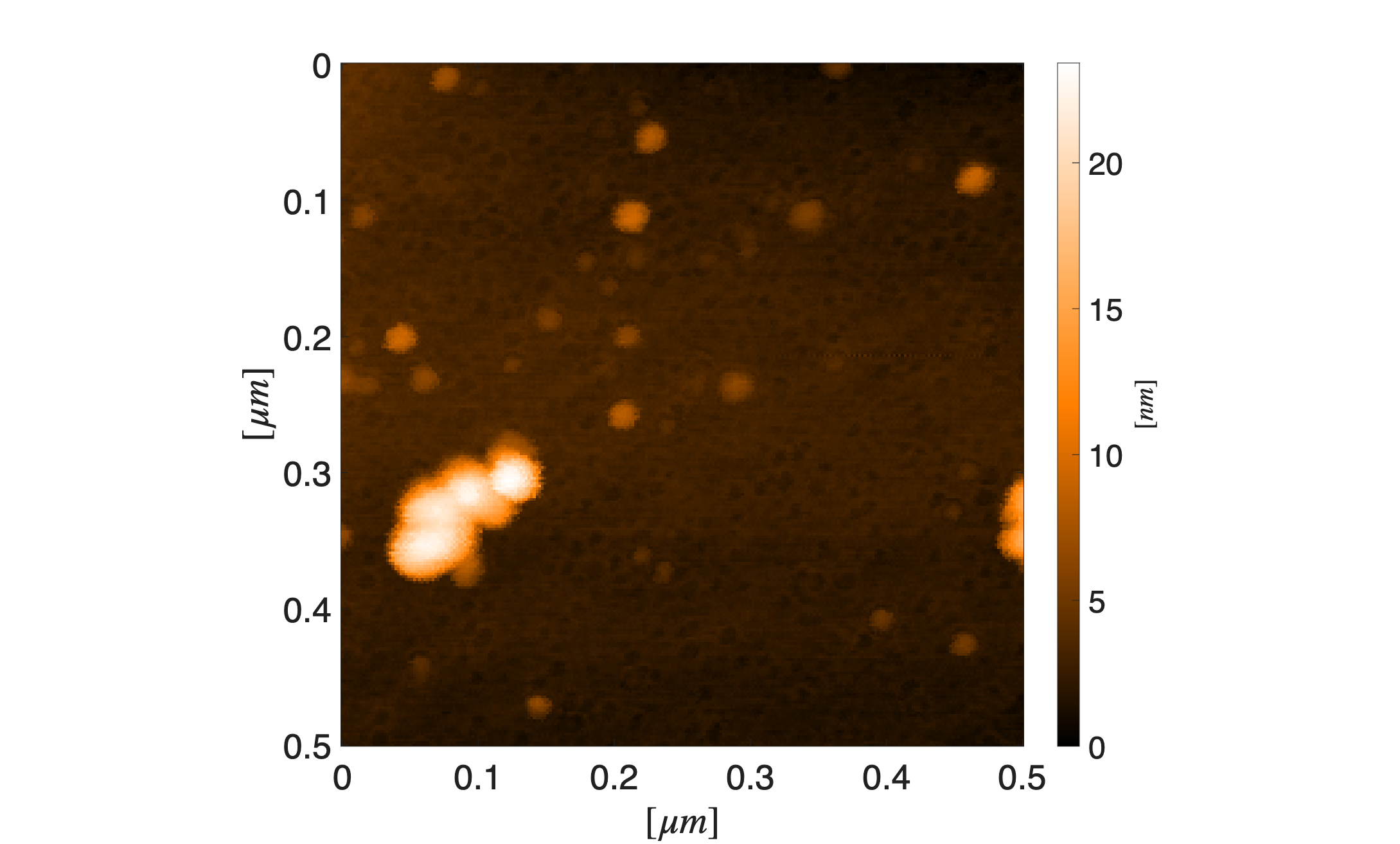}
    \caption{Top surface AFM image of Quartz sample (lab grown) irradiated with \SI{15}{MeV} \ce{Au^{5+}} ions at UM. 
    }
    \label{fig:AFM image}
\end{figure}

Atomic Force Microscopy (AFM) is a surface technique that typically measures mechanical properties at the nm scale with \si{pN} sensitivity. Atomic-sized defects are imaged convoluted with the probe, typically \SI{50}{nm} in diameter. It is a low throughput technique ($\sim$\SI{10}{ \um ^2.hr^{-1}} to map strain with the required sensitivity).  AFM is a real space imaging technique that requires samples that are relatively smooth. UNF has multiple AFM setups that are used for surface characterization of samples that are relevant for mineral detection.  Fig.~\ref{fig:AFM image} displays an AFM image of lab grown quartz irradiated with \SI{15}{MeV} \ce{Au^{5+}} ions at UM.  We are continuing to characterize these samples to determine if we can identify the tracks created by the \ce{Au^{5+}} ions, but equipment failures have severely slowed our progress.

\subsection{EBSD}
\label{sec:EBSD}
\begin{figure}
    \centering
    \includegraphics[width=0.6\linewidth]{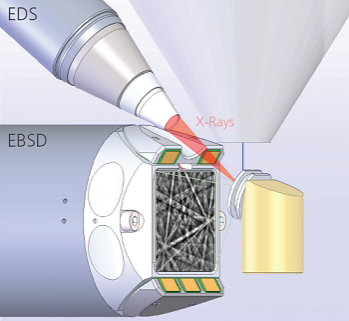}
    \caption{Schematics of the Oxford Instruments AZtec EBSD system, with illustrative diffraction pattern shown on the EBSD detector.}
    \label{fig:oxford_EBSD}
    \end{figure}

We have developed procedures and collected data on ion-irradiated olivine and quartz using electron backscatter diffraction (EBSD) imaging at UNF.  Elastically scattered electrons are imaged on a fluorescent film, forming a diffraction pattern reflecting the relative orientation of the crystalline planes.  Fig.~\ref{fig:oxford_EBSD} shows a schematic diagram with a source of electrons incident on a sample, along with an EBSD detector and an energy-dispersive X-ray spectroscopy (EDS) detector.  The spatial resolution of this technique is sample-dependent ($\approx$ \si{nm}); the angular resolution is geometry dependent (field of view).

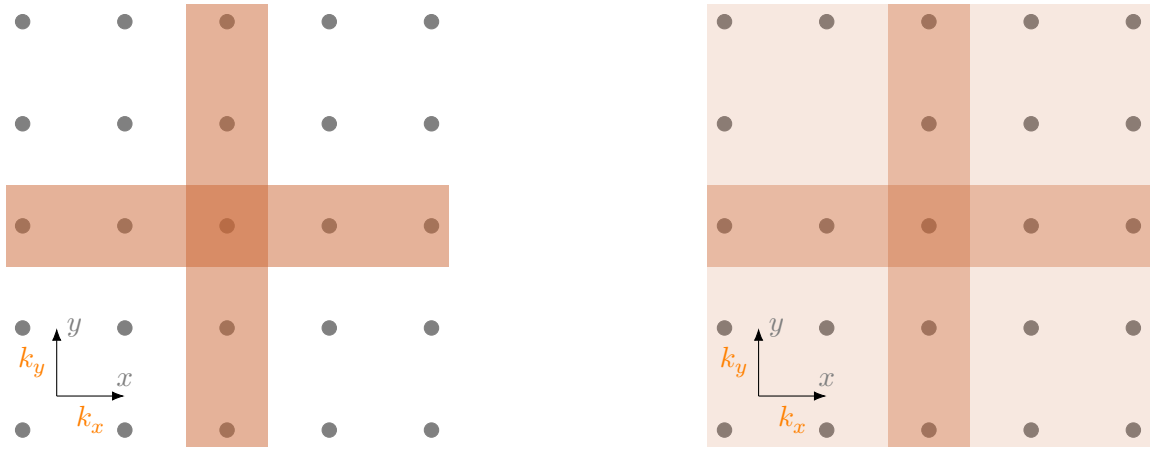
\begin{figure}[h]
    \centering
    \begin{tikzpicture}[>=Latex, scale=0.45]
                \def\alpha{34} 
                \def\n{1.5} 

                  \foreach \i in {-2,...,2} {
                        \foreach \j in  {-2,...,2} {
                            \filldraw[gray] (\i*30mm,\j*30mm) circle (6pt);
                          }
                          }
                \fill[orange!80!blue, opacity=0.5] (-1.2,-6.5) rectangle (1.2,6.5);
                \fill[orange!80!blue,opacity=0.5] (-6.5,-1.2) rectangle (6.5,1.2);

                \draw[thin, -Latex] (-5,-5) -- (-3,-5);
                \node[color=gray, above] at (-3,-5) {$x$};
                \node[color=orange,below] at (-4,-5) {$k_x$};

                \draw[thin, -Latex] (-5,-5) -- (-5,-3);
                \node[color=gray, right] at (-5,-3) {$y$};
                \node[color=orange,left] at (-5,-4) {$k_y$};

    \end{tikzpicture}
    \hfill
    \begin{tikzpicture}[>=Latex, scale=0.45]
                \def\alpha{34} 
                \def\n{1.5} 

                 \foreach \i in {-2,...,2} {
                            \filldraw[gray] (-60mm,\i*30mm) circle (6pt);
                          }
                \foreach \i in {-2,...,2} {
                            \filldraw[gray] (0,\i*30mm) circle (6pt);
                          }
                \foreach \i in {-2,...,0} {
                            \filldraw[gray] (-30mm,\i*30mm) circle (6pt);
                          }
                \foreach \i in {2,...,2} {
                            \filldraw[gray] (-30mm,\i*30mm) circle (6pt);
                          }
                \foreach \i in {-2,...,2} {
                            \filldraw[gray] (30mm,\i*30mm) circle (6pt);
                          }
                \foreach \i in {-2,...,2} {
                            \filldraw[gray] (60mm,\i*30mm) circle (6pt);
                          }    
                \fill[color=orange!80!blue, opacity=0.35] (-1.2,-6.5) rectangle (1.2,6.5);
                \fill[orange!80!blue,opacity=0.35] (-6.5,-1.2) rectangle (6.5,1.2);
                \fill[orange!80!blue,opacity=0.15] (-6.5,-6.5) rectangle (6.5,6.5);

                \draw[thin, -Latex] (-5,-5) -- (-3,-5);
                \node[color=gray, above] at (-3,-5) {$x$};
                \node[color=orange,below] at (-4,-5) {$k_x$};

                \draw[thin, -Latex] (-5,-5) -- (-5,-3);
                \node[color=gray, right] at (-5,-3) {$y$};
                \node[color=orange,left] at (-5,-4) {$k_y$};
    \end{tikzpicture}
    \caption{Diffraction pattern from a crystal with no defect (left). Diffraction pattern from a crystal with a point-like defect (right).}
    \label{fig:crystal with point-like defect}
\end{figure}

To help explain the output of the diffraction pattern in EBSD, Fig.~\ref{fig:crystal with point-like defect} plots the $1^{st}$-order diffraction pattern for the smallest crystalline indexes as an orange overlay on the crystal lattice, shown in gray in the left frame. The presence of the defect takes light away from the $1^{st}$-order diffraction bands into all directions, resulting in a signal more similar to what is shown in the right frame.

Plotting the diffusivity of the diffraction pattern as a function of position is expected to reveal crystal defects. One method to accomplish this is to perform a raster scan across a sample, collecting an EBSD image at each point.  A cross-correlation of these images will show differences in crystal structures at each point.  A simple implementation of this procedure would be to subtract a reference image (one of the diffraction patterns from one of the points in the scan) from the diffraction patter collected at each point in the scan.  This ``difference image'' will show a ``signal'' when there are differences in the crystal structure (defects/tracks) between the image at a point in the raster scan and the reference image. This technique, or something similar, would also provide an opportunity to apply machine learning algorithms to extract the track signals from the diffraction patterns collected through EBSD.

\begin{figure}[htbp]
    \centering 
    \includegraphics[trim=0 70 20 55,clip,width=0.4\textwidth]{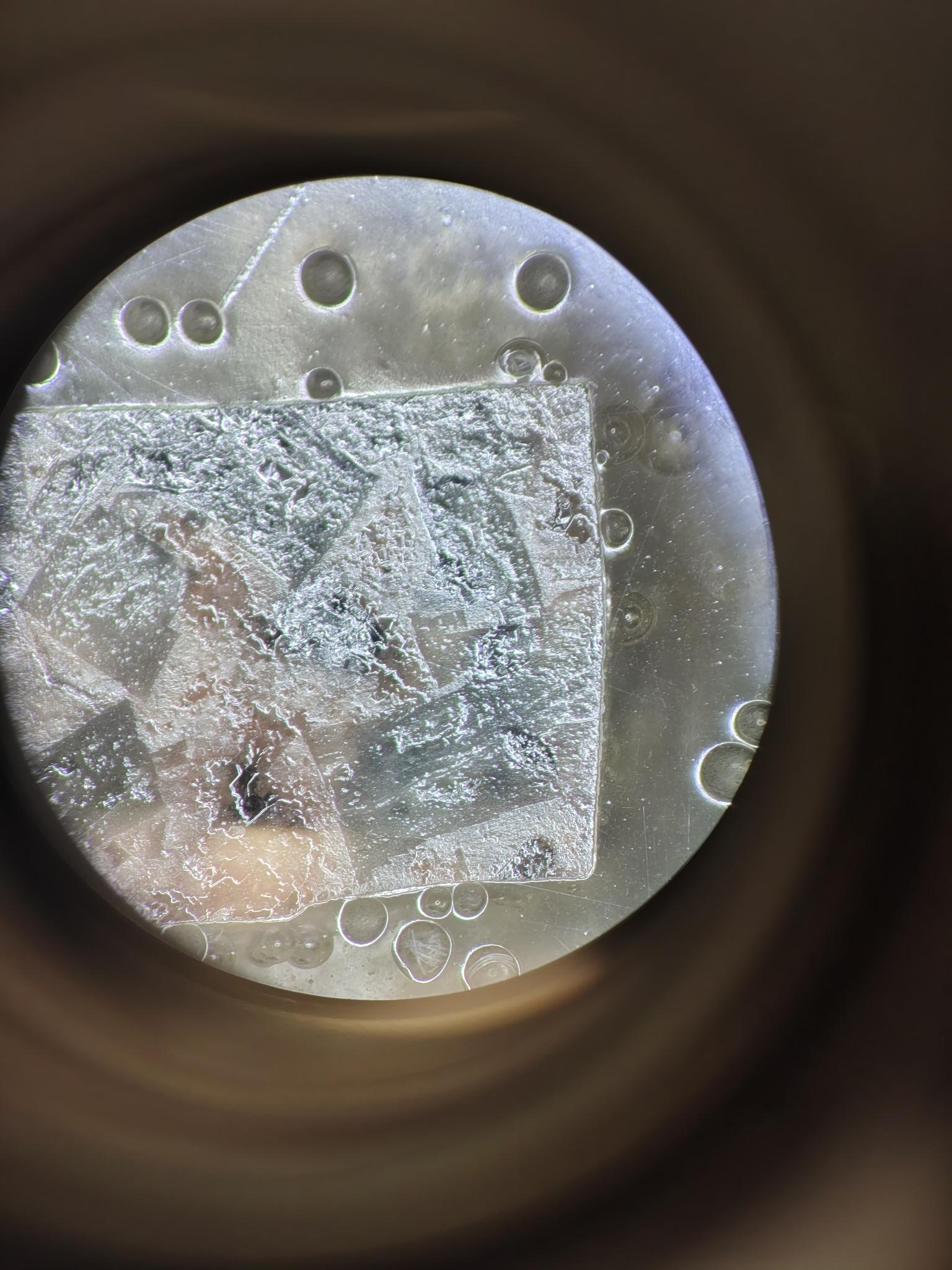}
    \hfill
    \includegraphics[width=0.5\textwidth]{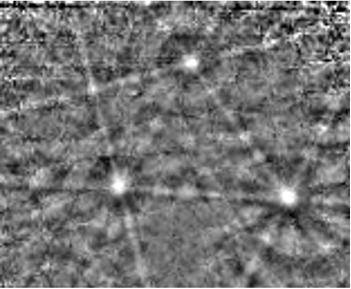}
    \caption{Optical image of polished sample UM-2024-Unknown-UPGR-8700-SAMP-0.0.1 (left).  This sample was prepared for EBSD analysis at UNF  with one of the collected diffraction patterns shown (right).}
    \label{fig:ebsd_at_UNF}
    \end{figure}

Fig.~\ref{fig:ebsd_at_UNF} shows an optical image of a polished sample (UM-2024-Unknown-UPGR-8700-SAMP-0.0.1) that was prepared for EBSD analysis at UNF in the left frame.   The collected diffraction pattern of a point in this sample is shown in the right frame.  The bright line-like features are called Kikuchi bands and will likely play important roles in the analysis that is searching for crystal defects in the samples described previously.
   
This work will continue at UNF, in which we will measure the crystal orientation of grains within the samples (e.g., \ce{Au^{5+}} irradiated quartz) using EBSD.  Both the results from our EBSD analysis and the samples will be sent to the APS Center for Nanoscale Materials (CNM) for preliminary analysis by APS staff. The highly focused, coherent X-ray nanoprobe beamline at APS-CNM Sector 26 is expected to enable nanoscale characterization of strain induced by scattering from dark matter particles with the crystal lattice via Bragg ptychography. This preliminary data will support the submission of a full beam time proposal to CNM.

\clearpage

%
%
%
%

\bibliographystyle{JHEP.bst}
\bibliography{theBib}

\end{document}